\documentclass[a4paper,onecolumn,11pt,accepted=2023-04-19]{quantumarticle} 
\pdfoutput=1
\usepackage{amsmath}
\usepackage{amssymb}
\usepackage[english]{babel}
\usepackage{bbold}
\usepackage{booktabs}
\usepackage{braket}
\usepackage{caption} 
\usepackage{comment}
\usepackage{dsfont}
\usepackage[T1]{fontenc}
\usepackage{graphicx}
\usepackage{hyperref}
\usepackage[utf8]{inputenc}
\usepackage{lipsum}
\usepackage{subcaption} 
\usepackage{tcolorbox}  
\usepackage{tikz}
\usepackage{xifthen} 

\usetikzlibrary{arrows,shapes,positioning} 
\usetikzlibrary{decorations.markings} 
\usetikzlibrary{topaths,calc}

\tikzstyle arrowstyle=[scale=1] 
\tikzstyle directed=[postaction={decorate,decoration={markings,
		mark=at position .65 with {\arrow[arrowstyle]{stealth}}}}]
\tikzstyle reverse directed=[postaction={decorate,decoration={markings, mark=at position .65 with {\arrowreversed[arrowstyle]{stealth};}}}]  
\graphicspath{{../../Figures/}}

\newcommand{\oneline}[1]{%
  \newdimen{\namewidth}%
  \setlength{\namewidth}{\widthof{#1}}%
  \ifthenelse{\lengthtest{\namewidth < \textwidth}}%
  {#1}% do nothing if shorter than text width
  {\resizebox{\textwidth}{!}{#1}}% scale down
} 
\providecommand{\keywords}[1]{\textbf{\textit{Keywords:\ }} #1}   

\begin{document} 
 
\title{Unitary Evolutions Sourced By Interacting Quantum Memories: Closed Quantum Systems Directing Themselves Using Their State Histories}

\author{Alireza Tavanfar} 
\email{alireza.tavanfar@research.fchampalimaud.org}
\affiliation{Champalimaud Research, Champalimaud Center for the Unknown, 1400-038 Lisboa, Portugal}
\affiliation{Institute of Neuroscience, University of Oregon, Eugene, OR 97403, USA}

\author{Aliasghar Parvizi} 
\email{a.parvizi@ut.ac.ir}
\affiliation{Department of Physics, University of Tehran, 14395-547, Tehran, Iran} 
\affiliation{School of Particles and Accelerators, Institute for Research in Fundamental Sciences (IPM), P.O. Box 19395-5531 Tehran, Iran}

\author{Marco Pezzutto}
\email{marco.pezzutto@uni.lu} 
\affiliation{Complex Systems and Statistical Mechanics, Physics and Materials Science Research Unit, University of Luxembourg, L-1511 Luxembourg}
%\orcid{https://orcid.org/0000-0003-1850-340X}

\maketitle

\begin{abstract} 
We propose, formulate and examine novel quantum systems and behavioral phases in which momentary choices of the system's memories interact in order to source the internal interactions and unitary time evolutions of the system. In a closed system of the kind, the unitary evolution operator is updated, moment by moment, by being remade out of the system's `experience', that is, its quantum state history. The `Quantum Memory Made' Hamiltonians (QMM-Hs) which generate these unitary evolutions are Hermitian nonlocal-in-time operators composed of arbitrarily-chosen past-until-present density operators of the closed system or its arbitrary subsystems. The time evolutions of the kind are described by novel nonlocal nonlinear von Neumann and Schr\"odinger equations. We establish that nontrivial Purely-QMM unitary evolutions are `Robustly Non-Markovian', meaning that the maximum temporal distances between the chosen quantum memories must exceed finite lower bounds which are set by the interaction couplings. After general formulation and considerations, we focus on the sufficiently-involved task of obtaining and classifying behavioral phases of one-qubit pure-state evolutions generated by first-to-third order polynomial QMM-Hs made out of one, two and three quantum memories. The behavioral attractors resulted from QMM-Hs are characterized and classified using QMM two-point-function observables as the natural probes, upon combining analytical methods with extensive numerical analyses. The QMM phase diagrams are shown to be outstandingly rich, having diverse classes of unprecedented unitary evolutions with physically remarkable behaviors. Moreover, we show that QMM interactions cause novel purely-internal dynamical phase transitions. Finally, we suggest independent fundamental and applied domains where the proposed `Experience Centric' Unitary Evolutions can be applied natuarlly and advantageously.

\end{abstract}
\newpage

\tableofcontents
\newpage

\section*{\center{Table of Acronyms}}

\begin{table}[h]
\begin{center}
\begin{tabular}{ll}
	\toprule
	Acronym	&	Meaning									\\
	\midrule
	\\
	QM \hspace{3.8cm}		&	Quantum Memory			\\
	\\
	QMM		&	Quantum Memory Made						\\
	\\	
	QMM-UE	&	Quantum Memory Made Unitary Evolution 	\\
	\\
	QMM-H	&	Quantum Memory Made Hamiltonian			\\
	\\
	QMD		&	Quantum Memory Distance					\\
	\\
	QM-TPF &	Quantum Memory Two Point Function		\\
	\\
	QM-TPO	&	Quantum Memory Two Point Operator		\\
	\\
	QMM-MF	&	Quantum Memory Made Magnetic Field		\\
	\\
	LHS		&	Left Hand Side							\\
	\\
	RHS		&	Right Hand Side						\\	
	\\	
	\bottomrule
\end{tabular}
\end{center}
\end{table}

\newpage

%%%%%%%%%%%%%%%%%%%%%%%%%%%%%%%%%%%%%%
%%%%%   1 . Introduction
%%%%%%%%%%%%%%%%%%%%%%%%%%%%%%%%%%%%%%%
\section{Introduction}\label{introduction} 
In conventional treatments of the unitarily-evolving quantum systems, it is typically assumed that many-body interactions and the dynamical equation of the system are (as physically postulated and mathematically formulated,) independent of the system's quantum-state history. In this discipline, the total Hamiltonian generating the system's unitary time evolution is a (time-independent or time-dependent) Hermitian operator that is `implanted' inside the system, independent of which trajectory of quantum states the system has been taking up until the present moment. But, this is not necessarily the most general framework to be taken, as one can envision. In fact, one naturally tends to investigate general formulations and concrete realizations of \emph{fundamental-or-emergent closed quantum many body systems which function beyond this established discipline}. Intending this investigation in the present work, and regarding intrinsic connections between the state-history of a quantum system and the system's quantum memories, we begin with a very selective and brief overview on the conventional perspective and framework of Non-Markovian dynamics in quantum many body systems.\\\\
Time evolution of a quantum system is typically described through one of three main types of processes~\cite{BreuerPetr,AlickiLendi,NielsenChuang}: i) the unitary evolution of a closed system generated by a Hamiltonian, be it time-independent or time-dependent subject to temporal locality; ii) the non-unitary evolution which is induced by a quantum measurement process; iii) the dissipative, irreversible dynamics of an open system undergoing interactions with an environment, as for example formulated by a Lindblad Markovian master equation. At least for two reasons, processes of type ii) and iii) are not completely independent: on the one hand, the action of a quantum measurement repeated a sufficient number of times can lead to an effective description in terms of a master equation for the reduced dynamics of the system; on the other hand, the action of an environment interacting with the system can sometimes be interpreted as a series of repeated measurements.\\\\
However, processes of type i) and iii) are opposite in characteristics. The unitary evolution of every closed system under a time-independent Hamiltonian (with discrete spectrum) is quasi-periodic, due to the quantum Poincaré recurrence theorem \cite{qpt:a}, while an open system evolving under a Lindblad master equation can eventually evolve to attractor asymptotic states independent of the initial state, such that the system's initial-state information is not preserved. Furthermore, while the interaction with the environment brings dissipative effects, and so irreversibility in the dynamics, it can also produce coherent effects developed by modifications to the Hamiltonian dynamics \cite{BreuerPetr,Kossakowski72,Lindblad76,GorKossSud76}.
Between these two opposite settings, there comes a variety of more complex scenarios in which the interacting environment does not simply act as a sink in which the information is lost. Instead, 
the system-environment interactions and the internal dynamics of the environment may be such that the information from earlier quantum states of the system can be dynamically fed back to the system. This interplay results in Non-Markovian profiles for the system's time evolution, because of becoming by nature state-history dependent.
Indeed, memory dependencies in quantum dynamics have been acknowledged since the birth of the theory of open quantum systems. A step taken often in deriving master equations is the Born-Markov approximation in which the environment quickly `forgets' any trace of its interactions with the system. Hence, the system's time evolution is described by a local-in-time differential equation which is independent of the system's (long or short) state-history. Such a decoupling can be reasonable only under certain conditions, like for sufficiently-weak system-environment couplings, or if the characteristic timescale of the environment dynamics is sufficiently shorter than the one of the system.\\\\
However there are many important physical settings in which the approximation of system's memory-independent dynamics breaks down significantly, leading to complex Non-Markovian evolutions. Despite their relevance and significance, only in recent times a promising deal of coherent research has been conducted on analyzing these much-more-challenging settings. Indeed, in an increasingly more diverse spectrum of domains in open quantum systems, finding out the many impacts of strong memory effects has turned into a doorway to many body physics scenarios with remarkable novelties from the standpoints of theoretical and experimental quantum physics, and even mathematics. See, merely as few examples, \cite{Floquet18,Zhang19,PollockModi19}. In this realm, developing more
inclusive definitions, characterizations and measures of Non-Markovian quantum processes is still an evolving endeavour, hosting a variety of approaches, proposals and concrete formulations ~\cite{Rivas14,Breuer16,deVega15}. These distinct approaches are based upon a variety of different but compatible standpoints such as, 
the backflow of quantum information from the environment to the system~\cite{Breuer09,Laine10,Wissmann12}, the divisibility of the dynamics~\cite{Rivas10}, the evolution of the geometric volume in the state space accessible to the system~\cite{Lorenzo13}, and several more. Increasing efforts are devoted to develop a comprehensive framework which sharpens similarities and discrepancies between all these approaches~\cite{Chruscinski11,Chruscinski14,buscemi16,acin16,Pollock15,Pollock18,Lili18}.\\\\ 
Most known methods of quantifying non-Markovianity are formulated from the very perspective of the system's dynamics alone: access to the reduced density operator of the system during its time evolution should be sufficient to accurately decode the required signatures of Non-Markovianity. This dominant standpoint has the double advantage of being minimalistic in formulation, while also more practical form both theoretical and experimental viewpoints, but is incapable of satisfactorily capturing the root causes of the emergence of Non-Markovianity and memory effects in open quantum systems. However, manifesting directly in the formulation, the actual role of the system-environment correlations in sourcing and mediating the memory effects is attracting increasing attention~\cite{Mazzola12,Smirne13,fanchini14,DArrigo14,Campbell18}. To this aim, a variety of scenarios have been developed in recent years where the interacting environment and its interplays with the system are modelled explicitly~\cite{Ciccarello13,Kretschmer16,Lorenzo17}, allowing us to understand better the physics of crossovers between Markovian and Non-Markovian evolutions~\cite{Rodriguez08,GonzalesTudela10,Man15a,Man15b,Man15c,Brito15,LoFranco15}.\\\\
Finally, recent promising advancements in experimental techniques for controlling and manipulating quantum systems have led to the first direct observations of Non-Markovian dynamics ~\cite{chiuri,cinesi,cinesi2,brasiliani,recover,brasiliani2,Xu13}, while its impact in a whole variety of areas has been also addressed. These important areas, to name a few, include topological phases of matter~\cite{Zambrini19,GeomPhase18,Palma17}, quantum optics~\cite{QED18,Biref18,Paspa19,Meystre19}, thermodynamics~\cite{Addis15,Kutvonen14,Goold15,Guarnieri16} and entanglement dynamics~\cite{Marcantoni19,Chru18,Subhash19}.\\\\
\emph{The present work}, however, aims to take the study of quantum-memory effects to the very extreme: \emph{we develop an idealization of a Non-Markovian quantum dynamics which is completely coherent, free from any dissipative effect, and retains the totality of the specifying qubits of the system's states at past times}. Stated more specifically, we study the time evolution of general closed quantum systems under a completely new kind of Hamiltonians which are (functionally and explicitly) made from arbitrary choices of the system's states in the course of its history from an initial time up to the present time. Therefore, the physical context is that of a unitarily-evolving quantum many body system whose many-body interactions and Hamiltonian emerge, moment by moment, in an intrinsically Non-Markovian way: at every present moment, the closed quantum system itself reconstructs its internal interactions and its Hamiltonian by `connecting' or `linking' (a number of) its quantum memories. In effect, we develop an original general type of quantum dynamics which is marked with two complementary faces, at once and consistently: \emph{the system's evolution is 
Non-Markovian and Unitary.}\\\\ 
We propose and work out the most direct definition and formulation of the promised theories for generally-defined closed quantum systems. One first  envisions a unitary quantum state history of a closed system beginning from some initial moment $t_0$ up to the present moment $t$. We now identify the density operator of the system, $\rho(t')$, at each moment $t^\prime \in [t_0 , t]$, with the \emph{quantum memory} of the closed system at that moment. Henceforth, the quantum state history of the system evolves onwardly, on a moment-by-moment basis, using the available quantum memories. In this way, an entire unitary quantum state history can be constructed from $t_0$ all the way to $t = \infty$.
Concretely speaking, the closed system `selects' and combines its available quantum memories at every present moment to construct a composite (generically nonlocal in time) state-history-dependent Hamiltonian, being called \emph{`Quantum-Memory-Made Hamiltonian' (QMM-H)}, which generates its unitary time evolution operator. \\\\ 
But, mathematically formulating a QMM-H is one thing, establishing that the resulted dynamical equations can develop solutions which are consistent physically is another thing. In particular, physical consistency of the QMM unitary evolution implies that the quantum memories which construct the Hamiltonian polynomial are \emph{not} actually independent variables. That is, any pair of system's density operators at times $t'$ and $t''$ which come in the 
above proposed construction of the Hamiltonian at the present time $t$, must be themselves obtainable from one another under the accordingly-constructed time evolution operator in between the times $t'$ and $t''$.\\\\ 
The upshot becomes the statement that the physical solutions to the nonlocal-in-time nonlinear Schr\"odinger equation resulted from the QMM-H must realize, one by one, a unitarily self-consistent state-history for the system. After proving the consistency of the theory, it is important to establish that the physical moduli space of the QMM Schr\"odinger equations, namely that which corresponds to its 
unitarily self-consistent solutions, is a large and rich landscape which, compared to the conventional settings, accommodates novel phases of unitary state-histories, and also novel phase transitions. \emph{This demanding task} consists the focus of a large part of the present paper which, by means of analytical 
and numerical analyses in combination, is done successfully.\\\\
A brief outline of the paper is now presented with the selected highlights. We commence with a general theory of `$(N,L)$ QMM-Hs', formed as order-$L$ polynomials made of $N$ quantum memories, together with their highly-novel Schr\"odinger equations. There, we also introduce the `Quantum Memory Characters', specially the `QMM Two-Point Functions', in terms of which QMM Unitary Evolutions (QMM-UEs) are reformulated and characterized naturally and usefully. Moreover, the general (2,2) QMM-UEs are detailed for further clarifications. Finally, we establish that all purely-QMM unitary dynamics are
`Robustly Non-Markovian', meaning that non-constant wavefunctions are developed only when the maxima of the `quantum-memory distances' exceed the finite thresholds which depend on the QMM-H couplings.\\\\
In the course of Sections \ref{888}, \ref{iml} and \ref{SV}, given total novelty and significant theoretical complexity of QMM-UE, we focus on exploring the most elementary case studies: the closed system of one qubit evolving under (1,1), (2,2), (2,3) and (3,3) QMM-Hs. First, the dynamical equations of all these cases are detailed with important explanations, in Section \ref{888}. Section \ref{iml} presents a variety of analytic solutions and results for the (2,2) QMM-UE. In Section \ref{SV}, phase diagrams of one-qubit wavefunctions made purely by quantum memories are analyzed for each one of the above QMM-Hs, finding wide classes of qualitatively unprecedented unitary quantum state histories. Moreover, we present several selected examples of Dynamical Phase Transitions which are unitarily developed in the course of the evolutions generated by QMM-Hs (in even the most elementary closed quantum system, namely) the 
one-qubit closed system, as a result of the higher order interactions between the system's quantum memories. Moreover, we present highlighted lessons, selected from all these analyses, to complete Section \ref{SV}.\\\\ 
The Proposal is revisited, elucidated and also evaluated in Section \ref{visions}. On one hand, we highlight several interesting open questions to be addressed in the future works. On the other hand, we briefly present our current visions on the diverse categories where QMM-UEs can be realized and utilized advantageously, in future.
%%%%%%%%%%%%%%%%%%%%%%%%%%%%%%%%%%%%%%
%%%%%  2 . General QMMUEs 
%%%%%%%%%%%%%%%%%%%%%%%%%%%%%%%%%%%%%%%
\section{Unitary State-History-Based Quantum Behavior: \\ The Hamiltonian Formulation}
\label{gfp} 
\subsection{The Central Idea}
%\subsection{Interlinked Quantum Memories And Resulted Quantum-Memory-Made Hamiltonians (`QMM-Hs')}\label{gfp}
The central mission of the present work is to address and investigate in details the  question phrased below, which is significant and intriguing, but largely unexplored. The roadmap we use to answer it, is first presenting the general formulation of the merging envisioned in it, followed by extensive (analytical and numerical) explorations of the resulted novel quantum behaviors, in the simplest specific models and in some general cases.\\\\
\emph{Can one unify, in the context of a closed quantum system and in a consistent way, the principle of system's time evolution unitarity with one's request for the system's many-body interactions and total Hamiltonian being (entirely or partially) made from the system's state history?}\\\\
We now begin with presenting a quite general formulation of self consistent unitary evolutions whose generating Hamiltonians are being made (partially or entirely) by the past-until-present quantum-state-history of a general closed system, and subsequently realizing the main implied features of such proposed systems upon undertaking diverse analyses. Visions related to (the inquiry of) \emph{what kinds of natural or artificial systems} can be described by these novel Hamiltonians is indeed an intriguing and important question which we postpone to \emph{Section} \ref{visions}.\\\\
On general grounds, the defining operator of the total Hamiltonian of a closed quantum many body system can  have a \emph{dichotomic} structure (effectively or fundamentally) as follows,
\begin{equation}
\begin{split}
& H_{{\text{(The Most General)}}}^{\text{(Total)}}(t) \; =\; H_{(\text{State-History Independent})}(t) \; + \; H_{(\text{State-History Dependent})}(t) \; \equiv \; \\ & \hspace{1.5 cm} \equiv \;  H_{(\text{Conventional})}(t) \; + \; H_{(\text{Quantum Memory Made})}(t) \; \equiv \;  H_{{\rm C}}(t) \; + \; H_{{\rm QMM}}(t). 
\label{H-CQS} 
\end{split} 
\end{equation} 
Given a closed quantum many body system at hand, the conventional part $H_C(t)$ in definition (\ref{H-CQS}) is the most general Hamiltonian one can formulate out of the complete set of operators whose definitions do not depend on the system's state history.  The `Quantum-Memory-Made' (QMM) contributions to ($\ref{H-CQS})$, collectively named $H_{{\rm QMM}}(t)$, is the most general Hamiltonian one can formulate as a Hermitian composite operator whose constituting components have intrinsic dependencies on the state history of the system which is developed from an initial time to the present moment. Conventional Hamiltonians,  collectively denoted by $H_C(t)$, can be either constant operators or time dependent operators. In contrast, QMM-Hs do always have time dependencies which are \textit{induced} by their being made from the ongoing state histories of the closed systems. 
In what follows from now, we develop a general formulation of $H_{{\rm QMM}}(t)$ made directly from momentarily-chosen past-until-present quantum  states of the closed system.
\subsection{Quantum-Memory-Made Hamiltonians: A General Formulation}
Suppose we have a completely arbitrary closed quantum system and want to consider its 
Hamiltonian-generated unitary evolution all the way from an initial time to infinity, $t \in [t_0, \infty)$. One straightforward way to represent the system's state-history up to the present moment $t$ is the continuos set of density
operators which represent the system's quantum states at the corresponding times in the course of this history,
\begin{equation}
\mathcal{P}_{[t_0 , t]} \; \equiv \; \{ ~ \rho(t^\prime)  \;,\; \forall ~ t^\prime \in [t_0, t]  ~ \} \; \subset \; \lim_{t \to \infty} \mathcal{P}_{[t_0 , t]} \; \equiv \; \mathcal{P}_{[t_0 , \infty)}. \label{densityOp}
\end{equation} 
It is clear that every single bit of quantum information which is contained in each one $\rho(t^\prime) \in \mathcal{P}_{[t_0 , t]}$, namely in every system's past-or-present state, is one `qubit memory' of the system. In other words, the qubit memories of a quantum system up until a present time $t$ correspond, in a one to one manner, to all the quantum information bits necessary and sufficient to specify the system's quantum-state-history $\mathcal{P}_{[t_0 , t]}$ uniquely. Likewise, we refer to every density operator in the 
quantum-state-history $\mathcal{P}_{[t_0 , t]}$ as \emph{one `quantum memory' of the system}. Note that by a slight extension of terminology, the system's momentary density operator is likewise to be called a quantum memory. We, now highlight a point which is conceptually essential to all the formulation to be put forth: a contrast between \eqref{densityOp} for a closed quantum system and its counterpart in an open quantum system. The quantum systems whose state-histories 
$\eqref{densityOp}$ we address here are isolated, undergoing time evolutions which are reversible by being unitary. By this virtue, the totality of the quantum information by which the system's states, $\rho(t^\prime) \in \mathcal{P}_{[t_0 , \infty)} \; \forall t^\prime$, are uniquely specified, are enforced to stay inside the entirety of the quantum system and be eternally preserved. Hence, what plays the role of an `storage', containing all the memories to be utilized in the construction of $H_{{\rm QMM}}$, is the entire quantum system itself.
Re-emphasizing that the closed quantum many-body system under consideration has an arbitrary nature in terms of its defining degrees of freedom, and can have a wide range of defining structures, we now move on to formulate the envisioned evolution.\\\\  
\emph{Our aim is} to formulate a unitary evolution operator whose generating Hamiltonian $H_{{\rm QMM}}(t)$, at each present moment $t$, emerges from a chosen assembly of the system's available qubit memories, being arbitrarily selected. That is, $H_{{\rm QMM}}(t)$ is (re-)formed, moment by moment, as a nonlocal-in-time Hermitian operator which features structural dependence on the information content of some chosen elements of $\mathcal{P}_{[t_0 , t]}\;$. Therefore, the qubit memories contained in system's density operators at arbitrary past-to-present moments, $\rho( t^\prime_{{\rm chosen}}) \equiv \rho_{t^\prime_{{\rm chosen}}} \in \mathcal{P}_{[t_0 , t]}, \; \forall\; t^\prime_{{\rm chosen}} \in [t_0, t] $, participate in the momentary re-construction of the Hamiltonian $H_{{\rm QMM}}(t)$. To make it more clear, we first determine a subset of $\mathcal{P}_{[t_0 , t]}$, denoted by $\mathcal{P}^{_{{\rm chosen}}}_{[t_0 , t]}$, which contains (nothing but all) the system's `participant quantum memories' out of which $H_{{\rm QMM}}(t)$ is to be made, 
\begin{equation}\label{clh} \;\;\;\;\;\;\;\;\;\;\;\;\;\;\;\;\;\;\;\; \mathcal{P}^{{\rm chosen}}_{[t_0 , t]} \; \equiv \; \{ ~ \rho_{t^\prime_{{\rm chosen}}}  \;,\; \forall ~ t^\prime_{{\rm chosen}} \in [t_0, t]  ~ \} \; \subset \; \mathcal{P}_{[t_0 , t]}. \end{equation}
We now state the general definition of a \emph{QMM Hamiltonian (QMM-H)} as follows,
\begin{equation}\label{VMV} H_{{\rm QMM}}(t) \; =\; \mathcal{O}[\;\mathcal{P}^{{\rm chosen}}_{[t_0 , t]} \;]. \end{equation}
In definition (\ref{VMV}), $\mathcal{O} = \mathcal{O}^\dagger$ denotes a Hermitian operator and the brackets (in the R.H.S of it) mean \emph{`explicit structural dependence'}. Overall, the definition (\ref{VMV}) states that a QMM-H at present time $t$ is a Hermitian operator with explicit structural dependence on a chosen subset of the past-to-present quantum-state-history $\mathcal{P}^{{\rm chosen}}_{[t_0 , t]} $. Obviously, there are infinitely-diverse ways to introduce operators $\mathcal{O}$ which realize (\ref{VMV}), mainly based on: a) how one extracts the qubit memories contained in the chosen quantum memories $\rho_{t^\prime_{{\rm chosen}}} \; \forall t^\prime_{{\rm chosen}}$, and b) how one assembles those qubit memories to construct a Hermitian operator as the QMM-H. Among all possibilities, in what follows, we choose an infinitely-large family of QMM unitary evolutions (QMM-UEs) whose QMM-Hs are formulated by taking the straightforward way to realize (\ref{VMV}). \emph{We introduce QMM-Hs which, as Hermitian composite operators, are functions, specially polynomials, of the density operators which consist $\mathcal{P}^{{\rm chosen}}_{[t_0 , t]}$} \eqref{clh}. That is, $\mathcal{O}[\mathcal{P}^{{\rm chosen}}_{[t_0 , t]}]$ is a nonlocal-in-time Hermitian operator whose constituting operators \emph{include} all the chosen past-or-present density operators of the system, namely all the chosen quantum memories themselves: $\rho_{t^\prime_{{\rm chosen}}} \; \forall t^\prime_{{\rm chosen}}$. Furthermore,
we can let $\mathcal{P}^{{\rm chosen}}_{[t_0 , t]}$ be itself a countable or uncountable set. For concreteness, we take the former possibility in what follows, returning to the latter possibility in Section \ref{visions}.
\subsection{A Polynomial Family of Quantum-Memory-Made Hamiltonians}
Having made all these specifications, one concrete family of QMM-Hs is now presented according to the following \emph{recipe}.\\\\ 
 The recipe begins with choosing a doublet of positive integers, $(N , L)$. $N$ equals $|\mathcal{P}^{{\rm chosen}}_{[t_0 , t]}|$ and is the total number of the system's past-or-present density operators out of which the present-time QMM-H, $H_{{\rm QMM}}(t)$, is constructed. $L$ is the maximum algebraic length of the QMM operator monomials, made of the $\mathcal{P}^{{\rm chosen}}_{[t_0 , t]}$ elements, which combine linearly to form $H_{{\rm QMM}}(t)$. Now, given $N$ and looking backwardly from the present moment $t$, we take a choice of $N$ `quantum-memory distances' (QMDs), $\Delta^{(t)}_{i = 1 \cdots N}$, which uniquely identify all the Hamiltonian-maker quantum memories as the density operators of the system at $N$ past-to-present moments given by  
$t_i \equiv t - \Delta^{(t)}_i$: $\{\rho_{t_i}\}$. The chosen QMDs  are constrained and ordered as follows, 
\begin{equation}\label{ds} (\Delta^{(t)}_1 \;, \; \cdots \;,\; \Delta^{(t)}_N) \;\;\;;\;\;\;  0 \; \leq \; \Delta^{(t)}_N \; < \; \cdots \; < \; \Delta^{(t)}_1 \; \leq \; (t - t_0).  \end{equation} We note that with the above selections of $N$ and the spectrum of QMDs in (\ref{ds}), the QMM-H argument in \eqref{VMV} is completely fixed as follows, \begin{equation}\label{cps} \mathcal{P}^{{\rm chosen}}_{[t_0 , t]} = \{\; \rho_{t_1} \;,\; \cdots \;,\, \rho_{t_N} \; \}. \end{equation} \\ The recipe continues with specifying the QMM monomials which combine linearly to become the QMM-H. A Hamiltonian monomial is denoted by $\Phi^{s_{i_1} \; \cdot \cdot \cdot \; s_{i_r}}_{ i_1 \;  \cdot \cdot \cdot \; i_r }(t)$, where $r_{\; \geq 1}^{\; \leq L}$ is the monomial's length as a density-operator string, and the lower and upper indices specify respectively the monomial's variables picked up from $\mathcal{P}^{{\rm chosen}}_{[t_0 , t]}$ and their algebraic powers. Every QMM monomial  $\Phi^{s_{i_1} \; \cdot \cdot \cdot \; s_{i_r}}_{ i_1 \;  \cdot \cdot \cdot \; i_r }(t)$ is structured as follows. 
First, we choose an $r$-tuple $(i_1, \cdots, i_r) \subset (\{ 1, \cdots, N\})^r$ which fixes $(t_{i_1}, \cdots, t_{i_r}) \equiv (t -\Delta^{(t)}_{i_1} , \cdots , t - \Delta^{(t)}_{i_r})$. Second, we have the `option' to take a choice of $r+1$ \emph{state-history-independent} operators $\mathcal{A}^{(r)} = ( A^{(r)}_1, \cdots, A^{(r)}_{r+1} )$. Third, we choose the monomial-forming powers $(s_{i_1}, \cdot \cdot \cdot , s_{i_r})$. Now, the \emph{`QMM-monomial operator'} $ \Phi^{s_{i_1} \; \cdot \cdot \cdot \; s_{i_r}}_{ i_1 \;  \cdot \cdot \cdot \; i_r }(t)$ is conceived in the following form, 
\begin{equation}\label{QMMmonml} \Phi^{s_{i_1}  \cdots \; s_{i_r}}_{ i_1 \;  \cdot \cdot \cdot \; i_r }(t) \; \equiv \; A^{(r)}_{1} \; ( \rho_{t_{i_1}} )^{s_{i_1}}  A^{(r)}_{2}\; ( \rho_{t_{i_2}} )^{s_{i_2}} \; \cdots \; ( \rho_{t_{i_{r-1}}} )^{s_{i_{r-1}}}  A^{(r)}_{r} \;( \rho_{t_{i_r}} )^{s_{i_r}} A^{(r)}_{r + 1}. \end{equation}
Finally, we form a linear combination of these QMM monomials, following a choice of complex-or-real (constant-in-time or time-varying) `coupling' coefficients $\lambda_{i_1  \cdots i_r}^{s_{i_1} \cdots s_{i_r}}(t)$, and sum it up with its Hermitian conjugate. We now have, in result, a Hermitian nonlocal-in-time operator integrating some degree-$l^{(\leq L)}$ monomials of $N$ past-or-present system's quantum states, chosen from a specified $\mathcal{P}^{{\rm chosen}}_{[t_0 , t]}$ as (\ref{cps}). Therefore, the recipe outputs a broad realization of QMM-Hs (\ref{VMV}), denoted by $H^{(N,L)}_{({\rm QMM} | \mathcal{A})}(t)$ with $\mathcal{A} \equiv \{ \mathcal{A}^{(r)} , \forall r \}$, in the following form,
\begin{equation}\label{MMH} H^{(N,L)}_{({\rm QMM} | \mathcal{A})}(t) \;\; \equiv  \sum^{( \text{must include}\; r = L )}_{\substack{ \text{all chosen} \;r \\ \\  1 \; \leq \;  r \; \leq \; L  }} \; \sum^{(i_1, \cdots, i_r) \; \subset \; \{ 1, \cdots, N\}^r}_{\substack{\text{all chosen} \big( (i_1,s_{i_1}),\;  \cdots \;, (i_r , s_{i_r}) \big)}} \; \lambda_{i_1  \cdots i_r}^{s_{i_1} \cdots s_{i_r}} \;\;
\Phi^{s_{i_1} \; \cdot \cdot \cdot \; s_{i_r}}_{ i_1 \;  \cdot \cdot \cdot \; i_r}(t) \;\;+\;\; h.c.  
\end{equation} 
We must highlight three important points about the QMM-Hs \eqref{MMH}. \emph{First}, the above $\lambda$-couplings can be non-constants, furthermore, they can have explicit dependencies on \emph{scalar quantities} made from the elements of \eqref{clh}. However, we shall assume that all $\lambda$-couplings are state-history-independent constants. \emph{Second}, we obtain more complex QMM-Hs as linear combinations of $H^{(N,L)}_{({\rm QMM} | \mathcal{A})}$'s with distinct choices of integer powers, coupling profiles and the insertion operators in $\mathcal{A}$. \emph{Third,} depending on the context, one chooses or must employ as some $\mathcal{A}^{(r)}$ elements \emph{nontrivial} state-history-independent operators. Examples are projection or swap operators corresponding to the eigenstates of the relevant observables. However, we choose to focus on the behavioral phases of the \emph{unexplored and intriguing category in which the QMM-Hs are `entirely' made of the system's state-history}: we carefully analyze, in Sections \ref{888}, \ref{iml} and \ref{SV}, the intriguing subclass of QMM-Hs \eqref{MMH} given by constant couplings and the uniform choice: $A^{(r)}_l = {\mathbb 1}$.\\\\ 
The resulted \emph{Entirely-QMM-Hs}, denoted by  $H^{(N,L)}_{{\rm QMM}}(t)$, read as follows,
\begin{equation}\label{MMHwa1} H^{(N,L)}_{{\rm QMM}}(t) \;\; \equiv  \sum^{( \text{must include}\; r = L )}_{\substack{ \text{all chosen} \;r \\ \\  1 \; \leq \;  r \; \leq \; L  }}  \sum^{(i_1, \cdots, i_r) \; \subset \; \{ 1, \cdots, N\}^r}_{\substack{\text{all chosen} \big( (i_1,s_{i_1}),\;  \cdots \;, (i_r , s_{i_r}) \big)}} \lambda_{i_1  \cdots i_r}^{s_{i_1} \cdots s_{i_r}} \;
( \rho_{t_{i_1}} )^{s_{i_1}}  \; \cdots \; ( \rho_{t_{i_r}} )^{s_{i_r}} \;\;+\;\; h.c.  
\end{equation} 
One sees that additional care must be taken concerning the unitary time evolution in a sufficiently-early time interval which we call the `quantum-memory storing period'. Given $N$ and a choice of the largest QMD in (\ref{ds}), $\Delta^{(t)}_1$, we define $t_*$ to be the earliest moment as of which, or right after which, all the $N$ participant quantum memories in (\ref{cps}), have become available for the system. For example, if one selects a constant value $a$ for the largest QMD, $ \Delta^{(t)}_1 = a $, we have: $t_* = t_0 + a$. However, if one chooses a time-dependent linear profile, 
$ \Delta^{(t)}_1 = c_1\; (t - t_0) $ with $|c_1| \leq 1$, then: $t_* = t_0$. In any case, the `quantum-memory storing period' is always the time interval from $t_0$ to $t_*$. This is because before going past this period, the system has not yet developed enough 
state-history resources to build up $H_{({\rm QMM}|\mathcal{A})}^{(N,L)}(t)$ as defined in (\ref{MMH}). Independently, even when $t_* = t_0$, because the initial-time Hamiltonian $H^{(N,L)}_{({\rm QMM}|\mathcal{A})}(t_0)$ becomes a polynomial in only $\rho_{t_0}$, no nontrivial dynamics is generated by it, as the system is at state $\rho_{t_0}$. Therefore, unless $t_0 = -\infty$, we must employ an \emph{alternative} Hamiltonian to be at work in the quantum-memory storing period. This initial `Kicker Hamiltonian', $H^{({\rm Kicker})}(t)$,  i) generates a nontrivial dynamics at the initial time of the state-history, $t_0$, and ii) builds up dynamically a `quantum-memory pool', extended from $t_0$ to the moment $t_*$, using which the closed system can hand all its afterward evolution to the QMM-H (\ref{MMH}). As such, the total Hamiltonian of the closed quantum system is given by,
\begin{equation} \label{MMHpure} 
H^{({\rm Total})}_{({\rm QMM}|\mathcal{A})}(t) \; \equiv \; \theta(t_*-t)\; H^{({\rm Kicker})}(t)  \; + \;\theta(t - t_*) \;  
 H^{(N,L)}_{({\rm QMM}|\mathcal{A}) }(t). \end{equation}
It is worth to remark the following points regarding the Kicker Hamiltonian. \emph{First}, it is easily possible to construct a large variety of Kicker Hamiltonians $H^{({\rm Kicker})}(t)$ which themselves are (Purely) QMM, as we detail in Appendix \ref{appendix1}. \emph{Second}, as explained above, there are infinitely many time dependent choices of the largest QMD, that is: $a \, = \,a(t)$, for which the quantum-memory storing period vanishes, $t_* = t_0$, such that no Kicker Hamiltonian is necessary.
\emph{Third}, choosing $H^{({\rm Kicker})} = H_{{\rm C}}$ as was characterized in (\ref{H-CQS}), namely a state-history independent operator as the Kicker Hamiltonian, is sufficiently good for us in investigating the examples of Sections \ref{888}, \ref{iml} and \ref{SV}.\\\\  
%%%%%%%%%%%%%%%%%%%%%
The QMM time evolution operator corresponding to a time interval $[t', t'']$, $U^{(\text{Total})}_{({\rm QMM}|\mathcal{A})}(t'' , t')$, generated by a continuous sequence of momentary Hamiltonians $H^{\text{(Total})}_{({\rm QMM}|\mathcal{A})}(t \in [t', t''])$ defined in (\ref{MMHpure}), is unitary. The unitary evolution operator, $U^{(\text{Total})}_{({\rm QMM}|\mathcal{A})}(t'' \;,\; t')$, equals the (limit) product of infinitely many unitary operators corresponding to infinitesimal time evolutions covering the entire interval $[t',t'']$, which is resummed in the form of a Dyson expansion using time ordering product operator,  $\mathcal{T}$, as follows,
\begin{equation}\label{mue}
\begin{split}
& U^{(\text{Total})}_{({\rm QMM}|\mathcal{A})}(t'' \;,\; t') \;\; = \;\lim_{M \to \infty} \; \prod_{m = 1}^{ M } \; U_m^{(\text{Total})} \bigg(t' + \frac{t'' - t'}{M}\; m \;,\; t' + \frac{t'' - t'}{M}\; (m-1) \bigg) \equiv \\
& \hspace{3cm} \;  \equiv \; \lim_{ M \to \infty} \; \prod_{m = 1}^{ M } \; \exp \bigg[ \; i \;  \frac{t'' - t' }{ M } \; H_{({\rm QMM}|\mathcal{A})}^{(\text{Total})} \Big(t' + \frac{t'' - t'}{M}\; (m-1) \Big) \; \bigg] = \\
& \hspace{3cm} \; = \, \lim_{ M \to \infty} \; \prod_{m = 1}^{ M } \;  \bigg[\; \mathbb{1}\; + \; i \;  \frac{t'' - t' }{ M } \; H_{({\rm QMM}|\mathcal{A})}^{(\text{Total})} \Big( t' + \frac{t'' - t'}{M}\; (m-1) \Big) \; \bigg] = \\
& \hspace{3cm} \; = \; \mathcal{T} \exp \bigg( -i \, \int_{t'}^{t''} d\tau \; H_{({\rm QMM}|\mathcal{A})}^{(\text{Total})} (\tau) \; \bigg) .\\
\end{split}
\end{equation}
As such, the system's quantum states, which themselves are the constructional resource for the
QMM-H, must evolve consistently-and-unitarily as described by,
\begin{equation}\label{ues} \rho_{t''} = U^{(\text{Total})}_{({\rm QMM}|\mathcal{A})}(t'' \;,\; t') \; \rho_{t'} \; U^{(\text{Total}) \dagger}_{({\rm QMM}|\mathcal{A})}(t'' \;,\; t').   \end{equation}
%%%%%%%%%%%%%%%%%%%%% \hspace{2.54 cm} U
\subsection{Pure State Histories Composing QMM Unitary Evolutions}
The QMM-Hs (\ref{MMH}) can be made from arbitrary mixed quantum states. But, from now on, it suffices to focus on investigating the QMM-UEs which are made by \emph{the pure states} of the system, on the account of their sufficient richness.
Assuming state purity at the initial moment, given that the whole QMM time evolution is ensured to be unitary (\ref{mue}, \ref{ues}), the corresponding 
state-history of the system (\ref{densityOp}) will be developed as a continual sequence of pure states: it will be a `pure-state-history'. Being so, as of now, we only focus on 
pure-state-histories of closed systems, that is, 
\begin{equation}\begin{split}\label{hps}
& \mathcal{P}_{[t_0 , \infty)} \; = \; \{ ~ \rho_{t^\prime} \;= \; | \Psi_{t'} \rangle \langle \Psi_{t'} | \;\;\;,\; \quad \forall ~ t'  \in [ t_0 , \infty )  ~ \}, \\
& \hspace*{1.33cm}\;\;\; \rho_{t^\prime} \;= \; \rho^{s}_{t'}\;\;\;,\;\;\; \forall s \in \mathbb{N}  \;,\; \quad \forall ~ t'  \in [ t_0 , \infty ).
\end{split}\end{equation}
Now, merging (\ref{MMHwa1}) and (\ref{hps}), we arrive at the following family of \emph{pure-state-history QMM-Hs}, denoted by $H^{(N|L)}_{{\rm QMM}}(t)$,
\begin{equation}\begin{split}
& H^{(N,L)}_{\text{QMM.PureStateHistory}}(t) \; \equiv \; \\ & \hspace{.0000213cm} \equiv \; H^{(N|L)}_{{\rm QMM}}(t) \; = \; \sum^{( \text{must include}\; r = L )}_{\substack{ \text{all chosen} \;r \\ \\  1 \; \leq \;  r \; \leq \; L  }}\;\;  \sum^{(i_1, \cdots, i_r) \; \subset \; \{ 1, \cdots, N\}^r\;;\;i_l \neq i_{l+1} , \forall\; l}_{\substack{\text{all chosen} \big( (i_1,s_{i_1}),\;  \cdots \;, (i_r , s_{i_r}) \big)}}\;\;\; \big[ \; \lambda_{i_1 \cdots i_r} \; (\; \rho_{t_{i_1}} \;\cdots \; \rho_{t_{i_r}} \;)  \; + \; \\ & \hspace{3.18cm} \hspace{5 cm} \hspace{1.99 cm} \;\;\;\;\; + \; \lambda^{\star}_{i_1 \cdots i_r} \; \; (\; \rho_{t_{i_r}} \;\cdots \; \rho_{t_{i_1}} \;) \; \big]  \label{MMHparts}.
\end{split} \end{equation}
The QMM-UEs generated by the above family of Hamiltonians will be the very subject of all our analytical and numerical findings in the rest of this work. Some results are general: they hold for the whole family (\ref{MMHparts}), either precisely or at least qualitatively. But, many exact results are the outcomes of detailed analyses in a number of chosen concrete examples. Specifically, in what follows, we write down the explicit choices of the QMM-Hs (\ref{MMHparts}) whose one-qubit-system realizations we will detail in Section \ref{888}, and map the phase diagrams of their QMM wavefunctions in Section \ref{SV},
\begin{equation}\label{NL}\begin{split} & 
H^{(1|1)}_{{\rm QMM}}(t) \; = \; \mu \; \rho_{t-a} \hspace{1.99 cm} \hspace{5.0000399 cm} \;\;\;\;\;\;\;\;\;\;\;\;\;\;\;\;\;\;\;\;\;\;\;\;\;\;\;\; \Delta^{(t)}_1 = a,\;\;\;\;\;\;\;\; \hspace{.399 cm} \; \hspace{.165 cm} %\;\;  (N,L) = (1,1), 
\\ & H^{(2|2)}_{{\rm QMM}}(t) \; = \; \mu \; \rho_{t-a} \; + \; \lambda \; \rho_{t-a} \; \rho_{t} + \lambda^{\star}\; \rho_{t} \; \rho_{t-a}\;\;\; \hspace{2.99 cm} \;\;\;\;\; (\Delta_2^{(t)} , \Delta_1^{(t)}) = (0 , a),\;\; \hspace{.0165 cm} %\;\;\;\;\;\; (N,L) = (2,2), 
\\ & H^{(2|3)}_{{\rm QMM}}(t)\; = \; \lambda \; \rho_{t-a} \; \rho_{t} + \lambda^{\star}\; \rho_{t} \; \rho_{t-a} \; + \; \eta \; \rho_{t-a} \; \rho_{t} \; \rho_{t-a}\;\;\;\;  \hspace{1.86001118999 cm} \;\;\;  (\Delta_2^{(t)} , \Delta_1^{(t)}) = (0,a), \;\; \hspace{.165 cm} % (N,L) = (2,3), 
\\ & H^{(3|3)}_{{\rm QMM}}(t) \;= \; \kappa \; \rho_{t-a} \; \rho_{t-b} \; \rho_{t} + \kappa^{\star} \; \rho_{t} \; \rho_{t-b} \; \rho_{t-a} \;\;\;\;\;\;\; \hspace{1.88 cm} (\Delta_3^{(t)} , \Delta_2^{(t)} , \Delta_1^{(t)}) \;=\; (0 ,b, a ). \;\;\;\; % (N,L) = (3,3)
 \end{split} \end{equation}
The QMM-Hs (\ref{MMHparts}) together with their QMM Schr\"odinger equations (to be coming very soon) and their QMM evolving wavefunctions, can be recast in \emph{an alternative way} which we highlight specially for its elegance and effectiveness. One defines a family of `Quantum-Memory Characters and Operators' which are natural to the proposed systems. 
Given one's choice of all the participant quantum memories shown in (\ref{cps}), these quantum-memory characters are as follows: the \emph{`Quantum-Memory Two-Point Functions'} $m_{ t_{i_l} t_{i_r} }$ (QM-TPFs); all their QM $n$-point-function descendants $m_{ t_{i_{r_1}} \cdots t_{i_{r_n}} }$; the QM Two-Point Observables, in other words, the pairwise \emph{QM-Fidelities}, $w_{t_{i_l} t_{i_r}}^2$. They are defined as such,
\begin{equation}\begin{split}\label{ctmcs} 
& m_{t_{i_l} t_{i_r}} \equiv \; 
\braket{\Psi_{t_{i_l}} | \Psi_{t_{i_r}}}  \;\;\;;\;\;\; m_{t_{i_1} t_{i_2} \cdots t_{i_{r-1}} t_{i_r}} \equiv \; m_{t_{i_1} t_{i_2}} m_{t_{i_2} t_{i_3}} \cdots m_{t_{i_{r-1}} t_{i_r}}  =  (\; m_{t_{i_r} t_{i_{r-1}} \cdots t_{i_2} t_{i_1}} \; )^\star,  \\
& w_{t_{i_l} t_{i_r}}^2  \equiv  | m_{t_{i_l} t_{i_r}} |^2 = {\rm Tr} [\; \rho_{t_{l}} \; \rho_{t_{r}} \; ] = \big( \text{Tr}\big[( \sqrt{\rho_{t_{l}}}\;\rho_{t_{r}}\;\sqrt{\rho_{t_{l}}} )^{\frac{1}{2}}\big] \big)^2 = F (\;\rho_{t_{l}} \;,\; \rho_{t_{r}}\; ).
\end{split}\end{equation}
Moreover, we define the \emph{`Quantum-Memory Two-Point Operators'} (QM-TPOs), $M_{t_{i_r} t_{i_l}}$,
\begin{align}
\label{ctmtpos} 
\hspace{.0000199 cm} 
\;\;\; M_{t_{i_r} t_{i_l}}\; &\equiv \; \ket{\Psi_{t_{i_r}}} \bra{\Psi_{{t_{i_l}}}}\; =\; M^\dagger_{t_{i_l} t_{i_r}}\;\;\:;\;\;\;\; 
M_{t_{i_r} t_{i_r}}\; =\; \rho_{t_{i_r}} \;\;\;;\;\;\: \text{Tr}[\; M_{t_{i_r} t_{i_l}} \;] = m_{t_{i_l} t_{i_r}} = m^\star_{t_{i_r} t_{i_l}} \nonumber \\ 
\end{align}
% \emph{`Quantum Memory Assemblers'}. 
Now, in terms of the above quantum-memory measures and operators, the QMM-Hs formulated in (\ref{MMHparts}) are recast as follows,
\begin{equation}\label{mmhtpswmctc}\begin{split} & H^{(N|L)}_{{\rm QMM}}(t) \; = \; \\ & \hspace{1.25cm} = \; \sum^{( \text{must include}\; r = L )}_{\substack{ \text{all chosen} \;r \\ \\  1 \; \leq \;  r \; \leq \; L  }}\;\;  \sum^{(i_1, \cdots, i_r) \; \subset \; \{ 1, \cdots, N\}^r\;;\;i_l \neq i_{l+1} , \forall\; l}_{\substack{\text{all chosen} \big( (i_1,s_{i_1}),\;  \cdots \;, (i_r , s_{i_r}) \big)}}\;\;\; \big( \; \lambda_{i_1 \cdots i_r} \; m_{t_{i_1} t_{i_2} \cdots t_{i_{r-1}} t_{i_r}} \; M_{t_{i_1} t_{i_r}}  \; + \; \\ & \hspace{9.3330999cm} \; + \; \lambda^{\star}_{i_1 \cdots i_r} \; \; m_{t_{i_r} t_{i_{r-1}} \cdots t_{i_2} t_{i_1}} \; M_{t_{i_r} t_{i_1}}\;\; \big). \end{split} 
\end{equation}
Although obvious, it would be worth emphasizing that the preservation of the total measurement probabilities does hold in the course of QMM-UE, that is:  ${\rm Tr} [\rho(t)] = 1$, or for the pure states: $<\psi_t|\psi_t> = 1$, if the system's state at $t_0$ is normalized to one. This dynamical preservation is an automatic and direct implication of the \emph{unitarity} of the system's evolution under the QMM-Hs (\ref{MMH},\ref{MMHparts}), manifested in (\ref{MMHpure},\ref{mue},\ref{ues}). Moreover, detailing $H^{(2|2)}_{{\rm QMM}}$ in Subsection \ref{msc}, we even explicitly demonstrate that $|\ket{\Psi_t}|$ is the invariant of QMM-UEs, using their QMM Schr\"odinger equations \eqref{pmmscheqwmctc}. \\\\
 Finally, we highlight an important point. Apart from the QMM-H with $(N,L) = (1,1)$, in examples (\ref{NL}) to be analyzed in Sections \ref{888}, \ref{iml} and \ref{SV}, we have taken the smallest QMD to be zero, such that $\rho_t \in \mathcal{P}^{{\rm chosen}}_{[t_0 , t]}$. But, this is optional: a `working-frame' choice. That is, in definition (\ref{MMH}), we do \emph{not} demand that the present moment state of the system, $\rho_t$, should be necessarily used in the making of $H_{{\rm QMM}}(t)$. Indeed, whether or not the chosen $\mathcal{P}^{{\rm chosen}}_{[t_0 , t]}$ contains $\rho_t$ is an arbitrary option under which the formulation of $H_{{\rm QMM}}(t)$ remains fully covariant.
%%%%%%%%%%%%%%%%%%%%%%%%%%%%%%%%%%%%%%
%%%%%%%%%%%%%%%%%%%%%%%%%%%%%%%%%%%%%%%
\section{The Dynamical Equations Of Unitary State-History-Based Quantum Behavior}\label{gtp}
\subsection{QMM von Neumann and Schr\"odinger Equations}
As (\ref{MMHpure},\ref{mue}) demonstrate, evolution under QMM-Hs (\ref{QMMmonml},\ref{MMH}), or its pure-state-history version (\ref{MMHparts}), is unitary. The time-integrated evolution of the states is given by (\ref{ues}) which combined with (\ref{mue},\ref{MMHparts}) leads to a highly complex structure for the differential dynamical equation corresponding to it. The complexity originates from the intrinsic quality of QMM-UEs: that, \emph{states evolve under states}, and that for this process to run, there must be a consistent interplay between the system's states at distinct moments. This interplay imposes a nonlocal-in-time nonlinearity on the differential dynamical equations of QMM-UEs. Here we derive these equations and write their expressions in alternative useful forms.  Before moving on, let us note that we employ conventional Hamiltonians for the Kicker Hamiltonian which is at work during $t \leq t_*$, throughout the paper. Hence, the unitary evolution of the states, given by (\ref{MMHpure},\ref{mue},\ref{ues}), leads to a known conventional Schr\"odinger equation during $[t_0, t_*]$. Being so, we only need to work out the QMM delay-differential dynamical equation which is on role as of having the quantum-memory pool filled up, namely for 
$t \geq t_*$.\\\\ 
The equations (\ref{mue}) and (\ref{ues}) once combined immediately result at the well-known general structurally-unpacked form of quantum dynamical equation for the system's states. The (Liouville–)von Neumann equation for QMM-H $(\ref{MMHparts})$ during $ t \in [t_*, \infty)$, is given by,
\begin{equation}
\dot{\rho}_t \;\; = \; i \; [\; \rho_t \;,\; H^{(N|L)}_{{\rm QMM}}(t) \; ]. \label{evolution}
\end{equation}
Now, applying the explicit expression of $H^{(N|L)}_{{\rm QMM}}(t)$ in \eqref{MMHparts} to general equation (\ref{evolution}), the dynamical equations for the corresponding family of pure-state-history QMM-UEs is yield as the following von Neumann equation,
\begin{equation}\label{evo-HMMpure}\begin{split} &
i \; \dot{\rho}_t =
\sum^{( \text{must include}\; r = L )}_{\substack{ \text{all chosen} \;r \\ \\  1 \; \leq \;  r \; \leq \; L  }}\;  \sum^{(i_1, \cdots, i_r) \; \subset \; \{ 1, \cdots, N\}^r\;;\;i_l \neq i_{l+1} , \forall\; l}_{\substack{\text{all chosen} \big( (i_1,s_{i_1}),\;  \cdots \;, (i_r , s_{i_r}) \big)}}\;\; \big(\;\lambda_{i_1 \cdots i_r} ~ [~ \rho_{t_{i_1}} \; \cdots \; \rho_{t_{i_r}} \;,\; \rho_t ~] \;\; + \; \\ & \hspace{3.54 cm} \hspace{3.99 cm} \;\;\;\;\;\;\;\;\; + \; \lambda_{i_1 \cdots i_r}^\star ~ [~ \rho_{t_{i_r}} \; \cdots \; \rho_{t_{i_1}} \;,\; \rho_t ~] \;\; \big). \;\;\;\;\;\;\;\;\; 
\end{split} \end{equation}
We have highlighted that it is insightful and useful to have the physics of a QMM-UE also expressed in terms of the corresponding QM-TPFs together with their descendants, and QM-TPOs, defined in (\ref{ctmcs},\ref{ctmtpos}). Now, applying those alternative expressions of the QMM-Hs \eqref{mmhtpswmctc} to \eqref{evolution}, we obtain an alternative form of the von Neumann equations (\ref{evo-HMMpure}),
\begin{equation}\label{prhoue}\begin{split} & i\; \dot{\rho}_t  
=  \sum^{( \text{must include}\; r = L )}_{\substack{ \text{all chosen} \;r \\ \\  1 \; \leq \;  r \; \leq \; L  }}\; \sum^{(i_1, \cdots, i_r) \; \subset \; \{ 1, \cdots, N\}^r\;;\;i_l \neq i_{l+1} , \forall\; l}_{\substack{\text{all chosen} \big( (i_1,s_{i_1}),\;  \cdots \;, (i_r , s_{i_r}) \big)}} \big( \; \lambda_{i_1 \cdots i_r} \; m_{t_{i_1} t_{i_2} \cdots t_{i_{r-1}} t_{i_r}} [\; M_{t_{i_1} t_{i_r}} \;,\; \rho_t \;] \; +  \\ & \hspace{8.3338999 cm} + \; \lambda^{\star}_{i_1 \cdots i_r} \; m_{t_{i_r} t_{i_{r-1}} \cdots t_{i_2} t_{i_1}} \; [\; M_{t_{i_r} t_{i_1}} \; , \; \rho_t \;]\; \big). \end{split}
\end{equation}
The above equation can be rewritten in a nicer form. Remembering that $\rho_t = \ket{\Psi_t} \bra{\Psi_t}$ and using (\ref{ctmcs}) and (\ref{ctmtpos}), we immediately validate the following relations,
\begin{equation}\label{simpleidentity} m_{t_a ... t_b} \; m_{t_b t} \;=\; m_{t_a ... t_b t},
 \end{equation}
and,
\begin{equation}\label{beingused} 
[\;M_{t_{i_1} t_{i_r}}\;,\;\;\rho_t\;]\; = \; m_{t_{i_r} t}\;M_{t_{i_1} t}\;-\; m_{t t_{i_1}}\;M_{t t_{i_r}}. 
\end{equation} 
Upon applying (\ref{simpleidentity},\ref{beingused}) to (\ref{prhoue}), the \emph{`QMM von Neumann equations'} are recast as, \begin{equation}\label{ase} \begin{split} & i\; \dot{\rho}_t\;=\; i\;\dot{M}_{tt} =  \\ & = \sum^{( \text{must include}\; r = L )}_{\substack{ \text{all chosen} \;r \\ \\  1 \; \leq \;  r \; \leq \; L  }}\; \sum^{(i_1, \cdots, i_r) \; \subset \; \{ 1, \cdots, N\}^r ; i_l \neq i_{l+1} , \forall\; l}_{\substack{\text{all chosen} \big( (i_1,s_{i_1}),\;  \cdots \;, (i_r , s_{i_r}) \big)}} \lambda_{i_1  \; ... \; i_r } \;( m_{t_{i_1} \; \cdot \cdot \cdot \; t_{i_r} t} \; M_{t_{i_1} t} -  m_{t t_{i_1} \; \cdot \cdot \cdot \; t_{i_r}} \; M_{t t_{i_r}} )\;+ \\ & \; \hspace{3.118 cm} \hspace{1.99 cm} \hspace{1.99 cm}  + \; \lambda^\star_{i_1 \; ... \; i_r }\;( m_{ t_{i_r} \; \cdot \cdot \cdot \; t_{i_1} t} \; M_{t_{i_r} t} -  m_{t t_{i_r} \; \cdot \cdot \cdot \; t_{i_1} } \; M_{t t_{i_1}} ).  \end{split} 
\end{equation}
Moreover, given the imposed purity of the system's quantum-state-history (\ref{hps}), one recasts the original form of Schr\"odinger equation for the unitary evolution under $H^{(N | L)}_{{\rm QMM}}(t)$,
\begin{equation}\label{pmmscheq}   i\;\ket{\dot{\Psi}_t} \; = \; H^{(N | L)}_{{\rm QMM}}(t)\;\ket{\Psi_t} . 
\end{equation} 
It can be recast by extraction from (\ref{ase}), or applying (\ref{mmhtpswmctc}) to (\ref{pmmscheq}), and using,
\begin{equation} M_{t_a t_b} \ket{\Psi_t} \;=\; m_{t_b t}\;\ket{\Psi_{t_a}}.
\end{equation} 
Either way, one finally arrives at `\emph{QMM Schr\"odinger Equation' for unitary evolution generated by $H^{(N | L)}_{{\rm QMM}}(t)$},
\begin{equation}\label{pmmscheqwmctc}\begin{split}    i \;\ket{\dot{\Psi}_t} \; = &  \sum^{( \text{must include}\; r = L )}_{\substack{ \text{all chosen} \;r \\ \\  1 \; \leq \;  r \; \leq \; L  }}\;\;  \sum^{(i_1, \cdots, i_r) \; \subset \; \{ 1, \cdots, N\}^r\;;\;i_l \neq i_{l+1} , \forall\; l}_{\substack{\text{all chosen} \big(\; (i_1,s_{i_1}),\;  \cdots \;, (i_r , s_{i_r}) \big)}}\;\;\; \big( \;\; \lambda_{i_1 \cdots i_r}\;\;m_{t_{i_1} \; \cdots \; t_{i_r} t}\; \ket{\Psi_{t_{i_1}}} \;  \; \\ & \hspace{3.99 cm} \hspace{3.599cm} \; +\; \lambda^\star_{i_1 \cdots i_r}\;\; m_{t_{i_r}  \; \cdot \cdot \cdot \; t_{i_1} t}\; \ket{\Psi_{t_{i_r}}} \; \big) . \end{split} \end{equation}
Equation (\ref{pmmscheqwmctc}) is one of the main results of the present paper in three complementary ways. First, it encodes the physical content of (\ref{MMHparts},\ref{mue},\ref{ues}), all in all. Second, it is a mathematically novel-and-elegant nonlinear delay-differential equation with extremely complex and resourceful structure and indeed moduli space. Third, all the rest of this paper is devoted to exploring and understanding the novel physics, specially the phase diagrams, of the most elementary closed system under the simplest QMM-UEs, upon solving the corresponding Purely-QMM Schr\"odinger Equations (\ref{pmmscheqwmctc}). Furthermore, we highlight in here three characteristic features of (\ref{pmmscheqwmctc}). First, notice that solving this equation is also determining $H^{(N | L)}_{{\rm QMM}}(t)$. Because before that, the QMM-H (\ref{mmhtpswmctc}) is an unidentified operator, as it is made out of the yet-unsolved state-history of the closed system (\ref{VMV}). This is intrinsically unlike conventional closed systems where the Hamiltonian is an identified operator and we only solve the Schr\"odinger equation for the wavefunctions. But for QMM-UEs, the triple equations (\ref{mmhtpswmctc},\ref{mue},\ref{ues}) are to be solved jointly, upon solving (\ref{pmmscheqwmctc}), with solutions identifying the system's unitary time evolution operators and the corresponding quantum-state-histories, all together.\\\\ 
Second, QMM Schr\"odinger equation is nonlocal-in-time. The `temporal nonlocality' of (\ref{pmmscheqwmctc}) is a direct implication of $H^{(N | L)}_{{\rm QMM}}(t)$ being made of the states at different times. In fact, nonconstant solutions to (\ref{pmmscheqwmctc}) are developed only if $\Delta^{(t)}_1$ exceeds some finite threshold: the temporal nonlocality of (\ref{pmmscheqwmctc}) cannot be even `softened' arbitrarily. \\\\
Third, QMM Schr\"odinger equation is inherently nonlinear. Physically, the nonlinearity is caused by the interactions which the closed system must turn on between some of its quantum memories, to develop unitarity evolution based on (\ref{mmhtpswmctc}). Mathematically, it is sourced by all the QM-TPFs $m_{t_i t_j} = \braket{\psi_{t_i}|\psi_{t_j}}$ which enter into (\ref{pmmscheqwmctc}). Hence, the temporal nonlocalities and the nonlinearities of (\ref{pmmscheqwmctc}) are merged to the core, and in a specific form. Indeed, QMM Schr\"odinger equation (\ref{pmmscheqwmctc}) and their solutions (to be shown) are, to the best of our knowledge \cite{acronlse:a,acronlse:b,acronlse:c,NLSEQI20189,KS2019}, physically and mathematically distinctive from all nonlinear Schr\"odinger equations available in the literature. 
%%%%%%%%%%%%%%%%%%%%%%%%%%%%%%%%%%%%%%%%%%%%%%%%%%%%%%%%%%%%%%%%%%%%%%%%%%%%%%%%%%%%%%%%%%
\subsection{Detailing A Family of QMM-Hs With Two Quantum Memories
}\label{msc}
We now unpack the detailed structures of the QMM-H and its QMM Schr\"odinger equation for an arbitrary closed system whose pure-state-history realizes the unitary evolution generated by the QMM-H \eqref{MMHparts} based on two quantum memories. The momentary Hamiltonian of the system, $H^{(2|L)}_{{\rm  QMM}}(t)$, is a Hermitian polynomial of arbitrary degree $L$ in the system's states at the two past-and-present moments identified with the following choice of the QMDs: $(\Delta^{(t)}_2 ,\Delta^{(t)}_1) = (0,a) $ where $a$ is an arbitrary positive constant. As before, $H^{({\rm Kicker})}(t) = H_{C}$, such that we only focus on $H^{(2|2)}_{QMM}$ and its QMM Schr\"odinger equation during $t > t_*$. Given the purity of the state history, the most general QMM-H characterized with $N = 2$, arbitrary $L$, and the above-made choices of QMDs is formed with the following structure of QMM monomials,
\begin{equation}\label{oec}\begin{split} &
H^{(2|L)}_{{\rm QMM}}(t) \; \equiv \; H^{(2|L)}_{({\rm odd})}(t), \;+\; H^{(2|L)}_{({\rm even})}(t), \\ & H^{(2|L)}_{({\rm odd})}(t) \; \equiv \; \big(\;  \mu_1 \; \rho_{t-a} \; + \; \mu_3 \; \rho_{t-a} \rho_t \rho_{t-a} \; + \; \mu_5 \; \rho_{t-a} \rho_t \rho_{t-a} \rho_t \rho_{t-a} \; + \; \cdots \;\big)_{{\rm up\;to\;degree}\; =\; L} \\ & \hspace{ 6.99 cm} \hspace{ 5.33 cm} ; \; \mu_{r = 1, 3, 5, \cdots} \in \mathbb{R}, \\ & H^{(2|L)}_{({\rm even})}(t)  \equiv  \big(\;  \lambda_2 \; \rho_{t-a} \rho_t  +  \lambda_4 \; \rho_{t-a} \rho_t \rho_{t-a} \rho_t  +  \lambda_6 \; \rho_{t-a} \rho_t \rho_{t-a} \rho_t \rho_{t-a} \rho_t +  \cdots  +  \\ & \hspace{1.330099 cm} \;\;\;\;\; + \lambda_2^\star \; \rho_t \; \rho_{t-a} + \lambda_4^\star \; \rho_t \rho_{t-a} \rho_t \rho_{t-a}  +  \lambda_6^\star \; \rho_t \rho_{t-a} \rho_t \rho_{t-a} \rho_t \rho_{t-a}  + \cdots \big)_{{\rm up\;to\;degree}\; = \; L} .
\end{split} \end{equation}
Notice that all the monomials beginning with the $\rho_t$ factor have been ignored in the `odd' part of the QMM-H (\ref{oec}). This is because, by the purity of the states, all such monomials commute with the system's density operator at the present time, $\rho_t$. Thus, the system's state gains no dynamics under their inclusion in $H^{(2|L)}_{{\rm QMM}}(t)$.\\\\ 
For $N=2$, and with the chosen spectrum of the QMDs $(\Delta^{(t)}_2, \Delta^{(t)}_1) = (0,a) $, the only QM-TPFs and nonlocal-in-time QM-TPOs are: $m_{t-a,t} = m_{t,t-a}^\star = \braket{\Psi_{t-a}|\Psi_t}$, $M_{t-a,t} = M_{t,t-a}^\dagger = \ket{\Psi_{t-a}}\bra{\Psi_t}$, respectively.
Moreover, the central observable to probe the dynamics of the wavefunctions of $N = 2$ QMM-UE is the norm of the QM-TPF,
\begin{equation}\label{nmtpf}
	w_{t-a \; t} \;\equiv\; w_{(a|t)} \; \equiv \; |m_{t-a,t}| \;=\; |\braket{\Psi_{t-a}|\Psi_t}| \;= \;
	( \text{Tr} [\; \rho_{t-a} \; \rho_t ] )^{\frac{1}{2}}.
\end{equation} 
Now, we introduce the following twin polynomials in terms of the QM observable $w_{(a,t)}$ whose coefficients are given by the Hamiltonian couplings $\mu$'s and $\lambda$'s, respectively,
\begin{equation}\label{wpoly}\begin{split} &
P_{L}^{(\mu)}(w_{(a|t)})\;\;\; \equiv \sum_{s}^{\;\;\;\; s \; \in \{ 1, \cdots, \frac{L+1}{2 }\}} \mu_{2s-1} \; w_{(a|t)}^{2(s-1)}, \\ & P_{L}^{(\lambda)}(w_{(a|t)})\;\;\; \equiv \sum_{s}^{\;\;\;\;\;\;\; s \; \in \{ 1, \cdots, \frac{L}{2}\}} \lambda_{2s} \; w_{(a|t)}^{2(s-1)}. \end{split}
\end{equation}
By means of the polynomials (\ref{wpoly}), and the QM-TPs and QM-TPOs themselves, the QMM-H (\ref{oec}) takes its final expression as follows,
\begin{equation}\label{oecms}\begin{split} &
H^{(2|L)}_{{\rm QMM}}(t) \; = \; P_{L}^{(\mu)}(w_{(a|t)})\; \rho_{t-a} \;+\;  P_{L}^{(\lambda)}(w_{(a|t)})\; m_{t-a, t} \; M_{t-a,t} \;+\; P_{L}^{(\lambda^\star)}(w_{(a|t)})\; m_{t-a, t}^\star \; M_{t-a,t}^\dagger .
\end{split} \end{equation}
Now, applying (\ref{oecms}) to (\ref{pmmscheq}), or by working out the right hand side of (\ref{pmmscheqwmctc}), the QMM Schr\"odinger equation corresponding to 
$H^{(2|L)}_{{\rm QMM}}(t)$ is organized in an efficient way in terms of two particular quantum-memory functions $\mathcal{S}^{(2|L)}_{R}$ and $\mathcal{S}^{(2|L)}_{N}$, as follows,
\begin{equation}\label{tqmmse}\begin{split} &
i~ |\dot{\Psi}_t \rangle \; = \; \mathcal{S}^{(2|L)}_{R} \ket{\Psi_{t-a}} \;+\; \mathcal{S}^{(2|L)}_{N} \ket{\Psi_t}, \\ & \hspace{1.65 cm} \; \mathcal{S}^{(2|L)}_{R} \; \equiv \; P_{L}^{(\mu + \lambda)}(w_{(a|t)}) \; \braket{\Psi_{t-a}|\Psi_t} \; = \; (\; P_{L}^{(\mu)}(w_{(a|t)}) \;+\;  P_{L}^{(\lambda)}(w_{(a|t)})\;) \; \braket{\Psi_{t-a}|\Psi_t}, \\ & \hspace{1.65 cm} \; \mathcal{S}^{(2|L)}_{N} \; \equiv \; P_{L}^{(\lambda^\star)}(w_{(a|t)})\;w_{(a|t)}^2 .
\end{split}\end{equation} 
The above QMM functions $\mathcal{S}^{(2|L)}_{R}$ and $\mathcal{S}^{(2|L)}_{N}$ are denoted here as the `Retarded' and `Now' (Shr\"odinger equation) Functions of the $(2,L)$ QMM-UE, respectively. Finally, let us further write down the fully-unfolded forms of these two functions for the choices of $L = 2$ and $L =3$, given that the one-qubit-wavefunction solutions to (\ref{tqmmse}) will be analyzed later for both of these cases. Upon renaming the couplings according to (\ref{NL}), here they come, 
\begin{equation}\label{ltt}\begin{split} &
L \; = \; 2\;: \;\;\; \mathcal{S}^{(2|2)}_{R} \; = \; (\mu + \lambda) \; \braket{\Psi_{t-a}|\Psi_t} \;\;\;;\;\;\; \mathcal{S}^{(2|2)}_{N} \; = \;  \lambda^\star \; |\braket{\Psi_{t-a}|\Psi_t}|^2 , \\ & L \; = \; 3\;: \;\;\; \mathcal{S}^{(2|3)}_{R} \; = \; (\mu + \lambda  + \eta \; |\braket{\Psi_{t-a}|\Psi_t}|^2 ) \; \braket{\Psi_{t-a}|\Psi_t} \;\;\;;\;\;\; \mathcal{S}^{(2|3)}_{N} \; = \;  \lambda^\star \; |\braket{\Psi_{t-a}|\Psi_t}|^2 .
\end{split}\end{equation}
%Now $\mu$ 1 \\
Before closing this subsection, we would like to explicitly check and make sure that the QMM Schr\"odinger equation (\ref{tqmmse}) preserves the norms of its wavefunction solutions. This explicit confirmation would be done here despite that the norm-conservation of the wavefunctions $\ket{\Psi_t}$ is guaranteed by the very fact that the QMM time evolution operator is unitary (\ref{mue}). But it is still worth showing, as a mere consistency check, that the corresponding QMM Schr\"odinger equation also by itself realizes this feature. Indeed, it would be enough to do this check for the case of $L=2$. Applying the $(R,N)$ Schr\"odinger operators (\ref{ltt}) to (\ref{tqmmse}), one can easily check that,
\begin{equation}\begin{split} \frac{d}{dt} \braket{\Psi_t | \Psi_t}  % =  \; \braket{ \dot{\Psi}_t | \Psi_t} \; + \; \braket{\Psi(t) | \dot{\Psi}_t} \; 
=  2\; \text{Re} \big[\braket{\Psi_t | \dot{\Psi}_t} \big]  =  2\;\text{Im} \big[\bra{\Psi_t}  \mathcal{S}^{(2|L)}_{R}\ket{\psi_{t-a}}  + \bra{\Psi_t} \mathcal{S}^{(2|L)}_{N} \ket{\Psi_t} \big] \;=\; 0 . %\text{several cancelling terms} \; = \; 0  
\end{split} \end{equation}  
Being so, the QMM Schr\"odinger equation does by itself secure that the total probability of the measurements is preserved to be one by the wavefunction solutions which evolve under the corresponding QMM-H. In fact, this preservation remains always true in the entire course of the state-history of the closed quantum system $\mathcal{P}_{[t_0 , \; \infty)}$,
\begin{equation} \hspace{1.33 cm} {\rm Tr} [\; \rho_t \;]  =  1 \;\;\;,\;\;\; \forall\; \rho_t \in \mathcal{P}_{[t_0 , \; \infty)} . \end{equation}
%%%%%%%%%%%%%%%%%%%%%%%%%%%%%%%%%%%%%%
%%%%%%%%%%%%%%%%%%%%%%%%%%%%%%%%%%%%%%%%%%
\subsection{The Robust Non-Markovianity of Purely-QMM Unitary Evolutions}\label{S:IV:Strong}
One important aspect in understanding the physics of a memory-dependent quantum many body system is probing its `Near-Markovian limit', namely a limit of the control parameters where QMM contributions to the system's time-evolution and observables  can be perturbatively computed as some sufficiently-small modifications to a Markovian version of the system. Clearly, the exact identification of this limit and the formulation of the perturbation scheme with mathematical and physical rigour can vary depending on the specifics of the system. We now formulate and probe the Near-Markovian limit of the QMM-UE (\ref{mue},\ref{ues},\ref{MMHparts}), for a general closed system whose entire evolution (as of $t_*$) is generated by its interacting quantum memories, uncombined with any conventional Hamiltonian. This is surely not the general case, as the system's total Hamiltonian can be in its hybrid form (\ref{H-CQS}). These hybrid settings will be also briefly analyzed in  the present paper, but the main idea of our analyses in this work is that one naturally begins with exploring and understanding the unprecedented aspects of the most extreme case of Unitary Non-Markovian physics, that is, closed quantum systems whose unitary evolutions are completely made of their quantum memories. \\\\ 
Fortunately the definition of the Near-Markovian limit of QMM-UE is straightforward. This is because, the Hamiltonian $H^{(N|L)}_{{\rm QMM}}(t)$ is by construction a direct composite of the system's quantum memories which define its state-history $\mathcal{P}_{[t_0 , t]}$ (\ref{hps}). Being so, all the QMDs 
$\Delta_{i = 1, \cdots, N}^{(t)} \equiv t - t_i$ (\ref{ds}), by which the building-block density operators are singled out, form a set of variables to which the QMM-H (\ref{MMHparts}) explicitly depend as a function. Given this feature, the above-mentioned Near-Markovian limit of $H^{(N|L)}_{{\rm QMM}}(t)$ corresponds to taking the largest QMD to be a 
sufficiently small parameter, $ \Delta^{(t)}_1 \ll 1$, in terms of which all the observables of the system can develop perturbative expansions. With this definition, recollecting that the full content of the equations (\ref{mue},\ref{ues},\ref{MMHparts}) is contained in the QMM Schr\"odinger equation (\ref{pmmscheqwmctc}) or in its equivalent form 
(\ref{evo-HMMpure}), the central question we are addressing in this subsection is the following:\\\\ 
In a perturbative scheme based on the sufficient smallness of the largest QMD $ \Delta^{(t)}_1 \equiv a$, $a \ll 1$, what perturbative solutions $\rho_{t}^{({\rm pert})}$ to the QMM dynamical equations (\ref{evo-HMMpure}) are developed by a closed quantum system which unitarily evolves under $H^{(N|L)}_{{\rm QMM}}(t)$? Are the solutions to (\ref{evo-HMMpure}) analytic functions of $a$, specially at the Markovian point $a = 0$, and of all the other $\Delta^{(t)}_i$s likewise? Moreover, one wants to know from a more general mathematical perspective, what functions of the QMDs the typical solutions to the equations (\ref{evo-HMMpure}) or equivalently to QMM Schr\"odinger equations (\ref{pmmscheqwmctc}) are.\\\\
The solutions to the $H^{(N|L)}_{{\rm QMM}}$ dynamical equation (\ref{evo-HMMpure}) can be generally thought as functions $\rho(t,a) \equiv \rho_t (a)$, with the QMD $a$ playing the role of an independent variable. We now look for the near-Markovian dynamical solutions $\rho^{({\rm pert})}(t,a) \equiv \rho_{t}^{({\rm pert})}(a)$ to (\ref{evo-HMMpure}) which, as functions of a sufficiently small variable $a \ll 1$, are analytic at the Markovian point $a = 0$. In other words, we look for the solutions $\rho^{({\rm pert})}(t,a)$ which can be recast as a full Taylor expansion over the second variable $a$ at the point $a = 0$,
\begin{equation}\label{nmp} 
\hspace{1.999 cm} \rho^{({\rm pert})}(t,a) \; = \; \sum_{j=0}^{\infty} \;  a^j \rho^{(j)}(t) \; \equiv \; \sum_{j=0}^{\infty} \;  a^j \rho_t^{(j)} \;\;\;\;\;\;\;;\;\;\;\;\; % {\rm in\; which}\;:\rho_t^{(j)} \; \equiv \; \frac{\partial^j}{\partial a^j} \; \rho(t,a)_{|_{a = 0}} \;\;\;{\rm s.t.}\;\;\;\;  
\; \rho_t^{(j)}: a-\text{independent}. \; 
\end{equation}
Now, to capture the near-Markovian solutions to (\ref{evo-HMMpure}) in a way which is consistent order by order in the perturbation parameter $a$, we must also implement the QMM `temporal' Taylor expansions, as follows. For every past-time quantum memory that comes into the making of $H^{(N|L)}_{{\rm QMM}}(t)$, namely for every state corresponding to $\Delta^{(t)}_i > 0$,  the right hand side of the dynamical equation (\ref{evo-HMMpure}) contains one time-shifted density operator, $\rho_{t - \Delta^{(t)}_i}$. Given the perturbation bound $a \ll 1$ and the analyticity assumption, every such entry $\rho^{({\rm pert})}(t-\Delta^{(t)}_i,a)$ additionally admits a temporal Taylor expansion corresponding to its tiny time shift, $- \Delta^{(t)}_i$, away from the present moment $t$,
\begin{equation}\begin{split} &
 \rho^{({\rm pert})}(t-\Delta^{(t)}_i,a) = \sum_{m=0}^{\infty} \frac{(-1)^m}{m!}\; [\Delta^{(t)}_i]^m\; \frac{d^m}{dt^m} \rho^{({\rm pert})}(t,a) \; = \; \\ & \hspace{3 cm} \; = \;
\sum_{m = 0}^{\infty} \; \sum_{n = 0}^{\infty}\;\frac{(-1)^m}{m!}\; a^n\; [\Delta^{(t)}_i]^{m}\; \rho_t^{(n|m)} \;\;\;\;\;\;;\;\;\;  \rho_t^{(n|m)} \; \equiv \; \frac{d^m}{dt^m}\; \rho_t^{(n)} . \label{pertb+a}
\end{split}\end{equation}
Merging (\ref{evo-HMMpure},\ref{nmp},\ref{pertb+a}) and organizing the result order by order in the perturbation parameter $a$, we are led to a hierarchically-coupled system of differential equations for the unknown functions $\rho_t^{(j)}$. The outcomes are the time-dependent operator-coefficients of the Taylor expansion (\ref{nmp}) which yield the Near-Markovian analytic solutions to (\ref{evo-HMMpure}). We now work out the details and the result of this procedure for one specific case which is nevertheless sufficient to clearly draw the general result which holds for all the $(N,L)$ QMM-Hs defined in \eqref{MMHparts}. \\\\
One can see that it is sufficient to address the most minimal case: $(N,L) = (1,1)$ specified in (\ref{NL}), because similar patterns develop, leading to the same end result, in the cases with larger $N$ or $L$, but only more involved algebraically. For this take, the QMM-H and the QMM dynamical equation (\ref{evo-HMMpure}) resulting from it read as follows,
\begin{equation}
H^{(1|1)}_{{\rm QMM}}(t) = \mu \; \rho_{t-a} = \mu \; \rho(t-a,a) \;\;\;;\;\;\; i\; \frac{\partial}{\partial t}\rho(t,a) \; = \; \mu \; [\;\rho(t-a,a) \;,\; \rho(t,a) \;] . \label{EQ-W-SCHRO-11}
\end{equation}
The near-Markovian solutions of \eqref{EQ-W-SCHRO-11} are of the form (\ref{nmp}), in which we take $a$ to be an arbitrary constant, for mere simplicity. For $N=1$, there can be only one past-time quantum memory, and therefore one time-shifted density operator entering in \eqref{EQ-W-SCHRO-11}. Being so, \eqref{pertb+a} is given as follows,
\begin{equation}\begin{split} &
\rho^{({\rm pert})}(t-a,a) \hspace{.000003 cm} \; = \;
\sum_{m = 0}^{\infty} \; \sum_{n = 0}^{\infty}\;\frac{(-1)^m}{m!}\; a^{n+m}\; \rho_t^{(n|m)} .  \label{pertba}
\end{split}\end{equation}
Applying (\ref{nmp},\ref{pertba}) to \eqref{EQ-W-SCHRO-11}, we obtain the following system of differential equations, with $j \; \geq \; 0$ denoting the perturbation order in terms of the QMD $a$,
\begin{equation}
i \; \dot{\rho}^{(j)}_t = i \; \rho_t^{(j|1)} = \mu \; \sum_{m,n = 0}^{m+n = j} \; \frac{(-1)^m}{m!} \;\; [\; \rho_t^{(n|m)} \;,\; \rho_t^{(j-n-m)}] . \label{eq:sch-pertb}
\end{equation}
Now we look for the functions $\rho^{(j)}_t$ which solve the system \eqref{eq:sch-pertb} order by order in the Near-Markovian perturbation. At the zeroth order, $j=0$, we find,
\begin{equation}
i \;\dot{\rho}_t^{(0)} \; = \; 0 .
\end{equation}
That is, $\rho^{(0)}_t$ is constant. At the first order of perturbation, we extract the differential equation for $\rho_t^{(1)}$ from \eqref{eq:sch-pertb}, and apply the zeroth-order result to it. Thus, we obtain,
\begin{equation}
i \;\dot{\rho}_t^{(1)} = \; \mu \; [\;  \rho^{(0)}_t , \dot{\rho}^{(0)}_t \;] \; = \; 0 .
\end{equation}
Therefore, $\rho^{(1)}_t$ is also found to be constant. Next, extracting from \eqref{eq:sch-pertb} the equation for the second-order entry $\rho^{(2)}_t$ given in (\ref{nmp}), and using the constancy of $\rho^{(0)}_t$ and $\rho^{(1)}_t$, yields,
\begin{equation}\begin{split}
i \;\dot{\rho}_t^{(2)} & = - \mu \; (\; [\; \dot{\rho}^{(0)}_t,\rho^{(1)}_t\;] - \dfrac{1}{2} \; [\; \ddot{\rho}^{(0)}_t, \rho^{(0)}_t\;] \; + [\; \dot{\rho}^{(1)}_t, \rho^{(0)}_t\;] \;) \; = \; 0 .
\end{split}\end{equation}
That is, $\rho^{(2)}_t$ is also a constant solution of time. Continuing this procedure to all the higher orders in the perturbation over the QMD variable $a$, we obtain, 
\begin{equation}\begin{split} & \dot{\rho}_t^{(m \; \geq \; 0)}  = 0 \;\;\; \Longrightarrow \;\;\;  \rho^{({\rm pert})}(t,a) \;  = \; \text{constant} .
\end{split}\end{equation}
Therefore, one concludes as follows. The only solutions to the $(1,1)$-QMM dynamical equations \eqref{EQ-W-SCHRO-11} which are analytic at the Markovian point $a=0$ are constant-in-time density operators. In other words, the Near-Markovian limit of the equations \eqref{EQ-W-SCHRO-11} does not admit any time-dependent perturbative solutions.\\\\
It is easy to see that by following the aforementioned procedure, detailed for the $(1,1)$ QMM-UE, in the case of larger values of $N$ and $L$, we are lead to the same conclusion. For each case, the computation based on (\ref{evo-HMMpure},\ref{nmp},\ref{pertb+a}) is straightforward and similar, but just becomes increasingly more involved, with the final result remaining the same. Robust Non-Markovianity is stable under interactions between quantum memories.\\\\
\emph{The purely-QMM-UEs are `Robustly non-Markovian'}. They are robustly non-Markovian in the following senses: i) Their Near-Markovian limits, namely when the largest QMD becomes sufficiently small, are perturbatively trivial, developing only constant-in-time states as perturbative solutions to their QMM Schr\"odinger equations. ii) Given any $H^{(N|L)}_{{\rm QMM}}(t)$, the analytic and nontrivial quantum-state-histories are developed only when the largest QMD, $\Delta_1^{(t)} \equiv a$, surpasses (or is not less than) some finite threshold $a^\star$ which depends on the couplings. That is, collectively denoting the couplings $\lambda_{i_1 \cdots i_{r^{(\leq N)}}}$ introduced in \eqref{MMHparts} by $\vec{\lambda}$, there is the following `Robust Non-Markovianity Condition' to be satisfied for developing state-histories which are dynamically non-trivial and solve the QMM Schr\"odinger equation (\ref{pmmscheqwmctc}),
\begin{equation}\label{threshholds} a \; \geq \; a^* = a^* (\vec{\lambda}) . \end{equation} 
Three highlights are in order. First, typical state-histories solving the QMM Schr\"odinger equations of $H^{(N|L)}_{{\rm QMM}}(t)$s are highly dynamical and develop giant landscapes which are extremely rich qualitatively. The many solutions of the QMM Schr\"odinger equations \eqref{pmmscheqwmctc} which we shall obtain analytically and numerically in the next sections establish the fact. Moreover, for each one of those solutions, we will confirm that the Robust Non-Markovianity Condition \eqref{threshholds} is validated.\\\\ 
Second, the threshold $a^\star$ of the purely-QMM-UE of is generally a complicated function of the $\lambda_{i_1 \cdots i_{r}}$ couplings. In some elementary examples, we compute $a^\star$, or lower bounds to it, analytically. For example, for a specially interesting class of  unitary state-histories developed by $H^{(2|2)}_{{\rm QMM}}(t)$ in \eqref{NL}, the right hand side of \eqref{threshholds} is realized as,
\begin{equation}
	a^\star\:\ge\;\frac{1}{|\lambda|} .
\end{equation}
Third, hybrid QMM-UEs are threshold-free, as we will show later. For them, the conventional parts of the hybrid QMM-Hs add source terms to \eqref{evo-HMMpure}, such that $\rho^{(m)}_t$s in \eqref{nmp} become non-constant, yielding dynamical solutions for arbitrarily small $a$.
%%%%%%%%%%%%%%%%%%%%%%%%%%%%%%%%%%%%%%
%%%%%   3 . One-Qubit
%%%%%%%%%%%%%%%%%%%%%%%%%%%%%%%%%%%%%%%
\section{Purely-QMM Unitary Evolutions of The One-Qubit Closed System}\label{888}
\subsection{State-History Representations And Quantum Memory Characters}
Having formulated a very general theory of QMM-UEs, we now focus on chracterizing, solving and analyzing, the unitarily-evolving wavefunctions of the closed system of a single qubit, to be called the `one-qubit closed system', as generated by its quantum memories, in the concrete forms specified  in (\ref{NL}). Physically and mathematically, the one-qubit closed system is the most fundamental setting to be explored in the novel context of QMM-UE, for three (two very basic and one rather surprising) reasons. First, its two dimensional Hilbert space is the smallest state-space capable of hosting genuinely-quantum features. As a result, given a choice of $H^{(N,L)}_{{\rm QMM}}$, the corresponding QMM Schr\"odinger equations \eqref{pmmscheqwmctc} which are very complex and unprecedented (highly nonlocal-in-time, highly-nonlinear) differential equations find their simplest possible mathematical profiles in this very setting. Despite significant simplifications, we will see that handling, analyzing and solving the one-qubit QMM Schr\"odinger equations is still a very demanding task, not only analytically, but also numerically. Second, one finds many physical situations in which the actual larger (even infinite dimensional) quantum system can be effectively or temporarily approximated by the 
one-qubit closed system, given that the quantum degrees of freedom which are dynamically active, or are relevant to the observables we are interested in, become effectively that of a qubit.
Third, we will find out that the unitary-evolving one-qubit wavefunctions which solve the QMM Schr\"odinger equations develop phase diagrams which are remarkably rich, in the two cases of 
purely-QMM-UE and hybrid-QMM-UE. In fact, as we will observe in Section \ref{SV}, these phase diagrams host universality classes of closed-system quantum states which (to the best of our knowledge) are qualitatively unprecedented, moreover are likely useful for various intriguing applications. Being so,  the mathematical forms and physical aspects of the purely-QMM Schr\"odinger equations \eqref{pmmscheqwmctc} of the one-qubit closed system unitarily evolving under \eqref{NL} are worked out in Section \ref{888} in all details. The solutions and phase diagrams will be obtained in Sections \ref{iml} and \ref{SV}.\\\\
The one-qubit Hilbert space, spanned by the states $\{| - \rangle , | +  \rangle\}$, is isomorphic to $\mathbb{CP}^1$. Qubit pure states, parametrised by the Bloch angles $\theta_t \in [0,\pi]$ and $\phi_t \in [0,2\pi)$ are,
\begin{equation}\label{density-B}\begin{split} &
\rho_t  =  \ket{\Psi_t} \bra{\Psi_t} = \sin^2 \frac{\theta_t}{2} \; A_-  +  \frac{1}{2}
\sin \theta_t e^{-i\phi_t} \; C_{-} + \frac{1}{2} \sin \theta_t e^{i\phi_t}  \; C_{+}  +  \cos^2 \frac{\theta_t}{2}  \; A_ + ,\\
 &  C_{\pm} \; \equiv \; |\mp \rangle \langle \pm | \;\;\;;\;\;\; A_{\pm} \; \equiv \; |\pm\rangle \langle \mp | ,
 \\ & | \Psi_t \rangle =   \sin\dfrac{\theta_t}{2} \; e^{i\phi_t} \; | - \rangle \; + \; \cos \dfrac{\theta_t}{2} \; |+\rangle . \end{split}\end{equation}
Representing by matrices, $(\ket{-} , \ket{+}) \equiv ( \begin{pmatrix} 0  \\ 1 \\ \end{pmatrix} , \begin{pmatrix} 1  \\ 0 \\ \end{pmatrix} ) $, the one-qubit pure states (\ref{density-B}) are rewritten as follows,
\begin{equation}\label{ppp} \ket{\Psi_t} \; = \;
\begin{pmatrix}
\cos \frac{\theta_t}{2}         \\
\sin \frac{\theta_t}{2} \; e^{i \phi_t}  \\
\end{pmatrix}
\;\;\;;\;\;\; \rho_t\; = \; \frac{1}{2} \;
\begin{pmatrix}
1 + \cos \theta_t \; &\; \sin \theta_t \; e^{- i \phi_t}  \\
\sin \theta_t \; e^{ i \phi_t} \; & \; 1 - \cos \theta_t \\ \end{pmatrix} . \end{equation}
In spin representation where one employs the three $SU(2)$ Pauli matrices $\vec{\sigma} = 2 \vec{S}$, the basis states $\ket{\pm}$ are mapped to the eigenstates of $\sigma^3$ with eigenvalues $\pm 1$, respectively. Moreover, we have, 
\begin{equation} \sigma^1 \ket{\mp} \; = \; \ket{\pm} \;\;\;;\;\;\; A_{\pm} \; = \; \frac{\mathbb{1} \pm \sigma^3}{2} \; = \; \frac{\mathbb{1}}{2} \pm S^3 \;\;\; ; \;\;\; C_{\mp} \; = \; \frac{ \sigma^1 \; \pm \; i \sigma^2 }{2} \; \equiv \; \frac{\sigma^{\pm}}{2} \; = \; S^{\pm} . \end{equation} 
Now, upon introducing a time-dependent Bloch vector of norm one, $\vec{r}_t$, the one-qubit pure states (\ref{ppp}) are represented in terms of the Pauli matrices $\sigma^\mu \equiv (\mathbb{1} , \vec{\sigma})$ as follows, 
\begin{equation}\label{spinlyro}\begin{split} & \rho_t \; \equiv \; \frac{1}{2} \; r^\mu_t \; \sigma^\mu \; = \; \frac{1}{2} \; ( \; \mathbb{1} \; + \; 
\vec{r}_t\; .\; \vec{\sigma}  \;), % \; = \; \frac{\mathbb{1}}{2} \; + x_t \; S^1 \; + \; y_t \; S^2 + \; z_t \; S^3 
\\  & \text{Pure-State Condition}, \; \forall\; t \;:\;\;\; \; |\vec{r}_t| \; = \; 1 \;\;\;\text{s.t.}\;  \\ & \vec{r} \; \equiv \; ( x_t \; , \; y_t \; , \; z_t ) \; = \; \; ( \sin \theta_t \; \cos \phi_t \; ,\; \sin \theta_t \; \sin \phi_t \;,\; \cos \theta_t) . \\   \end{split} \end{equation}
Alternatively, the one-qubit closed system can be reformulated fermionically, a la Schwinger. One begins with introducing two constrained fermion fields, say each one corresponding to either of the coloured flavours: red and green, $(f_r, f_r^\dagger)$ and $(f_g, f_g^\dagger)$.
The one-qubit Hilbert subspace is identified with the sub-Hilbert space of this pair of fermions for which the following constraint of total flavour-number conservation holds: $ n_r + n_g = 1$. In effect, one has the following correspondence,
\begin{equation}\label{mapping}
\sigma_z \; \longleftrightarrow \; n_g \;\;\; ; \;\;\; \sigma^{-} \;  \longleftrightarrow \; f_r^\dagger f_g \;\;\; ; \;\;\;  \sigma^{+} \; \longleftrightarrow \; f_g^\dagger f_r  .
\end{equation}
Applying this map to \eqref{spinlyro}, and introducing a pair of related coordinates $x_t^{\pm} \equiv x_t \pm iy_t$, one finds the following Schwinger fermionic representation of the one-qubit pure states,
\begin{equation}
\rho_t \; = \; \dfrac{1-z_t}{2} \; \mathbb{1} \; + \; z_t \; n_g \; + \; \frac{x^+_t}{2} \; f^\dagger_r f_g \; + \frac{x^-_t}{2} \; f^\dagger_g f_r . \label{den-fermion2}    
\end{equation}
Now, representations of the state-histories of the one-qubit closed system are clear. Given the unitarity of the evolution, every continuous-in-time succession of the above representations of the one-qubit pure quantum states, \eqref{spinlyro} 
or 
\eqref{den-fermion2}, collected from the initial time $t_0$ to the present moment $t$, must be identified with a representation of one (in-principle) possible pure-state-history $\mathcal{P}_{[t_0 , t]}$ of the one-qubit closed system (\ref{hps}). That is, given continuous functions $ (f^\theta(t) , f^\phi(t) ) \equiv (f^\theta_t  ,  f^\phi_t ) = (\theta_t , \phi_t)$, 
the map is,
\begin{equation}\label{ashinsr}  \;   \{ \;  (\; \sin(f^\theta_t)  \cos(f^\phi_t)  \;,\;  \sin(f^\theta_t)  \sin(f^\phi_t) \;,\;  \cos(f^\theta_t) \;) \; \equiv\;   \vec{r}_{t}  \;;\; \forall \; t \in [t_0 , t] \; \} \; \longleftrightarrow \; \mathcal{P}_{[t_0 , t]} .
\end{equation}
Now, we compute the QM assemblers, that is, QM-TPF \eqref{ctmcs} and QM-TPO \eqref{ctmtpos}, for any pair of moments $(t',t'')$ in the pure-state-history of the one-qubit closed system. It is natural to begin with computing the fundamental QM observable, that is, the (square of the) norm of the QM-TPF as defined in \eqref{nmtpf},  
\begin{equation}\label{nmtpfoqu}
	w_{t' t''}^2 \;= \; |m_{t' t''}|^2 \;=\; |\braket{\Psi_{t'}|\Psi_{t''}}|^2 \;= \; \frac{1}{2} \;
	\text{Tr} [\;\{ \rho_{t'} \; , \; \rho_{t''} \} \; ] .
\end{equation} 
Using the representation \eqref{spinlyro} and the algebra of Pauli matrices, we obtain,
\begin{equation}\label{acrhos} \{ \rho_{t'} \; , \; \rho_{t''} \} \; = \; \frac{1}{2} \; [\; (1 + \vec{r}_{t'} . \vec{r}_{t''}) \; \mathbb{1} + (\; \vec{r}_{t'} + \vec{r}_{t''} \;) . \; \vec{\sigma} \;] . \end{equation}
Therefore, from \eqref{nmtpfoqu} and \eqref{acrhos}, one finds as follows the fundamental QM observable of the one-qubit closed system,
\begin{equation}
\label{nmtpfoquis}\begin{split}
& w_{t' t''}^2 \; = \; \frac{1}{4} \; |  \vec{r}_{t'} + \vec{r}_{t''} |^2 \; =\; \frac{1}{2} \;
	(1 \;+\; \vec{r}_{t'} \;.\; \vec{r}_{t''}) \; = \; \\ & \;\;\; \hspace{.592cm} =\; \frac{1}{2}\; \big( \cos \theta_{t'} \cos \theta_{t''} \;+\;  \sin \theta_{t'} \sin \theta_{t''} \cos (\phi_{t'}-\phi_{t''}) \;+\; 1 	\big) .
\end{split}\end{equation} 
Indeed, the norm of the one-qubit QM-TPF reaches its maximum value, $w_{t' t''} = 1$, when $\vec{r}_{t''} = \vec{r}_{t'}$, for which $\rho_{t''} = \rho_{t'}$, and its minimum value, $w_{t' t''} = 0$, when $\vec{r}_{t''} = - \vec{r}_{t'}$, corresponding to $\rho_{t''} = 
1 -\rho_{t'}$. Now, the one-qubit QM-TPF $m_{t' t''}$ differs from its norm $w_{t' t''}$ by a phase factor which is removable by unobservable phase transformations on the QM wavefunctions. Therefore, the dynamical equations (\ref{evo-HMMpure}) and (\ref{ase}) do not depend on the to-be-determined phase of $m_{t_i t_j}$, because density operators are by construction insensitive to the phase redundancies of the wavefunctions. But, the total QM-TPFs, including their phases, do enter into the QMM Schr\"odinger equations (\ref{pmmscheqwmctc}) which are written in terms of the state-history wavefunctions. Using definition \eqref{ctmcs} and the Bloch parametrization of the wavefunctions \eqref{density-B}, we obtain the one-qubit QM-TPF,
\begin{equation}\label{qmtpfica} 
m_{t' t''} = \cos \frac{\theta_{t'}}{2} \; \cos \frac{\theta_{t''}}{2} \; +\; \sin \frac{\theta_{t'}}{2} \; \sin \frac{\theta_{t''}}{2} \; e^{i (\phi_{t''} - \phi_{t'})}  . \end{equation}
Finally, we must compute the QM-TPO $M_{t' t''}$ which can be done in different ways. 
Using representation \eqref{spinlyro} and also the algebra of Pauli matrices, one obtains the following expression for the product of any two one-qubit density operators,
\begin{equation}\label{prdoftdms} \rho_{t'} \rho_{t''} \;=\; \frac{1}{4}\;[\; (1 + \vec{r}_{t'} . \vec{r}_{t''}) \; \mathbb{1} + (\; \vec{r}_{t'} + \vec{r}_{t''} + \; i \; \vec{r}_{t'} \times \vec{r}_{t''}  \;) .\; \vec{\sigma} \;] . \end{equation}
Now, using (\ref{ctmcs},\ref{ctmtpos}), the one-qubit QM-TPO is obtained as follows,
\begin{equation}\label{Mmrho}   M_{t' t''}\; = \; \frac{1}{m_{t' t''}} \;  \rho_{t'}  \rho_{t''} = \begin{pmatrix}
\cos \frac{\theta_{t'}}{2}\; \cos \frac{\theta_{t''}}{2} \; &\; \cos \frac{\theta_{t'}}{2}\; \sin \frac{\theta_{t''}}{2} \; e^{- i \phi_{t''}}  \\
\sin \frac{\theta_{t'}}{2} \; \cos \frac{\theta_{t''}}{2} \; e^{ i \phi_{t'}} \; & \; \sin \frac{\theta_{t'}}{2} \;\sin \frac{\theta_{t''}}{2} \; e^{i (\phi_{t'} - \phi_{t''})} \\ \end{pmatrix} . \end{equation}
In conclusion, for any pair of moments $(t_i,t_j)$ in the pure-state-history of the one-qubit closed system, its QM-TPO $M_{t_i t_j}$ is determined by putting together the results (\ref{qmtpfica}), (\ref{prdoftdms}) and (\ref{Mmrho}).
%%%%%%%%%%%%%%%%%%%%%%%%%%%%%%%%%%%%%%
%%%%%%%%%%%%%%%%%%%%%%%%%%%%%%%%%%%%%%%
\subsection{(1,1) QMM Unitary Evolution:
	The Simplest QMM-H Made Of One Quantum Memory }
	We naturally begin with the most minimal QMM-UE ever possible, the choice: $N=1$. Upon this choice, the purity of the state-history formulated in \eqref{hps} implies that $L=1$ is the longest length that a monomial string in the QMM-H can take. That is, $\mu \; \rho_{t - \Delta^{(t)}_1} $, with $\mu$ being an arbitrary real valued coupling, can be the only monomial which the Hamiltonian can contain. Furthermore, setting the QMD to be an arbitrary constant parameter denoted by $a$, we conclude the QMM-H specified in \eqref{NL}: $H^{(1,1)}_{\text{QMM}}  = \mu \; \rho_{t-a}$. Finally, picking the `spin' representation of the one-qubit density operators given in \eqref{spinlyro}, and upon removing the dynamically-trivial term $\propto \mathbb{1}$, the QMM-H becomes,
\begin{equation}\label{11h}
\bar{H}^{(1,1)}_{\text{QMM}}(t) \; = \; % \text{constant} \; \mathbb{1} 
 \frac{1}{2}\; \mu \; \vec{r}_{t-a}  \cdot \vec{\sigma} .
\end{equation}
In useful analogy with conventional one-qubit Hamiltonians, it is natural to define a formal magnetic field in terms of which the QMM-H \eqref{11h} can be re-expressed as the Hamiltonian of a magnetic field. The magnetic field, however, is purely made from the quantum memory of the one-qubit closed system. We call this nonlocal-in-time QMM field, the `QMM Magnetic Field' (QMM-MF). The QMM-MF written in terms of the QM Bloch vector, and the  
re-expressed QMM-H \eqref{11h} are,
\begin{equation}\label{QMMagnet-11}
\vec{\mathcal{B}}^{(1,1)}_{\text{QMM}}(t) \;\equiv\; \mu \; \vec{r}_{t-a} \;\;\;;\;\;\; \bar{H}^{(1,1)} \;=\; \vec{\mathcal{B}}^{(1,1)}(t) \cdot \vec{S} .
\end{equation}
Having defined \eqref{QMMagnet-11}, the QMM-H \eqref{11h} and the dynamical equations \eqref{evo-HMMpure}, or \eqref{ase} for the one-qubit closed system unitarily evolving under \eqref{11h}, take the following forms,
\begin{equation}\begin{split} &
\dot{\vec{r}}_t \; = \; \vec{\mathcal{B}}_{\text{QMM}}^{(1,1)}(t) \times \vec{r_t} \; = \;  \mu \; \vec{r}_{t-a} \times \vec{r}_{t} . \label{MM-evo-eqn0} 
\end{split}\end{equation}
Equivalently, the QMM Schr\"odinger equation \eqref{pmmscheqwmctc} for the QMM-H \eqref{11h} is given by,
\begin{equation}\label{11seqmm}
i \ket{\dot{\Psi}_t} \;=\; 2 \; \mu \; m_{t-a \; t} \ket{\Psi_{t-a}} .
\end{equation} 
Now, we present the system of \emph{delay-differential equations} or difference-differential equations \cite{ArinoHD,Roussel2019,Erneux2009,MBaniYghoub Bani-Yahoub M} for the evolution of Bloch angles $(\theta,\phi)$ of a qubit unitarily evolving under $H^{(1,1)}_{\text{QMM}}(t)$. 
It can be obtained by extracting it from the vector equations 
\eqref{MM-evo-eqn0}, or by directly applying the QM-TPF $m_{t-a\;t}$ given in \eqref{qmtpfica} to \eqref{11seqmm}. The result is,
\begin{equation}
\dot{\theta}_t = \mu \; \mathcal{G}_{t-a, t} \;\;\;;\;\;\; 
\sin\theta_t \; \dot{\phi}_t = \mu \; \mathcal{F}_{t-a, t} . \label{MM-evo-eq0}
\end{equation}
in which,
\begin{align}
\label{fgp}
\mathcal{F}_{t-a, t} &= \cos\theta_{t-a} \sin\theta_t - \sin\theta_{t-a} \cos\theta_t 
\cos(\phi_t - \phi_{t-a}) , \nonumber \\
\mathcal{G}_{t-a, t} &= - \sin\theta_{t-a} \sin(\phi_t-\phi_{t-a}) .
\end{align}
The one-qubit unitary state-histories which solve (\ref{MM-evo-eq0},\ref{fgp}) will be investigated in the next sections. Moreover, we will present their phase diagram in Subsection \ref{SV-I}.
%%%%%%%%%%%%%%%%%%%%%%%%%%%%%%%%%%%%%%
%%%%%%%%%%%%%%%%%%%%%%%%%%%%%%%%%%%%%%%
\subsection{(2,2) QMM Unitary Evolution: QMM-H Made By Two System's States With Monomial Interactions Up To Second Degree}\label{mcw}
We now work out in full details the Purely-QMM-UE of the one-qubit closed system made by two interlinked quantum states, one of them being the present moment state, $\rho_t$, and the other the state a distance $a$ back in time, $\rho_{t-a}$, forming strings of length one (linear in QMs) and two (quadratic in QMs) in the Hamiltonian. That is, we formulate a unitary two-state system whose evolution is re-fabricated moment by moment by nothing but a pair of its quantum memories, corresponding to $(\Delta^{(t)}_2 , \Delta^{(t)}_1 ) = (0 , a)$, building up by their monomial interactions the $(N,L) = (2,2)$ version of the QMM-H \eqref{MMHparts}. We spelled out in \eqref{NL} this QMM-H for any closed system. Here we move on to presenting the mathematical and physical structures of the QMM-H, and its dynamical equations, once applied to the one-qubit closed system. In doing so, however, we shall add a dynamically effectless term linear in $\rho_t$ to the 
QMM-H \eqref{NL} which gives more symmetric forms to some of the quantities and equations to come. We highlight that this Hamiltonian term which commutes with $\rho_t$ will make no observable effect on the final dynamical equations apart from unphysical deformations which only have to do with unobservable phases of the wavefunctions $\ket{\Psi_t}$ given in \eqref{density-B}. Therefore, denoting the real and imaginary parts of the $\lambda$ coupling in \eqref{NL} by $(\lambda^R,\lambda^I)$ and upon renaming the past-QM $\mu$ coupling as $\mu_{t-a}$, we now consider an isolated qubit whose Hamiltonian is made in the following form by two of its quantum memories,
\begin{equation}
H^{(2,2)}_{\text{QMM}}(t) \; =\; \mu_{t-a}\; \rho_{t-a} \;+\; \mu_t \; \rho_t \; +\; \lambda^R\; \{\; \rho_{t-a}, \rho_t \;\} \;+ \; i \lambda^I \; [\;\rho_{t-a} \;,\; \rho_t \; ] . \label{QMMH22}
\end{equation}
Applying the representation \eqref{spinlyro} of the one-qubit pure density operator, using the elementary properties of the Pauli matrices, and introducing the deformed $\mu$ couplings: $( \hat{\mu}_{t-a} , \hat{\mu}_t )  \equiv (\mu_{t-a} + \lambda^R \;,\; \mu_t + \lambda^R) $, the QMM-H \eqref{QMMH22} takes the following form,
\begin{equation}\begin{split}
& H^{(2,2)}_{\text{QMM}}(t)  \;=\; \\ & =  \frac{1}{2} \;[\; ( \mu_{t-a} +  \mu_t   +   \lambda^R  +  \lambda^I \; \vec{r}_{t-a} \; \cdot \; \vec{r}_t \;) \; \mathbb{1} + \; ( \hat{\mu}_{t-a}\; \vec{r}_{t-a} +  \hat{\mu}_t \; \vec{r}_t - \lambda^I \; \vec{r}_{t-a} \; \times \; \vec{r}_t \; ) \cdot  \vec{\sigma}\; ] .
\label{MMH-OQ-SPIN}
\end{split} \end{equation}
It is natural to introduce a QMM-MF $\vec{{\mathcal{B}}}^{(2,2)}_{\text{QMM}}(t)$ which in every moment $t$ is remade from the two interacting quantum memories of the qubit $(\rho_{t-a} , \rho_{t})$, $\vec{{\mathcal{B}}}^{(2,2)}_{\text{QMM}}(t)  = \vec{{\mathcal{B}}}^{(2,2)}_{\text{QMM}}(\vec{r}_{t-a},\vec{r}_t)$. Being so, dropping the term $\propto \mathbb{1}$ in \eqref{MMH-OQ-SPIN} and renaming its dynamically relevant part as $\bar{H}^{(2,2)}_{\text{QMM}}(t)$, the (2,2) QMM-H of the one-qubit closed system is re-expressed as follows, 
\begin{equation}
\vec{{\mathcal{B}}}^{(2,2)}_{\text{QMM}}(t) \; \equiv \; \hat{\mu}_{t-a} \; \vec{r}_{t-a}  \;+ \; \hat{\mu}_t \; \vec{r}_t \; - \; \lambda^I\; \vec{r}_a \times \vec{r}_t \;\;\;;\;\;\; \bar{H}^{(2,2)}_{\text{QMM}}(t) \; = \; \vec{{\mathcal{B}}}^{(2,2)}_{\text{QMM}}(t) \cdot \vec{S} . \label{MMMagnet}	
\end{equation}
Accordingly, the dynamical equations (\ref{evo-HMMpure}) or (\ref{ase}) which are resulted from the above 
QMM-H come to the following familiar Bloch-vectorial form,
\begin{equation}\begin{split} & \hspace{.165 cm}
\dot{\vec{r}}_t  \;= \; \vec{\mathcal{B}}^{(2,2)}_{\text{QMM}}(t) \times \vec{r_t} \;\;\; \Longrightarrow \;\;\; \dot{\vec{r}}_t  \;= \; \lambda^I \; \vec{r}_{t-a} \; - \; \lambda^I \; (\vec{r}_{t-a} \cdot \vec{r}_t) \; \vec{r}_t \; + \; \hat{\mu}_{t-a} \;\;  \vec{r}_{t-a} \times \vec{r}_t .  \label{MM-evo-eqn3} \end{split}
\end{equation}
One now highlights, as an alternative, the insightful formulation of the (2,2) QMM-UE of the one-qubit closed system by means of Schwinger fermions. The purely-fermionic representation of the QMM-H, $H^{(2,2)}_{\text{QMM}}(t)$, can be done in two ways: either we work it out independently by going back to \eqref{QMMH22}, applying the fermionic representation \eqref{den-fermion2} to it and doing the algebra, or we simply apply the mapping \eqref{mapping} to the spin representation \eqref{MMH-OQ-SPIN}. We shall write the final result only. First, introducing the following compact notations are useful,
\begin{equation}\label{notations}\begin{split} &
v \; \equiv \; 1-z \;\;\;;\;\;\; (a \wedge b)_{t-a \; t} \; \equiv \; a_{t-a} \; b_t \; - \; a_t \; b_{t-a} \;,\; \forall \; (a,b) .
\end{split} \end{equation}
Now, the $(2,2)$ QMM-H of the one-qubit closed system finds the following alternative representation as a QMM purely-fermionic system,
\begin{equation}\label{hubbard33}\begin{split} &
H^{(2,2)}_{\text{QMM}}(t) \; = \; C_1 (t) \; \mathbb{1} \; + \; \bar{H}^{(2,2)}_{\text{QMM}}(t) ,  \\ &
2 \; C_1 (t) \; = \; \mu_{t-a} \; v_{t-a} \; + \; \mu_t \; v_t \; + \; \lambda^I \; ( x^-_{t-a} \; x^+_t )^I \; + \; \lambda^R \; [\; (x^-_{t-a} \; x^+_t )^R \; + \; v_{t-a} \; v_t \;] , \\ &
\bar{H}^{(2,2)}_{\text{QMM}}(t) \; = \; \zeta_{\text{QMM}}(t) \; n_g \; + \; \tau_{\text{QMM}}^\star (t) \; f^\dagger_r f_g \; + \; \tau_{\text{QMM}} (t) \; f^\dagger_g f_r .   
\end{split} \end{equation}
As we see in \eqref{hubbard33}, the dynamically relevant part of the Hamiltonian, $\bar{H}^{(2,2)}_{\text{QMM}}(t)$, takes the following couplings: a nonlocal-in-time real coupling $\zeta_{\text{QMM}}(t)$ which is similar to a QMM chemical potential, and a nonlocal-in-time complex coupling $\tau_{\text{QMM}}^\star (t)$ which is similar to a QMM tunneling amplitude. In terms of the Bloch coordinates of the system's states at $t-a$ and $t$, these QMM couplings are derived as follows,
\begin{equation}\begin{split} \label{hubbardcouplingsare} & \;\;\;
\zeta_{\text{QMM}}(t) \; = \; \hat{\mu}_{t-a} \; z_{t-a} \; + \; \hat{\mu}_t \; z_t \; - \; \lambda^I \; ( x^-_{t-a} \; x^+_t )^I ,    \\ &
2 \; \tau_{\text{QMM}} (t) \; = \; \hat{\mu}_{t-a} \; x^-_{t-a} \; + \; \hat{\mu}_t \; x^-_t \; + \; i \; \lambda^I \; (z  \wedge  x^-)_{t-a \; t} , \\ &
2 \; \tau_{\text{QMM}}^\star (t) \; = \; \hat{\mu}_{t-a} \; x^+_{t-a} \; + \; \hat{\mu}_t \; x^+_t + i \; \lambda^I \; (z \wedge x^+)_{{t-a} \; t}  .
\end{split} \end{equation}
Given (\ref{hubbard33},\ref{hubbardcouplingsare}), the $(2,2)$-QMM one-qubit closed system physically takes the form of a (nonlocal-in-time version of) Hubbard model \cite{hmr} with two sites, each one site for one of the Schwinger fermions. The unconventionality of the above Hubbard model  comes about by the QMM structure of the couplings: the tunneling amplitude and chemical potential are remade, moment by moment, by the state-history history of the system.  We also obtain an `unconventional-Hubbard-model' representation of the $(1,1)$ one-qubit closed system by turning off the $\lambda$ coupling in (\ref{hubbardcouplingsare}). Likewise, the $n$-qubit closed system yield a nonlocal-in-time $2n$-site Hubbard model in which all tunneling amplitudes, density-density interactions and chemical potentials are QMM. Finally, we note that unconventional fermionic-or-spin systems similar to (\ref{hubbard33},\ref{hubbardcouplingsare}) or \eqref{MMMagnet}, but albeit with increasingly more complex QMM many-body interactions, represent the $(N,L)$-QUEs of all finite-dimensional closed quantum systems.\\\\ 
Now using either of the equivalent representations (\ref{hubbard33},\ref{hubbardcouplingsare}) or \eqref{MMMagnet}, one can obtain the differential equations for the time evolution of the one-qubit Bloch coordinates. For example, these equations can be extracted from the dynamical equation \eqref{MM-evo-eqn3}, by using the angular parametrization of the two QM Bloch vectors $(\vec{r}_{t-a},\vec{r}_t)$, given in \eqref{spinlyro}. Alternatively, they can be obtained by applying the QM-TPF data \eqref{qmtpfica} and \eqref{nmtpfoqu}, together with the Bloch-angle parametrizations \eqref{density-B}, to the (2,2) QMM Schr\"odinger equation given in (\ref{tqmmse},\ref{ltt}). The result is the following coupled system of nonlocal-in-time differential equations,
\begin{align}
\label{MM-evo-eqn2}
\dot{\theta}_t \; = \; \hat{\mu}_{t-a} \; \mathcal{G}_{t-a, t} \; - \; \lambda^I \; \mathcal{F}_{t-a, t} , \nonumber  \\
\sin\theta_t \; \dot{\phi}_t \; = \; \hat{\mu}_{t-a} \; \mathcal{F}_{t-a, t} \; + \lambda^I \; \mathcal{G}_{t-a, t} .  
\end{align}
The two QMM functions $(\mathcal{F}_{t-a, t},\mathcal{G}_{t-a, t})$ are the same as given in (\ref{fgp}). In next sections, some analytic solutions to the above equations are obtained and analyzed. Moreover, we numerically work out the total phase diagram of the unitarily-evolving wavefucntions which solve \eqref{MM-evo-eqn2} is Subsection \ref{SV-II}. 
%%%%%%%%%%%%%%%%%%%%%%%%%%%%%%%%%%%%%%
%%%%%%%%%%%%%%%%%%%%%%%%%%%%%%%%%%%%%%%
\subsection{(2,3) QMM Unitary Evolution: QMM-H Made By Two System's States With Monomial Interactions Up To Third Degree} 
	Now, we consider, and later analyze and solve, the QMM-UE with $N = 2$ and $L = 3$. The motivation is more than just analyzing a QMM-H with more complex structure: the aim is to understand the effect of turning on higher-order monomial interactions (in this case: third order), while keeping the number of quantum memories fixed (in this case: two). The corresponding Hamiltonian, $H^{(2,3)}_{\text{QMM}}(t)$, is specified in \eqref{NL}, with linear terms being ignored in it, because (as we have seen) the linear coupling $\mu_{t-a}$ can be absorbed in a deformation of the coupling of the second-degree monomial interaction: $\lambda^R  \to \hat{\mu}_{t-a} = \mu_{t-a} + \lambda^R$. To this aim, as motivated before, we restrict to the fundamental example: the one-qubit closed system under QMM-UE generated by $H^{(2,3)}_{\text{QMM}}(t)$ in \eqref{NL}. For the one-qubit closed system,  using (\ref{QMMH22},\ref{nmtpfoquis},\ref{spinlyro}), the $(2,3)$ QMM-H in \eqref{NL} is,
\begin{equation}\begin{split}\label{avn} & H^{(2,3)}_{\text{QMM}}(t) \; = \; H^{(2,2)}_{\text{QMM}}(t)_{|_{(\mu_{t-a},\mu_t) = (0,0)}} \; + \; H^{(2, 3 | \Delta L = 1 )}_{\text{QMM}}(t) , \\ &
 H^{(2,3 | \Delta L = 1 )}_{\text{QMM}}(t) \; = \; \eta \;  \rho_{t-a} \; \rho_t \; \rho_{t-a} \; = \; \eta \; w_{t-a\;t}^2 \; \rho_{t-a} \; = \\ & \;\;\;\;\; \hspace{1.85cm} = \; \frac{\eta}{2} \; (\; 1 + \vec{r}_{t-a} \cdot \vec{r}_t \;) \; \rho_{t-a} \;= \;  \frac{\eta}{4} \; (\; 1 + \vec{r}_{t-a} \cdot \vec{r}_t \;) \; (\mathbb{1} + \vec{r}_{t-a} \cdot \vec{\sigma}) .
  \end{split} \end{equation}
As in (1,1) and (2,2) cases, we now define a QMM-MF $\vec{{\mathcal{B}}}^{(2,3)}_{{\text QMM}}(t)$. Introducing $\bar{H}^{(2,3)}_{\text{QMM}}(t)$ by setting aside the term $\propto \mathbb{1}$ in \eqref{avn}, and remembering \eqref{MMMagnet}, we immediately find,
\begin{equation}\begin{split}  & \bar{H}^{(2,3)}_{\text{QMM}}(t) \; = \; \vec{{\mathcal{B}}}^{(2,3)}_{{\text QMM}}(t) \; \cdot \; \vec{S} ,\\ & 
\vec{{\mathcal{B}}}^{(2,3)}_{{\text QMM}}(t) \; = \;    \lambda^R \; \vec{r}_t \; + \; [\; \lambda^R \;+ \; \dfrac{\eta}{2} \; (\; 1 \; + \; \vec{r}_{t-a} \cdot \vec{r}_t \;) \;]\; \vec{r}_{t-a} 
- \; \lambda^I \; \vec{r}_{t-a} \times \vec{r}_t .	\label{QMMagnet-23}	
\end{split} \end{equation}
Being so, the resulted dynamical equations (\ref{evo-HMMpure}) or (\ref{ase}) are given as follows,
\begin{equation}\begin{split} &
\dot{\vec{r}}_t \; = \; \vec{{\mathcal{B}}}^{(2,3)}_{{\text QMM}}(t) \times \vec{r}_{t} \;\;\; \Rightarrow \; \\ &  \dot{\vec{r}}_t \; = \; \lambda^I \; \vec{r}_{t-a} \; - \; \lambda^I \; (\vec{r}_{t-a} \cdot \vec{r}_t) \; \vec{r}_{t} \; + \; [ \; \lambda^R + \dfrac{\eta}{2} \; ( 1 + \vec{r}_{t-a} \cdot \vec{r}_t) \;]\; \vec{r}_{t-a} \times \vec{r}_t . \label{SE23VECTOR}
\end{split} \end{equation}
Finally, either by applying (\ref{qmtpfica}, \ref{nmtpfoqu}; \ref{density-B}) to QMM Schr\"odinger equation (\ref{pmmscheqwmctc}) for $H^{(2,3)}_{\text{QMM}}(t)$, or by working out \eqref{SE23VECTOR} in Bloch polar coordinates, a coupled system of nonlocal-in-time differential equations describing the time evolution of the Bloch angles under one-qubit $(2,3)$ QMM-UE is obtained. Defining functions $(\mathcal{A}^{\theta}_{t-a, t} , \mathcal{A}^\phi_{t-a, t})$ as,
\begin{equation}
\begin{split}
\label{ate:ma} 
& \mathcal{A}^{\phi}_{t-a, t} \;=\; 
\frac{1}{4}  \bigg(-\cos (\phi_t-\phi_{t-a}) \Big(\sin (2 \theta_t) \sin ^2(\theta_{t-a}) \cos (\phi_t-\phi_{t-a})  \\ &
\;+\; 2 \cos (\theta_t) \sin (\theta_{t-a})+\cos (2 \theta_t) \sin (2 \theta_{t-a}) \Big) \\ &
\;+\; \sin (2 \theta_t) \cos ^2(\theta_{t-a})+2 \sin (\theta_t) \cos (\theta_{t-a}) \bigg) .
\end{split} 
\end{equation} 
\begin{equation}
\begin{split}
\label{ate:mb} 
& \mathcal{A}^{\theta}_{t-a, t} \;=\; -\frac{1}{2} \; \;  \sin (\theta_{t-a}) \sin (\phi_t-\phi_{t-a}) \; \times  \\ &
\;\times\; \Big( \sin (\theta_t) \sin (\theta_{t-a}) \cos (\phi_t-\phi_{t-a})+\cos (\theta_t) \cos (\theta_{t-a})+1 \Big) .
\end{split}
\end{equation}
and recalling $(\mathcal{F}_{t-a, t}, \mathcal{G}_{t-a, t})$ in \eqref{fgp}, the dynamical equations are presented as follows,
\begin{equation}
\begin{split} 
\label{SE23POLAR} &
\dot{\theta}_t \;=\; \eta \; \mathcal{A}^{\theta}_{t-a, t} \; + \; \lambda^R \; \mathcal{G}_{t-a, t} \; - \; \lambda_{t-a, t}^I \; \mathcal{F}_{t-a, t}, \\ &
\sin\theta_t \; \dot{\phi}_t \;=\; \eta \; \mathcal{A}^{\phi}_{t-a, t} \; + \; \lambda^R \; \mathcal{F}_{t-a, t} \; + \; \lambda_{t-a, t}^I \; \mathcal{G}_{t-a, t}  .
\end{split} 
\end{equation} 
The phase diagram of the unitarily-evolving wavefunctions which solve the equation \eqref{SE23POLAR} is explored in Section \ref{SV-III}.
%%%%%%%%%%%%%%%%%%%%%%%%%%%%%%%%%%%%%%
%%%%%%%%%%%%%%%%%%%%%%%%%%%%%%%%%%%%%%%
\subsection{(3,3) QMM Unitary Evolution: A Specific 
	QMM-H Made By Three System's States With Third-Degree Monomial Interactions}
Finally, we detail the QMM-UE of the one-qubit closed system with $(N, L) = (3, 3)$. With this choice of (N,L), at every present moment $t$, the system's Hamiltonian $H^{(3,3)}_{\text{QMM}}(t)$ must be made of three time-varying quantum memories which interact in the form of monomials of maximum degree three. Furthermore, we fix these interacting quantum memories by choosing the following spectrum of QMDs: $(\Delta_3^{(t)},\Delta_2^{(t)},\Delta_1^{(t)}) = (0,b,a)$, with $a$ and $b < a$ being arbitrary real numbers. That is, the three quantum memories are identified with the system's states at the present time $t$, at a past moment $t-b$, and at a far past moment $t-a$, respectively: $(\rho_{t-a},\rho_{t-b},\rho_{t})$.\\\\ 
Finally, it suffices to include only interaction monomials of degree three in defining the corresponding QMM-H, given the sufficient complexity and richness of this choice, as will be seen later in the resulted phase diagram of the one-qubit closed system. Therefore, given the assumed purity of the states, the most general $(3,3)$ QMM-H with this criteria is $H^{(3,3)}_{\text{QMM}}(t)$ specified in \eqref{NL}. As in all the previous cases, the one-qubit QMM-H is naturally expressed in terms of a QMM-MF which in every moment $t$ is remade by the identifying Bloch vectors of the three interacting states: $(\vec{r}_{t-a}, \vec{r}_{t-b}, \vec{r}_{t})$. That is, using the representation \eqref{spinlyro} and defining the dynamically relevant part of the QMM-H $\bar{H}^{(3,3)}_{\text{QMM}}(t)$ which differs from the original Hamiltonian given in \eqref{NL} only by a dynamically irrelevant term  $\propto \mathbb{1}$, one has,
\begin{equation}
\label{QMMagnet-33}
\begin{split} 
&H^{(3,3)}_{\text{QMM}}(t) \; = \; \kappa \; \rho_{t-a} \; \rho_{t-b} \; \rho_{t} \; + \; \kappa^\star \; \rho_{t} \; \rho_{t-b} \; \rho_{t-a} \;=\; C(\vec{r}_{t-a}, \vec{r}_{t-b}, \vec{r}_{t}) \; \mathbb{1} \; + \; \bar{H}^{(3,3)}_{\text{QMM}}(t), \\ 
&\bar{H}^{(3,3)}_{\text{QMM}}(t) \;=\; \vec{\mathcal{B}}^{(3,3)}_{\text{QMM}}(t) \cdot \vec{S} , \\ 
&\vec{\mathcal{B}}^{(3,3)}_{\text{QMM}}(t) \;=\; \dfrac{\kappa^R}{2} \; \big[ \; (\vec{r}_{t-a} + \vec{r}_{t-b} + \vec{r}_t)  +  (\vec{r}_{t-a} \cdot \vec{r}_{t-b}) \vec{r}_t - (\vec{r}_{t-a} \cdot \vec{r}_t) \vec{r}_{t-b} + (\vec{r}_{t-b} \cdot \vec{r}_t) \vec{r}_{t-a} \; \big] - \\ 
& \;\;\;\;\;\;\; \hspace{1 cm} - \dfrac{\kappa^{I}}{2} \; (\;\vec{r}_{t-a} \times \vec{r}_{t-b} \; + \; \vec{r}_{t-b} \times \vec{r}_t \; + \; \vec{r}_{t-a} \times \vec{r}_t \;) .	 
\end{split}
\end{equation}
Dynamical equations (\ref{evo-HMMpure}) or (\ref{ase}) which correspond to the $(N,L) = (3,3)$ QMM-UE of the one-qubit closed system under the above Hamiltonian \eqref{QMMagnet-33} are now given by,
\begin{equation}\begin{split} &
\dot{\vec{r}}_t \;=\; \frac{1}{2} \; \kappa^R \; \big[ \; (1 + \vec{r}_{t-b} \cdot \vec{r}_t) \; \vec{r}_{t-a} \times \vec{r}_t \; + \; (1 - \vec{r}_{t-a} \cdot \vec{r}_t) \; \vec{r}_{t-b} \times \vec{r}_t \; \big] + \\ & \;\;\;
\;+\; \frac{1}{2} \; \kappa^I \; \big[ \; (1 + \vec{r}_{t-b} \cdot \vec{r}_t) \; \vec{r}_{t-a} \; + \; (1 - \vec{r}_{t-a} \cdot \vec{r}_t) \; \vec{r}_{t-b} \; - \; (\vec{r}_{t-a} \cdot \vec{r}_t \; + \; \vec{r}_{t-b} \cdot \vec{r}_t) \; \vec{r}_t \; \big] . \label{SE33VECTOR}
\end{split} \end{equation}
Finally, we obtain a coupled system of nonlocal-in-time differential equations describing the dynamics of the one-qubit wavefunction Bloch angles under \eqref{QMMagnet-33}. As before, this can be done either by applying  (\ref{qmtpfica}, \ref{nmtpfoqu}) and (\ref{density-B}) to QMM Schr\"odinger equation (\ref{pmmscheqwmctc}) corresponding to the one-qubit Hamiltonian $H^{(3,3)}_{\text{QMM}}(t)$ given in \eqref{QMMagnet-33}, or by detailing the vectorial equation \eqref{SE23VECTOR} in Bloch polar coordinates. The result is,
\begin{align}
\dot{\theta}_t &= \kappa^R \; \tilde{\mathcal{G}}_{t-a,t-b, t} \; - \; \kappa^I \;   \tilde{\mathcal{F}}_{t-a,t-b, t} , \nonumber \\
\sin\theta_t \; \dot{\phi}_t &= \kappa^R \;  \tilde{\mathcal{F}}_{t-a,t-b, t} \; + \; \kappa^I \; \tilde{\mathcal{G}}_{t-a,t-b, t} .  \label{SE33POLAR}
\end{align}
The pair of nonlocal-in-time QMM functions $(\tilde{\mathcal{F}}_{t-a,t-b, t} \; , \; \tilde{\mathcal{G}}_{t-a,t-b, t})$ in equations \eqref{SE33POLAR} are specified as follows,
\begin{equation} \begin{split} &
\tilde{\mathcal{F}}_{t-a,t-b, t} \;\;=\; \frac{1}{2} \; \big[ -\sin (\theta_{t-a}) \cos (\phi_{t}-\phi_{t-a})
\cos (\theta_{t-b}) \;+\;  % \hspace{1.85 cm} \hspace{1cm}
 \cos (\theta_{t-a}) \sin (\theta_{t-b}) \cos (\phi_{t}-\phi_{t-b}) \;+ \\ & \hspace{3 cm}
\; + \; \sin (\theta_{t}) \cos (\theta_{t-a}) \; -\; \cos (\theta_{t}) \sin (\theta_{t-a}) \cos (\phi_{t}-\phi_{t-a}) \;+ \\ & \; \hspace{3 cm} + \; \sin (\theta_{t}) \cos (\theta_{t-b}) 
\;-\; \cos (\theta_{t}) \sin (\theta_{t-b}) \cos (\phi_{t}-\phi_{t-b}) 
 \; \big] ,  \\ &  \\ &
\tilde{\mathcal{G}}_{t-a,t-b, t} \;=\; \frac{1}{2} \; [\;-\cos (\theta_{t}) \sin (\theta_{t-a}) \sin (\phi_{t}-\phi_{t-a}) \cos (\theta_{t-b})  \\ & \;\;\; \hspace{1.85 cm} \;\;\;\;\;\;+\; \sin (\theta_{t-b}) \cos (\phi_{t-b}) \; (\sin (\theta_{t}) \sin (\theta_{t-a}) \sin (\phi_{t-a})-\sin(\phi_{t})) \;+ \\ & \;\;\; \hspace{1.85 cm} \;\;\;\;\;\;+\; \cos (\theta_{t}) \cos (\theta_{t-a}) \sin (\theta_{t-b})
\sin (\phi_{t}-\phi_{t-b}) \;+  \\ & \;\;\; \hspace{1.85 cm} \;\;\;\;\;\;+\; \sin (\theta_{t-b}) \sin (\phi_{t-b})
(\cos (\phi_{t})-\sin (\theta_{t}) \sin (\theta_{t-a}) \cos (\phi_{t-a})) \;+  \\ & \;\;\; \hspace{1.85 cm} \;\;\;\;\;\;+\; \sin (\theta_{t-a}) \sin (\phi_{t-a} - \phi_{t}) \; ] . \end{split} \label{fgof33}  
\end{equation}
The complex phase diagram of the QMM unitary state-histories which solve equations \eqref{SE33POLAR} is probed in Section \ref{SV-IV}. Indeed, we will demonstrate that, in comparison with its $N=2$ counterpart whose unprecedented phase diagram is remarkably rich, $N=3$ QMM-UE of the one-qubit closed system does develop several phases of QMM unitary state-histories which are distinctive and unprecedented qualitatively.
%%%%%%%%%%%%%%%%%%%%%%%%%%%%%%%%%%%%%%
%%%%%   4 .  Analytical
%%%%%%%%%%%%%%%%%%%%%%%%%%%%%%%%%%%%%%%
\section{QMM Unitary Evolutions of The One-Qubit Closed System:\\ Analytical Results}\label{iml}
\subsection{The General Scope of Analytical Investigations}
In this section, a number of distinctive unitarily-evolving wavefunctions which, exactly or perturbatively, solve the QMM Schr\"odinger equations of the one-qubit closed system are obtained analytically. Moreover, we work out analytically selected features of these solutions, because next to solving the equations mathematically, we must develop an intuitively-clear understanding of the qualitatively-novel physics of the QMM unitary state-histories which are realized by them. To develop these analyses, it suffices to focus all through this section on the $(2,2)$ QMM-UE of the 
one-qubit closed system which was extensively detailed in Subsection \ref{mcw}. The main reason for this sufficiency is twofold. First, QMM Schr\"odinger equations \eqref{MM-evo-eqn2} are already very challenging mathematically. Second, as we will find out, the solutions to these equations do realize unitary-evolving one-qubit states which cannot be developed by conventional Hamiltonian counterparts, already before considering the more-involved $(N,L)$ QMM Schr\"odinger equations.
%%%%%%%%%%%%%%%%%%%%%%%%%%%%%%%%%%%%%%%%%%%
%%%%%%%%%%%%%%%%%%%%%%%%%%%%%%%%%%%%%%%%%%%
\subsection{Purely-QMM Unitary State-Histories On The `$\theta$-Orbit'}
In this subsection, the simplest (exact and `excitonic') analytic wavefunction solutions to the QMM Schr\"odinger equations corresponding to the QMM-H \eqref{QMMH22}, in the form presented in 
\eqref{MM-evo-eqn2}, are presented. Moreover, we analyze, determine or constrain the $a^\star$ thresholds, which mark the Robust Non-Markovianity of a family of related solutions. 
%%%%%%%%%%%%%%%%%%%%%%%%%%%%%%%%%%%%%%%%%
%%%%%%%%%%%%%%%%%%%%%%%%%%%%%%%%%%%%%%%%%
%\subsubsection{Attractor Solutions }
%%%%%%%%%%%%%%%%%%%%%%%%%%%%%%%
%%%%%%%%%%%%%%%%%%%%%%%%%%%%%%%
\subsubsection{The Profile-Invariant State-History, Stability, Dynamical Attractor} \label{SIV-II}
We now introduce the first special `orbit' of the state time evolution of the one-qubit closed system, and then obtain and analyze several (here the simplest) wavefunction solutions along it, realizing QMM unitary state-histories in correspondence with \eqref{QMMH22}. This special orbit, and its counterpart introduced in the next subsection, are obtained by imposing the simplest dynamically-consistent constraints on the time evolution of the state, and also on the Hamiltonian couplings. The motivation for considering these orbits is that they are instructionally insightful and useful, specially for one's analytic explorations. This is because the QMM Schr\"odinger equations are simplified along these special orbits considerably, nevertheless, they do
not miss the main features of the general-orbit equations, neither mathematically nor physically. Moving to Section \ref{SV}, we solve equations \eqref{MM-evo-eqn2} along generic one-qubit orbits, using numerical tools.\\\\ 
The first special orbit, \emph{the $\theta$-orbit}, refers collectively to all the unitary time evolutions of the 
one-qubit wavefunctions for which, stated in terms of the Bloch representation \eqref{density-B}, the $\phi$ angle is kept constant, but the $\theta$ angle is set free to evolve based on its interacting memories: $\dot{\theta}_t = f(\theta_{t-a}, \theta_{t})$. Along the $\theta$-orbit, where $\forall \;t$: $\phi_{t}=\phi_{t_0}$, one has,
\begin{equation}\label{fgotto}
\mathcal{G}_{t-a, t} = 0 \;\;\;;\;\;\; \mathcal{F}_{t-a, t} = \; \sin(\theta_t-\theta_{t-a}) .
\end{equation}
Moreover, imposing the dynamical constraint $\dot{\phi}_t \equiv 0$ to equations \eqref{MM-evo-eqn2} enforces the following constraint on the couplings of the QMM-H \eqref{QMMH22}: $\hat{\mu}_{t-a} = \mu_{t-a} + \lambda^R = 0$. These constraints, together with \eqref{fgotto}, yield \emph{the dynamical equation on $\theta$-orbit},
\begin{equation}
\dot{\theta_t}  =  - \lambda^I \sin \big( \theta_t- \theta_{t-a} \big) . \label{EXCT-THETA}
\end{equation}
The beautiful equation \eqref{EXCT-THETA} is concise, but infinite dimensional. It must be highlighted that the above equation inherits the essential characteristics of equations
\eqref{MM-evo-eqn2}. It is a nonlinear delay-differential equation \cite{ArinoHD,Roussel2019,Erneux2009,MBaniYghoub Bani-Yahoub M} with a simplified but similar form. Being so, our analytic analyses will be mostly based on equation \eqref{EXCT-THETA}. Applying the defining $\theta$-orbit constraint $\phi_t = \phi_{t_0}$ to \eqref{qmtpfica}, and using the corresponding equation \eqref{EXCT-THETA}, one finds these expressions for the QM-TPFs of the unitary state-histories on the $\theta$-orbit, \begin{equation}\label{wto} m_{t-a\;t} = w_{t-a\;t} = \cos \big( \frac{\theta_t - \theta_{t-a} }{2} \big) \;\;\;;\;\;\; w^2_{t-a\;t}\; \big( 1 - w^2_{t-a\;t} \big) = \frac{\dot{\theta}_t^2}{4 \; (\lambda^I)^2} . \end{equation} \\\\  
It is natural to begin with looking for those exact solutions to the delay-differential $\theta$-orbit equation (\ref{EXCT-THETA}) whose defining profiles, as expressed in terms of finitely-many elementary functions, remain invariant in the whole course of time evolution. These are solutions $\theta_t$ which preserve their defining initial-history profile, chosen for $[t_0,t_0+a]$, ever after the turning-on of all the interacting quantum memories: namely all through $[t_0 + a , \infty)$. Obviously, compared with generic solutions whose profiles can change  in $a$-dependent intervals, the `profile-invariant solutions' are the first to be obtained.\\\\  
First, we see that all the $(N,L)$ Purely-QMM-UEs of all closed quantum many-body systems admit constant-in-time density operators as their trivial exact solutions. This fact can be easily confirmed by noticing that upon substituting: $\rho_{t_l} = \rho_{t} ,\forall\; t_l$ into \eqref{MMHparts}, we get: $H_{\text{QQM}}^{(N,L)}(t) \propto \rho_t $, resulting at: $\dot{\rho}_t = \rho_{t_0}$. The perturbative of this static solution to \eqref{EXCT-THETA}, $\theta_t = \text{const.}$, will be investigated later in what comes in this section. Next, we ask: what is the simplest profile-invariant dynamical solution to the $\theta$-orbit equation? Equation \eqref{EXCT-THETA} is featured with the special structure that the time-derivative of its variable is sourced by the mere difference between the variable at (the quantum memory moments) $t-a$ and $t$. Being so, the delay differential equation turns into a generically-solvable algebraic equation upon substituting in it a linear-in-time ansatz. Likewise, equation \eqref{EXCT-THETA} realizes its the simplest dynamical solution as a one-qubit unitary state-history (\ref{hps}) whose evolving density operators are parametrized by a linear-in-time $\theta$ angle, modulo its identifications. That is, we look for the state-history, \begin{equation}\label{ts5}\begin{split} \; & \hspace{1.85 cm} \hspace{.18 cm}  
 \big( \theta_t \;, \;\phi_t \big) \; = \; \big( \alpha \; t +  \beta \;,\; \phi_{t_0} \big) , \\ & \;\;\;\;\;\;\;\;\; \hspace{.25 cm} \;\;\;\;\;\;  m_{t-a\;t}\; =\; w_{t-a\;t}\; =\; \cos \big( \frac{a \; \alpha}{2} \big) . \end{split} \end{equation} Indeed, \eqref{ts5} satisfies the equation \eqref{EXCT-THETA} for arbitrary $\beta$ if the slope $\alpha$  is a real root of, \begin{equation} \alpha +\lambda^I  \sin(a \; \alpha) = 0 . \label{eq:alphaII} \end{equation}
The
parameters of the above algebraic equation in terms of which the slope $\alpha$ is fixed are
$\lambda^I$, namely the Hamiltonian coupling introduced in (\ref{QMMH22}), and $a$, being the largest QMD. For example, as long as $\lambda^I $ is taken positive, $\alpha = 0$ is the only solution, but $a = 1$ and $\lambda^I = -2$ yields the following three solutions: $\alpha = 0$ and $|\alpha| \approx  1.89549$. Indeed, one solution to (\ref{eq:alphaII}) is always $\alpha = 0$, resulting at a static one-qubit state-history. But, even numbers of non-zero $\alpha$ solutions to (\ref{eq:alphaII}) appear (in $\pm$ pairs) only on a proper subspace of the $(a,\lambda^I)$ plane. Stated precisely, this subspace corresponds to the $(\lambda^I < 0 , a > a^\star)$ region for a threshold $a^\star=a^\star(\lambda^I)$ which will be determined in Subsection \ref{sub-threshold}.\\\\
Now, we advance to point out that the solution (\ref{ts5},\ref{eq:alphaII}), despite being a special one, has a very significant relation with all the solutions in the moduli space of the $\theta$-orbit. Let us consider any state-history $(\theta_t = \theta(t) ,\phi_t = \text{const.})$ which identifies a one-qubit unitarily-evolving wavefuncion on the $\theta$-orbit,  by solving \eqref{EXCT-THETA}. Clearly, an important characteristic of any generic solution is its \emph{late-time asymptotics}, also (getting to know) if this asymptotics by itself corresponds to any specific exact solution of \eqref{EXCT-THETA} as a (static or dynamical) \emph{attractor} of the system. We now analyze analytically all these questions. To this aim, we first introduce a natural dimensionless coordinate which is the time measured in the units of the largest QMD, that is: $\tau \equiv \frac{t}{a}$. Likewise, we introduce the dimensionless coupling $\xi \equiv \lambda^I a$ of the QMM-H, $H^{(2,2)}_{\text{QMM}}(\tau)$ by which the $\tau$-evolution is generated. Re-expressing the $\theta$-orbit dynamics in terms of $(\tau,\xi)$, equation \eqref{EXCT-THETA} takes the following form,
\begin{equation}\label{top} \dot{\theta}_\tau \;=\; - \; \xi \; \sin \big( \theta_\tau - \theta_{\tau -1} \big) . \end{equation} The objective is obtaining the leading-order solution to \eqref{top} at sufficiently-late times $\tau \gg 1$, which also is the general asymptotic solution to the above equation for $\tau \to \infty$. Now, to work out the dominant solution at late times by means of a well-defined Taylor series, it is good to do a second change of variable from $\tau$ to $\tilde{x} \equiv \tau^{-1}$, such that the asymptotics is captured by $\tilde{x} \ll 1$. Applying the change of variables, the largest QMD $\Delta_1^{(t)} = a$ is mapped respectively to $\Delta_1^{(\tau)} = 1$ and $\Delta_1^{(\tilde{x})} = - \tilde{x}^2 + \text{O}[\tilde{x}^3] \approx  - \tilde{x}^2$. Therefore, the difference between the $\theta$ angle at the two quantum-memory moments which appears in the right hand side of the above equation is recast, to the leading order in the $\tilde{x} \ll 1$ Taylor expansion, as follows: $\theta_{\tilde{x}} - \theta_{\tilde{x} - \Delta_1^{(\tilde{x})}} \approx - \tilde{x}^2 \; \dot{\theta}_{\tilde{x}}$. Being so, the late-time dynamics of the $\theta$-orbit equation is described by the following first-order differential equation,  \begin{equation}\label{xte} \big( \tilde{x}^2\; \dot{\theta}_{\tilde{x}} \big) + \xi \; \sin \big( \tilde{x}^2 \; \dot{\theta}_{\tilde{x}}  \big) = a \; \big[\; \dot{\theta}_t + \lambda^I \sin \big( a \; \dot{\theta}_{t}  \big) \; \big] =  0 . \end{equation}
Finally, comparing \eqref{xte} with \eqref{eq:alphaII} proves that the late-time dynamics is given by,
\begin{equation}\label{sap} \dot{\theta}_t^{(\frac{t}{a} \gg 1)} = \text{constant} = \alpha(\lambda^I , a )_{|_{ \text{a real root of}\; \eqref{eq:alphaII}}} . \end{equation}
This completes the proof that the profile-invariant unitary state-history (\ref{ts5},\ref{eq:alphaII}) is \emph{the dynamic-or-`curve' attractor} of all one-qubit unitary state-histories on the the $\theta$-orbit. Note that we did not demand the smallness of $a$: the proof holds for arbitrary $a^{(\ge a^\star)}$. Moreover, as anticipated, generic solutions do not wait for very long times (in $a$ units) to realize their dynamical attractor (\ref{ts5},\ref{eq:alphaII}) with good approximations. In Subsection \ref{wnetto}, both for re-establishing the result \eqref{sap}, and for examining the fast tendency of the $\theta$-orbit state-histories in approaching it, we present numerical solutions  to (\ref{EXCT-THETA}).\\\\
Now, we investigate the stability of the profile-invariant solution (\ref{ts5},\ref{eq:alphaII}) under general perturbations, noting that the stability analysis of the trivial solution is also included. The very statement that the solution (\ref{ts5},\ref{eq:alphaII}) is the dynamical attractor for all $\theta$-orbit solutions indicates its stability under the subclass of perturbations which are confined on the $\theta$-orbit. Here in what follows analytically, joined with the numerical results of Subsection \ref{wnetto}, we directly examine the most general perturbations of this solution. That is, we must look for sufficiently small deformations of the $\theta$-orbit special solution (\ref{ts5},\ref{eq:alphaII}) in the entire moduli space of solutions to the general-orbit equations \eqref{MM-evo-eqn2}. Introducing a perturbation parameter $\epsilon \ll 1$, and choosing any root $\alpha$ of (\ref{eq:alphaII}) for a given $(a,\lambda^I)$, we apply to \eqref{EXCT-THETA} the following ansatz,
\begin{equation}\begin{split}\label{abe}
& \theta_t \; =\; \theta^{(0)}_t \;+\; \epsilon \; \theta^{(1)}_t  \;+\; \cdots \; = \;  ( \alpha \;t + \beta ) \;+\;  \epsilon \; \theta^{(1)}_t \; +\; \cdots \\ & \phi_t \; =\; \phi^{(0)}_t \;+\; \epsilon \; \phi^{(1)}_t  \; + \; \cdots \; = \; \phi_{t_0} \; + \;  \epsilon \; \phi^{(1)}_t  \; + \; \cdots  \end{split}\end{equation} 
Expansing the expressions to first order in $\epsilon$ yields two linear first-order delay-differential equations for the time evolution of the fluctuation modes. For the $\theta^{(1)}_t$ modes we obtain, 
\begin{equation}\begin{split}\label{whl} & 
\dot{\theta}^{(1)}_t \;=\; - \lambda^I \;  \cos( a \;  \alpha ) \; \big( \theta^{(1)}_t - \theta^{(1)}_{t-a} \big) . \end{split}\end{equation} The $\phi^{(1)}_t $ modes are evolved under an independent similar equation which is nevertheless featured, in distinction with \eqref{whl}, with a time-dependent prefactor, \begin{equation}\begin{split}\label{whlphi}                                                                                                       
 \dot{\phi}^{(1)}_t \;=\; \alpha \; \big[ \cot ( a\; \alpha ) - \cot ( \alpha \; t + \beta ) \;\big] \; \big( \phi^{(1)}_t - \phi^{(1)}_{t-a} \big) . \end{split}\end{equation}  
The first thing to observe about the above equations is that, similar to their nonlinear ancestor equation \eqref{EXCT-THETA}, the time-derivatives are sourced by the mere differences of the functions at the quantum-memory moments $t-a$ and $t$. Given this characteristics, it is clear that the solution for the mode ever after the onset of quantum memories, namely: $t \ge t_0 + a$, vanishes identically, that is: $\theta^{(1)}_t = 0 $ or $\phi^{(1)}_t = 0$, if the initial-history profile of the perturbation during $t \in [t_0 , t_0 + a]$ is taken to be any constant-in-time function. Now, we must deal with the solutions for the modes for time-dependent initial histories of the perturbations. Given the decoupling of the dynamical equations for $\theta^{(1)}_t$ and $\phi^{(1)}_t$, one can investigate the two moduli spaces independently.\\\\
Let us begin with profile-invariant solutions to \eqref{whl}. It is immediate to see that by inserting in \eqref{whl} a linear-in-time profile for the mode, we get $\theta^{(1)}_t = \theta^{(1)}_c$: 
a static mode. Now, we must consider the exponential-in-time modes: $\theta^{(1)}_t \propto e^{\gamma \; t}$ \cite{ArinoHD,Roussel2019,Erneux2009,MBaniYghoub Bani-Yahoub M}. The `characteristic equation' which determines these modes is found from \eqref{whl} as follows, 
\begin{equation}\label{ats} \gamma + 
\lambda \; \cos(a \; \alpha ) \; (1 - e^{- a \; \gamma}) \;=\; 0  .  \end{equation} It is easy to see that the transcendental equation \eqref{ats} yields  $\gamma = 0$ as its single root. First, $\gamma = 0$ is always the trivial solution to the equation. Second, the derivative of the right hand side of \eqref{ats} with respect to $\gamma$ never change its sign on the entire real $\gamma$ axis. Being so, equation \eqref{ats} can have one real root, but not more. Given the reality of the modes, the complex roots of \eqref{ats} must be purely imaginary and must organize linear combinations with oscillatory profiles: $\theta^{(1)}_t = c_1 \sin(\gamma \; t  + \beta) + c_2 \cos(\gamma \; t  + \beta)$. But then it can be seen directly that demanding these profiles to solve the $\theta^{(1)}_t$ equation \eqref{whl} enforces $\gamma$ to vanish, again giving us the static $\theta^{(1)}_t$ mode as the only solution. Similarly, we get only static modes for all other types of profile-invariant fluctuations, for example, those with arbitrary polynomial-in-time profiles. The conclusion is that there cannot be any profile-invariant mode fluctuating over the solution (\ref{ts5},\ref{eq:alphaII}).\\\\ 
We now extend our analysis to the most general family of solutions to \eqref{whl}: these are the solutions which are developed out of arbitrary initial histories: $\theta^{(1)}_{t_{|_{(t_0 \le t \le t_0 + a)}}} \equiv \theta_{t}^{(1|0)}$, and being so, cannot be profile-invariant generically. Because general analysis of these generic solutions cannot be done using the known analytical methods, we must utilize numerical methods. This numerical investigation is to be presented in Subsection \ref{wnetto}. The upshot is that all the $\theta^{(1)}_t$ fluctuations which solve \eqref{whl} undergo ever-shrinking oscillations whose attractors, $\theta^{(1)}_c$, are history-dependent constants: all $\theta^{(1)}_t$ modes are destined to become static. The values of these fixed-point attractors are generically non-zero, nevertheless they correspond to unphysical modes. This is because the sole effect of a static modes is a perturbative shift in the $\beta$ parameter which appears in \eqref{ts5}, an effect which in turn can be gauged away by resetting $t_0$. Finally, given the decoupling of the $\theta^{(1)}_t$ and $\phi^{(1)}_t$ modes, we conclude that the profile-invariant solution (\ref{ts5},\ref{eq:alphaII}) is stable against all the perturbations which are confined to the $\theta$-orbit.\\\\ 
Next, we must probe generic $\phi^{(1)}_t$ modes. It is easy to find out that the evolution of these modes is significantly different from that of $\theta^{(1)}_t$ modes.  The reason is that $\phi^{(1)}_t$ modes evolve according to equation \eqref{whlphi} whose structure, although similar to \eqref{whl}, is marked by a diverging time-dependent factor in its right hand side. Indeed, precisely due to this factor, generic $\phi^{(1)}_t$ modes are rapidly driven to grow nonperturbatively, destabilizing their background $\theta$-orbit solution (\ref{ts5},\ref{eq:alphaII}). To see this fact concretely, we set a known but arbitrary initial-history $\phi^{(1)}_{t_{|_{(t_0 \le t \le t_0 + a)}}} \equiv \phi_{t}^{(1|0)}$ for the mode. Now, one traces the $\phi^{(1)}_{t}$ evolution across time intervals of length $a$ consecutively, using  ``method of steps''  \cite{ArinoHD,Roussel2019,Erneux2009,MBaniYghoub Bani-Yahoub M}. Denoting the $m$th-step solutions by $\phi^{(1|m)}_{t}$, every next-step solution $\phi^{(1|n+1)}_t$ satisfies this ordinary differential equation in terms of the already-known $\phi^{(1|n)}_{t}$, \begin{equation}  \dot{\phi}^{(1|n+1)}_t \;=\; \alpha \; \big[ \cot ( a\; \alpha ) - \cot ( \alpha \; t + \beta ) \;\big] \; \big( \phi^{(1|n+1)}_t - \phi_{t-a}^{(1|n)} \big) . \end{equation} As we observe, with $t$ approaching any root of $\sin(\alpha \; t + \beta) $ = 0,  $\dot{\phi}^{(1|n+1)}_t$ is boosted up boundlessly, unless
$\phi^{(1|n+1)}_t = \phi_{t-a}^{(1|n)} $ exceptionally. For example, taking $t_0 = \beta = 0$, the first divergence happens at $t = \frac{\pi}{\alpha}$ which, given \eqref{eq:alphaII}, is located in one of the time steps exceeding $t = \frac{\pi}{|\lambda^I|} $. Supposing again $(a,\lambda^I) = (1,-2)$, this first divergence is reached at the very first time interval after the outset of the quantum memories. Numerical simulations of the solutions to \eqref{whlphi} re-confirm the destabilizing behavior of the  $\phi^{(1)}_t$ modes, due to the role played by the time-dependent prefactor on its RHS.\\\\ 
This concludes the stability analysis of the profile-invariant state-history (\ref{ts5},\ref{eq:alphaII}) which is the dynamical attractor for all the state-histories realizable on the $\theta$-orbit. The result is that this special state-history is stable against perturbations confined to its orbit, but is destabilized by `general orbit perturbations', namely those launched by turning on a non-constant initial quantum-history of the $\phi$-angle fluctuations:  $\phi^{(1)}_{t_{|_{(t_0 \le t \le t_0 + a)}}} = \phi_{t}^{(1|0)}$. In fact, the above instability is qualitatively very fruitful. This is so, because it is a cause for the phenomenon that the state-histories which solve the equations \eqref{MM-evo-eqn2} form a  rich phase diagram, featured with generic late-time behavior of unitarily-evolving wavefunctions which is significantly different from the linear-in-time $\theta$-orbit solution. In Section \ref{SV}, using numerical methods, we numerically map the remarkably-rich phase diagram of the one-qubit closed system under $(2,2)$ Purely-QMM unitary evolution.  %%%%%%%%%%%%%%%%%%%%%%%%%%%%%%%%%%%%%%%%%%%%%%%%%%%%%%%%%%%%%%%%%%%%%%%%%%%%%%%%%%%%%%%%%%%%%%%%%%%%%%%%%%%%%%%%%%%%%%%%%%%%%%%%%%%%%%%%%%%%%%%%%%%%%%%%
\subsubsection{Generic Purely-QMM Unitary State-Histories On The $\theta$-Orbit }\label{vms}
Generic $\theta$-orbit state-histories, namely the solutions to the delay-differential equation \eqref{EXCT-THETA} which are developed out of a generic initial quantum-history $\theta_{t_{|_{(t_0 \le t \le t_0 + a)}}}  \equiv \theta_{t}^{[0]}$ are `profile-changing' (in the sense defined in Subsection \ref{SIV-II}). As established, all such state-histories must tend to (\ref{ts5},\ref{eq:alphaII}), being the only profile-invariant state-history on the $\theta$-orbit, as their dynamical attractor. Furthermore, in Subsection \ref{wnetto}, we  expose numerically examples of generic solutions to the $\theta$-orbit equation \eqref{EXCT-THETA}. Nevertheless, it remains worthwhile to develop an analytic understanding of these state-histories. It is the objective of the present subsection to fill this gap.\\\\
To work out such solutions, as used in probing the $\phi_t^{(1)}$ modes, the best method one utilizes is the method of steps to consecutively integrate the delay-differential equation across time intervals of length $a$, beginning from an initial quantum-history $\theta_{t}^{[0]}$ \cite{ArinoHD,Roussel2019,Erneux2009,MBaniYghoub Bani-Yahoub M}. The implemented procedure is iterative: the solution at each new $(n+1)$th-step, $\theta^{[n+1]}_{t}$, is obtained based on the already-obtained previous-step solution $\theta_t^{[m]}$, by solving this nonlinear first-order ordinary differential equation, \begin{equation}\label{mosfto}  \dot{\theta}^{[n+1]}_t \;=\; - \lambda^I \; \sin \big( \theta^{[n+1]}_t - \theta_{t-a}^{[n]} \big) . \end{equation}
To integrate iteratively the chain of ordinary differential equations which encode the full-time solution to a given delay-differential equation, one takes, as initial condition, an input-function which serves as the initial history in the time interval $[t_0,t_0+a]$. Depending on what input-function one chooses as the initial history, exact solutions to the resulted differential equations at (some of the) subsequent length-$a$ time-steps may be obtained analytically. In the case of the nonlinear delay-differential equation \eqref{mosfto}, if we take a linear-in-time initial quantum history $\theta_{t}^{[0]} \propto t$, being the simplest choice, then the first-step differential equation for $\theta_{t}^{[1]}$ becomes integrable in terms of trigonometric and inverse-trigonometric functions. However, because this integrability does not further even to the second time-step for $\theta_{t}^{[2]}$, it is not particularly instructional to write it down explicitly. Indeed, it is well-known to mathematicians that nonlinear delay-differential equations are very rarely integrable by available analytical methods. Therefore, we need to take an alternative approach to carry out the integration of the sequential differential equations \eqref{mosfto} analytically. The alternative is a perturbative approach in which the resulted integrability comes with the following three merits. First, it works with very diverse choices of the initial quantum history $\theta_{t}^{[0]}$. Second, it extends up to arbitrary orders in the perturbation, in a way which is entirely covariant. Third, it continues in a highly similar fashion to arbitrary length-$a$ time-steps.\\\\ 
Let us consider an arbitrary function $h^{[0]}_{t \in [t_0 + t_0+a]}$ and marginalize it by the action of a sufficiently small prefactor $\epsilon \ll 1$. We identify the initial quantum history with this combination. Now, we aim to construct each one of the fixed-step state-histories $\theta_{t}^{[n \; \geq \; 1]}$ in the orders of perturbation in terms of the $\epsilon$ parameter. That is, we set a perturbative scheme as follows to consecutively integrate the sequential differential equations \eqref{mosfto},  \begin{equation}\label{pmsfto} \hspace{1.33 cm}\;\;\;\;\;\;  \theta_{t}^{[0]} \; \equiv \; \epsilon \;  h^{[0]}_t \;\;\;\;\;\;;\;\;\;\;\;\; \theta_{t}^{[n]} \; = \; \sum_{m=1}^\infty \; \epsilon^m \; \theta_{t}^{[n|m]} .  \end{equation}
Substituting \eqref{pmsfto} into the first-step equation \eqref{mosfto} for $\theta_{t}^{[1]}$, and organizing the resulted expression as a series expansion in powers of $\epsilon$, we obtain a generically-inhomogeneous first-order differential equation of the following form for every one of the functions $\theta_{t}^{[1|m]}$,
\begin{equation}\label{mexpaneqswjsrc}\begin{split} & \dot{\theta}_{t}^{[1|m]} \; + \; \lambda^I\; \theta_{t}^{[1|m]} \;=\; \lambda^I \; J_t^{[1|m]} \big( h^{[0]}_{t-a}\;,\; \theta_{t}^{[1|r < m]} \big)  .  \end{split}\end{equation}  
To be concrete, here are the explicit forms of the first few `source' functions  $J_{t}^{[1|m]}$,
\begin{equation}\label{ste}\begin{split} & J_{t}^{[1|1]} = h^{[0]}_{t-a} \;;\; J_{t}^{[1|2]} = 0 \;;\; J_{t}^{[1|3]} = \frac{1}{6}\;
\big(  \theta_{t}^{[1|1]} - h^{[0]}_{t-a} \big)^3 \:;\; J_{t}^{[1|4 ]} = \frac{1}{2}\; \theta_{t}^{[1|2]} \; \big( \theta_{t}^{[1|1]} - h^{[0]}_{t-a} \big)^2 , \\ & J_{t}^{[1|5]} \;=\; \frac{1}{2}\; \big( \theta_{t}^{[1|1]} - h^{[0]}_{t-a} \big)^2 \; \theta_{t}^{[1|3]} \;+\; \frac{1}{2}\;\big(  \theta_{t}^{[1|1]} - h^{[0]}_{t-a} \big) \;  (\theta_{t}^{[1|2]})^2 \;+\; \frac{1}{120}\; \big(  \theta_{t}^{[1|1]} - h^{[0]}_{t-a} \big)^5 .   \end{split}\end{equation}
The general solution to differential equations \eqref{mexpaneqswjsrc} for $\theta_{t}^{[1|m]}$ is analytically well-known, and for example using the method of `variation of parameters', can be written as follows,
\begin{equation}\label{gefoinh} \hspace{1 cm} \theta_{t}^{[1|m]} =\; e^{- \lambda^I t} \; \big(\; c^{[1|m]} \;+ \int dt\;e^{\lambda^I t}\;J_{t}^{[1|m]}\; \big) \;\;\;;\;\;\; \forall \; m \in \mathbb{N} .  \end{equation}
The coefficients $c^{[1|m]}$ are determined by imposing the following boundary conditions,
\begin{equation} \;\;\;\;\;   \theta_{t_0 + a}^{[0]} \; = \; \theta_{t_0 + a}^{[1]} \;\;\Longrightarrow\;\; \theta_{t_0 + a}^{[1|m >1]} = 0 \;\;\;;\;\;\; h_{t_0 + a}^{[0]} \;=\; \theta_{t_0 + a}^{[1|1]}  . \end{equation}
The above patters goes on, almost identically, to the subsequent length-$a$ time-steps. That is, one obtains the same expressions for all fixed-step $\theta$-orbit state-histories $\theta_{t}^{[1|n > 1]}$. The only change that we need to implement at each time-step is to re-set the initial quantum history, being the state-history in the previous step, and re-compute the source functions $J_{t}^{[n|m]}$ accordingly. To be clear, let us now consider the $\theta$-orbit state-history in the second interval $[t_0 + 2 a, t_0 + 3 a]$, $\theta_{t}^{[2]}$. The corresponding initial quantum history is similarly marginalized by the $\epsilon$ parameter, but also receives the higher order terms in $\epsilon$ in the following form,
\begin{equation}\label{fstoshbm}\begin{split} 
\theta_{t}^{[1]} \; \equiv \; \epsilon \; h_t^{[1]} \;\;\;;\;\;\; h_{t}^{[1]}  \;=\; \theta_t^{[1|1]} \;+\; \sum_{m=1}^{\infty} \epsilon^m \; \theta_t^{[1|m+1]} .
\end{split}\end{equation}
Being so, the solutions $\theta_{t}^{[2|m]}$ will be found to be of the same form as \eqref{gefoinh}, with the only change being in the expressions of the source functions $J_{t}^{[2|m]}$, in comparison with their counterparts (\ref{ste}), given all the higher order terms in $\epsilon$ which enter into \eqref{fstoshbm}. In conclusion, the fixed-step $\theta$-orbit state-histories are all in the following form,
\begin{equation}\label{gefoinh} \hspace{3 cm} \theta_{t}^{[n|m]} =\; e^{- \lambda^I t} \; \big(\; c^{[n|m]} \;+ \int dt\;e^{\lambda^I t}\;J_{t}^{[n|m]}\; \big) \;\;\;;\;\;\; \forall\; (n,m) \in \mathbb{N}^2  . \end{equation}
The boundary conditions which fix the coefficients $c^{[n |m]}$, for every $n > 1$, are as follows,
\begin{equation} \;\;\;\;\;   \theta_{t_0 + n a}^{[n-1]} \; = \; \theta_{t_0 + n a}^{[n]} \;\;\Longrightarrow\;\; \theta_{t_0 + n a}^{[n - 1|m \geq 1]} = \theta_{t_0 + n a}^{[n|m \geq 1]} .  \end{equation}
%%%%%%%%%%%%%%%%%%%%%%%%%%%%%%%%%%%%%%%%%%%%%%%%%%%%%%%%%%%%%%%%%%%%%%%%%%%%%%%%%%%%%%%%%%%%%%%%%%%%%%%%%%%%%%%%%%%%%%%%%%%%%%%%%%%%%%%%%%%%%%%%%%%%%%%%%%%%%%%%%%%%%%%%%%
\subsubsection{Determining The Thresholds Along The $\theta-$Orbit} \label{sub-threshold} \label{sub-threshold}
One intriguing characteristic of all Purely-QMM-UEs is that they are developed only if their corresponding largest QMDs are larger than (or at least equal to) some thresholds whose finite values depend on the defining couplings of the QMM-H: $ a \ge a^\star= a^\star(\vec{\lambda})$. This general aspect was established in Subsection \ref{S:IV:Strong} and we will furthermore verify it in all numerical solutions to the QMM Schr\"odinger equations presented in Section \ref{SV}. Now, one asks: what functions of the Hamiltonian couplings $\vec{\lambda}$ are the thresholds $a^\star$? The fact is, given the mathematical significant complexity of generic delay-differential QMM-Schr\"odinger equations even for dimensionally-very-small quantum systems, the above dependencies are generically too complicated to be analytically computable by the available methods. But, there are rare cases where we can analytically determine $a^\star(\vec{\lambda})$. The simplest example of these rare cases is presented in this subsection, mainly in order to shed analytical and quantitative light on the `Robust Non-Markovianity' of Purely-QMM-UEs.\\\\ 
We determine in what follows the threshold $a^\star$ of the $(2,2)$ QMM-UE of the one-qubit closed system along the $\theta$-orbit. In Subsection \ref{SIV-II}, the profile-invariant solution to the $\phi$-orbit equation \eqref{EXCT-THETA} was obtained. The solution, given by (\ref{ts5},\ref{eq:alphaII}), corresponds to the one-qubit unitary state-histories with constant $\phi$ angles and linear-in-time $\theta$ angles. According to \eqref{eq:alphaII}, the slopes $\alpha$ of the linear-in-time $\theta$ angles must be a real root of the algebraic equation,   
\begin{equation}
a\; \alpha + \arcsin \big( \frac{\alpha}{\;\lambda^I} \big) \;= \; 0 . \label{eq:alpha-arc}
\end{equation}
Now, we analyze the \emph{full} Taylor-series expansion of the LHS of equation \eqref{eq:alpha-arc} as a function of $\alpha$, around $\alpha = 0$ as its trivial root. Notice that, only the convergence of the Taylor series concerns us in this analysis (rather than its perturbative utility), therefore, there is not any small-$\alpha$ restriction involved in here. Equation \eqref{EXCT-THETA} implies that every $\theta$-orbit state-history must validate the dynamical inequality: $|\dot{\theta}_t| < |\lambda^{I}|$. Remembering the convergence of the Taylor-series expansion of $\arcsin(x)$  around any point on the interval $|x| \leq 1$, we deduce that the Taylor series of the LHS of equation \eqref{eq:alpha-arc} is convergent for any profile-invariant unitary state-history (\ref{ts5},\ref{eq:alphaII}), resulting at the following strict inequality, 
\begin{equation}\begin{split} 
& a_{\,|_{\;\text{profile-invariant solution}\;\eqref{ts5}}} \,=\,  \frac{1}{- \lambda^I} \; \big\{\, 1+ \sum_{n = 1}^{\infty} \; \dfrac{(2n!)}{(2^n\;n!)^2 \; (2n+1)} \; (\frac{\alpha}{\lambda^I})^{2n} \; \big\} \;\;\;\; \text{   s.t.}  \\ & \; a_{\,|_{\;\text{profile-invariant solution}\;\eqref{ts5}}} > \; \dfrac{1}{|\lambda^I|} \; \equiv \; a^\star_{\;|_{\;\text{profile-invariant solution}\;(\ref{ts5},\ref{eq:alphaII})}} . \label{eq:lowerb} \end{split}
\end{equation}
Result (\ref{eq:lowerb}) is indeed satisfied by every nontrivial solution of (\ref{eq:alphaII}) and quantifies the Robust Non-Markovianity of the non-constant profile-invariant solutions on the $\theta$-orbit.
We remind that for positive values of $\lambda^I$, the $\theta$-orbit \eqref{EXCT-THETA} admits only trivial constant solutions, such that the negativity of $\lambda^I$ is assumed in resulting the inequality \eqref{eq:lowerb}.
Before going on, it is worth knowing quantitatively how the Robust Non-Markovianity feature, manifested by the strict inequality \eqref{eq:lowerb}, is satisfied by the `highest-rate' profile-invariant solution: $|\dot{\theta}_t| = |\alpha| = - \lambda^{I}$. Either by looking into the corresponding principal value of the RHS of the equation in the relation \eqref{eq:lowerb}, or directly from \eqref{eq:alphaII}, the measure of the Robust Non-Markovianity of the highest-rate solution is obtained to be: $a = \frac{\pi}{2} \frac{1}{|\lambda^{I}|} > \frac{1}{|\lambda^{I}|}$. As proved, result \eqref{ts5} places a strict lower bound on the largest QMDs of all the non-constant profile-invariant solutions on the $\theta$-orbit.\\\\ 
Moreover, one sees that the validity of the above inequality extends much beyond the solutions (\ref{ts5},\ref{eq:alphaII}) for which we laid out the given proof. There are two independent ways in which a much wider coverage of the lower bound \eqref{eq:lowerb} can be confirmed and also specified. First, as proved in Subsection \ref{SIV-II}, the profile-invariant solution is the late time asymptotics of all solutions on the $\theta$-orbit. As such, with the largest QMD $a$ being a constant, its late-time-asymptotics lower bound is necessarily imposed on the entire-time solution. Second, any smooth and differentiable solution along the $\theta$-orbit must become reducible in sufficiently time windows around every generic point on $\mathbb{R}_t$ to the profile-invariant solution (\ref{ts5}) corresponding to a real root of (\ref{eq:alphaII}). Given this requirement, the largest QMD $a$ of every nontrivial differentiable solution of the $\theta$-orbit dynamical equation (\ref{EXCT-THETA}) must admit $\frac{1}{|\lambda^I|}$ as a strict lower bound. We conclude that, \emph{a quantification of the Robust Non-Markovianity of all $\theta$-orbit unitary state-histories of the one-qubit Purely-QMM-UE under $H_{{\rm QMM}}^{(2|2)}(t)$} is stated by the following lower bound,
\begin{equation}\label{sct}
a^*_{\,|_{\;\text{all}\;\theta \text{-orbit solutions}}} \; \geq \; a^*_{\,|_{\;\text{profile-invariant solution}\;(\ref{ts5},\ref{eq:alphaII})}} \, = \;\dfrac{1}{|\lambda^I|} .
\end{equation}
%%%%%%%%%%%%%%%%%%%%%%%%%%%%%%%%%%%%%%%%%%%%%%%%%%%%%%%%%%%%%%%%%%%%%%%%%%%%%%%%%%%%%%%%%%%%%%%%%%%%%%%%%%%%%%%%%%%%%%%%%%%%%%%%%%%%%%%%%%%%%%%%%%%%%%%%%%%%%%%%%%%%%%%%%%%%%%%%%%%%%%%
\subsubsection{The $\theta$-orbit Purely-QMM-UEs In Their Infancy}
We have shown that the asymptotic behavior of all $\theta$-orbit state-histories, developed in the late-times regime $\frac{t}{a} \gg 1$, is the profile-invariant solution (\ref{ts5},\ref{eq:alphaII}). Naturally, one now examines the period when the Purely-QMM-UEs are in their `infancy', $\frac{t}{a} \ll1$, and asks what analytic solutions the $\theta$-orbit state-histories approximate in these very early times. We have shown that under the $(N, L)$ purely-QMM-Hs, there always are finite thresholds ${a}^{\star}$, such that non-constant wavefunctions are developed only if $a \geq {a}^{\star}$. Hence, an interesting subset of infant unitary state-histories are constructible by taking $a^\star$ thresholds which are sufficiently large and lead to the infancy periods $\frac{t}{a} \le \frac{t}{a^\star} \ll 1$ which are sufficiently long. For $\theta$-orbit state-histories which obey (\ref{sct}), the sufficient largeness of $a^\star$ is secured by taking sufficiently small values of the interaction coupling $|\lambda^I|$. Moreover, we remember that the unitary evolution under any $H^{(N,L)}_{{\rm QMM}}(t)$ is to be measured from the outset of having the quantum-memory-pool filled completely: $t = t_*$. All in all, if one introduces the dimensionless time variable: $\tau \; \equiv \; |\lambda^I| \; (t-{t}_{\star})$, rewrites the $\theta$-orbit equation (\ref{EXCT-THETA}) as a delay-differential equation in terms of $\tau$, and obtains polynomial solutions $\hat{\theta}( \tau)$ to it as an order-by-order perturbation in terms of $\tau \ll 1$, an interesting family of $\theta$-orbit infant unitary state-histories are constructed. Finally, because (for constant $a$,) ${t}_{\star} = t_0 + a$, and that with no loss of generality one can choose $t_0 = - a$, we set $\tau \; = \; |\lambda^I|\;t$ in what follows.\\\\
We now implement the recipe prescribed above. First, the QMD measured in the new time variable $\tau$ which parametrizes the post-$t^{\star}$ evolution in $\frac{1}{|\lambda^I|}$ units, reads as: $\hat{a} \equiv |\lambda^I|\; a \geq  \frac{a}{a^{\star}} \geq 1$. Redefining Bloch variable: $\hat{\theta}(\tau) \equiv \theta(-\lambda^I\;t)$, (\ref{EXCT-THETA}) takes the form,
\begin{equation}
\frac{d \hat{\theta}(\tau)}{d \tau} \;\equiv\; \dot{\hat{\theta}}(\tau) \;=\; - \sin \big(\; \hat{\theta}(- \hat{a} + \tau ) - \hat{\theta}(\tau) \; \big) . \label{eq:sch:thetahat}
\end{equation}
We confine to the regime $\tau \ll 1$ which given (\ref{sct}) guarantees the evolution infancy: $\frac{t}{a^\star} \leq \tau \ll 1$. Next, for solving the above QMM delay-differential equation, we expand the time-shifted $\hat{\theta}(-a + \tau)$, around the moment $\tau_0 = - a$, as a Taylor series for $\tau \ll 1$,
\begin{equation}
\hat{\theta}(-\hat{a} \;+\; \tau)  = \hat{\theta}(-\hat{a}) \; + \; \hat{\mathcal{R}}_0^{(-\hat{a})}(\tau) \; = \; \hat{\theta}_{-\hat{a}} \; +\; \dot{\hat{\theta}}_{-\hat{a}} \; \tau \;+\; \dfrac{1}{2}\; \ddot{\hat{\theta}}_{-\hat{a}} \; \tau^2 \;+\; \dfrac{1}{3!}\; \dddot{\hat{\theta}}_{-\hat{a}} \; \tau^3 \;+\; \cdots \; % \equiv \;\; \hat{\theta}_{-\hat{a}} \;+ \; \hat{\mathcal{R}}_0^{(-\hat{a})}(\tau)      
\label{def:exp:thetahat}
\end{equation}
Before working out the perturbative solutions as described, we ensure the convergence of the Taylor series (\ref{def:exp:thetahat}), by obtaining an upper bound on the remainder $\hat{\mathcal{R}}_0^{(-\hat{a})}(\tau)$. As equation \eqref{eq:sch:thetahat} implies, during the entire QMM-UE $\tau  \geq \tau_* = 0$, one has: $|\dot{\hat{\theta}}(\tau)| \leq 1$. Hence,  using the Remainder Estimation Theorem, we conclude that at every moment $\tau \in  [0, \infty)$, the zeroth-order absolute value of the remainder function $\hat{\mathcal{R}}_0^{(-\hat{a})}(\tau)$ in $(\ref{def:exp:thetahat})$ respects the upper bound,
\begin{equation}
 |\hat{\mathcal{R}}_0^{(-\hat{a})}(\tau)| \; \leq \; \frac{\tau^{1}}{1!} \cdot {\rm Max} \big(\;|\dot{\hat{\theta}}(\tau^{\prime}-a)| \;\big)_{\;\big|_{\tau^{\prime} \in \;[0, \tau]}} \;= \; \tau \; \cdot \; 1 \; = \; \tau .
\end{equation}
Hence, imposing $\tau \ll 1$ which secures the QMM-UE infancy, we obtain: $\hat{\mathcal{R}}_0^{(-\hat{a})}(\tau) \ll \hat{1}$. Therefore, the convergence of the Taylor series (\ref{def:exp:thetahat}) is guaranteed.\\\\
Now, everything is set for us to combine \eqref{eq:sch:thetahat} and \eqref{def:exp:thetahat}, and solve the following first-order differential equation for $\hat{\theta}(\tau)$,
\begin{equation}\label{38}\begin{split}
& \dot{\hat{\theta}}(\tau) \; = \; - \sin \big(\;\;\hat{\mathcal{R}}_0^{(-\hat{a})}(\tau)\;-\; \hat{\theta}(\tau) \; \;+\;\hat{\theta}_{-\hat{a}}\;\big) , \\
& \hat{\mathcal{R}}_0^{(-\hat{a})}(\tau) \; = \; \dot{\hat{\theta}}_{-\hat{a}} \; \tau \;+\; \dfrac{1}{2}\; \ddot{\hat{\theta}}_{-\hat{a}} \; \tau^2 \;+\; \dfrac{1}{3!}\; \dddot{\hat{\theta}}_{-\hat{a}} \; \tau^3 \;+\; \cdots \;
\end{split}\end{equation}
It is natural to impose polynomial-in-time profiles on the solutions ${\hat{\theta}}(\tau)$,
\begin{equation}
{\hat{\theta}}(\tau) \; = \; {\hat{\theta}}_0 \; + \; {\hat{\theta}}_1 \; \tau  \;+\; {\hat{\theta}}_2 \; \tau^2 \;+\; {\hat{\theta}}_3 \; \tau^3 \;+\; \cdots \label{def:exp:thetahat2}
\end{equation}
Applying \eqref{def:exp:thetahat2}, we solve \eqref{38} in perturbation over $\tau$, order by order. One notices that in the perturbative fixation of the solutions \eqref{def:exp:thetahat2}, the $n$-th order derivatives of $\hat{\theta}(\tau^\prime)_{|_{\tau^\prime = -a}}$, namely $({\hat{\theta}}_{-\hat{a}}, \dot{\hat{\theta}}_{-\hat{a}}, \ddot{\hat{\theta}}_{-\hat{a}}, \dddot{\hat{\theta}}_{-\hat{a}}, \cdots)$, play the role of the `boundary-conditions' constants, in terms of which the polynomial coefficients $(\theta_0, \theta_1, \theta_2, \theta_3, \cdots)$ are to be found as the equation variables. Assuming the perfect differentiability of the entire solution, these boundary constants $\frac{d^n \hat{\theta}(\tau^\prime)}{d \tau^{\prime\;n}}_{|_{\tau^\prime = -a}}$ must be read from the pre-$t_*$ solution developed by the Kicker Hamiltonian.
It is straightforward to obtain perturbative polynomial solutions of differential equation (\ref{38}) to arbitrary orders in the expansion \eqref{def:exp:thetahat2}. For example, one obtains the first three variables $(\hat{\theta_1}, \hat{\theta_2}, \hat{\theta_3})$ in terms of (the arbitrary constant) $\hat{\theta}_0$ and $({\hat{\theta}}_{-\hat{a}}, \dot{\hat{\theta}}_{-\hat{a}}, \ddot{\hat{\theta}}_{-\hat{a}})$ as follows,
\begin{equation}\begin{split}
& \hat{\theta}_1 = \sin(\hat{\theta}_0 - \hat{\theta}_{-\hat{a}}) ,\\& \hat{\theta}_2 = \frac{1}{2} \; \cos (\hat{\theta}_0 - \hat{\theta}_{-\hat{a}})\; \cdot \; [\;\sin(\hat{\theta}_0 - \hat{\theta}_{-\hat{a}}) - \dot{\hat{\theta}}_{-\hat{a}}\;] ,\\& \hat{\theta}_3 = \frac{1}{12} \; \big\{\; \sin[\; 3(\hat{\theta}_0 - \hat{\theta}_{-\hat{a}}) \;] - \sin(\hat{\theta}_0 - \hat{\theta}_{-\hat{a}})\; + \\ & \hspace{.0000199cm} \;\;\;\;\; + \; [\; 1-3\; \cos[2(\hat{\theta}_0 - \hat{\theta}_{-\hat{a}}) \;] - 2\; \sin(\hat{\theta}_0 - \hat{\theta}_{-\hat{a}}) \; \dot{\hat{\theta}}_{-\hat{a}}\; ]\; \dot{\hat{\theta}}_{-\hat{a}} - 2\; \cos(\hat{\theta}_0 - \hat{\theta}_{-\hat{a}}) \; \ddot{\hat{\theta}}_{-\hat{a}} \; \big\} . \end{split} \end{equation} 
%%%%%%%%%%%%%%%%%%%%%%%%%%%%%%%%%%%%%%%%%%%%%%%%%%%%%%%%%%%%%%%%%%%%%%%%%%
\subsection{Purely-QMM Unitary State-Histories On The `$\phi$-Orbit'}\label{tao}
The counterpart of the $\theta$-orbit is also a special orbit of the one-qubit Purely-QMM-UE. This second special orbit, the `$\phi$-orbit', is such that the $\theta$ angle is kept constant, but one lets the $\phi$ angle evolve purely based on its two interacting memories: $\dot{\phi}_t = f(\phi_{t-a}, \phi_{t})$. Although the defining constraint of the $\phi$-orbit is the exact counterpart to that of the $\theta$-orbit, it calls for an independent analysis in regard with its solutions. Indeed, the ways in which the two Bloch angles appear in the dynamical equations of the one-qubit closed system, in this case equations \eqref{MM-evo-eqn2}, are distinctive enough to make the moduli spaces of the purely-QMM unitary state-histories developed along these counterpart special orbits nonidentical. This analysis is undertaken carefully in what follows.\\\\ 
Along the $\phi$-orbit which is constrained by $\theta_t = \theta_{t_0}$, the functions $\mathcal{F}_{t-a,t}$ and $\mathcal{G}_{t-a,t}$  defined in \eqref{fgp} are given by,
\begin{equation}
\label{cpo}
\mathcal{F}_{t-a , t} =\; \sin(2 \theta_{t_0}) \; \sin^2(\frac{\phi_t - \phi_{t-a}}{2}) \;\;\;;\;\;\;  \mathcal{G}_{t-a , t} =\;- \sin(\theta_{t_0})\;\sin(\phi_t - \phi_{t-a}).
\end{equation}
Moreover, imposing the constancy of the $\theta_t$ angle on the general expressions (\ref{nmtpfoquis},\ref{qmtpfica}), the QM-TPF and its observable norm for a $\phi$-orbit unitary state-history are found as, 
\begin{equation} 
m_{t-a\;t} \;=\; \sin^2(\frac{\theta_{t_0}}{2}) \; \big(\; e^{i (\phi_t - \phi_{t-a})} \; - 1  \big) \;+ 1\;\;\;;\;\;\; w_{t-a\;t}^2 \;=\;- \sin^2(\theta_{t_0}) \; \sin^2 \big( \frac{\phi_t - \phi_{t-a}}{2} \big) \;+\;1 . \end{equation}
Now, we return to the one-qubit $(2,2)$-Purely-QMM delay-differential equations \eqref{MM-evo-eqn2}, apply to them the $\phi$-orbit traits $\dot{\theta}_t= 0$, and \eqref{cpo}, and work out all the implications. The outcomes depend on the couplings $\lambda^I$ and $\hat{\mu}_{t-a}$ and the Bloch-angle constant $\theta_{t_0}$. Let us present the final results as classified in the following five distinct categories.\\\\
\underline{$.$ \emph{Generic $(\hat{\mu}_{t-a} , \lambda^I , \theta_{t_0})$ Triplet}}\\
This is the generic category with the triplet $(\lambda^I$, $\hat{\mu}_{t-a},\theta_{t_0})$ being only constrained by: $\lambda^I \hat{\mu}_{t-a} \sin(\theta_{t_0}) \neq 0$. Modulo the arbitrariness in setting $\phi_{t_0}$, there is nothing but one unitary state-history realizable in this category along the $\phi$-orbit, given by the solution,
\begin{equation} 
\theta_t =\;\theta_{t_0}\;\;\;;\;\;\; \phi_{t} =\;\beta + \; \big[\;  \frac{2\;\cos(\theta_{t_0})\; \big(\; (\lambda^I)^2 + \hat{\mu}_{t-a}^2\; \big) \; 
\hat{\mu}_{t-a} }{(\lambda^I)^2 \; \cos^2(\theta_{t_0}) \; + \; \hat{\mu}_{t-a}^2} \;\big]\; t . 
\end{equation}
\underline{$.$ $\hat{\mu}_{t-a}$ and $\lambda^I$ \emph{Non-Vanishing, Fine-Tuned} $\theta_{t_0}:$ $\sin(\theta_{t_0}) = 0$}\\
This class is very special. The $\theta$-orbit dynamical equations are solved automatically, regardless of what function $\phi_t$ is. That is, one has a whole arbitrary family of unitary state-histories realizable on the $\phi$-orbit,
\begin{equation} 
\Big(\; \frac{\theta_t}{\pi} \;,\; \phi_t \;\Big) \;=\; \big(k \in \{0, 1 \} \;,\;f_{t} \; \big). 
\end{equation}
\underline{$.$ $\lambda^I = 0 \;$, \emph{Non-Vanishing} $\hat{\mu}_{t-a}$, \emph{Fine-Tuned} $\theta_{t_0}$}\\
If $\hat{\mu}_{t-a}$ is the only non-vanishing coupling of the $(2,2)$-Purely-QMM-H, modulo one's freedom in setting $\phi_{t_0}$, only a discrete family of non-constant unitary state-histories is realizable along the $\phi$-orbit, upon fine-tuning $\theta_{t_0}$. The solutions, being labeled by $k \in \mathbb{Z} $, are given by, 
\begin{equation}
\label{dscslnatphio}  
\theta_t =\; \arccos \; [\; \frac{(2 \; k + 1) \; \pi}{2\; a\;\hat{\mu}_{t-a}}\;] \;\;\;,\;\;\; \phi_t =\; \beta \;  + \;  \frac{(2 \; k + 1) \; \pi}{a} \;\; t . 
\end{equation}
Notice that although \eqref{dscslnatphio} is from \eqref{}, this is not a sub-class \\
\underline{$.$ \emph{Only One Non-Vanishing Coupling, Generic} $\theta_{t_0}$}\\
When either $\hat{\mu}_{t-a}$ or $\lambda^I$ is the only non-vanishing coupling, and $\theta_{t_0}$ is not fine-tuned to the values specified in the previous or the next category, the only $\phi$-orbit solutions are constant-in-times state-histories,
\begin{equation} 
(\;\theta_t \; ,\; \phi_t\;)\;=\; (\; \theta_{t_0} \;,\; \phi_{t_0} \;) . 
\end{equation}
\underline{$.$ $\hat{\mu}_{t-a} = 0\;$, \emph{Non-Vanishing} $\lambda^I$, \emph{Fine-Tuned} $\theta_{t_0}$}\\
Finally, when the only coupling which is turned on in the $(2,2)$-Purely-QMM-H given in \eqref{QMMH22} is $\lambda^I$, and furthermore one implements the fine-tuning: $\cos(\theta_{t_0}) = 0$, the unitary state-histories along the $\phi$-orbit are realized by solutions to the equation, 
\begin{equation}
\label{EXCT-PHI} 
\dot{\phi}_t \; =\; - \lambda^I \; \sin(\phi_t- \phi_{t-a}) . 
\end{equation}
Because the above delay-differential equation is mathematically identical to \eqref{EXCT-THETA}, it is immediate to draw three conclusions as follows.\\\\ 
First, the only profile-invariant $\phi$-orbit unitary state-history in this category is given by the following solution, 
\begin{equation}
\label{piposln} 
\big(\; \frac{\theta_t}{\pi} \; , \; \phi_t \; \big) \;=\; \big(\; \frac{k \in \{ 1, 3\}}{2} \;,\; \beta \;+\; \alpha \; t  \; \big)\;\;\;;\;\;\; \alpha \;+\; \lambda^I \; \sin(a \; \alpha)  = \; 0 \;\;\;;\;\;\; w^2_{t-a\;t} = \; \cos^2 (\frac{a \; \alpha}{2}) . 
\end{equation}
Second, the profile-invariant solution \eqref{piposln} is the dynamical attractor of all solutions to the $\phi$-orbit dynamical equation \eqref{EXCT-PHI}. This already secures that the profile-invariant solution \eqref{piposln} is stable against the perturbations which are confined to the $\phi$-orbit. Equation \eqref{EXCT-PHI} develops a large family of generic profile-varying unitary state-histories, some of which (modulo its different angular identifications) are given by the numerical solutions presented in Subsection \ref{SV} for its $\theta$-orbit counterpart.\\\\
Third, an analysis similar to what was done in Subsection \ref{SIV-II} demonstrates that the profile-invariant solution \eqref{EXCT-PHI} is however unstable against the more general family of perturbations with non-constant initial-history of $(\delta \theta)_t$ is also turned on.
%%%%%%%%%%%%%%%%%%%%%%%%%%%%%%%%%%%%%%%%%%%%%%%%%%%%%%%%%%%%%%%%%%%%%%%%%%%%%%%%%%%%%%%%%%
%%%%%%%%%%%%%%%%%%%%%%%%%%%%%%%%%%%%%%%%%%%
\subsubsection{The $\phi$-Orbit Thresholds}
Now, we move on to quantify analytically the Robust Non-Markovianity of the unitary state-histories along the $\phi$-orbit. The procedure is very straightforward and similar to its  
$\phi$-orbit counterparts. All it takes is to explicitly reconstruct only one section of the entire analysis we had to do in Subsection \ref{tao} in order to completely classify the $\phi$-orbit solutions. As such, in what follows, it suffices to this relevant part of the $\phi$-orbit impositions on the one-qubit dynamical system \eqref{MM-evo-eqn2}. Imposing the constancy of the $\theta_t$ angle on the $\dot{\theta}_t$ equation in the system \eqref{MM-evo-eqn2} states that $\mathcal{F}_{t-a,t}$ and $\mathcal{G}_{t-a,t}$ must be proportional: $\mathcal{F}_{t-a,t} = \frac{\hat{\mu}_{t-a}}{\lambda^I} \; \mathcal{G}_{t-a,t}$. Using this and the $\mathcal{G}_{t-a,t}$ in \eqref{cpo}, the $\dot{\phi}_t$ equation in \eqref{MM-evo-eqn2} yields,
\begin{equation} \;\;\;\;\;\;\;\;\;\;\; \dot{\phi_t}  \;=  -\; \big(\; \lambda^I  +\, \frac{\hat{\mu}_{t-a}^2}{\lambda^I} \; \big) \; \sin \big( \phi_t- \phi_{t-a}  \big) . \label{jphide}\end{equation}
Because the dynamical equation \eqref{jphide} is mathematically identical to equation \eqref{EXCT-THETA}, doing the same analysis as above, the Robust Non-Markovinaity of all the $\phi$-orbit unitary state-histories is immediately quantified likewise,
\begin{equation}
a^*_{\,|_{\;\text{all}\;\phi \text{-orbit solutions}}} \; \geq \;\; \frac{| \lambda^I |}{(\lambda^I)^2 \;  + \; \hat{\mu}_{t-a}^2} .
\end{equation}
%%%%%%%%%%%%%%%%%%%%%%%%%%%%%%%%%%%%%%%%%%%%%%%%%
%%%%%%%%%%%%%%%%%%%%%%%%%%%%%%%%%%%%%%%%%%%%%%%%%
\subsection{Hybrid Unitary Evolutions: General-Orbit Near-Markovian Solutions}\label{the hybrid versions}
We now look for one-qubit unitary state-histories made in the general context of \eqref{H-CQS}, when the dynamics of the qubit is generated by a linear combination of both QMM and conventional Hamiltonians, namely `Hybrid Hamiltonians'. Let us recall that by the term `conventional Hamiltonians', we refer collectively to all possible (local-in-time or nonlocal-in-time) Hamiltonians made only from the building-block operators which, in their definitions, are entirely independent of the system's state-history, as the discipline commonly used in formulating quantum systems. There are plenty of intriguing novel questions to address about the observables of closed quantum systems whose unitary evolutions are generated by the joining of conventional Hamiltonians and QMM-Hs. Here, we address only one such questions: the simplest one. We work out analytically the Near-Markovian limit of the hybrid-UE of the most elementary quantum system, namely the one-qubit closed system we have been considering from Section \ref{888}.\\\\ 
To be concrete, let us focus on the Hybrid-QMM-UE of the one-qubit closed system in which the $(2,2)$-Purely-QMM-H is deformed by the conventional Hamiltonian of a constant magnetic filed. Because the roles of the independent couplings $\mu_{t-a}$  and $\lambda^R$ in the QMM-Schr\"odinger equations \eqref{MM-evo-eqn2} are additively combined in a single coupling $\hat{\mu}_{t-a}$, we can safely restrict the monomial interactions of the two quantum memories to the second-order couplings $(\lambda^R,\lambda^I)$. The hybrid Hamiltonian of the qubit is given by,
\begin{equation}
H^{\text{Hybrid}}_{{\text{QMM}}}(t) \; \equiv \; H^{(2,2)}_{{\text{QMM}}}(t) \; + H_{\text{C}} \; = \; \lambda^R \; \{ \rho_{t-a}, \rho_t\} \;+\; i \; \lambda^I \; [\rho_{t-a}, \rho_t ] \;+\; \frac{1}{2} \; \vec{B}. \vec{\sigma} . \label{HybridHa}
\end{equation} %$(1, 1)$-QMM Hamiltonians. 
As we detailed in Subsection \ref{mcw}, and using the $(2,2)$-QMM magnetic field $\vec{{\mathcal{B}}}^{(2,2)}_{\text{QMM}}(t)$ as given in \eqref{MMMagnet}, %QMMB22 
the QMM-Schr\"odinger equation for the hybrid Hamiltonian \eqref{HybridHa} is,
\begin{equation}
\dot{\vec{r}}_t \; = \big[\; \vec{{\mathcal{B}}}^{(2,2)}_{\text{QMM}}(t) \;+\;  \vec{B}  \;\big] \;\times\; \vec{r}_t .
\end{equation}
Let us assume, with no loss of generality, that the memory-free magnetic field is coupled to only one of the Pauli matrices which we choose to be $\sigma^3$, $\vec{B} = B_z \; \hat{z}$. Recalling the general-orbit equations \eqref{MM-evo-eqn2} expressed in terms of the functions \eqref{fgp}, the one-qubit delay-differential dynamical system for the Bloch angles $(\theta_t,\phi_t)$ is given by,
\begin{align}
\label{Hybrid-evo-eqn2}
\dot{\theta}_t &= \lambda^R \; \mathcal{G}_{t-a, t} \; - \; \lambda^I \; \mathcal{F}_{t-a, t} \nonumber \\ \sin\theta_t \; \dot{\phi}_t  . &= \lambda^R \; \mathcal{F}_{t-a, t}  \;+\; \lambda^I \; \mathcal{G}_{t-a, t} + B_z \; \sin\theta_t . 
\end{align}
As explained, we only focus on the near-Markovian limit of the above Hybrid-QMM system which corresponds to arbitrarily-small values of the largest QMD. In this limit, we solve the system \eqref{Hybrid-evo-eqn2} order by order in terms of the parameter $a \ll 1$.\\\\ 
Being so, we introduce an infinite doublet-series of unknown angular functions $(\theta^{(n)}_t,\phi^{(n)}_t)$, as the to-be-obtained solutions at the Near-Markovian perturbation order: $n \leq N \to \infty$, such that the full solutions to the dynamical system \eqref{Hybrid-evo-eqn2} are so constructed, \begin{equation}\begin{split}\label{nme} &
\theta_t \; \equiv \; \sum_{n = 0}^{N \to \; \infty}\;a^n \; \theta^{(n)}_t \;\;\;;\;\;\;
\phi_t \;\equiv\; \sum_{n = 0}^{N \to \; \infty} \; a^n \; \phi^{(n)}_t  .
\end{split}\end{equation} 
Because \eqref{Hybrid-evo-eqn2} is a system of delay-differential equations involving functions $(\mathcal{F}_{t-a, t},\mathcal{G}_{t-a, t})$, the Near-Markovian expansions of the time-shifted functions $(\theta_{t-a},\phi_{t-a})$, being deduced from the above expansions, also come into play in every chosen order of perturbation. Applying the perturbative series of $(\dot{\theta}_t,\dot{\phi}_t)$ and $(\mathcal{F}_{t-a, t},\mathcal{G}_{t-a, t})$ to the LHS and the RHS of equations \eqref{Hybrid-evo-eqn2} respectively, we can analytically solve the system to arbitrarily-high order $N$ in the Near-Markovian perturbation \eqref{nme}, with the solutions $(\theta^{(n \leq N)}_t,\phi^{(n \leq N)}_t)$ being polynomials of maximum order $n$ in time. For concreteness, we have obtained the above Near-Markovian solution all the way to the order $N=6$. However, it suffices to detail the Near-Markovian solution to second order in the perturbation \eqref{nme}, as obtained according to the above procedure. The resulted $\mathcal{F}_{t-a, t}$ and $\mathcal{G}_{t-a, t}$ are given by,
\begin{equation}\label{sofandg}\begin{split}  & \mathcal{F}_{t-a, t}^{\;(\text{up to O}[a^3])} \;= \; a\; \dot{\theta}^{(0)}_t \; + \; a^2 \; \big[ \; \dot{\theta}^{(1)}_t  + \frac{1}{4}\; \sin(2 \theta^{(0)}_t) \;(\dot{\phi}^{(0)}_t)^2 - \;\frac{1}{2} \; \ddot{\theta}^{(0)}_t \;\big] , \\ & \mathcal{G}_{t-a, t}^{\;(\text{up to O}[a^3])} \;= - a\; \sin(\theta^{(0)}_t) \; \dot{\phi}^{(0)}_t \; + \\ & \; \hspace{2.1805cm} +\; a^2 \; \big[ \;  \cos(\theta^{(0)}_t)\;( \dot{\theta}^{(0)}_t\;-\;\theta^{(1)}_t)\;\;\dot{\phi}^{(0)}_t  + \frac{1}{2}\;\sin(\theta^{(0)}_t) \;\big( \;\ddot{\phi}^{(0)}_t - 2\; \dot{\phi}^{(1)}_t \big) \;\big] . \end{split} \end{equation} 
Given \eqref{sofandg}, to second order in the perturbation, the delay-differential system \eqref{Hybrid-evo-eqn2} is 
turned to the following chain of ordinary differential equations for the six unknown contributing functions
$\theta_t^{(0,1,2)} $ and $\phi_t^{(0,1,2)}$, 
\begin{equation}\label{sopeqs}\begin{split} & \dot{\theta}^{(0)}_t = 0 \;\;\;;\;\;\; B_z - \dot{\phi}^{(0)}_t = 0 ,
\\ & \dot{\theta}^{(1)}_t + \lambda^I  \dot{\theta}^{(0)}_t + \lambda^R \sin(\theta^{(0)}_t) \dot{\phi}^{(0)}_t  = 0 \;\;\;;\;\;\; \sin(\theta^{(0)}_t) \; \dot{\phi}^{(1)}_t  + \lambda^I \sin(\theta^{(0)}_t) \; \dot{\phi}^{(0)}_t  - \lambda^R \; \dot{\theta}^{(0)}_t  = 0  , \\ &  
    \dot{\theta}^{(2)}_t +  
   \lambda^I\;  \big[\; \dot{\theta}^{(1)}_t  + \frac{1}{4}  \sin(2 \theta^{(0)}_t)  (\dot{\phi}^{(0)}_t)^2 - \frac{1}{2}   \ddot{\theta}^{(0)}_t \; \big] \;+\; \\  & \hspace{.595cm} + \lambda^R \; \big[ \; \cos(\theta^{(0)}_t)  \theta^{(1)}_t  \dot{\phi}^{(0)}_t  -   \cos(\theta^{(0)}_t)  \dot{\theta}^{(0)}_t  \dot{\phi}^{(0)}_t  +  \sin(\theta^{(0)}_t)  \dot{\phi}^{(1)}_t - \frac{1}{2} \sin(\theta^{(0)}_t) \ddot{\phi}^{(0)}_t \;\big] = 0 , \\ & \sin(\theta^{(0)}_t) \;  \dot{\phi}^{(2)}_t   +  
   \lambda^I \; \big[\; 
   \sin(\theta^{(0)}_t)  \dot{\phi}^{(1)}_t - \cos(\theta^{(0)}_t)  \dot{\theta}^{(0)}_t  \dot{\phi}^{(0)}_t - \frac{1}{2} \sin(\theta^{(0)}_t)   \ddot{\phi}^{(0)}_t \; \big] + \\ & \hspace{1.786551cm} \;\;\;\; +  \lambda^R \; \big[\; \cot(\theta^{(0)}_t) \theta^{(1)}_t \dot{\theta}^{(0)}_t  - 
     \dot{\theta}^{(1)}_t  -  \frac{1}{4} \sin(2 \theta^{(0)}_t) (\dot{\phi}^{(0)}_t)^2  +  \frac{1}{2}  \ddot{\theta}^{(0)}_t \;\big]  = 0 .
\end{split} \end{equation} 
The general solutions to equations \eqref{sopeqs}, modulo angular identifications and upon ignoring the redundant constant parameters in the higher order contributions, are given by the following polynomial functions of the time variable,
\begin{equation}\label{psivbhl} \begin{split}&
\theta^{(0)}_t = \; c_1\;\;\;;\;\;\; \phi^{(0)}_t = \; B_z \; t + c_2,
 \\& \theta^{(1)}_t = - \sin
(c_1) \; \lambda^R \; B_z\; t  \;\;\;;\;\;\; \phi^{(1)}_t = \; - \;\lambda^I \;B_z \; t, \\&  \theta^{(2)}_t \; = \; \frac{1}{4}\; \sin(2 \; c_1) \; (\lambda^R)^2 \; B_z^2 \; t^2 \;+ \; \frac{1}{4}\; \big[\; 8 \;\sin (c_1) \; \lambda^I \;
\lambda^R  - \sin (2 \; c_1) \;  \lambda^I \; B_z \; \big] \; B_z \; t ,
 \\& \phi^{(2)}_t  = \frac{1}{2} \;\big[\; B_z \; \lambda_R \; \cos (c_1) \;+\; 2 \; \big( (\lambda^I)^2-(\lambda^R)^2 \big) \;\big]\; B_z \; t .
\end{split}\end{equation}
We remark that the polynomial structures of the corresponding higher order solutions, which we have obtained up to $\text{O}[a^7]$, remain as observed in \eqref{psivbhl}. For example, $\phi^{(6)}_t$ and $\theta^{(6)}_t$ are respectively found to be likewise-simple polynomials of orders five and six. Moreover, we note that the complete perturbative solution \eqref{nme}, being truncated to arbitrary order $N$ in the perturbation, is uniquely known by determining the constant doublet $(c_1,c_2)$ appearing in \eqref{psivbhl}. This doublet is in turn fixed by choosing the initial value of $(\theta_t,\phi_t)$ at $t = t_0 + a$, namely at the outset of the hybrid-QMM-UE.\\\\  
Let us end this subsection with highlighting the most important message we learn from the above investigation which comes in contrast with the corresponding aspect of purely-QMM-UE analyzed in Subsection \ref{S:IV:Strong}. The lesson is that the hybrid-QMM-UE breaks free from the lower bound of Robust Non-Markovianity: its Near-Markovian limit is dynamically non-trivial. In other worlds, the threshold of the largest-QMD $a^\star$ vanishes identically for generic hybrid-QMM-UE, such that no matter how infinitesimal the finite value of $a$ is set to be, solutions to the hybrid-QMM-Schr\"odinger equations realize unitary state-histories \eqref{densityOp} with non-constant $\rho(t')$. On the other hand, as we observe, upon setting the $|\vec{B}|$ in \eqref{HybridHa} zero which takes us back to purely-QMM-UE, the solutions \eqref{psivbhl} and all their high-order counterparts become constants-in-time. %QMM-UE \ref{Six}. 
%%%%%%%%%%%%%%%%%%%%%%%%%%%%%%%%%%%%%%
%%%%%   5 . Phases
%%%%%%%%%%%%%%%%%%%%%%%%%%%%%%%%%%%%%%%
\section{Mapping The Phase Diagrams of One-Qubit Pure States Evolving Under Purely-QMM-Hs: Numerical Investigations}\label{SV}  \subsection{Generalities About The Numerical Analyses And The Results}\label{nsg} 
The unitary state-histories which correspond to QMM-UEs are realized by solutions to QMM Schr\"odinger equations \eqref{pmmscheqwmctc} or (in their alternative equivalent form) \eqref{ase}. Mathematically, they are highly-nonlinear delay-differential equations with complicated forms. Thus, exact analytical solutions to these equations, or even `good' approximate analytic solutions to them which survive sufficiently-long times in units of the largest QMD, are rarely attainable using the known methods. This is a mathematically-known fact which we have concretely examined even in some simplest examples of the most elementary quantum setting: the closed system of only one qubit which, driven by its interacting quantum memories, evolves unitarily.\\\\ 
Therefore, we resort in the present section to (a couple of) numerical approaches which enable us to obtain generic unitarily-evolving wavefunction solutions to the one-qubit QMM Schr\"odinger equations developed by two or three interacting quantum memories. We numerically construct these generic solutions in a sufficiently-wide range of all the control parameters which quantitatively configure the setting: the QMM-H couplings, the one or the two QMM distance(s), and magnetic fields of the Kicker Hamiltonians. Carrying out such a broad numerical analysis, we succeed in mapping the complete phase diagram of the one-qubit unitary pure-state-histories generated by the least order polynomial Hamiltonians made by two interlinked quantum states chosen as $(\rho_{t-a},\rho_t)$. The specific cases, analyzed in their increasing complexities, are given as follows: the $(1,1)$, $(2,2)$ and finally $(2,3)$ QMM-H. Moreover, we implement one of these numerical methods in exploring the  example of a one-qubit closed system whose unitary evolution is generated by the simplest interactions of a triplet of its quantum memories: the $(3,3)$ QMM-UE. Mapping the phase diagram of this example is much more involved, however we are able to classify' the qualitatively-more-significant phases of the $(3,3)$ one-qubit QMM-UE. In each one of the above cases, we choose one of the quantum memories to be the present moment state of the system, and also take the QMDs to be constants. As such, we numerically solve equations \eqref{MM-evo-eq0}, \eqref{MM-evo-eqn2}, \eqref{SE23POLAR} and \eqref{SE33POLAR}.\\\\ 
The first numerical method which we employ relies on Mathematica's developed routine for numerically solving systems of coupled differential and delay-differential equations: NDSolve. The second approach is however an independent numerical solver. Here we remark briefly on the implementation and the specifics of both the approaches.\\\\ 
Let us begin with explaining the algorithms of the Mathematica codes we have used. The first remark is that the NDSolve routine has both a `standard' and a `generalized' version, and that, it is the \emph{standard version} which we utilize throughout Section \ref{SV}, except than Subsection \ref{wnetto} where, given its purpose, it suffices to use the alternative. The standard version construct numerical solutions to differential equations, using the given values of the unknown functions at an initial time, as the boundary condition. Similarly, the generalized version constructs numerical solutions to delay-differential equations, defined with a constant delay time, upon implementing the method of steps and using the given initial histories of the unknown functions as input of the procedure. As remarked above, the numerical code we have developed to construct the QMM-UEs of the one-qubit closed system uses the standard NDSolve routine, in a way that the method of steps is automatically, but only implicitly, realized by running it.\\\\ 
The code considers a single qubit whose density operator at $t_0 = 0$ is the arbitrarily chosen pure state identified with initial Bloch coordinates: $\ket{\rho_0} \longleftrightarrow (\theta_0,\phi_0)$, and then numerically the unitary time evolution of the state under the two-part Hamiltonian \eqref{MMHpure}. Unitary evolution preserves the purity of states. Therefore, the time-dependent state of the qubit remains to be uniquely coordinated by the two Bloch angles which, modulo their defining angular identifications and based on the two-part Hamiltonian \eqref{MMHpure}, must be both continuous functions of time: $(\theta_t,\phi_t) \longleftrightarrow \ket{\rho_t}$. In accordance with the total Hamiltonian \eqref{MMHpure}, the code divides the numerical construction of the time evolution of the qubit state into two successive distinctive parts. In the first part, which begins at $t_0 = 0$ and continues for a time duration of length $\Delta^{(t)}_1 = a$, the quantum memory pool is filled. In this part, the qubit-state time evolution is Markovian and is generated by a `Kicker Hamiltonian'  which we set to be that of constant magnetic field: $\vec{B} = B_y \; \hat{y}$. Immediately after having the quantum-memory pool filled, the unitary time evolution of the qubit state becomes Non-Markovian, generated by the $(N,L)$-QMM-H under consideration. This is the second part which begins from $t = a$ and goes on `forever', numerically meaning up to a sufficiently-long time $t_{\text{end}} \equiv K\; a$, with $K \gg 1$ being a few thousands in most examples.\\\\ 
This straightforward code gives us enough flexibility to range over sufficiently-wide spectra of all the control parameters of the QMM-UE of the qubit: the initial Bloch angles, the Kicker magnetic field $B_y$, the QMDs, and all the couplings of the QMM-H. As such, we have been able to efficiently explore the novel phases of the one-qubit $(N,L)$-Purely-QMM-UEs with $N \leq 3$ and $L \leq 3$. In fact, in the examples with $N < 3$, we have done the complete classification of the phases of one-qubit QMM-UEs.\\\\ 
Moreover, we have developed an independent numerical solver, in order to cross check the reliability of the above code, and also have a more direct control of the numerical solutions. It consists of a modified Runge-Kutta-4 algorithm with discrete fixed time step (see e.g.~\cite{NumericalRecipes}). To incorporate the QMM character of the evolution, the subroutines implementing a single Runge-Kutta step were modified so as to use the information of the dependent variables not only at the present time step, but also at an older time at fixed delay. We have tested both of these approaches in a wide range of configurations, especially with regard to the QMDs, and all the defining couplings of the QMM-Hs. We have found in all cases that, within acceptable numerical uncertainties, the two codes give same results, leading to the same phase diagrams of the one-qubit QMM-UEs.\\\\
Now, we begin to work out numerical constructions of the state-histories \eqref{hps} which are realized by the unitarily-evolving wavefunctions of a qubit whose quantum memories interact and compose each one of the four specific choices of the $(N,L)$-Purely-QMM-Hs \eqref{MMHparts} detailed in Section \ref{888}.
%%%%%%%%%%%%%%%%%%%%%%%%%%%%%%%%%%%%%% %%%%%%%%%%%%%%%%%%%%%%%%%%%%%%%%%%%%%%%% \subsection{Phase Diagram of Strong Non-Markovianity of Pure QMM Unitary Evolutions}%%%%%%%%%%%%%%%%%%%%%%%%%%%%%%%%%%%%%%%%%%%%%%%%%%%%%%%%%%%%%%%%%%%%%%%%%%%%%
\subsection{ The Phase Diagram of One-Qubit Wavefunctions Evolving Under The (1,1) 
Purely-QMM Hamiltonians} \label{SV-I}
\begin{figure} 
	\begin{subfigure}[b]{2.9in}
		\includegraphics[width=2.9in]{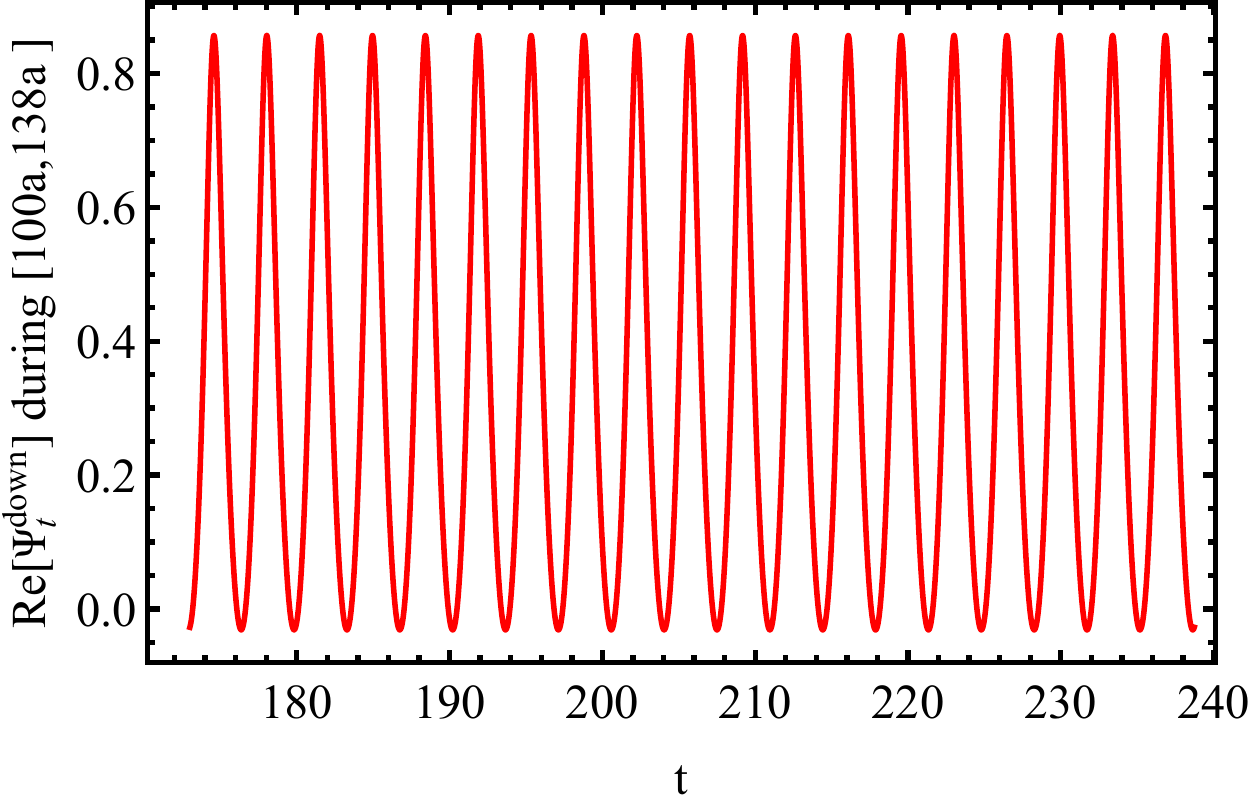}
	\end{subfigure}
	\quad
	\begin{subfigure}[b]{2.9in}
		\includegraphics[width=2.89in]{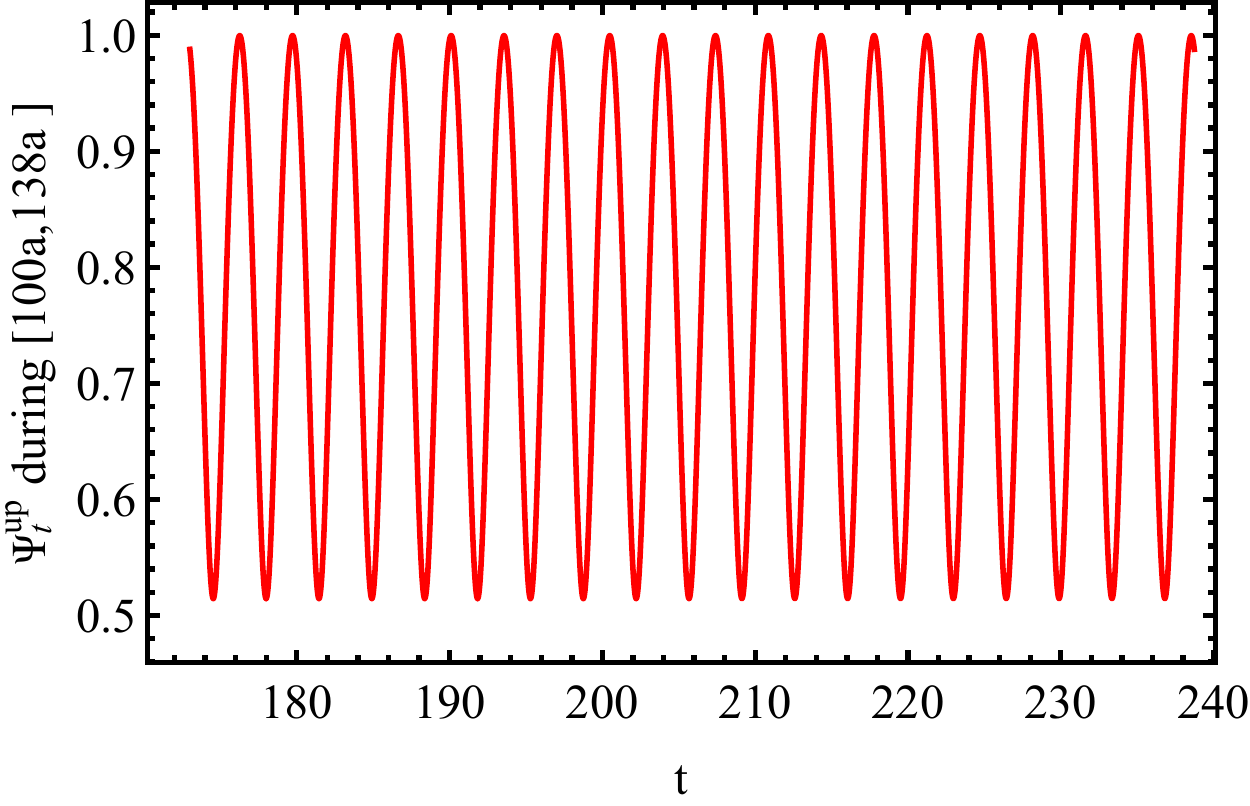}
	\end{subfigure} 
	\\
	\\
	\begin{subfigure}[b]{2.9in}
		\includegraphics[width=2.9in]{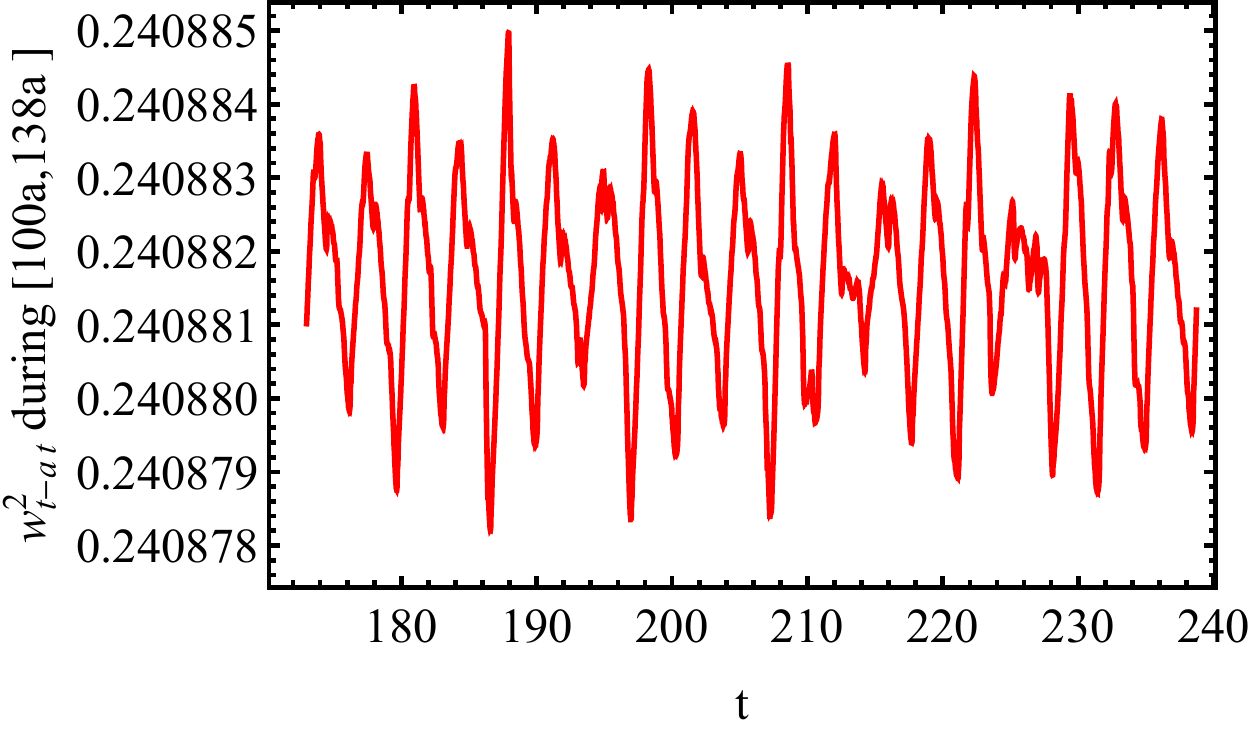}
\end{subfigure}
	\quad
	\begin{subfigure}[b]{2.9in}
		\includegraphics[width=2.9in]{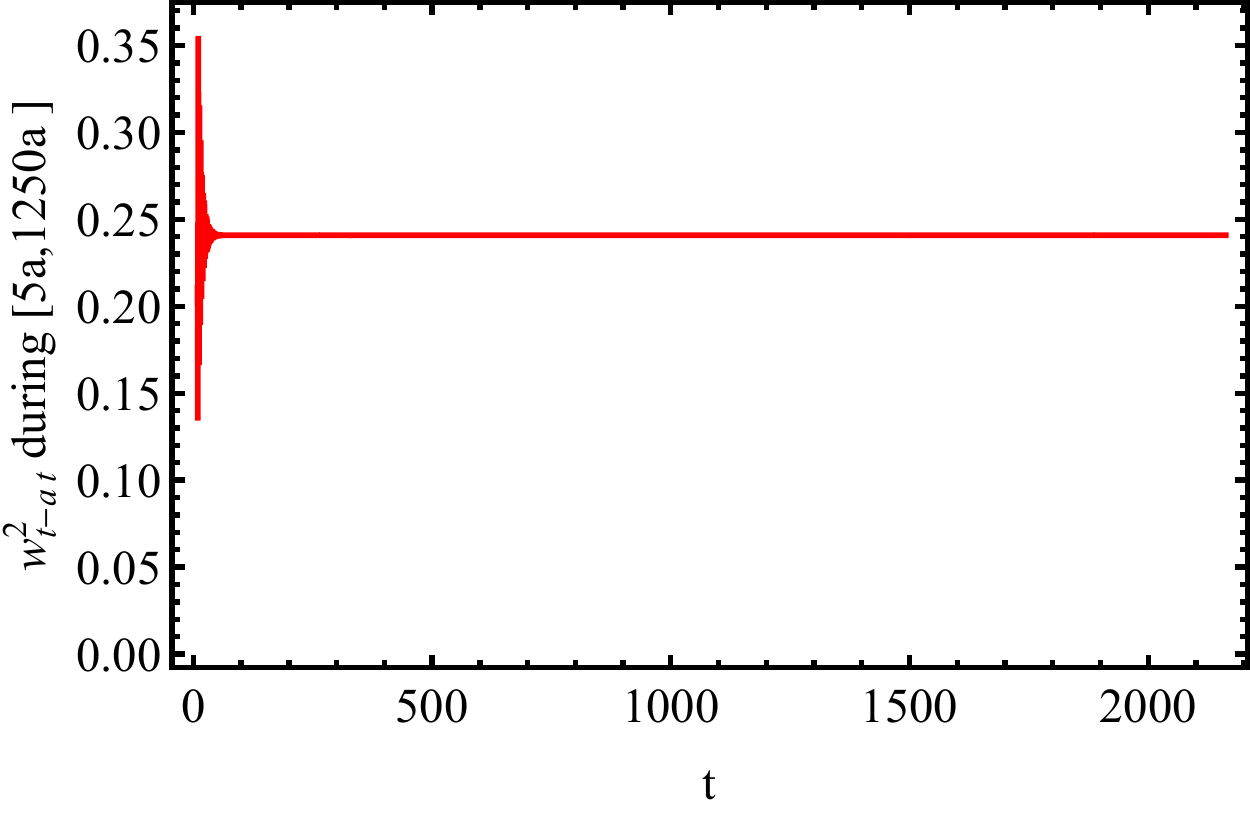}
	\end{subfigure}
	\caption{Numerical solution of the QMM-UE-System (\ref{MM-evo-eq0},\ref{fgp}) with parameters:\\
	($B^y_{\text{Kicker}} ; \theta_0 , \phi_0 | \mu \leftrightarrow \hat{\mu}_{t-a} | a  ; t_{\text{completion}} $)=($7\; ; 1.001 , 0.089 \;|\; 1.85 \; |\;  1.73 ;\; 2000 a$). %118\\
	 `Phase A' of the one-qubit $(1,1)$ QMM-UE $\cong$ `Phase Two' of the one-qubit (2,2) QMM-UE.}
	\label{fig-phaseA-11}
\end{figure}

\begin{figure} 
	\begin{subfigure}[b]{2.92in}
		\includegraphics[width=2.92in]{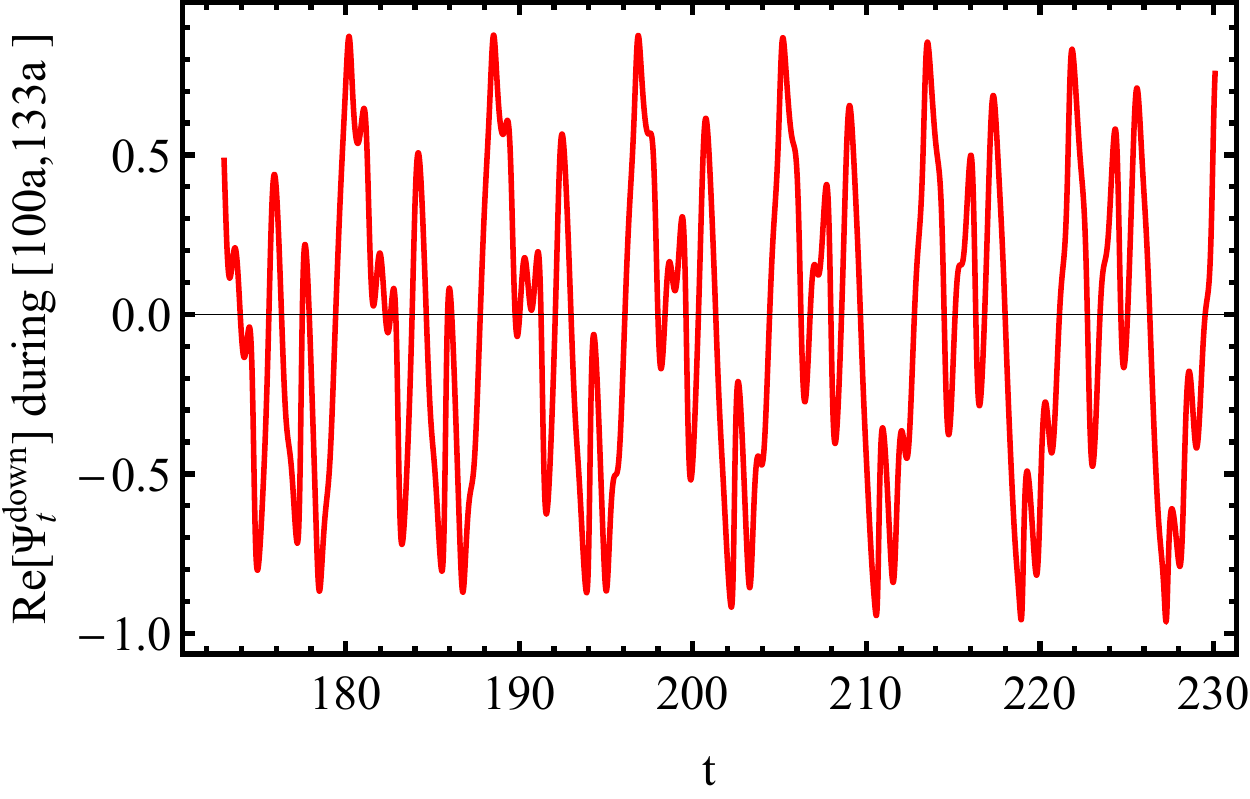}
		\caption{}
		\label{}
	\end{subfigure}
	\quad
	\begin{subfigure}[b]{2.88in}
		\includegraphics[width=2.88in]{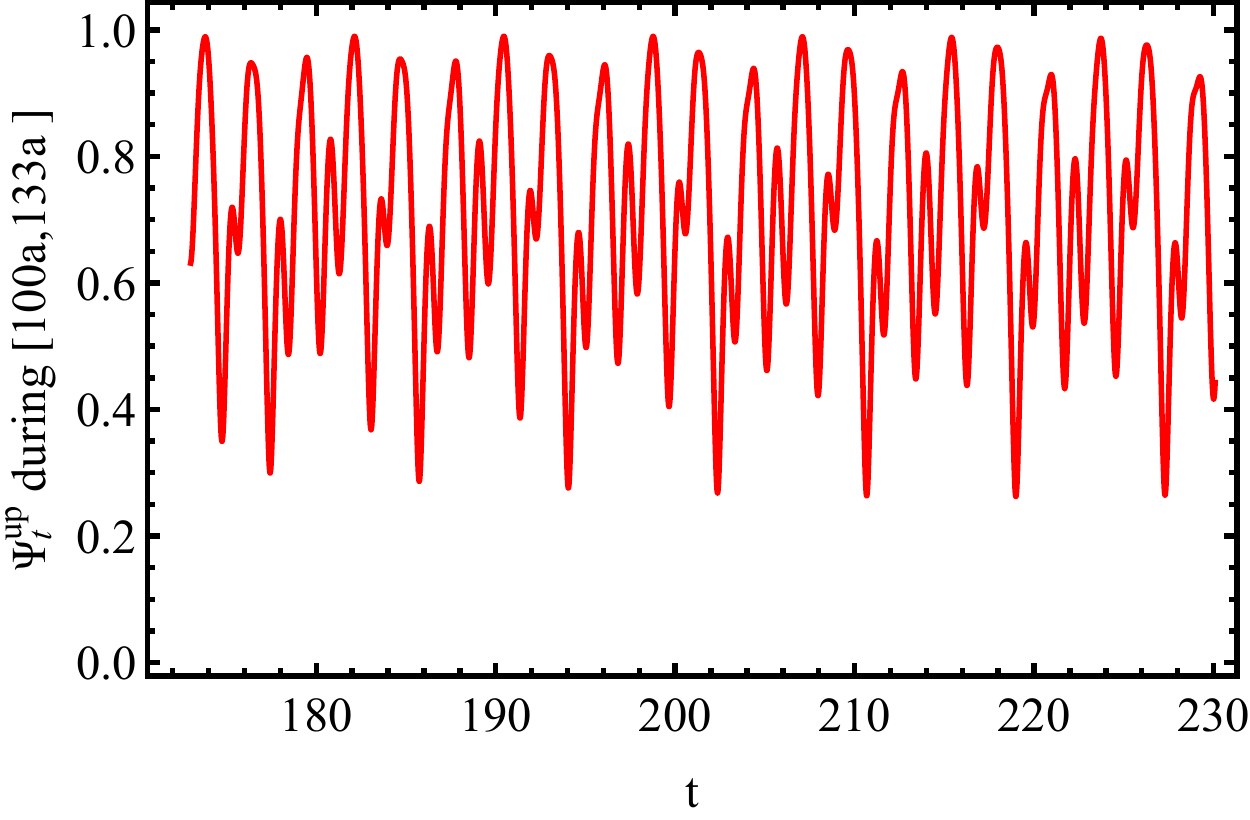}
		\caption{}
		\label{}
	\end{subfigure} 
	\begin{center}
	\begin{subfigure}[b]{2.9in}
		\includegraphics[width=2.9in]{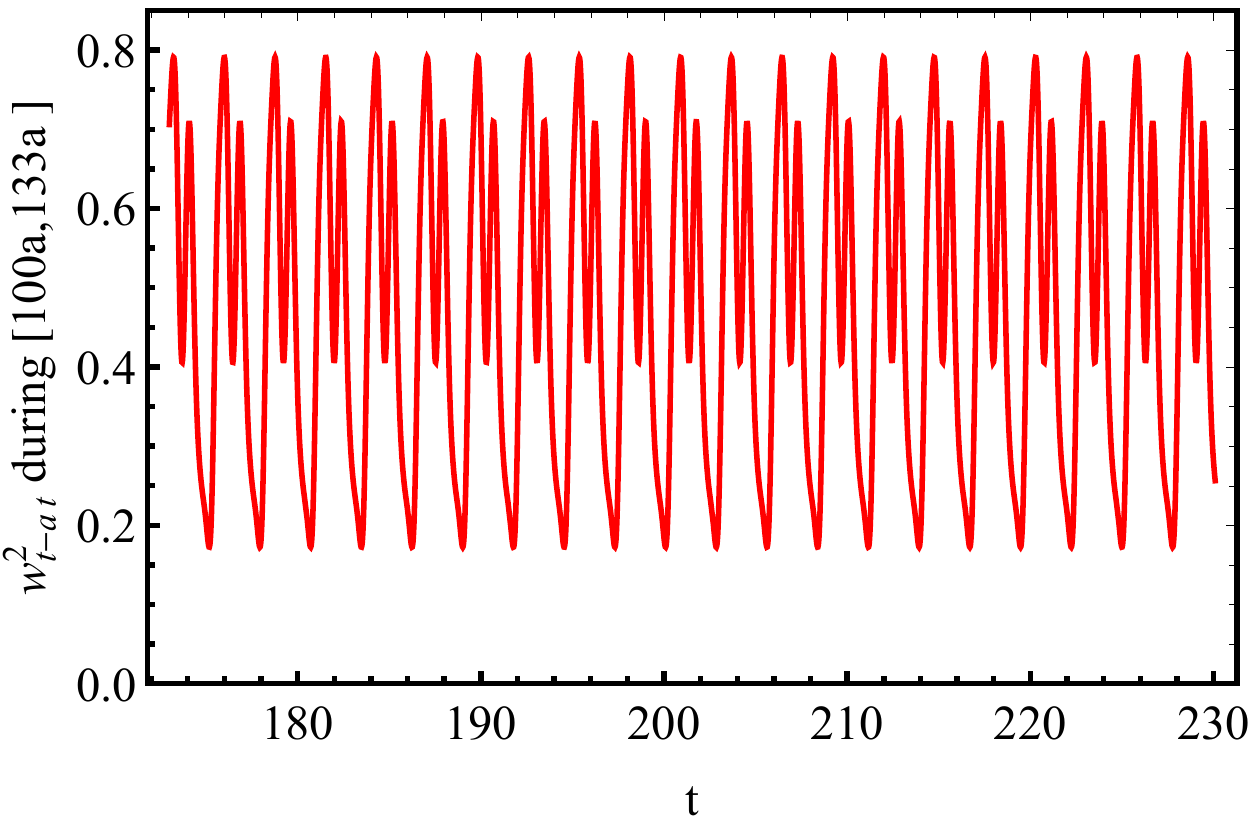}
		\caption{}
		\label{}
	\end{subfigure}
	\end{center}
	\caption{Numerical solution of the QMM-UE-System (\ref{MM-evo-eq0},\ref{fgp}) with parameters:\\
	($B^y_{\text{Kicker}} ; \theta_0 , \phi_0 | \mu \leftrightarrow \hat{\mu}_{t-a} | a  ; t_{\text{completion}} $)=($7\; ; 1.001 , 0.089 \;|\; 2.96 \; |\;  1.73 ;\; 2000 a$). %118\\
	 `Phase B' of the one-qubit $(1,1)$ QMM-UE $\cong$ `Phase Four' of the one-qubit (2,2) QMM-UE.}
	\label{fig-phaseB-11}
\end{figure}

\begin{figure}  		
\begin{subfigure}[b]{2.9in}
		\includegraphics[width=2.9in]{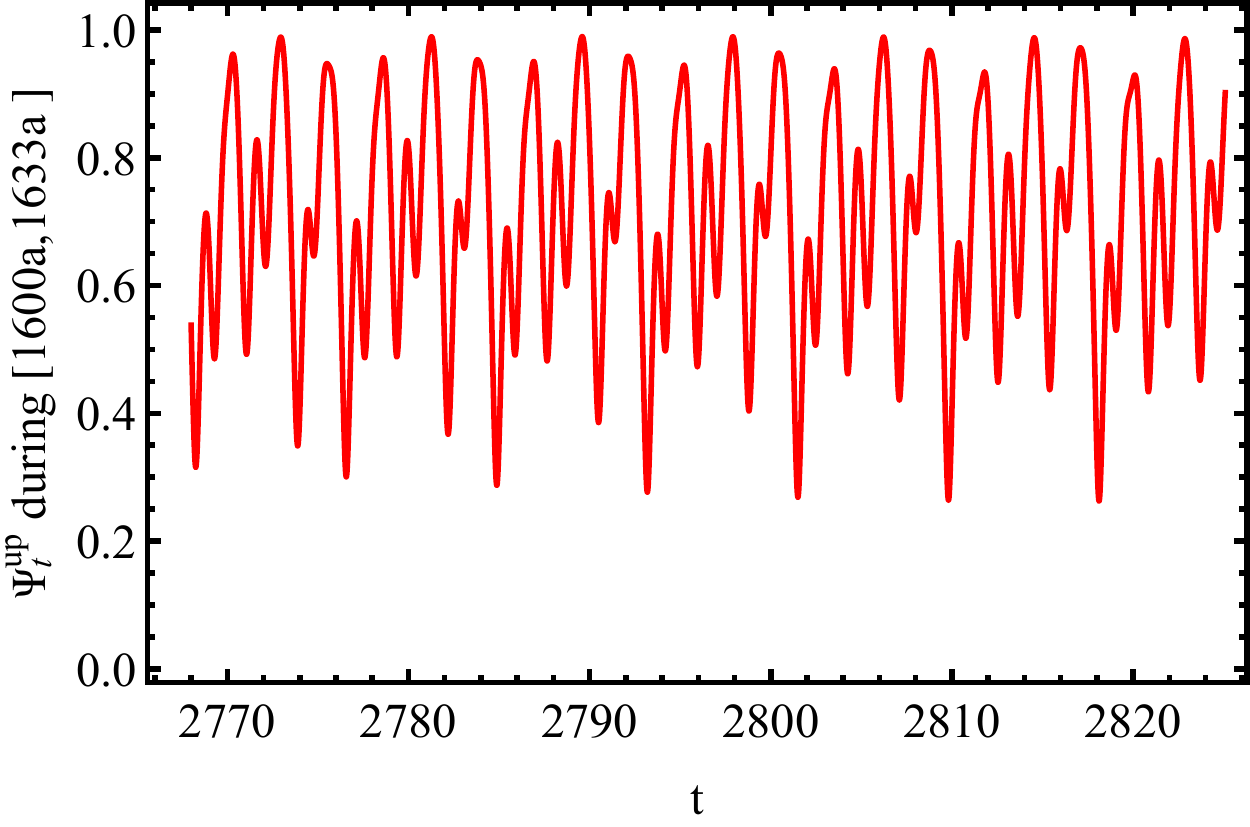}
		\caption{}
		\label{}
	\end{subfigure} 
	\hspace{.36in}
	\begin{subfigure}[b]{2.9in}
		\includegraphics[width=2.9in]{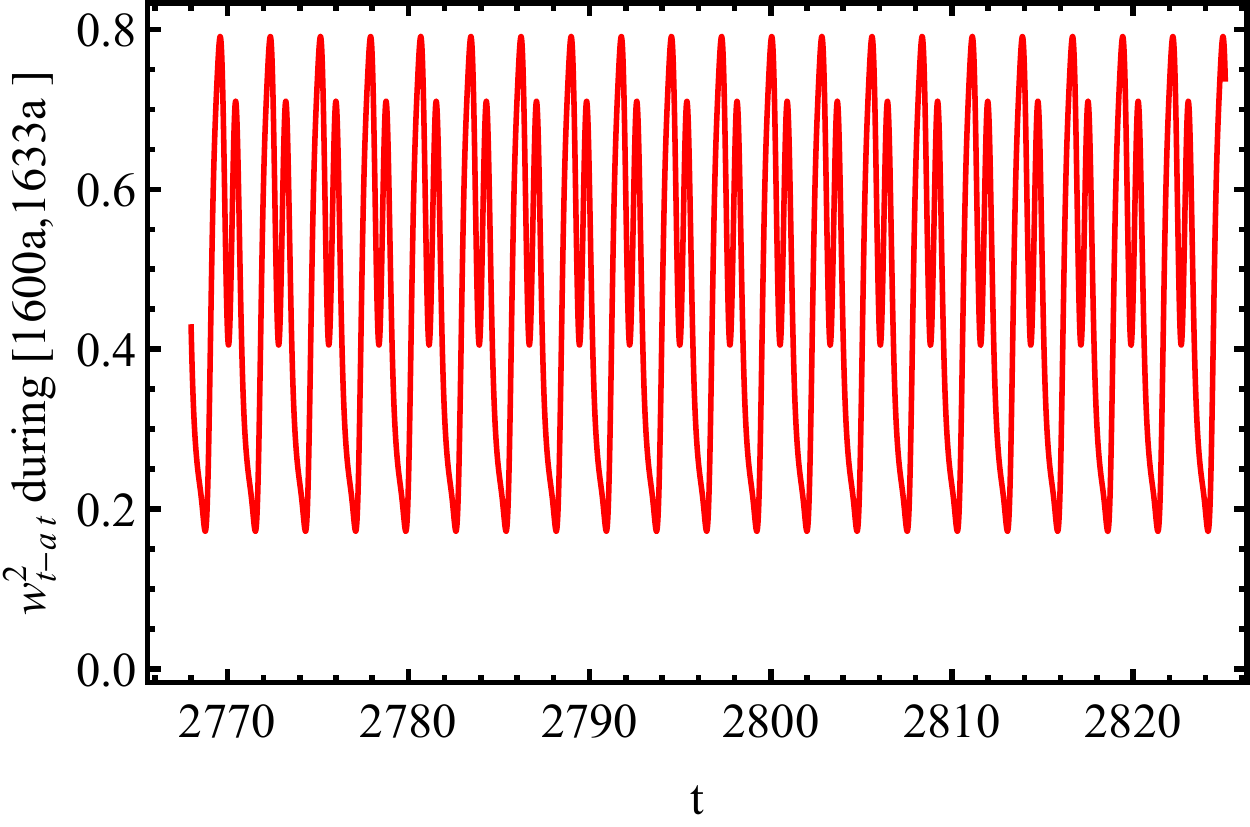}
		\caption{}
		\label{}
	\end{subfigure}
	\caption{Numerical solution of the QMM-UE-System (\ref{MM-evo-eq0},\ref{fgp}) with parameters:\\
	($B^y_{\text{Kicker}} ; \theta_0 , \phi_0 | \mu \leftrightarrow \hat{\mu}_{t-a} | a  ; t_{\text{completion}} $)=($7\; ; 1.001 , 0.089 \;|\; 2.96 \; |\;  1.73 ;\; 2000 a$). %118\\
	 `Phase B' of the one-qubit $(1,1)$ QMM-UE $\cong$ Phase Four' of the one-qubit (2,2) QMM-UE.}
	\label{fig-phaseBsz-11}
\end{figure}

\begin{figure}
	\begin{subfigure}[b]{2.88in}
		\includegraphics[width=2.88in]{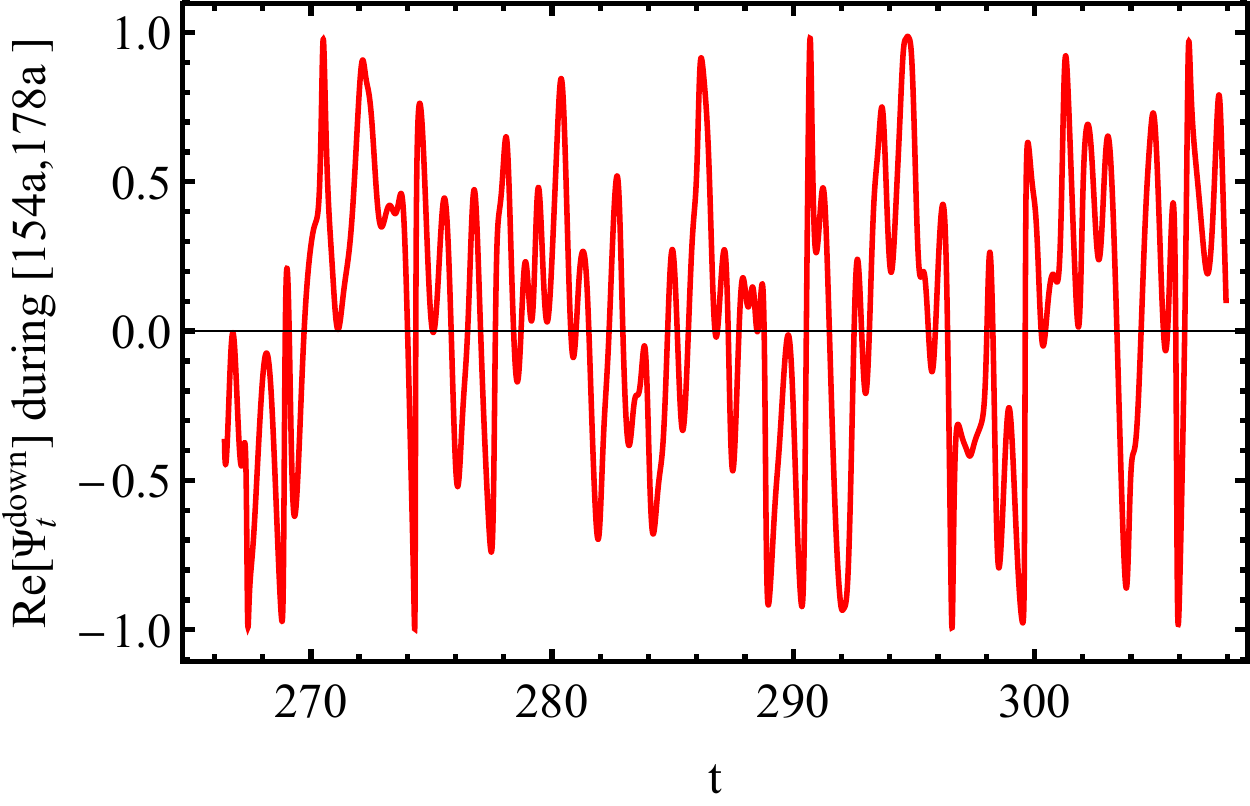}
		\caption{}
		\label{}
	\end{subfigure}
	\quad
	\begin{subfigure}[b]{2.88in}
		\includegraphics[width=2.88in]{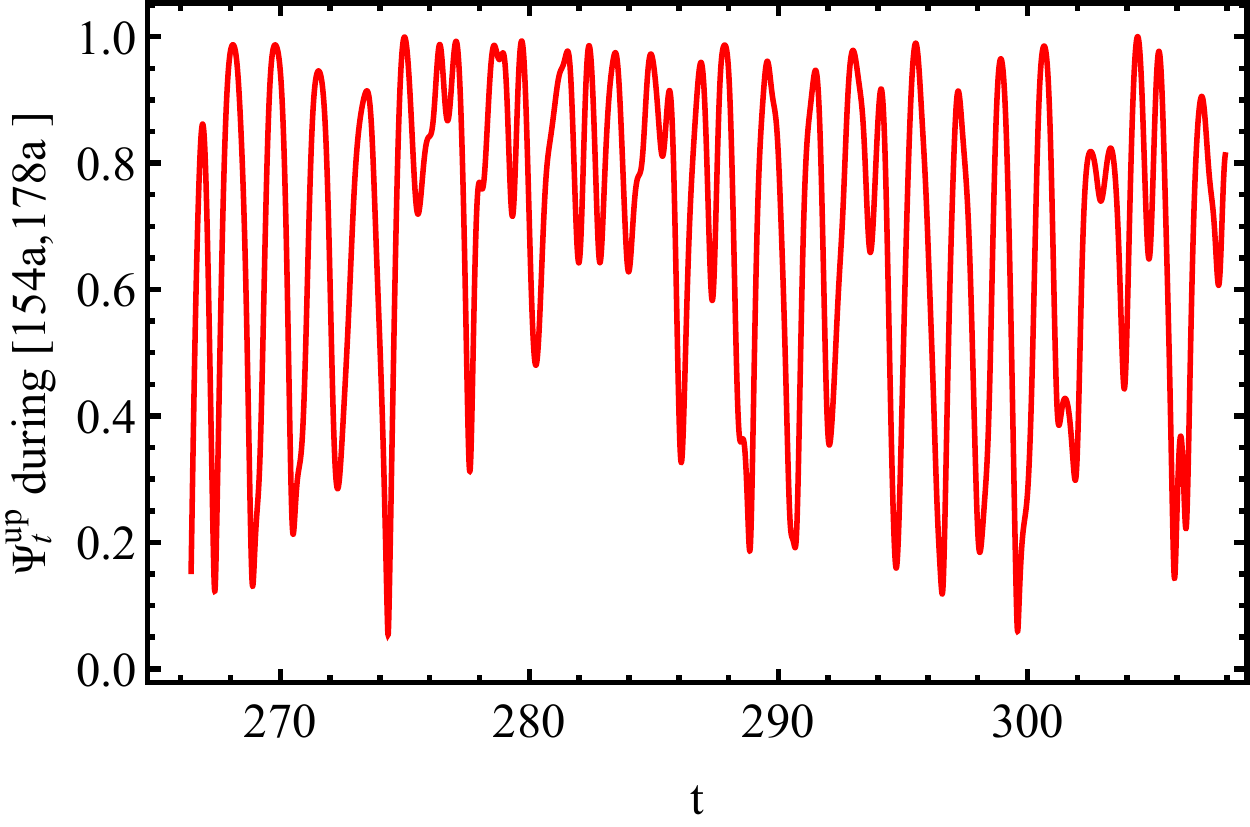}
		\caption{}
		\label{}
	\end{subfigure} 
	\begin{center}
	\begin{subfigure}[b]{2.9in}
		\includegraphics[width=2.9in]{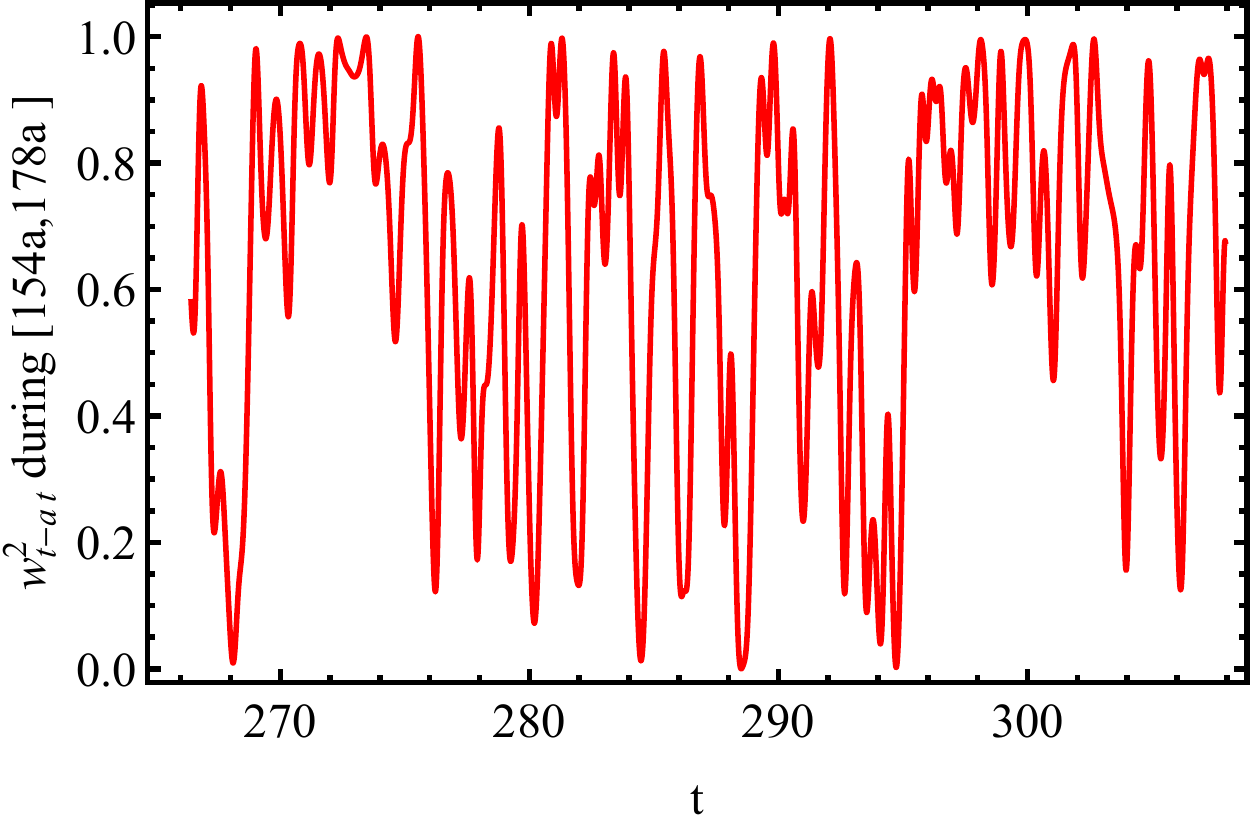}
		\caption{}
		\label{}
	\end{subfigure}
	\end{center}
	\caption{Numerical solution of the QMM-UE-System (\ref{MM-evo-eq0},\ref{fgp}) with parameters:\\ %where $\mu$ is positive.  
	($B^y_{\text{Kicker}} ; \theta_0 , \phi_0 | \mu \leftrightarrow \hat{\mu}_{t-a} | a  ; t_{\text{completion}} $)=($7 ; 1.001 , 0.089 \;|\; 5.124  \;|\;  1.73 ;\; 2000 a$). %118\\
	 `Phase C' of the One-qubit $(1,1)$ QMM-UE $\cong$ `Phase Three' of the one-qubit (2,2) QMM-UE.}
	\label{fig-phaseC-11}
\end{figure}

Let us begin with solving numerically the simplest setting, that is, the unitarily-evolving wavefunctions of the one-qubit closed system generated by the QMM-H of the type $(1,1)$. As we see in \eqref{NL}, the Hamiltonian at the present moment consists of the pure density operator of the qubit at a uniquely-specified moment in the past: that which is of a temporal distance $\Delta_t^{(t)} = a$ from the present moment. Moreover, because of the assumed purity of the unitary state-history \eqref{hps}, the Hamiltonian admits only one control parameter, being the real coupling $\mu$ introduced in \eqref{NL}. Although, in $(1,1)$ theory, one has a QMM-H which simply is a linear monomial of the system's quantum memory, the theory is actually nonlinear, interacting and also non-trivial dynamically. One really has a
physically-interacting theory because the QMM-H, although being linear in quantum memory, imposes observable interaction-type correlations between the two wavefunctions which participate in the dynamics. We can understand this fact most easily from the corresponding nonlocal-in-time and nonlinear QMM Schr\"odinger equation \eqref{MM-evo-eqn0}, or equivalently \eqref{11seqmm}, according to which $\ket{\psi_{t-a}}$ and $\ket{\psi_{t}}$ are made to interact due to the QM-TPF and its observable norm \eqref{nmtpfoquis} as explicitly given in \eqref{qmtpfica} and \eqref{nmtpfoquis}. Therefore, the phase diagram of the unitary state-histories which the $(1,1)$ theory develops is non-trivial and deserves its independent numerical exploration, as we present here in what follows. \\\\ 
We now present a careful classification of the phases by scanning over the profiles of the one-qubit-wavefunction dynamics, and the norm of its QMM-TPF, when $(\mu, a)$, the only coupling of the $(1,1)$ QMM-H and the QMD, are chosen freely in $\mathbb{R} \times \mathbb{R}^+$. For sufficiently-many choices of %$(\mu, a)$ 
the real coupling, the QMD, and all the other control parameters, the system (\ref{MM-evo-eq0},\ref{fgp}) is solved numerically for the Bloch angles $(\theta_t,\phi_t)$, applying the methods of Subsection \ref{nsg}. Briefing the procedure, the arbitrarily-chosen values of $(\mu, a ; \theta_0, \phi_0 ; B^y_{\text{Kicker}})$, and the numerically-filled quantum-memory pool of the two Bloch angles in the duration $t \in  [0 , a]$ as developed by the Kicker Hamiltonian, are fed to this nonlinear nonlocal-in-time infinite dimensional dynamical system as the input data. Henceforth, we construct the input-resulted numerical solution to \eqref{MM-evo-eq0}, beginning from the outset of the QMM-UE, that is $t = a$, up to a sufficiently-long time t = $t_{\text{completion}}$.\\\\ 
The complete solutions and the phase diagram resulting from the above procedure can be demonstrated optimally in three plots as identified now. Obviously, the outputs of solving the system (\ref{MM-evo-eq0},\ref{fgp}) are two dynamical plots: one for $\theta_t$ and one for $\phi_t$. But, the phases of the purely-QMM-UEs of the one-qubit closed system can be shown and understood in a more clear way, if (instead) two alternative plots are presented, together with a supplementary third one. As for the alternative independent plots, we choose two quantities which straightly represent the one-qubit wavefunctions \eqref{ppp}, so as to have the advantage of being more directly related to the physical observables. One-qubit wavefunctions $\ket{\Psi_t}$ have three real valued independent constituents in their vector parametrization \eqref{ppp}: the up component, $\Psi_t^{\text{up}} \equiv \cos(\frac{\theta_t}{2})$, the imaginary part of the down component: $\text{Im}[\Psi_t^{\text{down}}] \equiv \sin(\frac{\theta_t}{2}) \sin(\phi_t)$, and its counterpart, the real part of the down component: $\text{Re}[\Psi_t^{\text{down}} ] \equiv \sin(\frac{\theta_t}{2}) \cos(\phi_t)$. Because by unitarity, these three quantities satisfy: $|\braket{\Psi_t|\Psi_t}|^2 = 1$, there are two independent real valued constituents among them, with the third one being deducible. We have proved, and also specifically checked in Subsections \ref{msc} and \ref{gfp}, that QMM-Hs  generate unitary time evolutions. We choose the doublet $(\text{Re}[\Psi_t^{\text{down}}] , \Psi_t^{\text{up}})$ as the two alternative independent quantities which represent the QMM-UE of the one-qubit closed system. Moreover, the dynamics of the norm of the QM-TPF is the natural observable which serves as the discerning signature of the phases of the QMM-UE. Therefore, although being deducible from the other two quantities, it is crucial to present the plot of the QM-TPF observable for every solution.  
%one.\\\\  
On the whole, an optimal representation of the distinct phases of the wavefunction solutions to \emph{all} one-qubit QMM Schr\"odinger equations marked with delay spectrum $(\Delta^{(t)}_2,\Delta^{(t)}_1) = (0,a)$ is showing the numerical results in the figures each of which containing these three plots: one for $w_{t-a\;a}^2$, one for $\text{Re}[\Psi_t^{\text{down}} ]$ and one for $\Psi_t^{\text{up}}$. Indeed, from now on, for every numerical solution which represents one distinct phase of a one-qubit QMM-UE, we present the plots of these wavefunction quantities.\\\\ 
Now, we report and show the results of our scanning of the phase diagram of $(1,1)$ QMM-UE of the one-qubit closed system. For utter clarity, let us emphasize what we mean by `the phase diagram' of the QMM one-qubit wavefunctions for each one choice of $(N , L)$: \emph{the classification of the defining dynamical profiles of the `post-early-times' solutions to the associated QMM Schr\"odinger equations}. In the case at hand, the QMM Schr\"odinger equations correspond to the delay-differential equations (\ref{MM-evo-eq0},\ref{fgp}).\\\\ 
We begin with a specially remarkable phase of the one-qubit (1,1) Purely-QMM-UE whose representative is the solution shown in Fig.\ref{fig-phaseA-11}. The dynamical profiles of the triple $(\text{Re}[\Psi_t^{\text{down}} ],$ $\Psi_t^{\text{up}}, w_{t-a\;a}^2)$ that uniquely identify this phase are shown in the four, 3 + 1 = 4, plots of Fig.\ref{fig-phaseA-11} which, to enhance visibility, correspond to two sample temporal sub-zones, $[100 a ,138 a]$ and $[5a , 1250 a]$, of the entire numerical solution:  $ t \in[ 0 , 2000 a]$. As one sees, \emph{in this phase, the post-early-times wavefunction of the one-qubit closed system undergoes everlasting `regular' oscillations which are large and synchronized in such a way that the one-qubit QM-TP observable sticks effectively to a constant value}. All numerical solutions confirm that, in this phase, the dynamics of $w_{t-a\;t}$ is suppressed down to extremely narrow strips which are made by tiny fluctuations of almost-constant width around constant values, while the two independent components of the one-qubit wavefunction experience synchronized, large, regular oscillations. To distinguish it from the forthcoming distinct classes of all the solutions to the dynamical system (\ref{MM-evo-eq0},\ref{fgp}), this phase of unitary state-histories is to be called the phase `A' of the one-qubit (1,1) QMM-UE.  
%\begin{comment}
%\end{comment}
%\\\\ 
The numerical solution presented in Fig.\ref{fig-phaseB-11} shows another remarkable phase of the one-qubit (1,1) Purely-QMM-UE: the phase `B'.  As the representative solution of Fig.\ref{fig-phaseB-11} shows in a sample time zone of the entire numerical solution specified in the caption, the one-qubit wavefunction and its QM-TPF observable form \emph{large oscillations whose dynamical profiles are notably structured}. Specially, the hallmark of the phase B is the `pronounced' structure in the oscillations of the QM-TPF: the oscillations are \emph{formed in structural units of `bump modules' which are marked with robustly-ordered well-gapped local extrema}. In particular, in the third plot of Fig.\ref{fig-phaseB-11}, one sees that $w_{t-a\;a}^2$ picks up four ordered extrema over each bump module: the global maximum $\sim 0.79$, one local maximum $\sim 0.71$, one local minimum $\sim 0.4$ and the global minimum $\sim 0.17$. To observe how robustly these defining structures are preserved across different time scales of the corresponding QMM-UE, Fig.\ref{fig-phaseBsz-11} presents the two plots for $w_{t-a\;a}^2$ and $\Psi_t^{\text{up}}$ of the representative solution shown in Fig.\ref{fig-phaseB-11}, but in the time zone shifted by $+1500 a$.  
\\\\ 
Now, we come to the phase `C' as represented by the specific solution shown in the three plots for $\text{Re}[\Psi_t^{\text{down}} ], \Psi_t^{\text{up}}$ and $w_{t-a\;a}^2$ in Fig.\ref{fig-phaseC-11}. As in the previous cases, it suffices to show the plots in a sample zone $t \in [165 \; a , 178 \; a]$ of the total time span of the numerical solution, specified in the caption together with other control parameters, merely to enhance the fine structure visibility of the solution profile. As we see in the three plots of the Fig.\ref{fig-phaseC-11}, specifically that for $w_{t-a\;a}^2$, the first phase of the one-qubit $(1,1)$ Purely-QMM-UE is characterized by \emph{irregular oscillations} of the wavefunction and its QM-TPF observable. Indeed, all the numerical simulations performed in sufficiently wide temporal spans in this phase region show that, \emph{these unstructured oscillations of this phase go on likewise everlastingly}.
%One-qubit \\\ %\\  the one-qubit (1,1) Purely-QMM-UE \\\\ third
\\\\ 
Indeed, the fact that the $(1,1)$ QMM-UEs form a special sub-family of their $(2,2)$ counterparts is a clear statement, which is true more generally: given definition \eqref{MMHparts}, the phase diagram of any system under $(N,L)$-QMM-UE must be consistently locatable inside the phase-diagram of that system under $(N+1,L)$-QMM-UE. The other way around, one sensibly regards the phases of the $(N,L)$ system as \emph{`the building blocks'} of the `compositionally-reducible subsets' of the phases of the higher-$(N,L)$ systems. Henceforth, for clarity in both identifying and understanding the phases of all one-qubit $(N \leq 3,L \leq 3)$ QMM-UEs, we employ \emph{the five phases of $(2,2)$ one-qubit closed system}, specified in the next subsection, as a convenient `reference system'. Having it as a useful reference system at hand, we now summarize the statement of one-qubit (1,1) QMM-UE.\\\\ 
One indeed validates, by constructing sufficiently-many numerical solutions to (\ref{MM-evo-eq0},\ref{fgp}) upon varying $(\mu , a)$ and other control parameters in sufficiently-wide ranges, that the phase diagram of the (1,1) QMM-UE of the one-qubit closed system consists of the above \emph{three phases: A, B and C}. Moreover, insensitive to the sign of the real coupling $\mu$, these three phases take place in the alphabetic order: $A \to B \to C$, if one increases $a |\mu| \in \mathbb{R}^+$ monotonically. Finally, the phases A, B and C are equivalent with the three (to be identified) compositionally-reducible phases of the one-qubit (2,2) QMMUE, \emph{the phases two, four and three}, respectively. Accordingly, the representative solutions shown in Figs.\ref{fig-phaseA-11},\ref{fig-phaseB-11},\ref{fig-phaseC-11}, correspond respectively to the phases two, four and three of the one-qubit (2,2)-QMMUE.\\\\
As observed above, the particular QMM-UEs generated by (\ref{11h}) develop three classes of one-qubit state-histories with \emph{significant profile variability}. Certainly, the one-qubit $(1,1)$ system is \emph{the most elementary quantum system featuring the simplest QMM-UE}. In the landscape of all possibilities, it is the most minimalistic setting: A two-state degree of freedom is evolving unitarily as generated by momentary Hamiltonians which are essentially nothing but one of its quantum memories in its constant-distant past. To us, the fact that the most elementary implementation of QMM-UE leads to the above phase diagram, consisting of three phases with such level of distinctiveness in their defining qualities, is remarkable.
%%%%%%%%%%%%%%%%%%%%%%%%%%%%%%%%%%%%%%
%%%%%%%%%%%%%%%%%%%%%%%%%%%%%%%%%%%%%%%
\subsection{The Phase Diagram of One-Qubit Wavefunctions Evolving Under The (2,2) 
Purely-QMM Hamiltonians}\label{SV-II}
\subsubsection{The General-Orbit Phase Diagram}
\begin{figure}
	%\centering
	\begin{subfigure}[b]{2.9in}
		\includegraphics[width=2.9in]{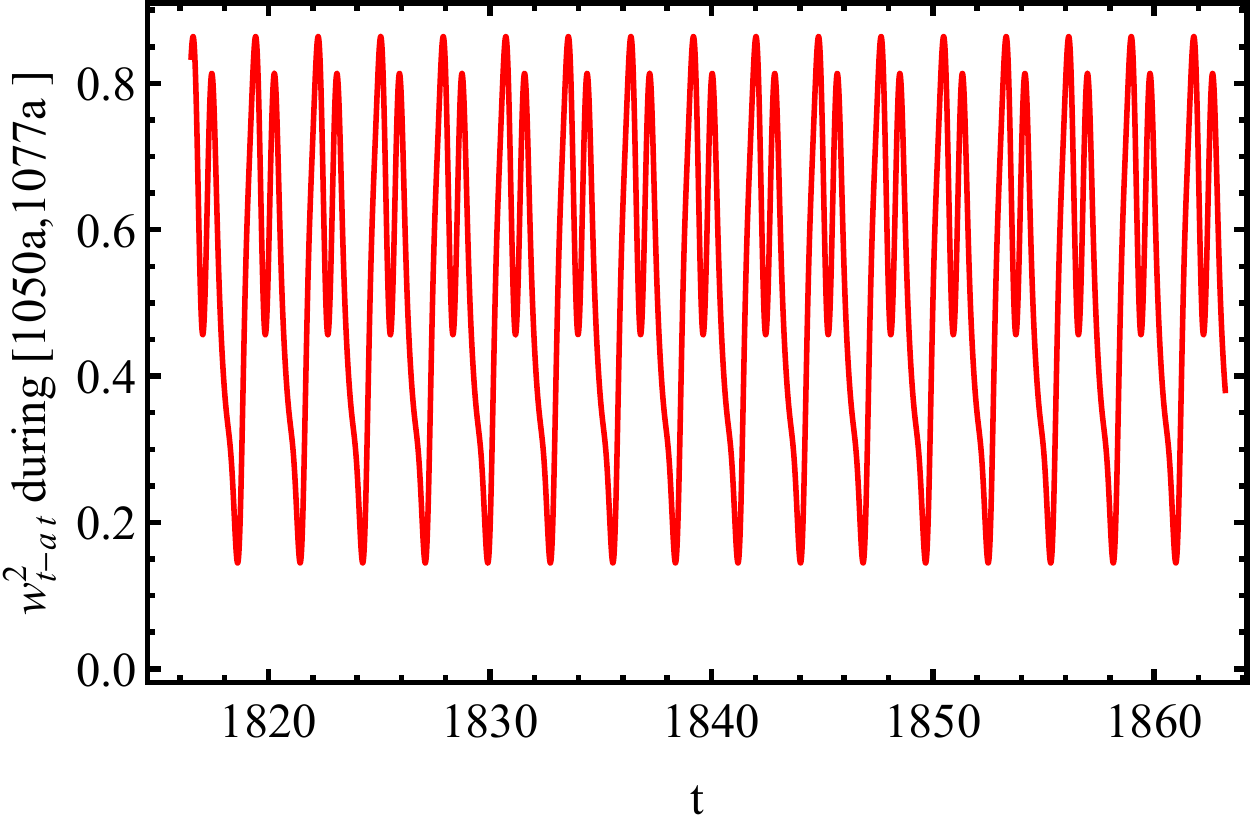}
		\caption{}
		\label{}
	\end{subfigure}
	\quad
	\begin{subfigure}[b]{3in}
		\includegraphics[width=2.9in]{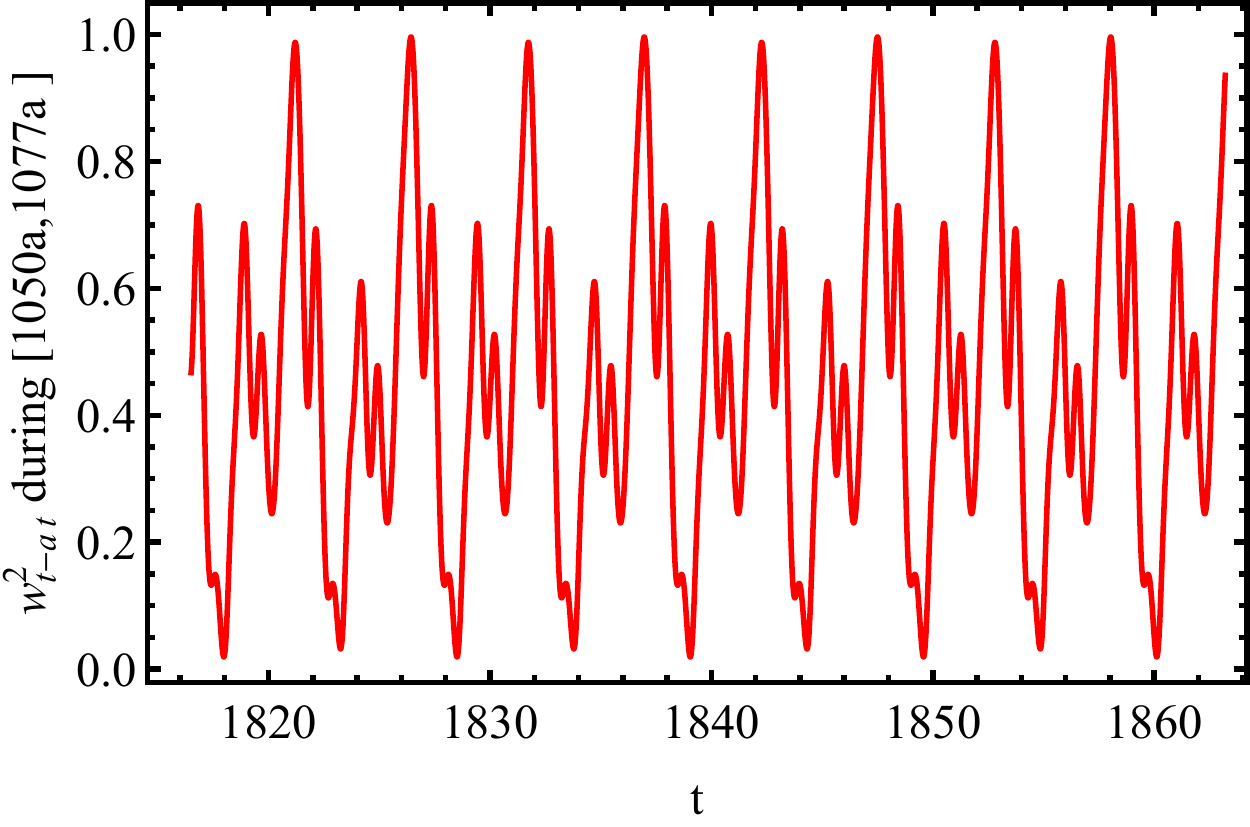}
		\caption{}
		\label{}
	\end{subfigure}
	\caption{QM-TPFs of the numerical solutions of (\ref{MM-evo-eq0},\ref{fgp}) with $(\lambda^I , \mu) = (0 , 3.202)$ in the left, and (\ref{fgp},\ref{MM-evo-eqn2}) with $(\lambda^I , \hat{\mu}_{t-a}) = ( - 0.35 , 3.202 )$ in the right, with the other parameters being given by ($B^y_{\text{Kicker}} ; \theta_0 , \phi_0 | a  ; t_{\text{completion}} $)=($2.75 ; 1.001 , 0.089 |  1.73 ; 2600\; a$). }
	\label{fig_11vs22_ABC}
\end{figure}

\begin{figure}
	%\centering
	\begin{subfigure}[b]{2.9in}
		\includegraphics[width=2.9in]{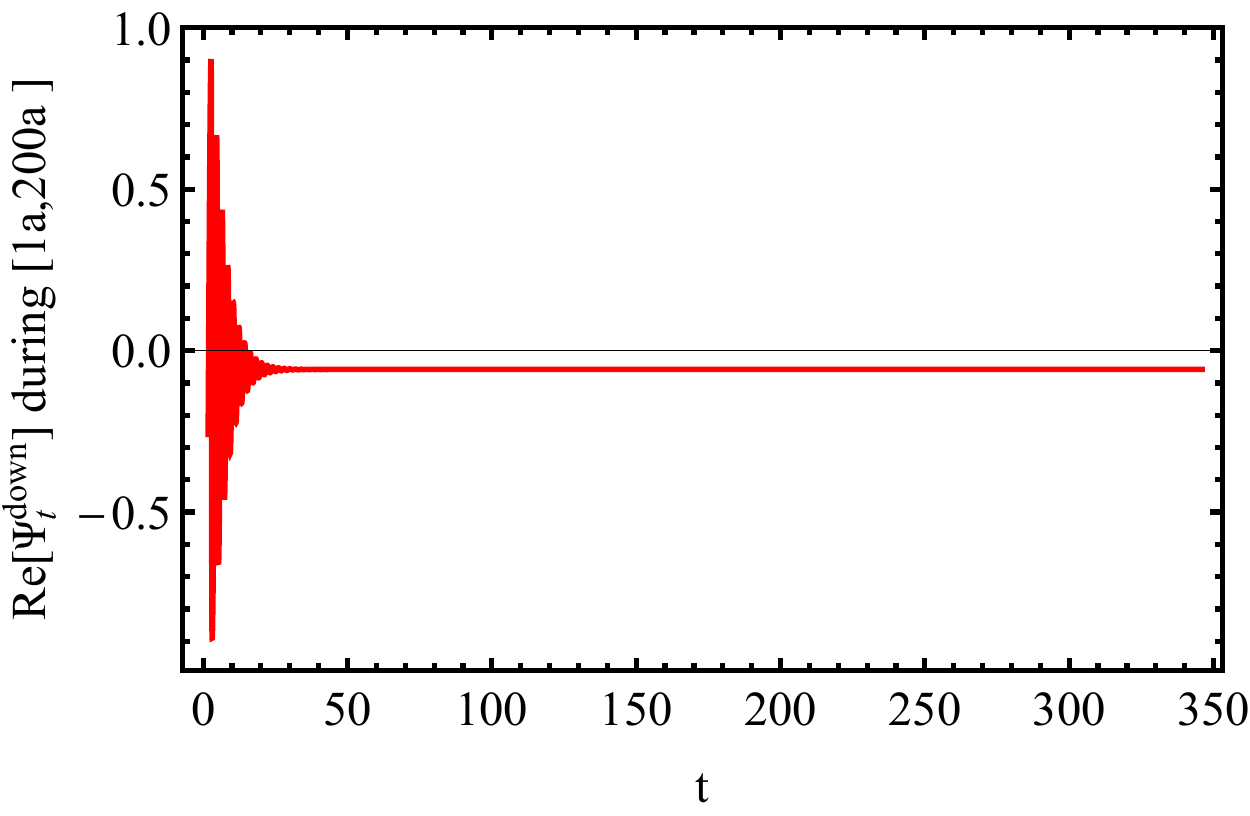}
		\caption{}
		\label{}
	\end{subfigure}
	\quad
	\begin{subfigure}[b]{3in}
		\includegraphics[width=2.9in]{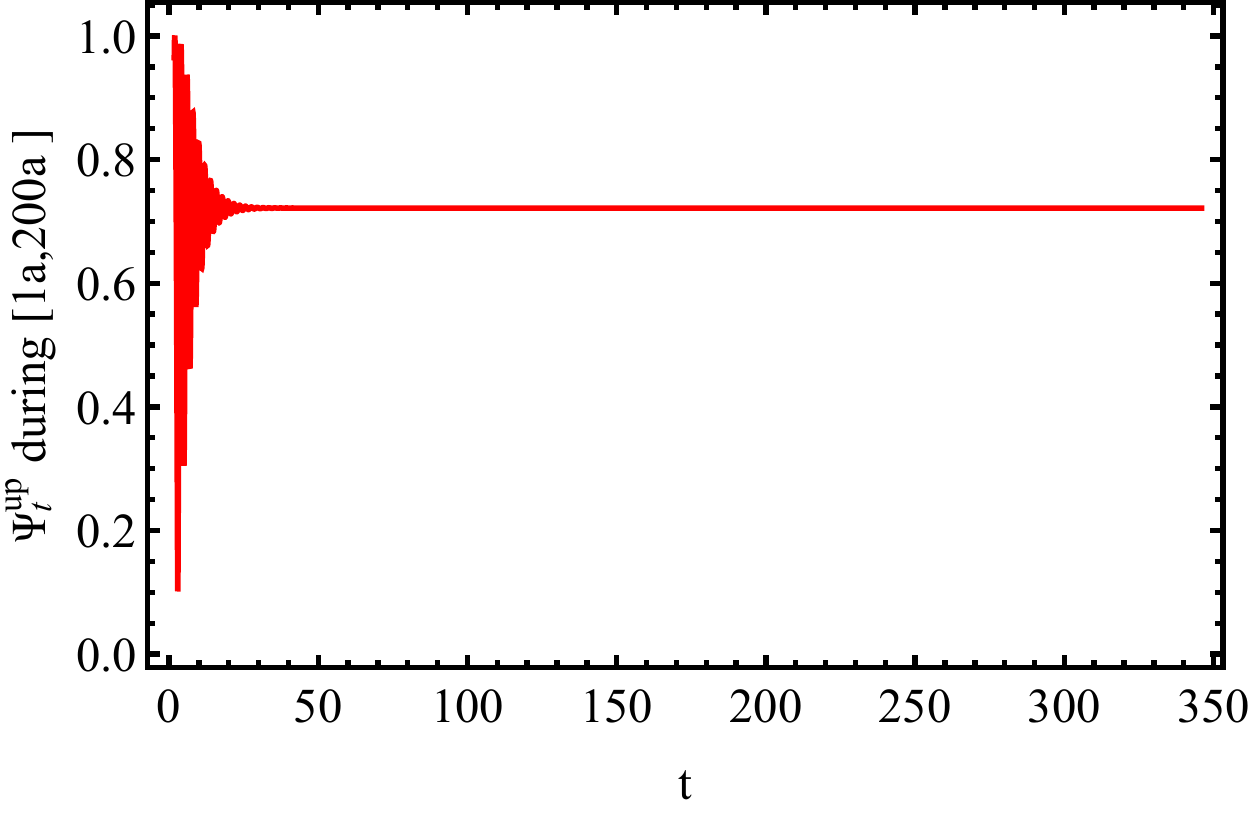}
		\caption{}
		\label{}
	\end{subfigure}
	\begin{center}
	\begin{subfigure}[b]{3.99in}
		\includegraphics[width=2.9in]{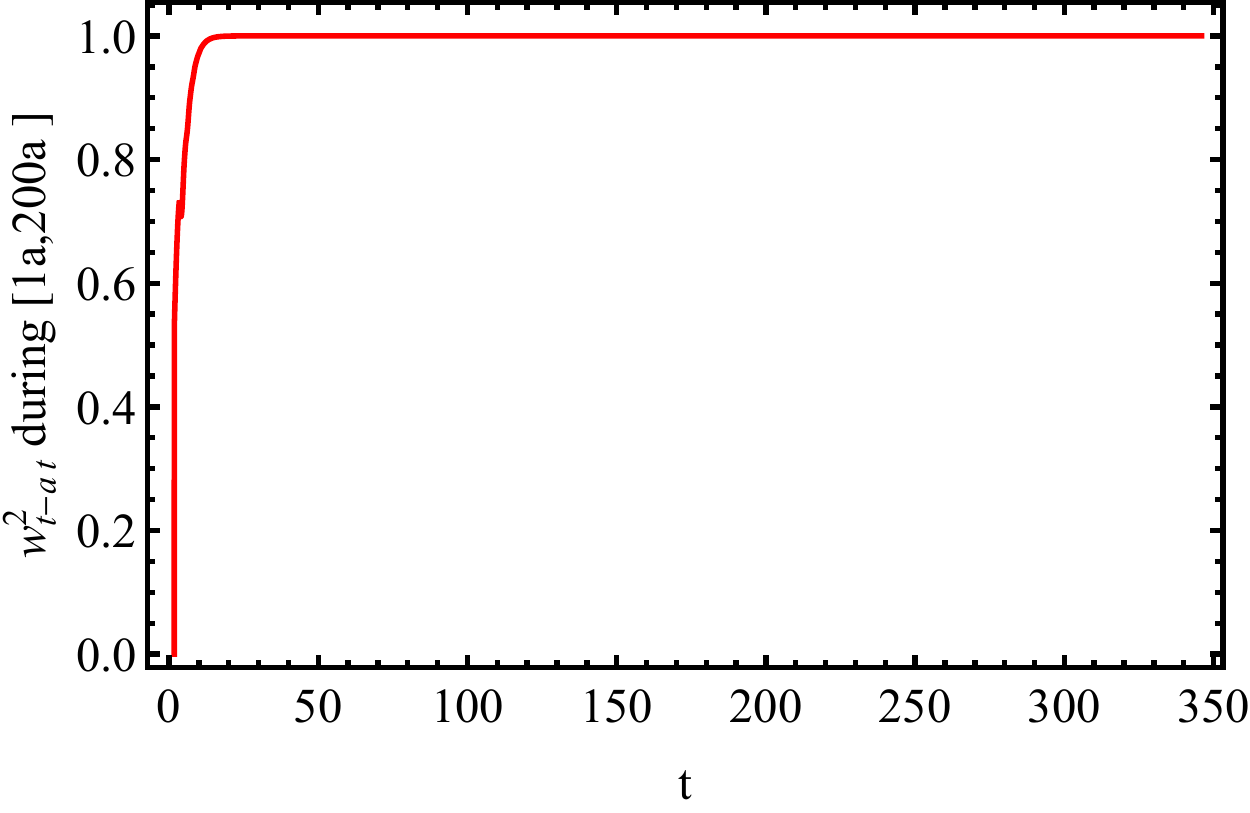}
		\caption{}
		\label{}
	\end{subfigure}
	\end{center}
	\caption{Numerical Solution of The QMM-UE system (\ref{fgp},\ref{MM-evo-eqn2}) with parameters:\\ ($B^y_{\text{Kicker}} ; \theta_0 , \phi_0 | \lambda^I , \hat{\mu}_{t-a} | a  ; t_{\text{completion}} $)=($2.75 ; 1.001 , 0.089 | 6.111 , 0.592 |  1.73 ; 2600\; a$).\\ 
`Phase 1' of one-qubit (2,2)-Purely-QMM-UE.}
	\label{fig-phase1}
\end{figure}

\begin{figure}
	%\centering
	\begin{subfigure}[b]{2.9in}
		\includegraphics[width=3.0in]{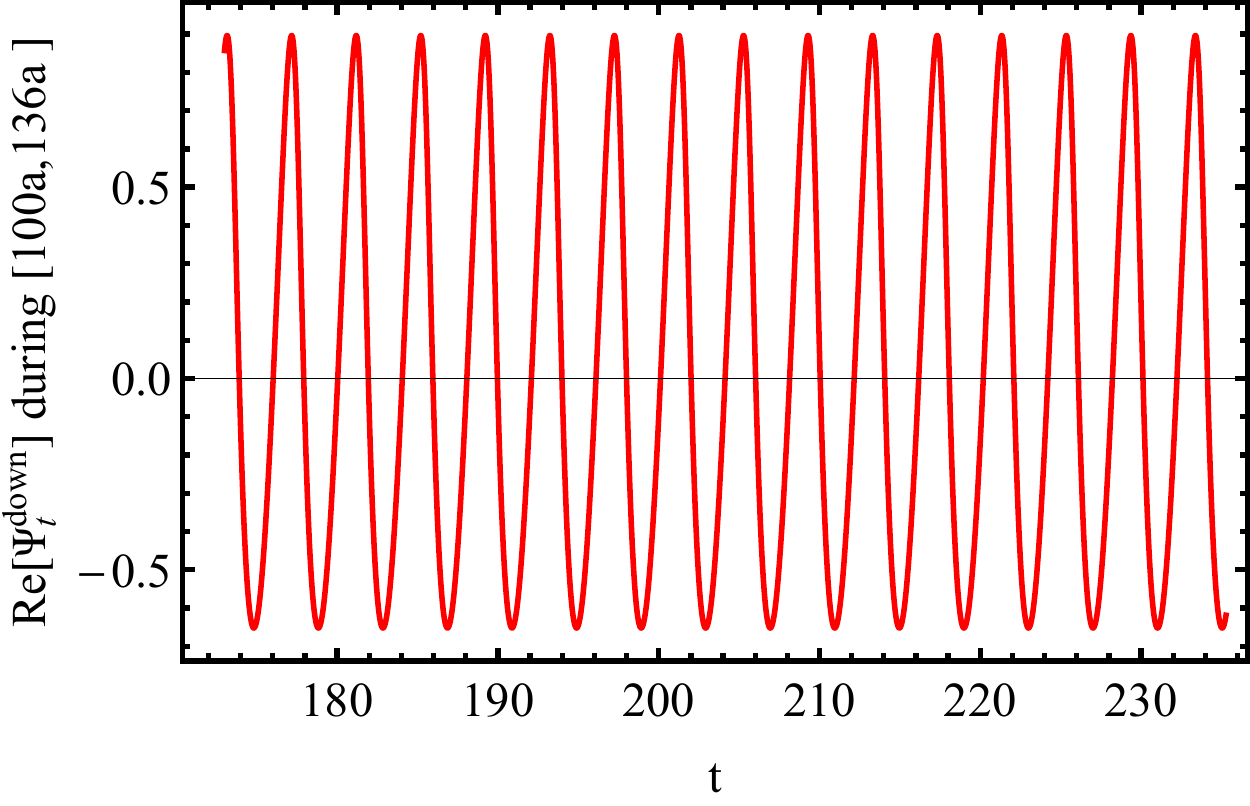}
	\end{subfigure}
	\quad
	\begin{subfigure}[b]{2.9in}
		\includegraphics[width=2.9in]{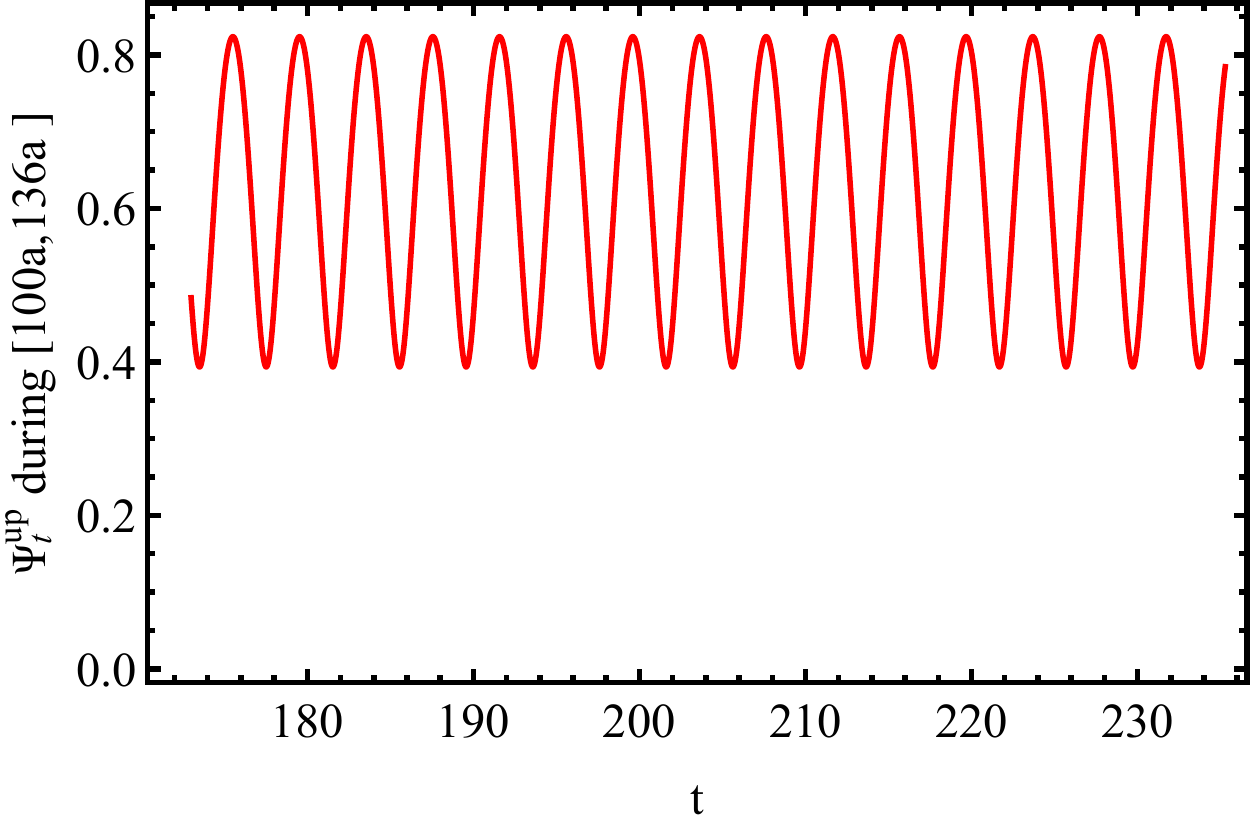}
	\end{subfigure}
	\\
	\\
	\begin{subfigure}[b]{2.9in}
		\includegraphics[width=2.7in]{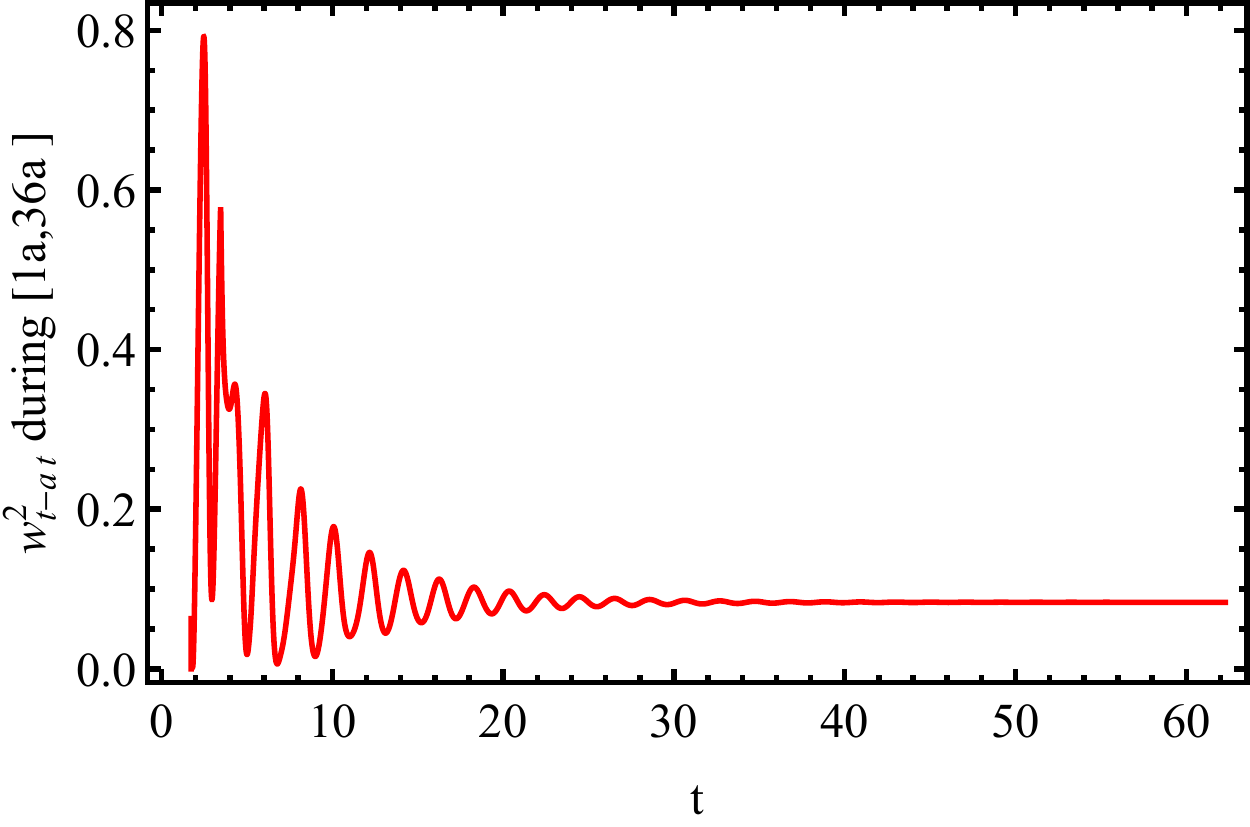}
	\end{subfigure}
	\begin{subfigure}[b]{2.9in}
		\includegraphics[width=3.0in]{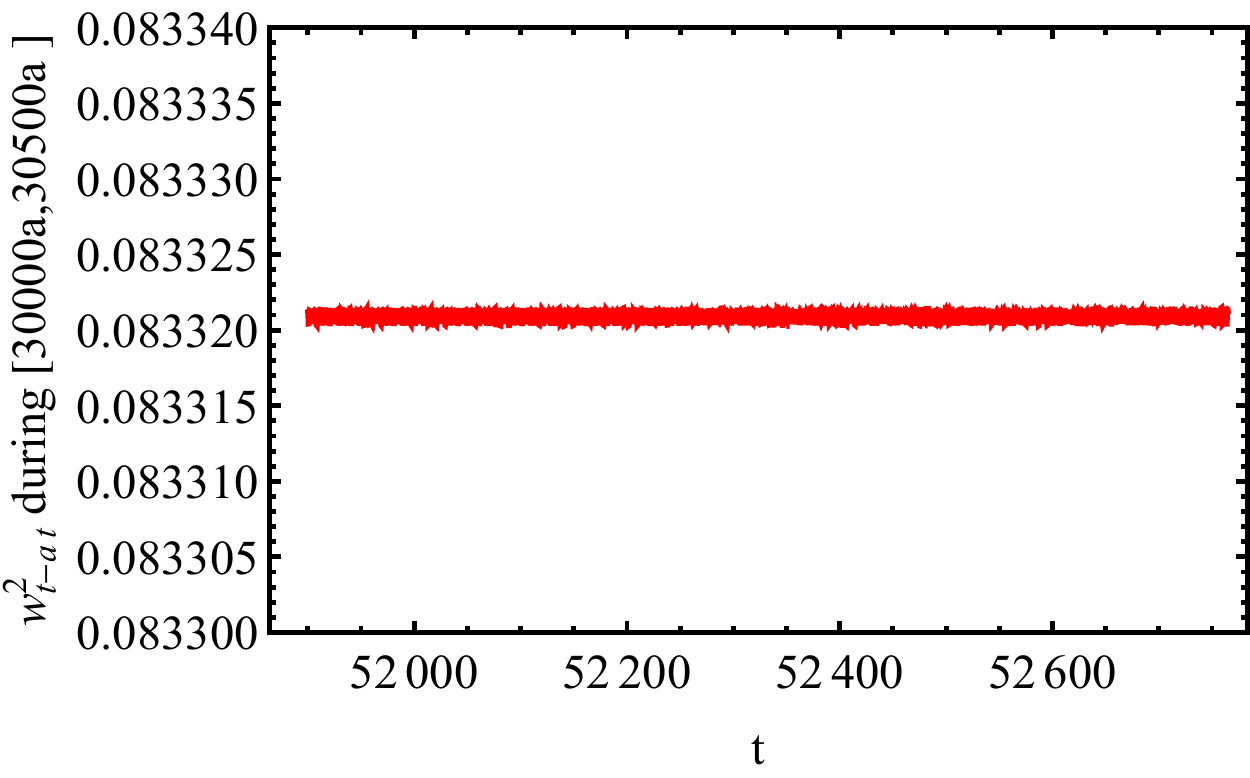}
	\end{subfigure}
	\caption{Numerical Solution of The QMM-UE system (\ref{fgp},\ref{MM-evo-eqn2}) with parameters:\\ ($B^y_{\text{Kicker}} ; \theta_0 , \phi_0 | \lambda^I , \hat{\mu}_{t-a} | a  ; t_{\text{completion}} $)=($5.75 ; 1.001 , 0.089 | -2.07 , 1.85 |  1.73 ; 6000\; a$).\\ 
`Phase 2' of one-qubit $(2,2)$-Purely-QMM-UE.}
	\label{fig-phase2}
\end{figure}

\begin{figure}
	%\centering
	\begin{subfigure}[b]{2.9in}
		\includegraphics[width=2.9in]{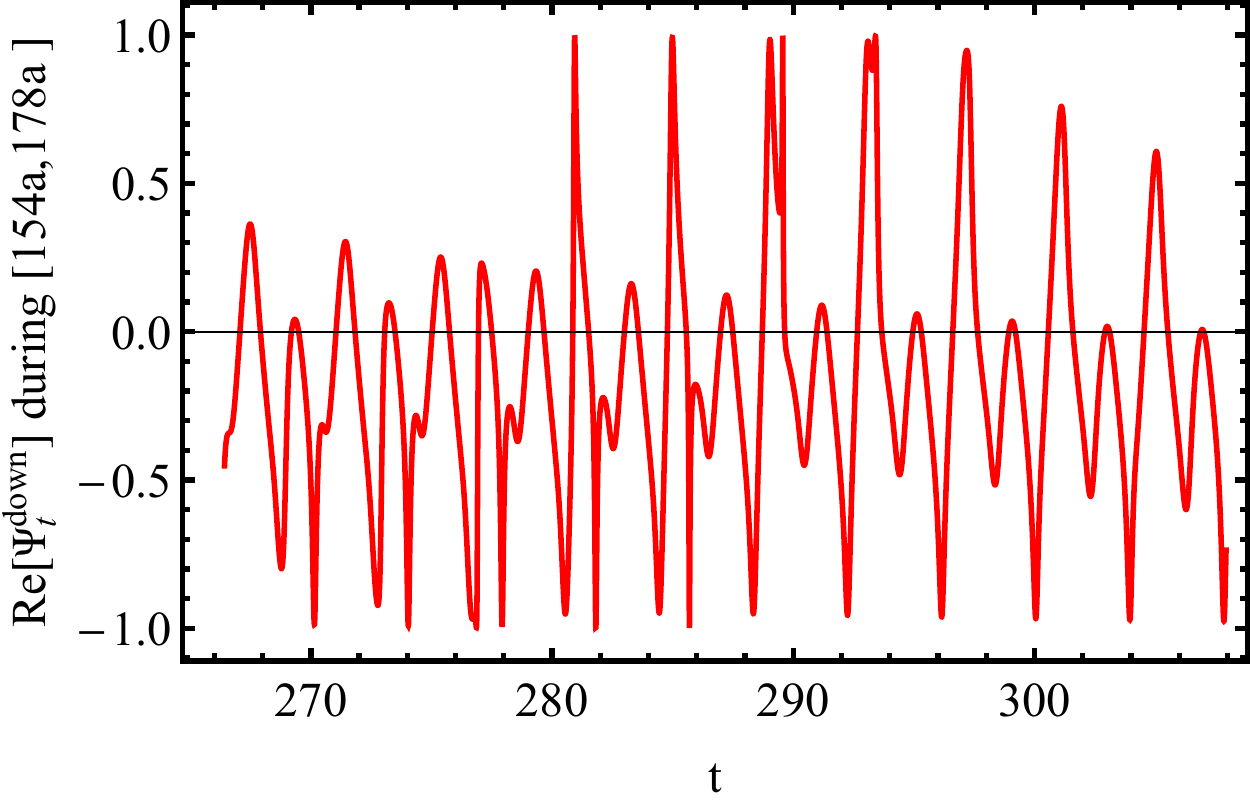}
		\caption{}
		\label{}
	\end{subfigure}
	\quad
	\begin{subfigure}[b]{2.9in}
		\includegraphics[width=2.8in]{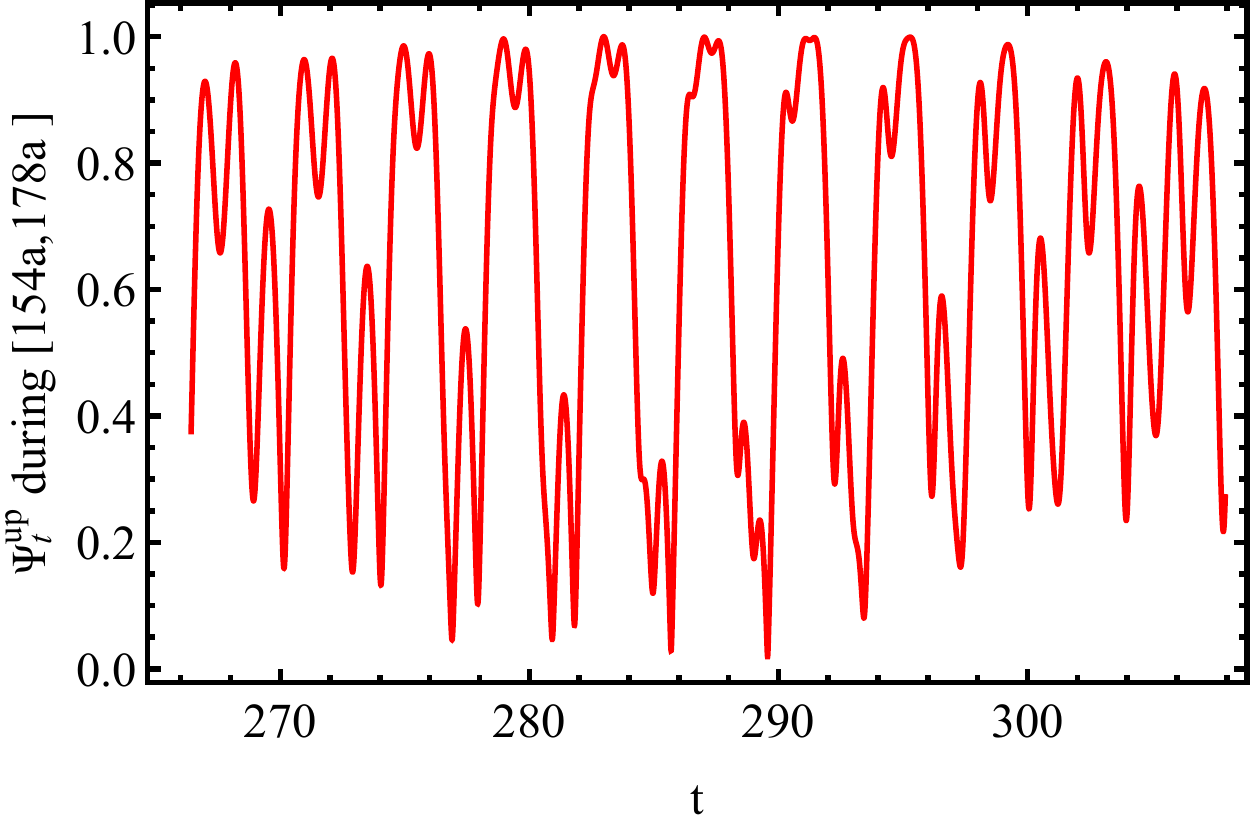}
		\caption{}
		\label{}
	\end{subfigure}
	\begin{center}
	\begin{subfigure}[b]{2.9in}
		\includegraphics[width=2.9in]{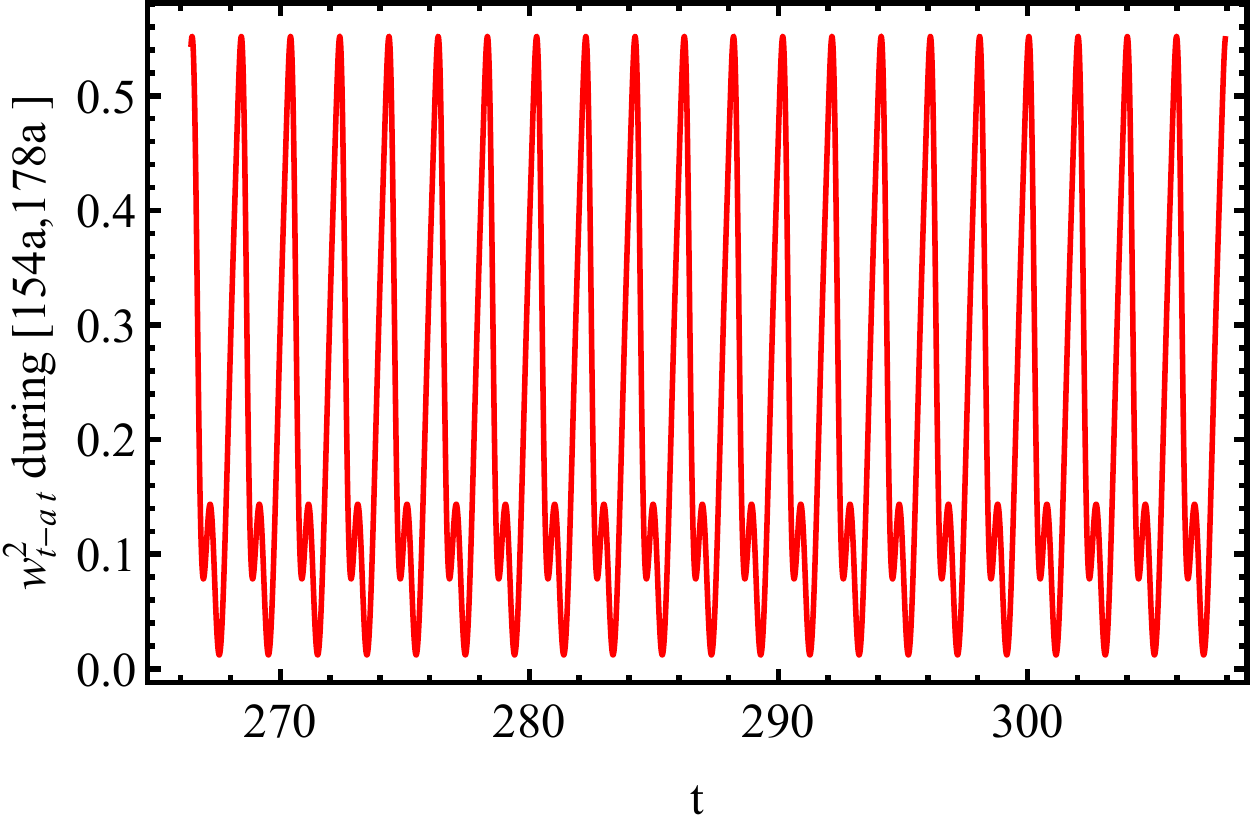}
		\caption{}
		\label{}
	\end{subfigure}
	\end{center}
	\caption{Numerical Solution of The QMM-UE system (\ref{fgp},\ref{MM-evo-eqn2}) with parameters:\\ ($B^y_{\text{Kicker}} ; \theta_0 , \phi_0 | \lambda^I , \hat{\mu}_{t-a} | a  ; t_{\text{completion}} $)=($7 ; 1.001 , 0.089 | -2 %6.111 
, 3.36 |  1.73 ; 2200\; a$).\\ `Phase 4' of one-qubit $(2,2)$-Purely-QMM-UE. }
	\label{fig-phase4 QMM}
\end{figure}

\begin{figure}
	\begin{subfigure}[b]{2.9in}
		\includegraphics[width=2.9in]{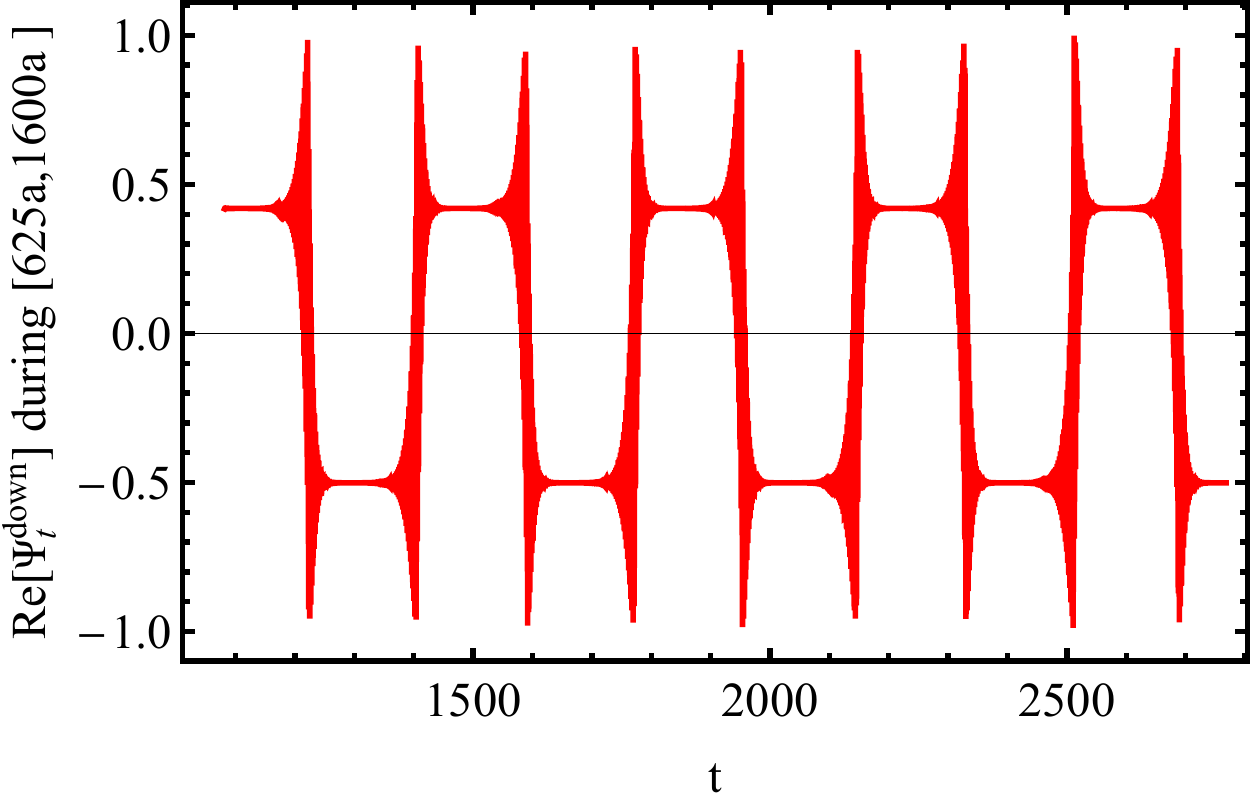}
		\caption{}
		\label{}
	\end{subfigure}
	\quad
	\begin{subfigure}[b]{2.9in}
		\includegraphics[width=2.8in]{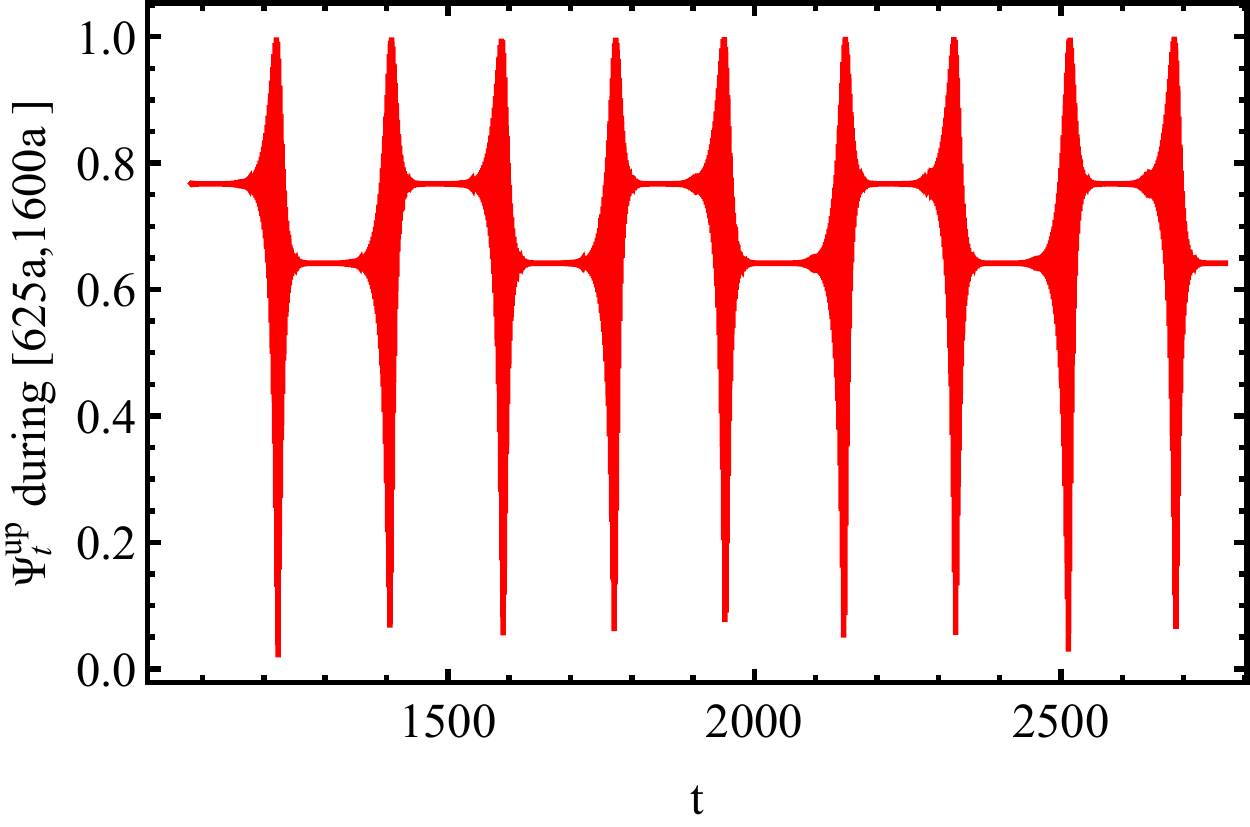}
		\caption{}
		\label{}
	\end{subfigure}
	\begin{center}
	\begin{subfigure}[b]{2.9in}
		\includegraphics[width=2.9in]{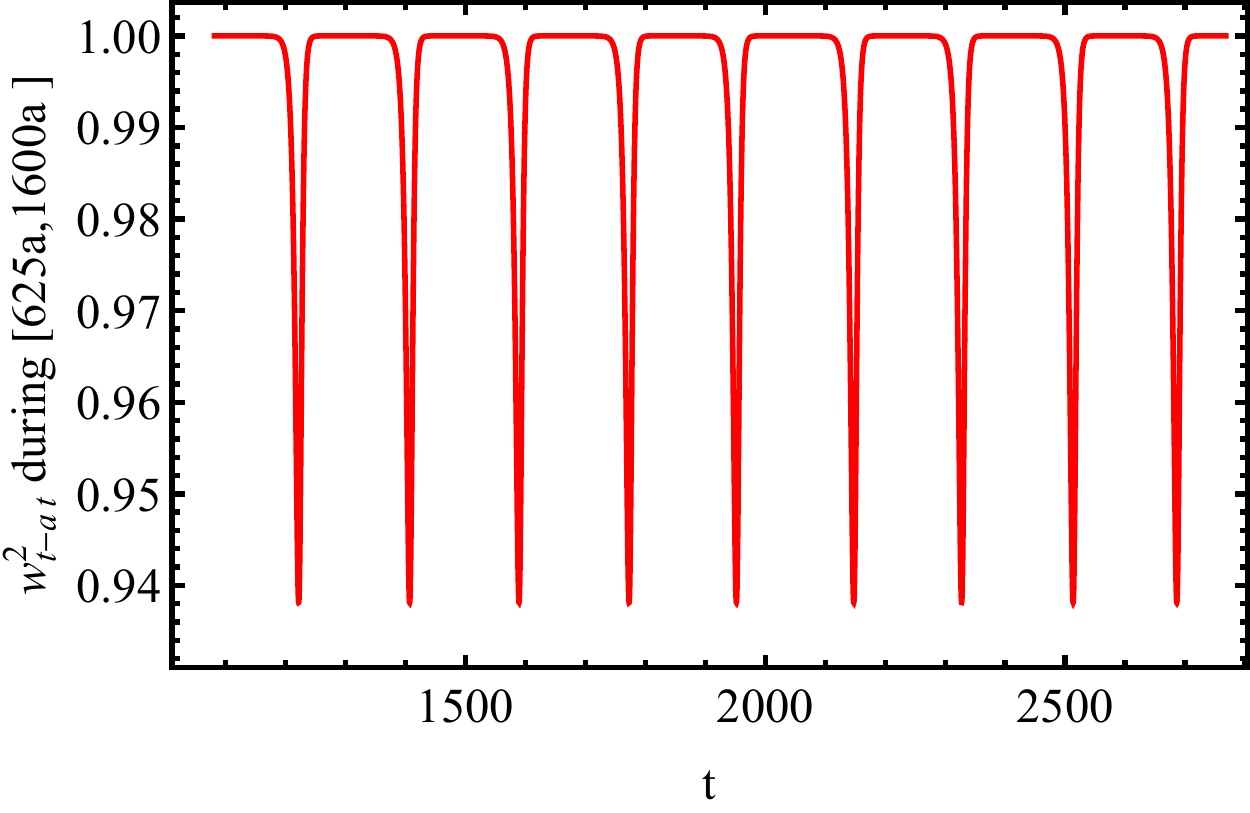}
		\caption{}
		\label{}
	\end{subfigure}
	\caption{Numerical Solution of The QMM-UE system (\ref{fgp},\ref{MM-evo-eqn2}) with parameters:\\ ($B^y_{\text{Kicker}} ; \theta_0 , \phi_0 | \lambda^I , \hat{\mu}_{t-a} | a  ; t_{\text{completion}} $)=($7.5 ; 1.001 , 0.089 | 5.921 , 3.786 |  1.73 ; 3600\; a$).\\ `Phase 5' of one-qubit (2,2)-Purely-QMM-UE.} \label{fig-phase5}
	\end{center}
\end{figure}

Now we turn to Unitary Non-Markovian dynamical system (\ref{fgp},\ref{MM-evo-eqn2}) and construct its numerical solutions, ranging over the spectra of its control parameters, to explore and make a complete map of its phase diagram. The novel dynamical system (\ref{fgp},\ref{MM-evo-eqn2}) is the nonlocal-in-time nonlinear Schr\"odinger equation for the (2,2)-Purely-QMM-UE which is given in (\ref{tqmmse},\ref{ltt}), applied to the closed system of one single qubit to become \eqref{MM-evo-eqn3}, and re-written as the Non-Markovian unitary time evolution of the one-qubit Bloch polar coordinates: $(\theta_t,\phi_t)$. To conclude, the delay-differential equations \eqref{MM-evo-eqn3} describe how pure states of a qubit unitarily evolve under the first-and-second degree monomials made of the chosen pair of its quantum memories: $\rho_{t-a}$ and $\rho_t$.\\\\ 
To construct solutions to the delay-differential equations (\ref{fgp},\ref{MM-evo-eqn2}), we follow the procedures detailed in Subsection \ref{nsg}. Moreover, we check and indeed confirm that, a) there always is a finite threshold $a^\star(\hat{\mu}_{t-a}, \lambda^I) $ as the strict lower bound of the largest QMD of any dynamical solution, b) if $\lambda^I$ is the only QMM-H coupling, the numerical threshold and the analytical determination \eqref{sct} are consistent. To characterize all one-qubit phases optimally, the numerical solutions are shown in three (`two + one') plots:  $(\text{Re}[\Psi_t^{\text{down}} ],\Psi_t^{\text{up}})$, supplemented with QMM signature of the corresponding phase of the system, being $w_{t-a\;a}^2$. Ranging over sufficiently-wide spectra of all the control parameters: $a, \theta_0, \phi_0, B^y_{\text{Kicker}}, \lambda^I, \hat{\mu}_{t-a} , t_{\text{completion}}$, the complete map of the phase diagram made by solutions to (\ref{fgp},\ref{MM-evo-eqn2}) is presented in what follows.\\\\ 
One firstly sees that, upon the following mapping of the associated couplings: $\hat{\mu}_{t-a} \leftrightarrow \mu$, the $(2,2)$ one-qubit QMM-UE becomes mathematically equivalent with one-qubit $(1,1)$ system: QMM-Hs (\ref{11h},\ref{QMMH22}), and their QMM Schr\"odinger equations (\ref{MM-evo-eq0},\ref{MM-evo-eqn2}) show this fact. Accordingly, by one's turning on a finite value of $|\lambda^I|$ which relative to $|\hat{\mu}_{t-a}|$ is sufficiently small, no sudden change is to be seen in the wavefunction behavior. Hence, imposing $\frac{|\lambda^I|}{|\hat{\mu}_{t-a}|} \ll 1$, every one-qubit (2,2) phase must fall into one of the  (1,1) phases: A, B, C. This fact can be also checked by numerical investigations when $|\hat{\mu}_{t-a}|$ becomes, typically by something like one order of magnitude, dominant: by turning on sufficiently small values of $|\lambda^I |$, the corresponding phase of the solution stays invariant, but it receives sub-structures which gradually become richer. Fig.\ref{fig_11vs22_ABC} shows an example of this fact: phase 4 $\sim B$ same but the effect of $|\lambda^I |$ multiplying the structured pattern extrema: in the left, only one minimum between every, but on the right, one global minimum and 8 local extrema between every subsequent pair of global maxima.\\\\  
As explained above, in the regions of the coupling space corresponding to $\frac{|\lambda^I|}{|\mu_{t-a} + \lambda^R|} \ll 1$, the $(2,2)$ system reproduces the three phases of the $(1,1)$ system as prescribed in the previous subsection. Therefore, we focus on describing the behavioral phases of the $(2,2)$ one-qubit QMM-UE in the `non-redundant' regions of the $(\mu, \lambda^R, \lambda^I)$-coupling space in which the effective $\mu$-coupling, $|\hat{\mu}_{t-a}| = |\mu_{t-a} + \lambda^R|$, is either sufficiently small or its strength is comparable with that of $| \lambda^I |$ but does not overwhelm it too much. We now briefly describe the \emph{`five'} behavioral phases which are observed, carefully examined and validated upon doing many simulations in our exhaustive numerical analysis.   
%%%%%
\\\\
\\\\
\\\\
\underline{\emph{PHASE 1: The Fixed-Point-Attractor Dynamics}}\\ 
In phase $1$ of one-qubit $(2,2)$-Purely-QMM-UEs, all the solutions develop \emph{fixed point attractors} at sufficiently late times. A representative solution of this phase is presented in Fig. \ref{fig-phase1}. As one observes in the $(\text{Re}[\Psi_t^{\text{down}} ] , \Psi_t^{\text{up}})$ plots, the one-qubit wavefunction becomes monotonically static. Moreover, the signature of the attractor phase is seen in the third plot of figure \ref{fig-phase1}: the fundamental QM-TPF observable, that is $w_{t-a\;a}$, evolves monotonically to its fixed-point attractor whose unique value can be nothing but $1$. As our numerical investigations confirm, irrespective of the initial control parameters $(B^y_{\text{Kicker}} ; \theta_0,\phi_0)$ and for arbitrarily values of the largest QMD which do satisfy $a \geq a^\star$, this attractor phase is formed in the following regime of the QM-couplings:  $\lambda^I \in \mathbb{R}^+$, while $|\hat{\mu}_{t-a}|$ is either zero or sufficiently smaller than $\lambda^I$.\\\\ 
Moreover, we highlight that in this phase, the (2,2)-QMM-UE can drive the one-qubit system to its fixed-point attractor within durations as short as several tens or several hundreds of length-$a$ time interval, as simulations show.  For example, in the specific solution shown in Fig. \ref{fig-phase1}, we see that the one-qubit wavefunction has already settled down to its attractor in about $ 35 \; a$ after $t_* = a$, namely the outset of QMM-UE.\\\\ 
Indeed, the fact that the QMM-UE of closed quantum systems can develop phases in which the global wavefunctions experience fixed-point attractor behavior, a typical hallmark of open quantum systems, is indeed remarkable. Indeed, going on with further explorations of the phase diagrams under consideration, we find that other celebrated features of open quantum system are naturally realized by (even the simplest versions of) QMM-UEs. Finally, we highlight in reference to the phase diagram of the $(1,1)$ system, that the phase one, namely the fixed-point attractor behavior, is new to the enlarged system of the $(2,2)$ QMM-UE. 
%%%%%
\\\\
\underline{\emph{PHASE 2: Regular Oscillations With (Almost-)Constant Attractors of The QM-TPF}}\\
The second phase of the one-qubit system under $(2,2)$ QMM-UE is equivalent with the phase `A' of its $(1,1)$ counterpart: the curious case of an \emph{`oscillatory-attractor mixed behavior'}. The system's QMM-UE suppresses the sufficiently-late-times oscillations of the QM-TPF down to infinitesimal fluctuations around initial-data-dependent constant values, while ordering the asymptotic dynamics of the global wavefunction in the form of large oscillations which are extremely regular.\\\\  
Interestingly enough, one finds that the second phase of the $(2,2)$ system can be easily formed even in the non-compact regions of couplings where the equivalence with the (1,1) system is broken completely. Specifically, simulations confirm the formation of the second phase with arbitrary \emph{negative} values of (the `new' coupling) $\lambda^I$ and with (zero or finite) $|\hat{\mu}_{t-a}|$ upper bounded as $\frac{|\hat{\mu}_{t-a}|}{- \lambda^I} \in [0 , 1 + \epsilon]$ where $\epsilon$ is a case-dependent parameter whose value is typically smaller than one.\\\\ 
A representative solution of the second phase, with its double-faced quality clearly manifested in the sample time windows, is presented in Fig. \ref{fig-phase2}. Indeed, the solution of Fig. \ref{fig-phase2} belongs to the non-compact region of the $(\mu, \lambda^R, \lambda^I)$-coupling-space where the one-qubit $(2,2)$ system is far beyond the reach of its $(1,1)$ counterpart. 
% % % % %
\\\\
\\\\
\underline{\emph{PHASE 3: Unstructured Oscillations}}\\ The third phase of the one-qubit $(2,2)$-Purely-QMM-UE is identical to the phase `C' of its $(1,1)$ counterpart in which the global wavefunction and the QM-TPF observable experience everlasting irregular oscillations. Fig. \ref{fig-phaseC-11} presents \emph{this `disordered' phase}. It is formed only when $\hat{\mu}$ has a finite value of either sign, sufficiently larger than $\lambda^I$.\\\\
%%%%%
\underline{\emph{PHASE 4: Oscillations which are clearly Structured or even Regular}}\\ This phase is equivalent with the phase `B' of the $(1,1)$ system. However, simulations show that it can be developed even going far beyond the regime of small $|\lambda^I|$ coupling. Specifically, choosing an arbitrary $\lambda^I \in \mathbb{R^-}$, while $\frac{|\hat{\mu}_{t-a\;t}|}{- \lambda^I}$ being neither less than one, nor too large, results in the formation of the phase four of the $(2,2)$-Purely-QMM-UE of the one-qubit closed system. %Fig. \ref{fig-phaseC-11}. 
The major hallmark, once more is the post-early-times behavior of QM-TPF: it develops \emph{repetitions of large `oscillatory patterns' which are robustly structured and are featured with several well-gapped extrema}. The wavefunction oscillations are not disordered likewise. In Fig. \ref{fig-phase4 QMM} a representative of the phase is shown.
%%%%%
\\\\
\underline{\emph{PHASE 5: BiState Meta-Stable Dynamics}}\\
This one is the most intriguing phase qualitatively. Moreover, among the five phases, it is the most structurally-ordered one: both in terms of the QM-TPF profile and the wavefunction oscillations themselves. Like other phases, the only condition for a generic solution of the dynamical system \eqref{MM-evo-eqn2} to be in this phase is one's choosing the `right' range of the couplings of the QMM-H \eqref{QMMH22}. As our numerical investigations confirm, the corresponding right range of the couplings for this phase is this: $\lambda^I \in \mathbb{R}^+$, while one sets a comparable strength for its competitor coupling $\hat{\mu}_{t-a\;t}$, that is, the ratio $\frac{|\hat{\mu}_{t-a\;t}|}{\lambda^I}$ is neither too small, nor too large. Hence, the (1,1) system cannot form this phase.\\\\  
One representative solution of the fifth phase is shown in the three plots of Fig. \ref{fig-phase5}. We highlight the temporal relations between the dynamical patterns of plots ((a),(b)) for the one-qubit wavefunction oscillations, and that of plot (c) for the oscillations of the fundamental QM observable: the emergent patterns are all synchronized robustly. Intriguingly, in the sufficiently-long duration of the representative solution of Fig. \ref{fig-phase5}, $\frac{t_{\text{completion}}}{a} = 3600$, one clearly sees a \emph{Two-State Metastable Dynamics}. This two-state meta-sable dynamics, resembling an `exhaling-holding-inhaling' pattern, emerges from the everlasting trails of fast dense oscillations, and is highly robust even quantitatively. Moreover, the one-qubit closed system switches the two metastable states by means of \emph{sharp transitions}. The time scale of the transitions is about one order of magnitude shorter than the duration of the metastable states. 
One discerns discerns the unique hallmarks of the phase five of the one-qubit $(2,2)$-Purely-OMM-UE, shown by the three plots of Fig. \ref{fig-phase5}. The pure state and the QM-TPF of the one-qubit closed system form highly-ordered temporally-correlated oscillatory behavior as such: (i) reach at one of their up-or-down levels quickly, (ii) settle down and very tinily fluctuate there for $\text{O}[100 a]$, (iii) swiftly transition to the switched-level round in succession. Remarkably, the QM-TPF transitions between a typically-deep minimum and its maximum being $1$, while holding to $1$ for $\text{O}[100 a]$.
\\\\ 
We \emph{conclude} that the one-qubit unitary state-histories which solve (\ref{fgp},\ref{MM-evo-eqn2}) yield five behavioral phases of Purely-QMM-Us generated by the second order polynomial made from two quantum memories, as explained above. In Section \ref{visions}, we come back to this surprisingly-rich five-region phase diagram, and its even richer generalizations in the next two subsections, to present a selection of interesting lessons.
%%%%%%%%%%%%%%%%%%%%%%%%%%%%%%%%%%%%%%%%%%
%%%%%%%%%%%%%%%%%%%%%%%%%%%%%%%%%%%%%%%%%%%
\subsubsection{Supplementary Numerical Solutions Along The $\theta$-orbit} \label{wnetto} 
\begin{figure}
	%\centering
	\begin{subfigure}[b]{2.9in}
		\includegraphics[width=2.9in]{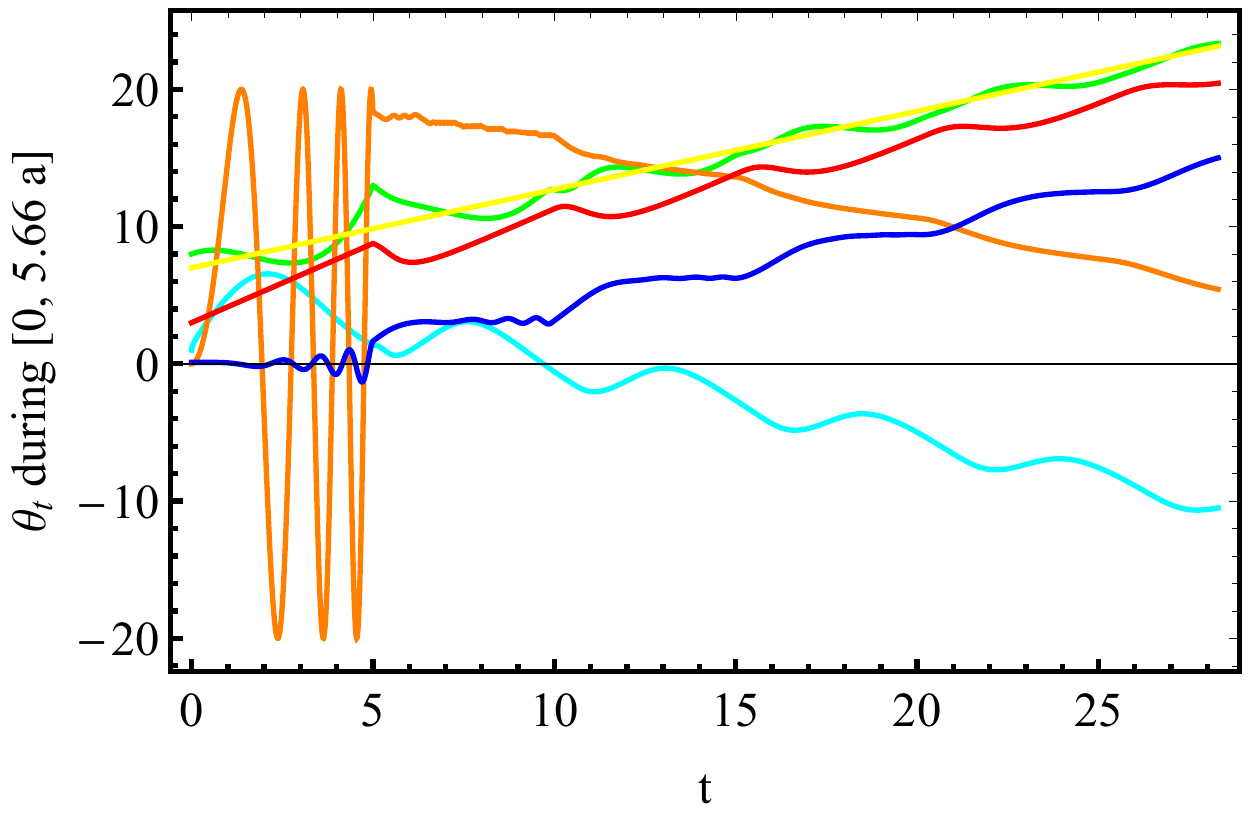}
		\caption{}
		\label{}
	\end{subfigure}
	\quad
	\begin{subfigure}[b]{2.9in}
		\includegraphics[width=2.9in]{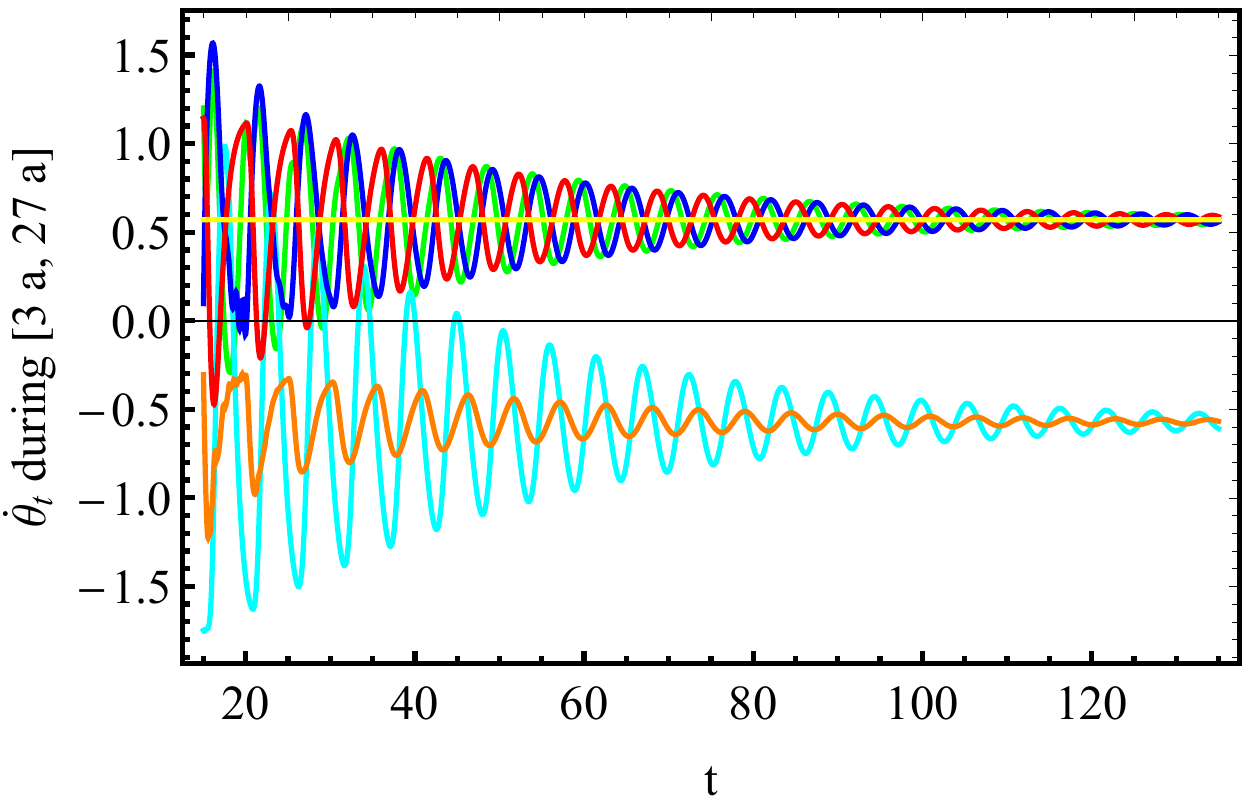}
		\caption{}
		\label{}
	\end{subfigure}
	\caption{Purely-QMM-UEs solving (\ref{EXCT-THETA}) with $(\lambda , a) = (-2 , 5)$, assuming the six initiations $\theta(t \leq a) = f(t) =$ $\frac{1}{5} t^3 - t^2 + t + 8$ (Green); $1.154\;t + 3$ (Red); $e^{\sqrt{3 t} - \frac{1}{7} t^2}$ (Light Blue); $20 \sin(\frac{5}{6}\; t^2)$ (Orange); $\frac{\cos(t^2)\;e^t}{t^2 + 11\; t \;+\; 9}$ (Strong Blue); $0.57\; t + 7$ satisfying (\ref{eq:alphaII})  (Yellow) .}
	\label{PIUSHBASYMPH}
\end{figure}

The unitary state-history (\ref{ts5},\ref{eq:alphaII}) features unique and significant characteristics among the theta-orbit solutions. These properties were analytically derived, and extensively discussed, in Subsection \ref{SIV-II}. First and foremost, although being the simplest of all nontrivial solutions to the dynamical system (\ref{EXCT-THETA}), this special unitary state-history has an interesting and important relation with \emph{all} the unitary state-histories in the $\theta$-orbit moduli space, as follows. The unitary state-history (\ref{ts5},\ref{eq:alphaII}) is the dynamic-or-curve attractor of all one-qubit unitary state-histories formed along the $\theta$-orbit. We now present a supplementary numerical study with double purposes. First, this numerical analysis re-validates, actually in a most general setting as explained below, the above statement. Second, as promised in Subsection \ref{SIV-II}, it demonstrates `the fast tendency' (in $a$ units) of the $\theta$-orbit unitary state-histories in approaching (\ref{ts5},\ref{eq:alphaII}).\\\\ 
Fig. \ref{PIUSHBASYMPH} presents six solutions $\theta_t$ to the dynamical system (\ref{EXCT-THETA}), numerically constructed based on six qualitatively-distinct functions $f(t)$ taken as their initial histories, $\theta(t \leq a)$. These exemplary functions are specified in the caption. The left and right plot show the dynamical behavior of $\theta_t$-and-$\dot{\theta}_t$ in the two time intervals of the Purely-QMM-UE  written on the vertical axes. For enhanced visibility and differentiation: (i) these six solutions are coloured: Light Blue (exponential $f_t$), Red (non-special linear $f_t$), Green (cubic polynomial $f_t$), Orange (oscillatory $f_t$), Strong Blue (complicated $f_t$) and Yellow (special linear $f_t$), see the caption; (ii) the curves are shown with exaggerated thickness. The yellow solution whose linear $f(t)$ realizes \eqref{eq:alphaII} is `a' profile-invariant one.\\\\ 
%%%%%%%%%%%%%%%%%%%%%%%%%
One clearly sees in the right plot of Fig. \ref{PIUSHBASYMPH}, showing the time derivatives of the six $\theta$-orbit solutions in an exemplary time duration of their Purely-QMM-UE: $\dot{\theta}_t\;;\;t \in [\;3 a\:, 27 a\;]$, that the five profile-non-invariant solutions all realize, precisely as the Subsection \ref{SIV-II} has prescribed, one of the two unitary-state-histories (\ref{ts5},\ref{eq:alphaII}) as their curve-attractor. As we see, the time-derivate $\dot{\theta}_t$ of every one of the five profile-non-invariant solutions has developed, after a while, a profile of monotonically-shrinking oscillations which swiftly converge to \emph{either} the slope of the profile-invariant yellow solution which satisfies \eqref{eq:alphaII}, \emph{or} to its sign-flipped mirror root. Indeed, the statement that every $\theta$-orbit solution, $\theta_t$, has a curve-attractor being a profile-invariant linear-in-time, $\theta^\star_t$, is translated to the statement that the time-derivates of the solutions, $\dot{\theta}_t$, have fixed-point attractors whose values are the roots $\alpha$ of the equation \eqref{eq:alphaII}. Accordingly, the right plot is a numerical manifestation of both the derived equation \eqref{sap} and the ever-shrinking simple-profile oscillations as fluctuations around it, realized by all the $\theta$-orbit unitary state-histories. \\\\ 
Moreover, we observe in the left plot of Fig. \ref{PIUSHBASYMPH} that even within the first few length-$a$ steps as of the outset of Purely-QMM-UE, all the five profile-non-invariant solutions have effectively taken profiles which approximate their dynamic attractors: the yellow (or its mirrored) solution. It must be further highlighted that we observe similar rapid tendencies, and the same curve-attractor behavior of $\theta$-orbit unitary state-histories shown in the right plot, using sufficiently-many other choices of the initiation functions for $\theta_{t \leq a}$, and in other sufficiently-late-times durations of their Purely-QMM-UE.\\\\  
Now, we come to the numerical investigation of the leading order `$\theta$-orbit-confined' perturbations over the profile-invariant unitary state-history (\ref{ts5},\ref{eq:alphaII}).  The aim is to investigate the asymptotic dynamics of \emph{general} leading order perturbation modes, $\theta_t^{(1)}$, as derived from equation \eqref{whl}. That is, we are interested in the late-time behavior of the solutions to the delay-differential equation \eqref{whl} which are resulted from general time-dependent functions, $f^{(1)}_t$, taken as their initial histories: $\theta_{t \leq a}^{(1)}$.\\\\ 
\begin{figure}[t]
	\centering
	\includegraphics[width=2.9in]{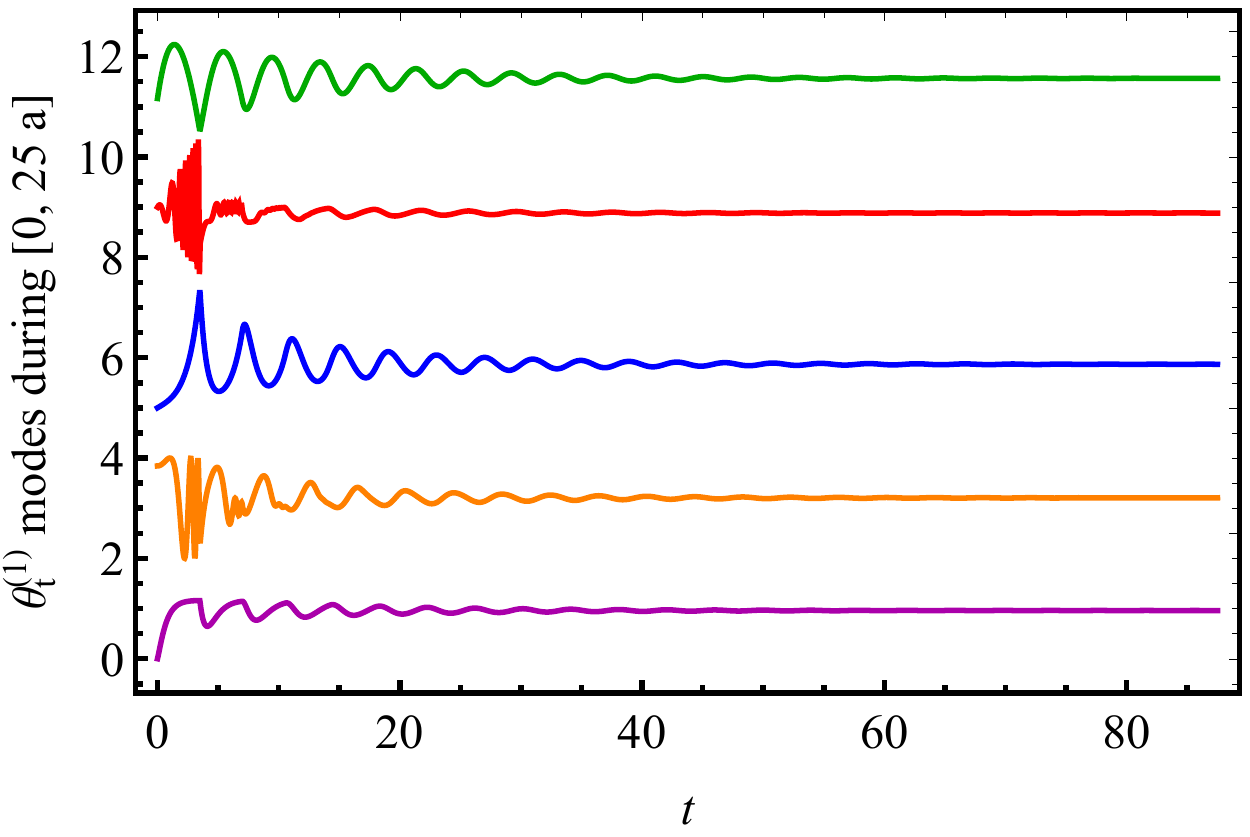}
	\caption{Evolving $\theta_t^{(1)}$ modes = excitons of the profile-invariant unitary state-history on the $\theta$-orbit, solving \eqref{whl} with $(\lambda , a) = (-2 , 3.5)$ s.t. $\alpha \sim 0.783$, and five initiations: $\theta^{(1)}(t \leq a) = f^{(1)}(t) =$ $\frac{1}{20} (t^3 - 14 t^2 + 33 t)$ (Green); $\frac{2}{5} t \cos(t^3 - t^2 + 5 t)$ (Red); $\frac{1}{7} \sinh(t)$ (Blue); $\sin(\cosh(t))$ (Orange); $\frac{1}{2}(t^{\;\frac{1}{1 + t}} + \tanh(t))$ (Magenta). For better visibility, the five curves have received appropriate amounts of relative vertical shifts.}
	\label{fvctuos:ne5}
\end{figure}
\\\\
%%%%%%%%%%%%%%%%%%%%%%%%%%%%
Fig. \ref{fvctuos:ne5} shows the solutions for $\theta_t^{(1)}$ modes evolving under \eqref{whl}, assuming the five initial histories $f^{(1)}_t$, specified in the caption, with diversely-different defining profiles. As one sees in Fig. \ref{fvctuos:ne5}, regardless of their initial histories, all $\theta_t^{(1)}$ modes have developed fixed-point attractors. The same behavior is observed from any other initial-history function taken for the simulation. Hence, as promised in Subsection \ref{SIV-II}, we observe that all $\theta_t^{(1)}$ excitons are asymptotically static and preserve the original slope of the profile-invariant solutions (\ref{ts5},\ref{eq:alphaII}): the $\theta$-orbit perturbations cannot destabilize the characteristic profile invariance of the unitary state-histories (\ref{ts5},\ref{eq:alphaII}).\\\\
We highlight that \emph{quite exceptionally} in this subsection, given the purposes of the above two numerical simulations, the quantum-memory pools were filled by means of initial histories chosen independent of Kicker Hamiltonians. That is, sufficient numbers of temporal functions with distinctive defining profiles were chosen as initial histories, to compare the $\theta$-orbit Purely-QMM unitary state-histories `initiated' quite differently. Mathematically viewed, in this way, we have effectively explored the (almost-) \emph{entire moduli spaces} of solutions to the delay-differential equations (\ref{EXCT-THETA}) and \eqref{whl}. These moduli spaces contain as their (proper or improper) sub-spaces the solutions generated by Kicker Hamiltonians. Such, we have come up with confirming and completing the analytical results of Subsection \ref{SIV-II} on the $\theta$-orbit one-qubit Purely-QMM-UEs, within an encompassing numerical approach.
%%%%%%%%%%%%%%%%%%%%%%%%%%%%
\subsection{The Phase Diagram of One-Qubit Wavefunctions Evolving Under A (2,3) 
Purely-QMM Hamiltonian }\label{SV-III}
\begin{figure}
	\begin{subfigure}[b]{2.9in}
		\includegraphics[width=2.9in]{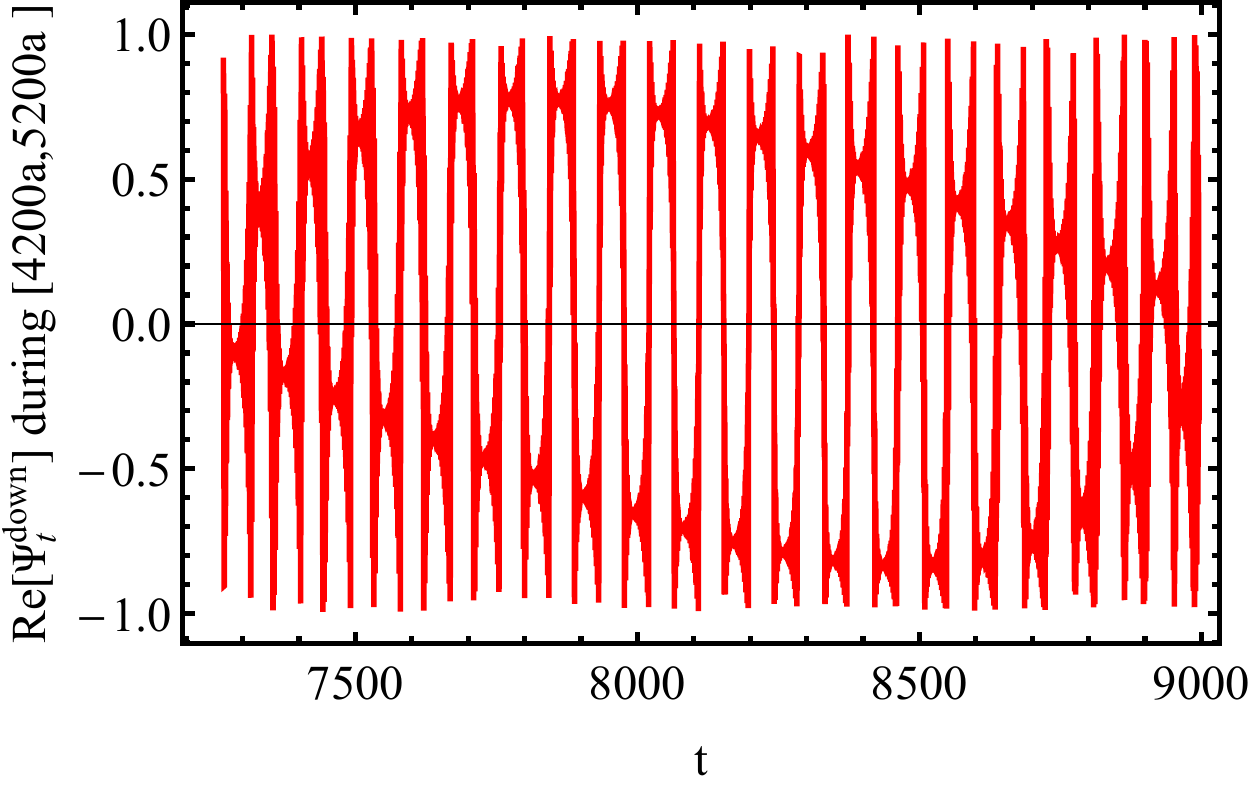}
		\caption{}
		\label{}
	\end{subfigure}
	\quad
	\begin{subfigure}[b]{2.9in}
		\includegraphics[width=2.9in]{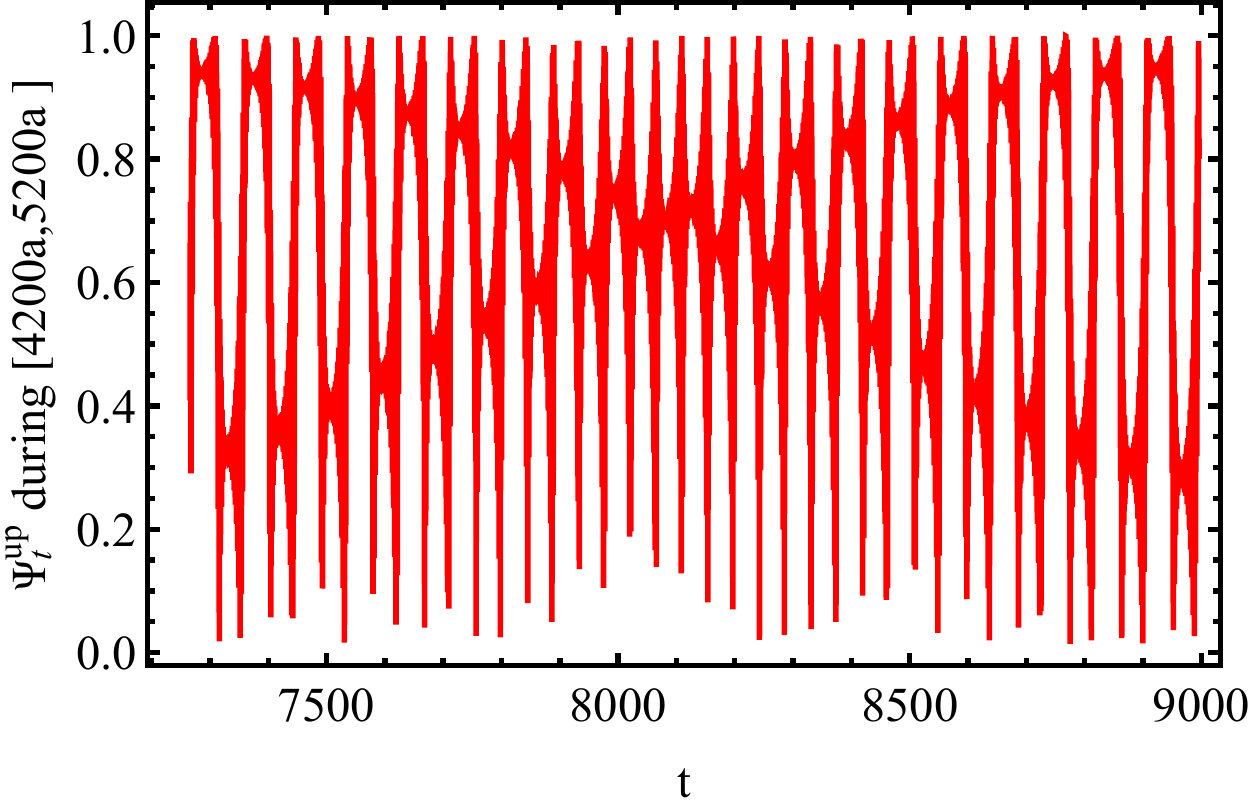}
		\caption{}
		\label{}
	\end{subfigure}
	\\
	\begin{subfigure}[b]{2.9in}
		\includegraphics[width=2.9in]{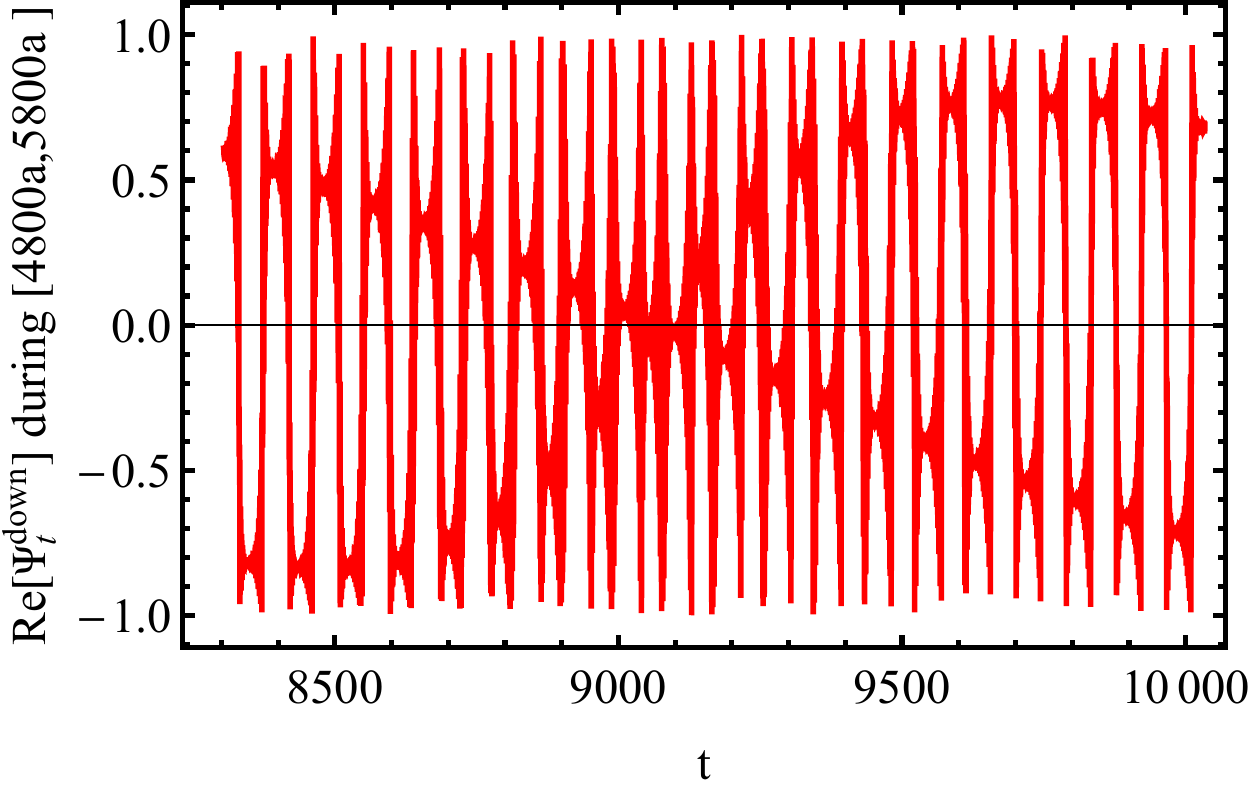}
		\caption{}
		\label{}
	\end{subfigure}
	\quad
	\begin{subfigure}[b]{2.9in}
		\includegraphics[width=2.9in]{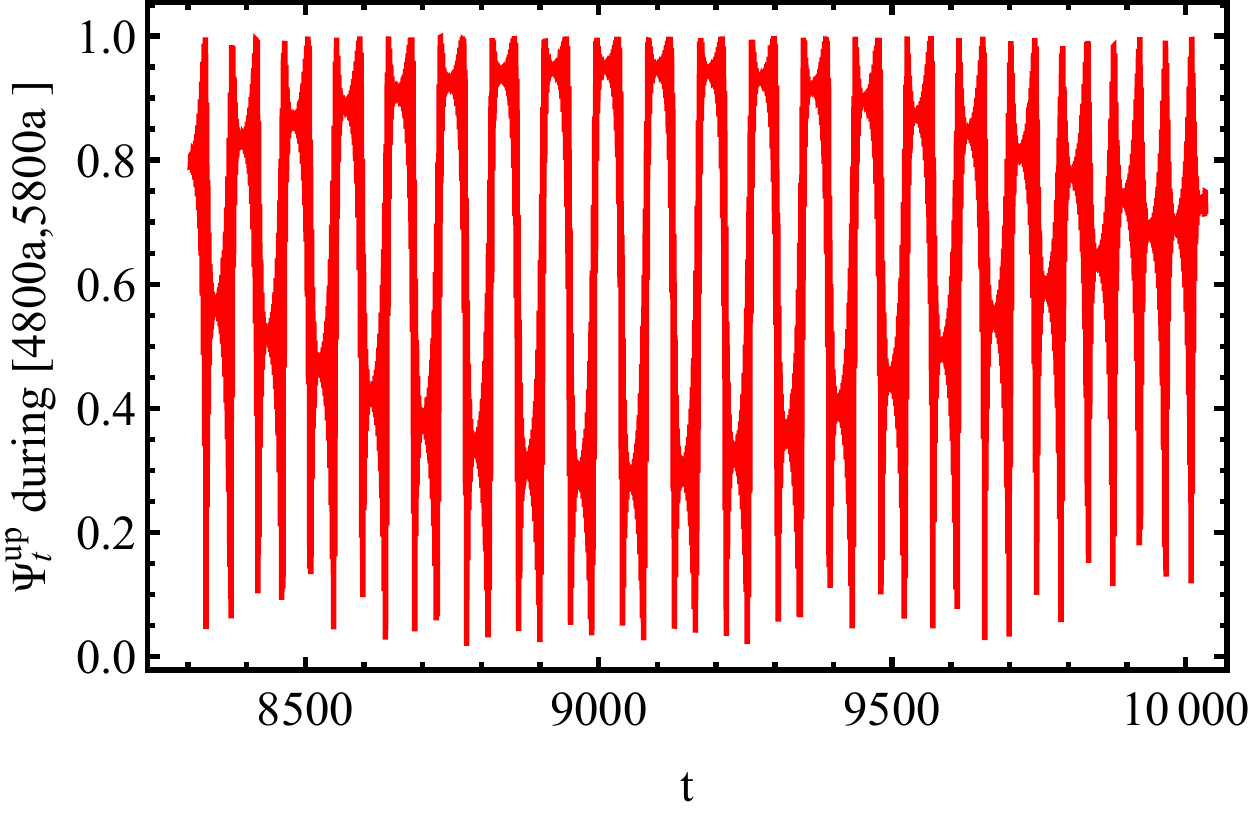}
		\caption{}
		\label{}
	\end{subfigure}	
	\begin{center}
	\begin{subfigure}[b]{2.9in}
		\includegraphics[width=2.9in]{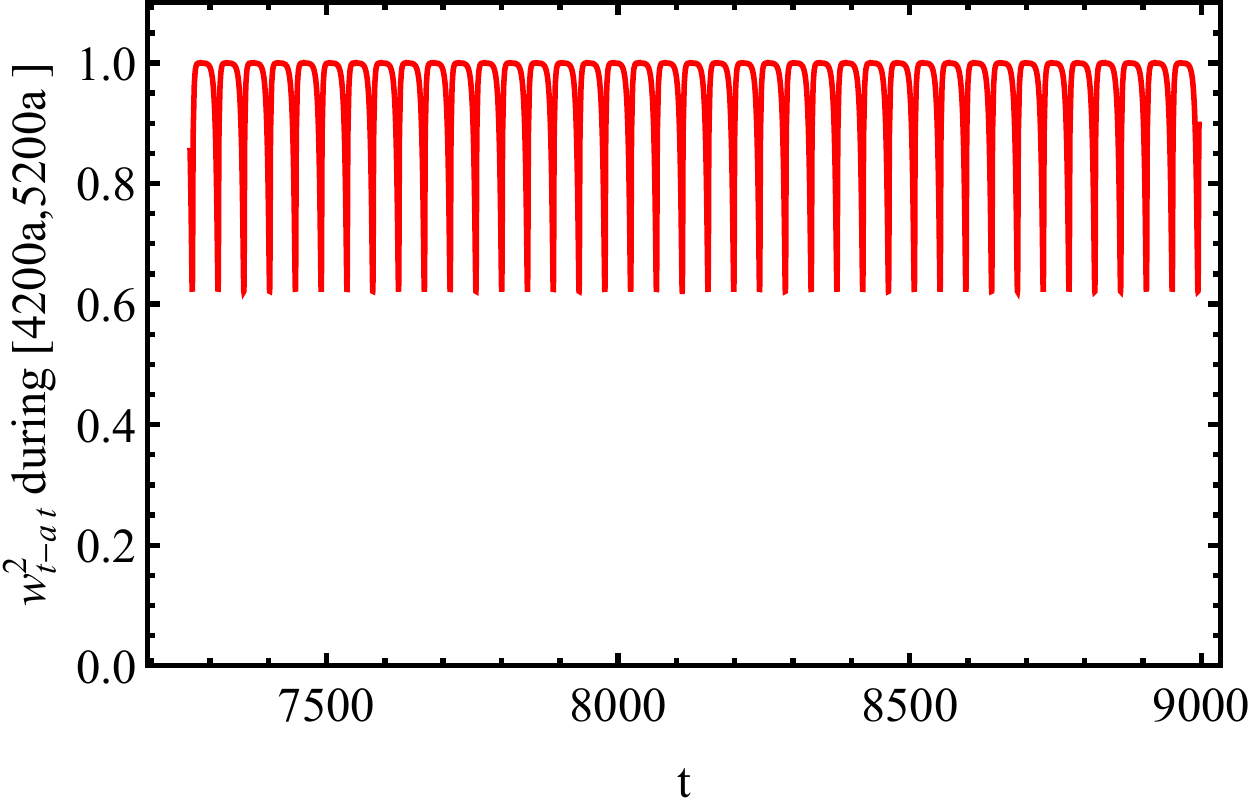}
		\caption{}
		\label{}
	\end{subfigure}
	\end{center}
	\caption{The Purely-QMM-UE described by (\ref{fgp},\ref{ate:ma},\ref{ate:mb},\ref{SE23POLAR}) with parameters:
		$(B_{Kicker}^{y} ; \theta_0; \phi_0 | \lambda^I, \eta | a; t_{completion}) = (7.5; 1.001, 0.089 | 2.6, 2.53 | 1.73; 8000 a)$.
		Oscillatory deformation of the `Phase 5' of one-qubit (2,2)-Purely-QMM-UE is featured.}
	\label{fig-23-S5C}
\end{figure}

\begin{figure}
	\begin{subfigure}[b]{2.8in}
		\includegraphics[width=2.8in]{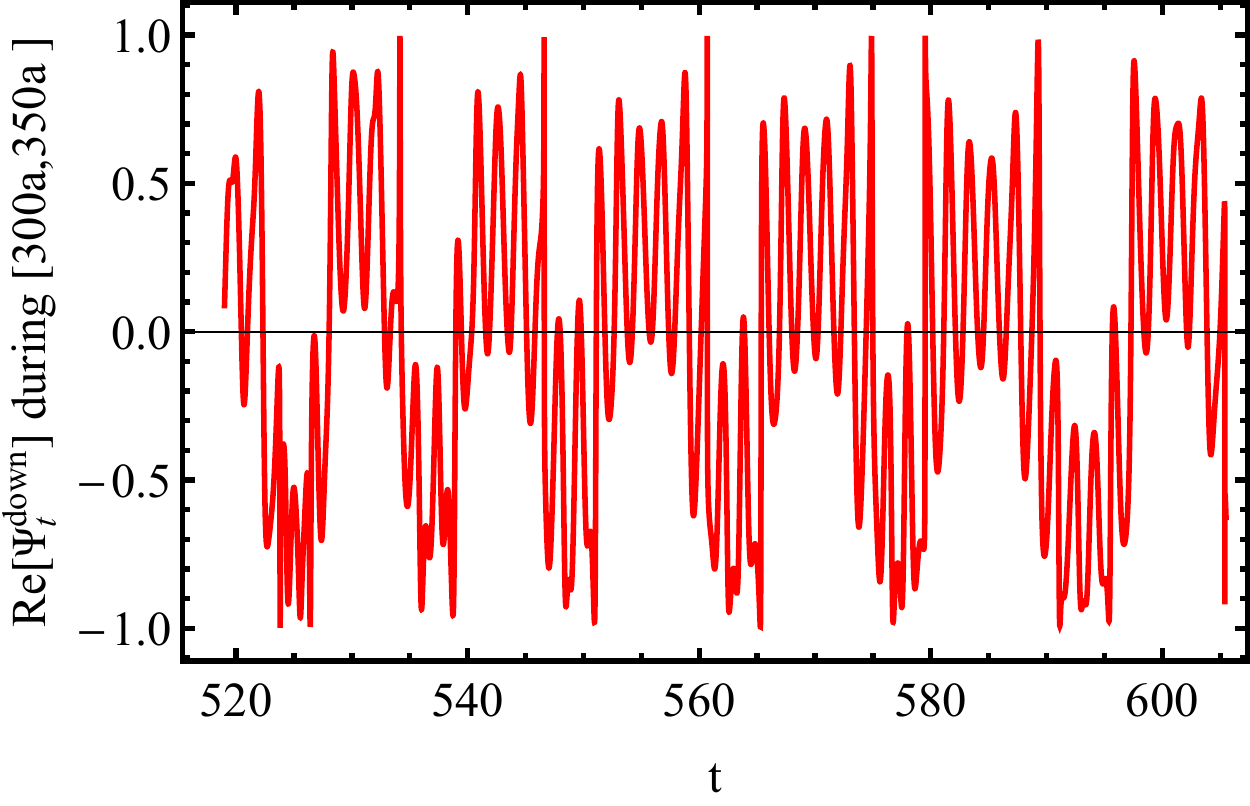}
		\caption{}
		\label{}
	\end{subfigure}
	\quad
	\begin{subfigure}[b]{2.8in}
		\includegraphics[width=2.8in]{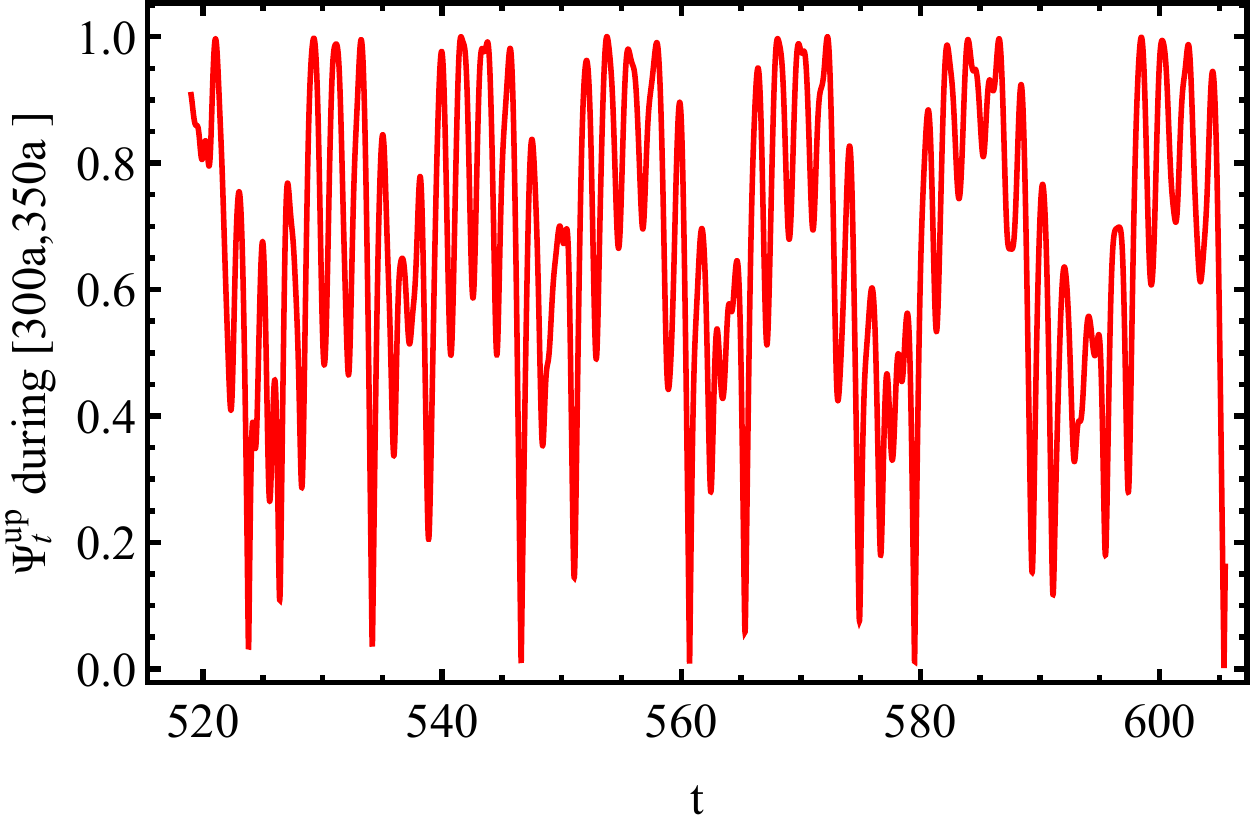}
		\caption{}
		\label{}
	\end{subfigure}
	\\
	\\
	\begin{subfigure}[b]{2.8in}
		\includegraphics[width=2.8in]{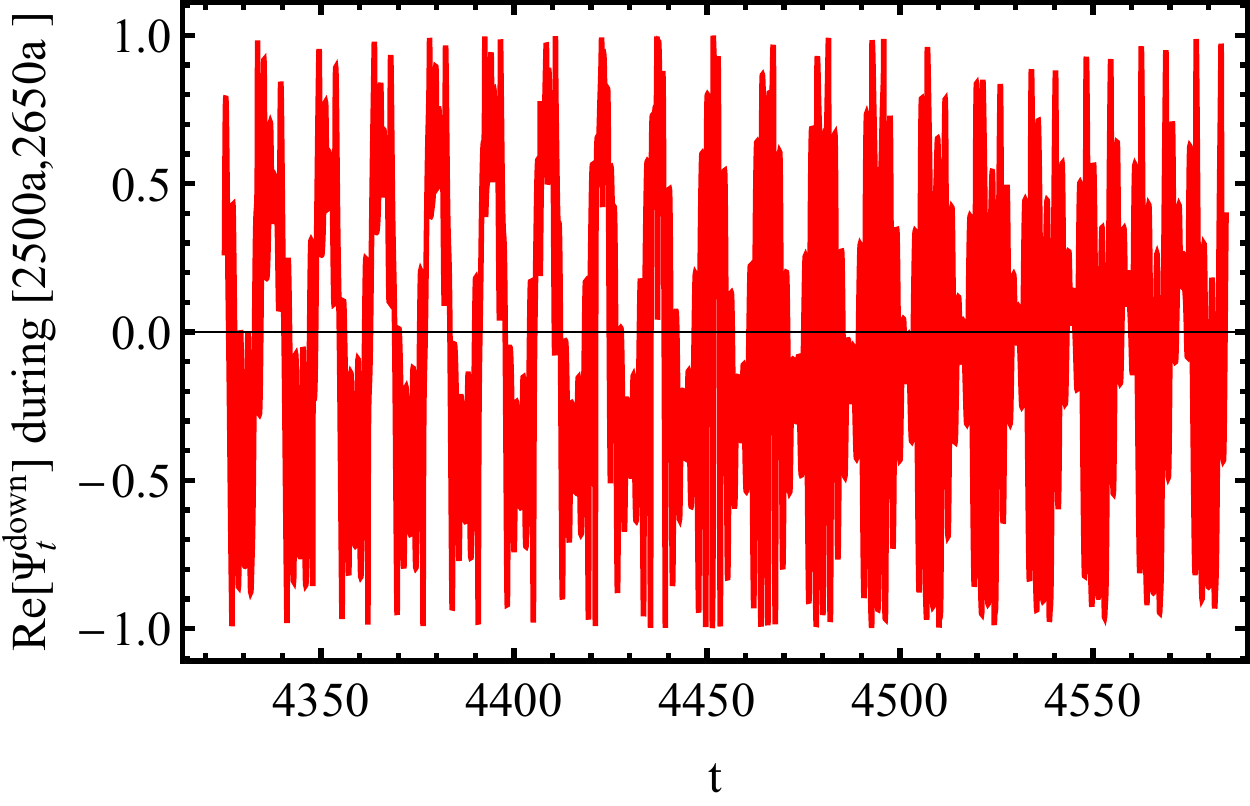}
		\caption{}
		\label{}
	\end{subfigure}
	\quad
	\begin{subfigure}[b]{2.8in}
		\includegraphics[width=2.8in]{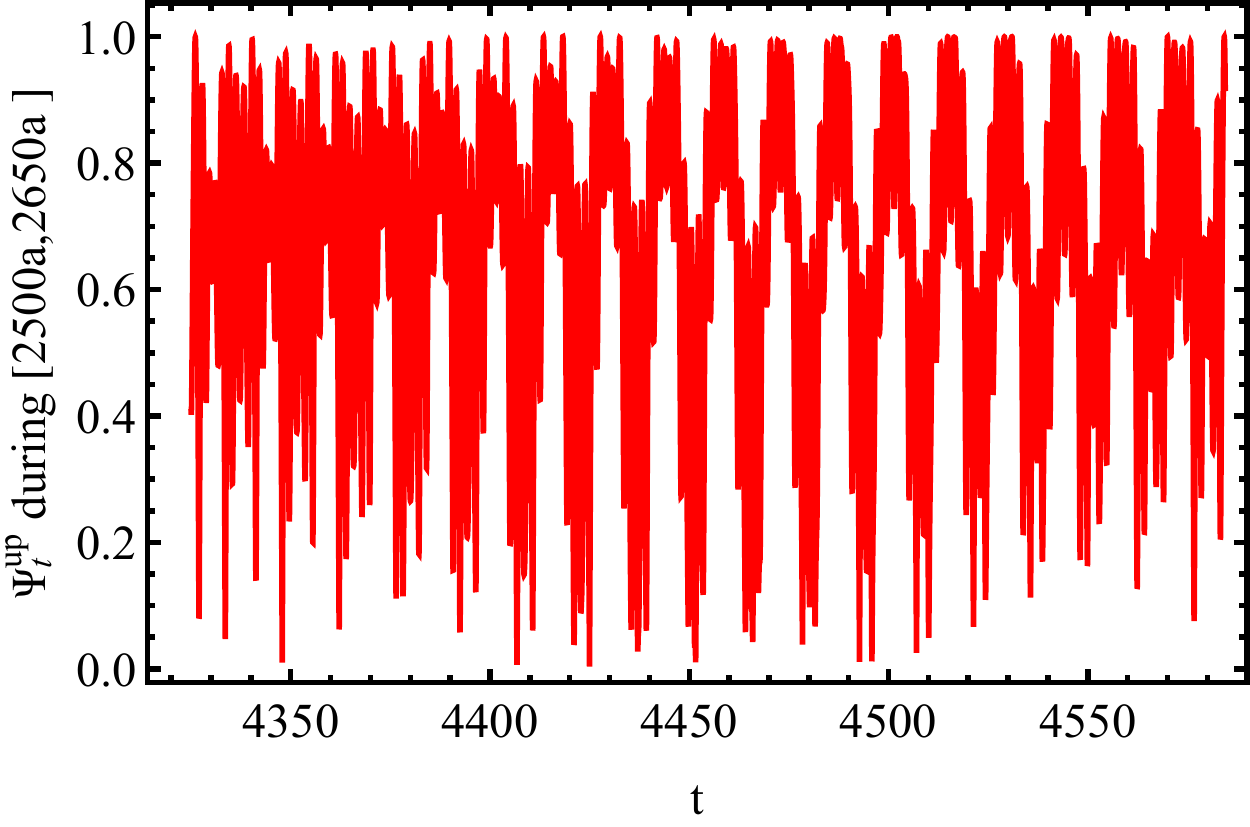}
		\caption{}
		\label{}
	\end{subfigure}
	\caption{The Purely-QMM-UE described by (\ref{fgp},\ref{ate:ma},\ref{ate:mb},\ref{SE23POLAR}) with parameters:
		$(B_{Kicker}^{y} ; \theta_0; \phi_0 \; | \; \lambda^I, \eta \;|\; a; t_{completion}) = (7.5; 1.001, 0.089 \; |\; 1.3, 4.93 \; | \; 1.73; 3000\; a)$.
		Dynamical Phase Transition of the one-qubit (2,3) system switching from a metastable `Phase 3' to a stable `Phase 5'.}
	\label{fig-23-S5}
\end{figure}

\begin{figure}
	\begin{center}
		\includegraphics[width=5.8in]{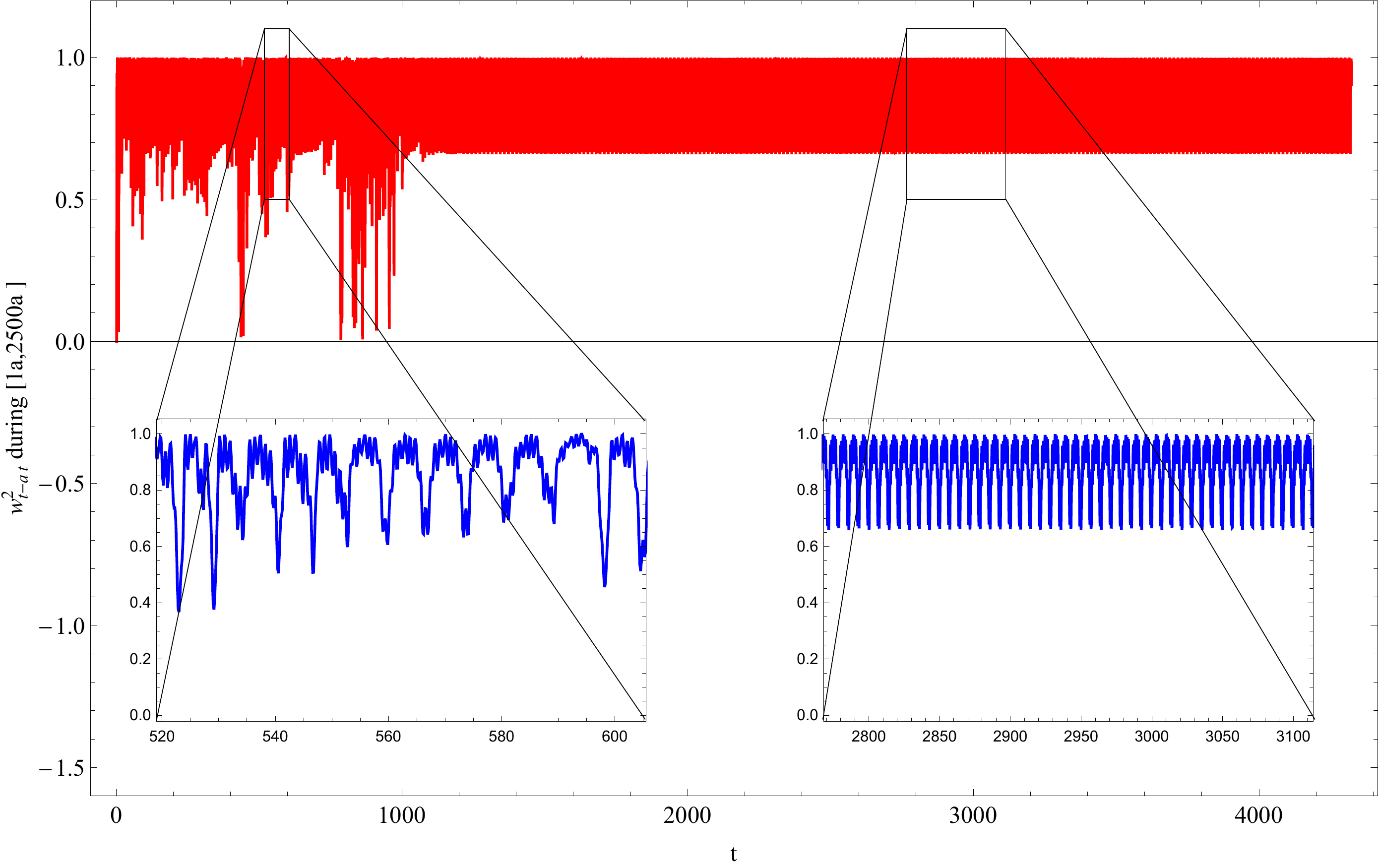}
		\caption{The Purely-QMM-UE described by (\ref{fgp},\ref{ate:ma},\ref{ate:mb},\ref{SE23POLAR}) with parameters:
			$(B_{Kicker}^{y} ; \theta_0; \phi_0 \; | \; \lambda^I, \eta \;|\; a; t_{completion}) = (7.5; 1.001, 0.089 \; |\; 1.3, 4.93 \; | \; 1.73; 3000\; a)$.
			Dynamical Phase Transition of the one-qubit (2,3) system switching from a metastable `Phase 3' to a stable `Phase 5'.}
		\label{fig-23-S5e}
	\end{center}
\end{figure}

We now set to understand how the first-to-come higher order monomial interactions between $N$ quantum memories impact the phase diagram of the one-qubit QMM-UE. To this aim, the same analyses is performed for $(N = 2, L= N+1 = 3)$ Purely-QMM system: we activate cubic links between a chosen pair of state-history-members of the one-qubit system,  $(\rho_{t-a}, \rho_t)$, and map the evolution phases generated by the QMM-H.\\\\ 
Specifically, we classify the Purely-QMM-UE phases generated by the $H^{(2,3)}_{\text{QMM}}(t)$ given in (\ref{avn}). Such unitary evolution of one-qubit states is described by the QMM Schr\"odinger equations (\ref{fgp},\ref{ate:ma},\ref{ate:mb},\ref{SE23POLAR}).  
This delay-differential system contains three couplings: $\lambda^R$ (whose value is kept zero for simplicity while maintaining a rich phase diagram), $\lambda^I$ and $\eta$ whose values we run over sufficiently-large domains to monitor the effects of the cooperation or competition between the second and third order interactions of the two quantum memories. In our numerical simulations, we safely choose the very same initial conditions as the (2,2) case, for comparison purposes. We highlight that the real coupling $\eta$ sets the strength of the third order interaction, $L = 3$, between the chosen quantum memories.\\\\
Indeed, higher order linking of the QMs are impactful. Numerical exploration proves that (2, 3) Purely-QMM-UE phases are physically richer. This is true despite the fact that the one-qubit (2, 3) system essentially develops the five principal phases, even when $|\frac{\eta}{\lambda^I}|$ is not small. First, one-qubit Purely-QMM-UE in the co-presence of the third order interactions develop \emph{dynamical phase transitions} (between those same five phases) in non-narrow regions of the $(\lambda^I,\eta)$-space. Moreover, even without dynamical phase transition taking place, the third order interactions can produce, without any fine-tuning, remarkably-deformed versions of the five phases of the $N=2$ system.\\\\
In order to make the effects of the third-order interactions crystal clear, we compare the numerically-constructed (2, 3) solutions with their (2,2) `reductions'. That is, for every one of the (2,3) numerical solutions, we additionally construct the solution of the (2,2) system with the same values of control parameters (aside from $\eta$), to make a more precise comparative analysis. 
But, it suffices to showing the (2, 3) solutions and explaining notable differences. To discern all the novel effects of the Purely-QMM third-order interactions, we choose sufficiently-many values of $\lambda^I$, run (for every one of them) the third-order interaction coupling $\eta$ from zero to sufficiently-large values, construct numerical solutions to the delay-differential system (\ref{fgp},\ref{ate:ma},\ref{ate:mb},\ref{SE23POLAR}), and classify the phases of the one-qubit unitary evolutions in correspondence. \\\\ 
\begin{figure}
	\begin{subfigure}[b]{2.8in}
		\includegraphics[width=3in]{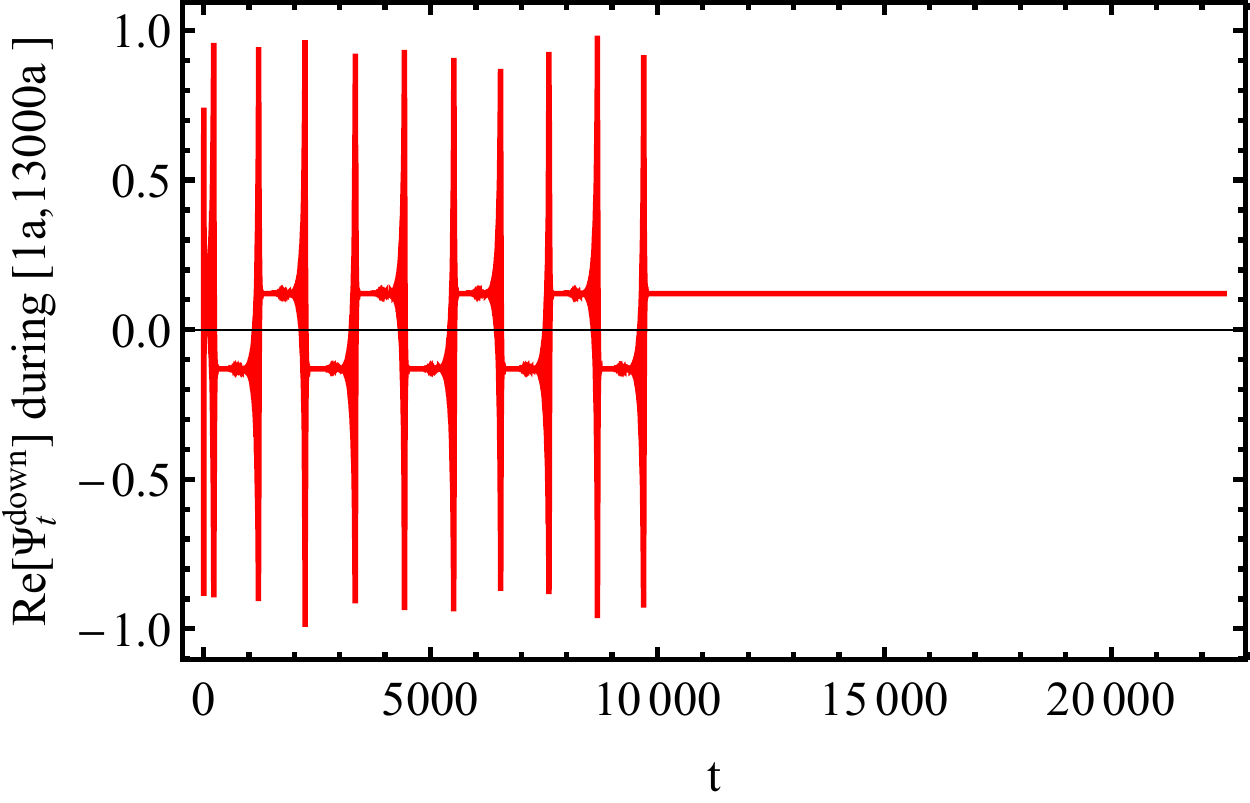}
		\caption{}
		\label{}
	\end{subfigure}
	\hspace{0.1in}
	\begin{subfigure}[b]{3.0in}
		\includegraphics[width=3in]{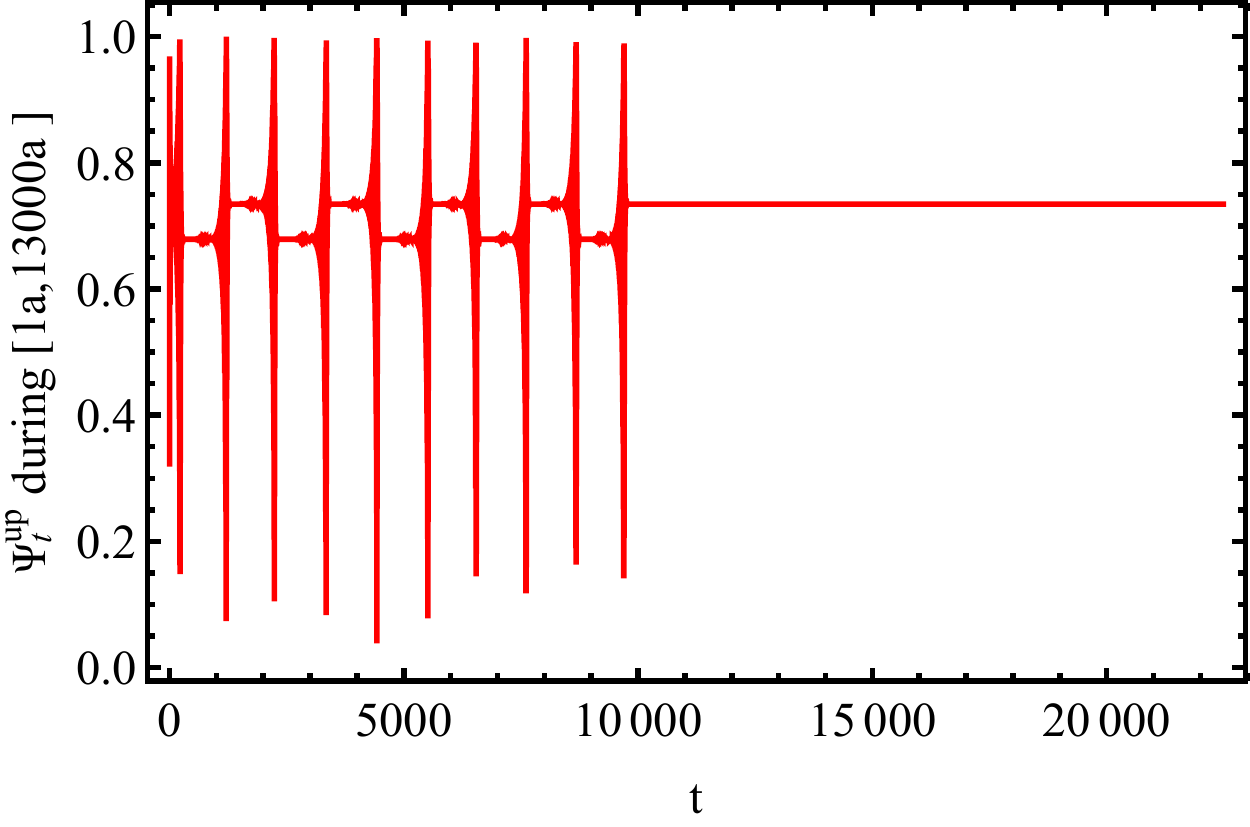}
		\caption{}
		\label{}
	\end{subfigure}
	\begin{center}
		\begin{subfigure}[b]{2.3in}
			\includegraphics[width=3in]{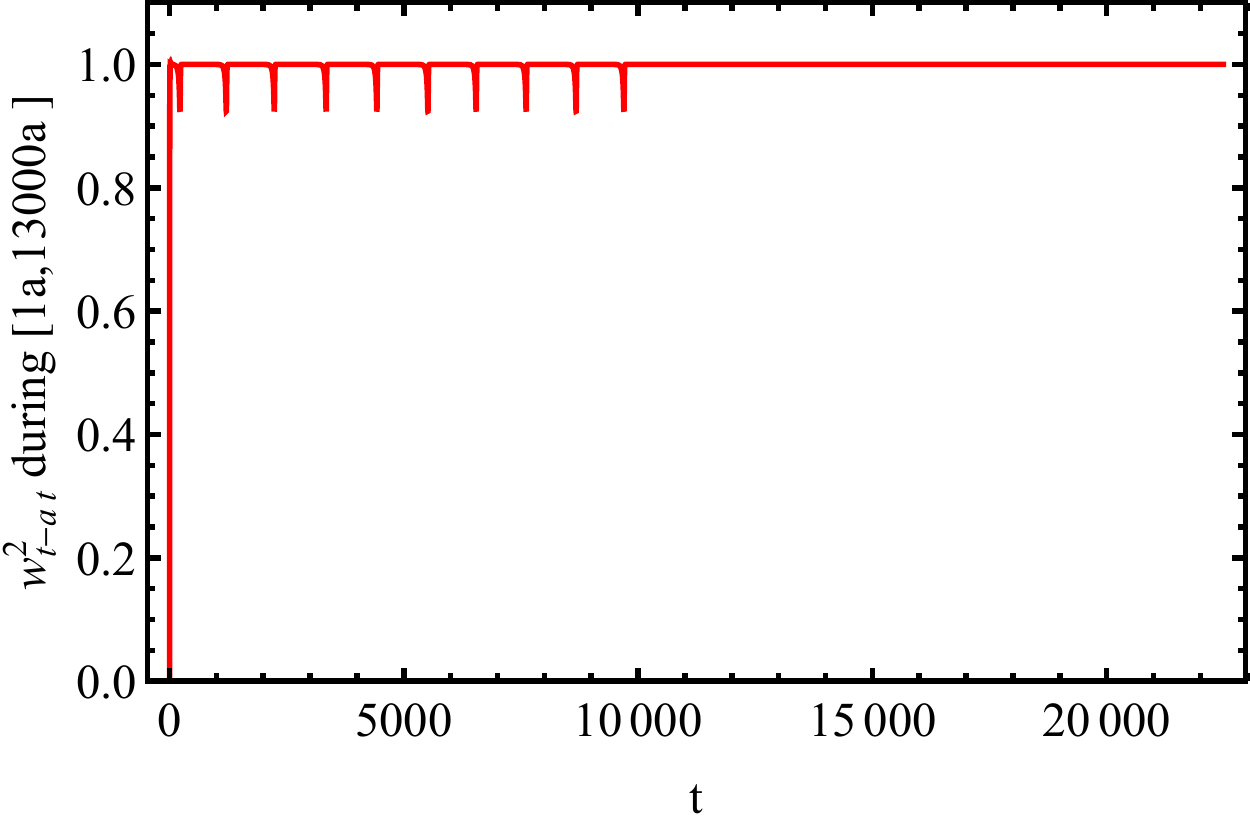}
			\caption{}
			\label{}
		\end{subfigure}
	\end{center}
	\caption{The purely-QMM-UE described by (\ref{fgp},\ref{ate:ma},\ref{ate:mb},\ref{SE23POLAR}) with parameters:
		$(B_{Kicker}^{y} ; \theta_0; \phi_0 | \lambda^I, \eta | a; t_{completion}) = (7.5; 1.001, 0.089 | 5.00, 2.00 | 1.73; 70000 a)$, showing
		dynamical phase transition of the one-qubit closed system under the (2,3) QMM-H: switching phase five to phase one.}
	\label{fig-23-A5}
\end{figure}
%%%%%%%%%%%%%%%%%%%%%%%%%%%%%%%%%%%%%%
One-qubit (2,3) phases are sensitive only to the sign of $\lambda^I$. We now begin with $\lambda^I \in \mathbb{R}^+$. Numerical investigations confirm that the range $\lambda^I \in  [0 , 5]$ covers all the interesting phases of the system. When $\lambda^I = 5$ at vanishing $\eta$, solutions show the attractor phase, as it must. Increasing $\eta$ slightly from zero, the phase 5 of the (2,2) system begins to be formed. Interestingly enough, before the establishment of the phase 5, in the whole range of $1.6 \lesssim \eta \lesssim 2.3$ with $\lambda^I = 5$, the (2,3) Purely-QMM-UE of the one-qubit states develops a \emph{dynamical phase transition} which switches, in the course of time, the two phases mentioned above. We must highlight the purely internal and dynamical nature of this `unitary' phase transition: it does not take any kind of external influences, or any change in the couplings or in the initial data, to occur. It happens all-internally to the one-qubit closed system, during its Purely-QMM-UE, and is sourced by nothing but the (higher order) interactions that the past-or-present density operators of the unitary-evolving qubit make. Also, in sharp contrast with (2,2) system, we notice the non-narrow bound for the formation of this dynamical phase transition: $\frac{\Delta \eta}{\eta} \sim 0.48 $. Fig. \ref{fig-23-A5} shows a solution of the kind during $[0, 13000\; a]$, but its numerical construction was performed for one order of magnitude longer time span, that is $[0, 70000\; a]$, for being ensured of the defining nature of this unitary phase transition. It is worthwhile to additionally mention that the time period for jumps between two stable states is almost the same as the corresponding (2, 2) system, but number of jumps before going into attractor phase is now larger. As we see in Fig. \ref{fig-23-A5} for the given values of $(\lambda^I, \eta)$, the one-qubit Purely-QMM-UE moves after $\sim 4000 a$ from phase 5 to phase 1.\\\\
%%%%%%%%%%%%%%%%%%%%%%%%%
\begin{figure}[t]
	\begin{subfigure}[b]{2.9in}
		\includegraphics[width=2.9in]{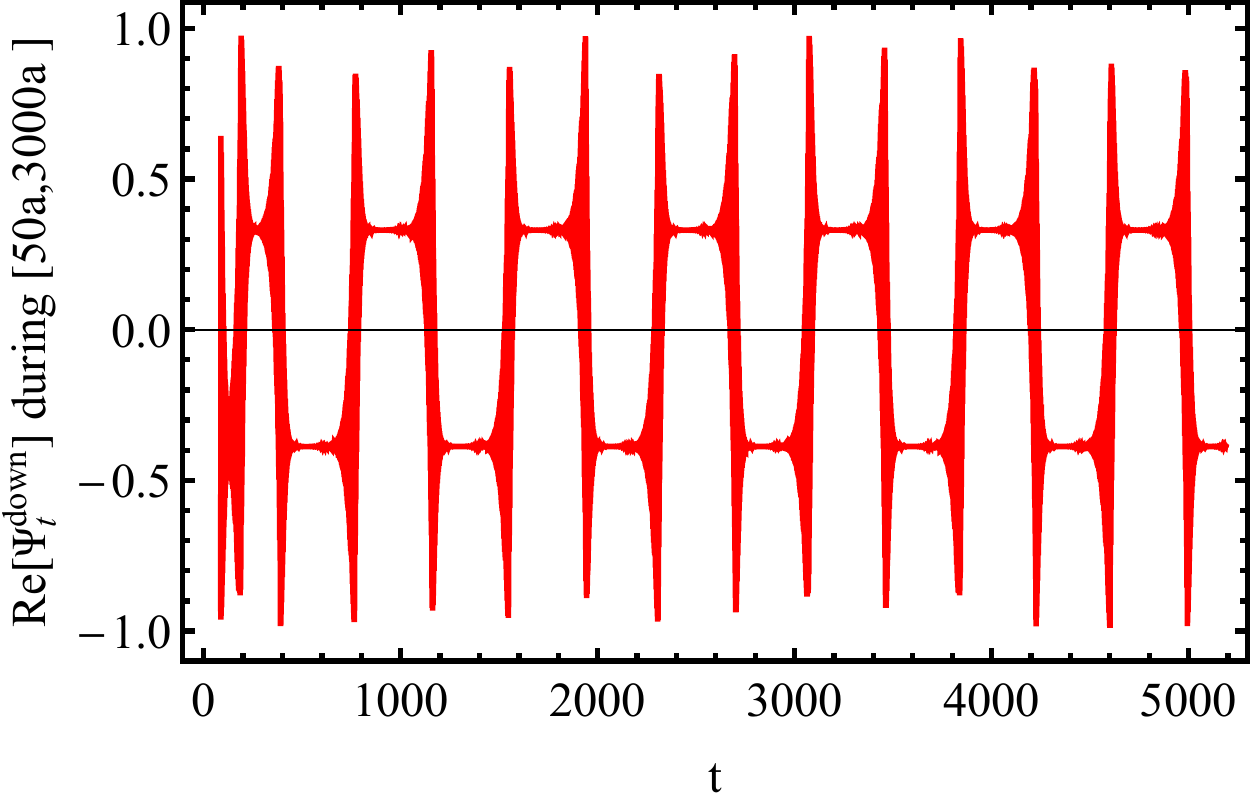}
		\caption{}
		\label{}
	\end{subfigure}
	\quad
	\begin{subfigure}[b]{2.9in}
		\includegraphics[width=2.9in]{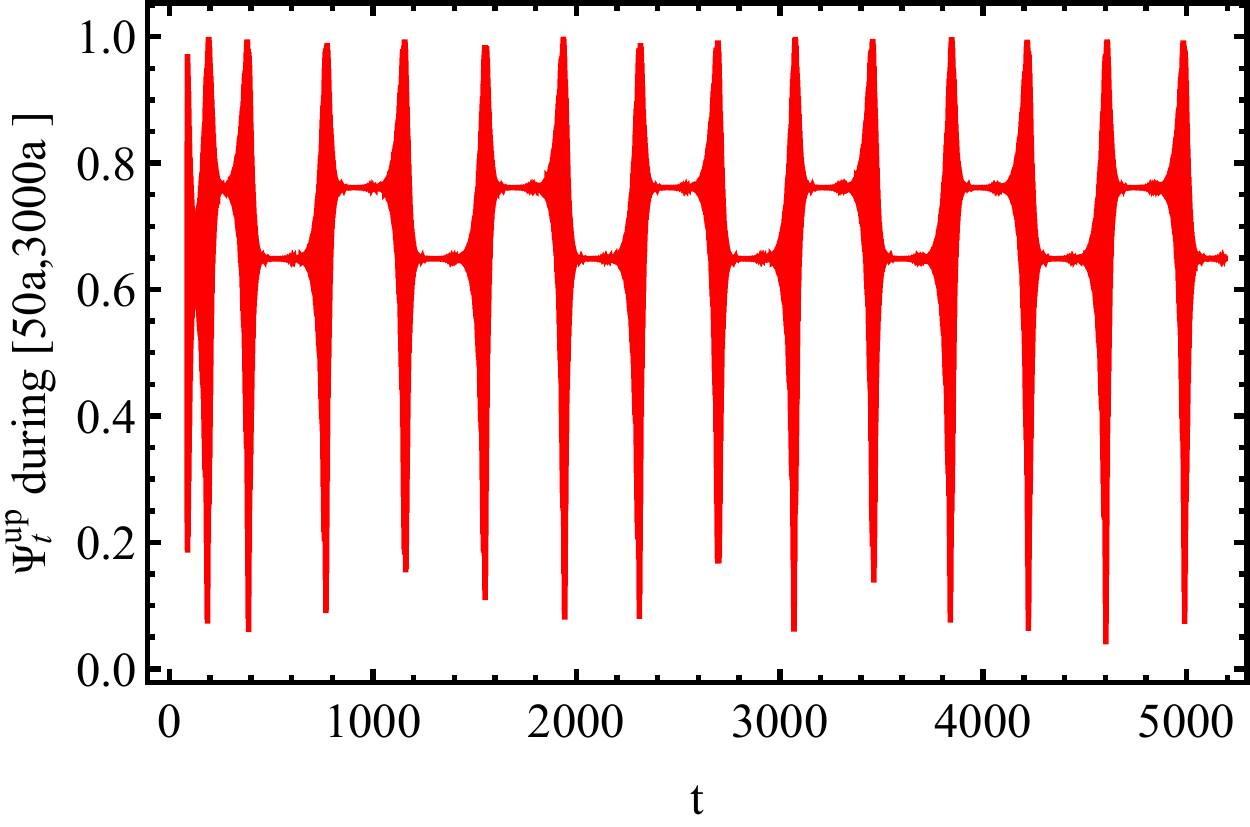}
		\caption{}
		\label{}
	\end{subfigure}
	\begin{center}
		\begin{subfigure}[b]{2.9in}
			\includegraphics[width=2.9in]{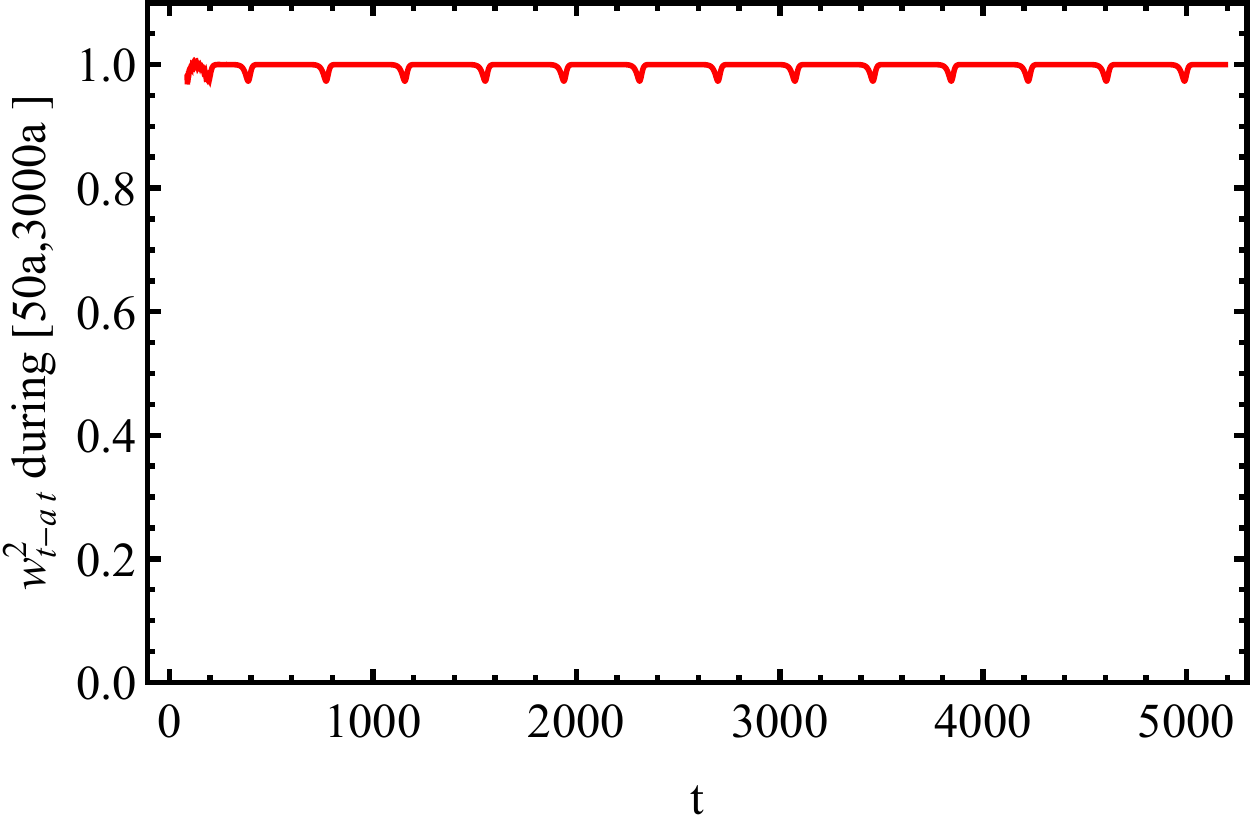}
			\caption{}
			\label{}
		\end{subfigure}
	\end{center}
	\caption{The Purely-QMM-UE described by (\ref{fgp},\ref{ate:ma},\ref{ate:mb},\ref{SE23POLAR}) with parameters:
		$(B_{Kicker}^{y} ; \theta_0; \phi_0 | \lambda^I, \eta | a; t_{completion}) = (7.5; 1.001, 0.089 | 10.00, 4.00 | 1.73; 5000 a)$.
		It equals the `Phase 5' of one-qubit (2,2)-Purely-QMM-UE.}
	\label{fig-23-5}
\end{figure}\\
Increasing $\eta$ above 2.3, we find an all-times-surviving two-state meta-stable phase which is identical to the phase 5 characterized for the (2, 2) system. A representative solution of this phase is presented in Fig. \ref{fig-23-5}. The time span of each meta-stable state equals its (2, 2) reduction, being $\sim 50\;a$, and $w^2_{t-at}$ variations are almost equal too. \\\\  
%%%%%%%%%%%%%%%%%%%%%%%%%
%%%%%%%%%%%%%%%%%%%%%%%%%%%%%%%%%%%%%% 
One also encounters in the phase diagram of the one-qubit $(2,3)$ Purely-QMM-UE altered versions of the standard phase 5. These are morphological deformations of the binary-metastable  QMM-UE which are interesting physically. One such deformation is presented in Fig. \ref{fig-23-S5C}: we see that the altered evolution features the standard profile of the QM-TPF in the phase 5, but now in such a way that the relative configurations of both metastable states undergo robust simple oscillations. The long wavelengths of these relative oscillations are determined primarily by the $\lambda^I$ and $\eta$ couplings. We explored these interesting deformations of the phase 5 in a sufficiently-wide range of the couplings. The numerical solution of Fig. \ref{fig-23-S5C} corresponds to $(\lambda^I , \eta) = (2.6, 2.53)$ as a representative, where a periodicity of about $T \sim 1000\;a$ is featured. For optimal clarity, the wavefunction is shown in two time intervals of the QMM evolution. One sees again that more complex phases are made because of the third order interactions.
%%%%%%%%%%%%%%%%%%%%%%%%%%%%%%%%%%%%%%
\\\\ 
\\\\
Moreover, we find that the one-qubit (2,3) Purely-QMM-UE can form dynamical phase transitions under which a metastable phase 3 is eventually switched with an stable phase 5. It has the same purely-internal nature that its previously mentioned counterpart has, where a dynamical switching from the phase 5 to the phase 1 was realized. Fig.\ref{fig-23-S5} and Fig.\ref{fig-23-S5e} present a representative solution of the kind where the unitary dynamical phase transition occurs at $t_{\text{dpt}} \simeq 1155\;a$, given the couplings $(\lambda^I, \eta) = (1.3. 4.93 )$. To clarify the defining criterion of this novel phase transition, we have exposed, and zoomed into, two sufficiently-distant time intervals as follows: $[\; 300\;a, 400\;a \;]$ and $[\;2200\;a, 2300\;a\;]$. Specially, one sees in our zooming into the QM-TPF profiles in the pre-vs-post phases of the purely-internal transition, Fig. \ref{fig-23-S5e}, that the (2,3) one-qubit Purely-QMM-UE has `itself' evolved from the phase-3 evolution to the phase-5 evolution. In the plots of Fig. \ref{fig-23-S5}, we see the transformed behaviors of $(Re[\Psi^{\text{down}}_t],\Psi^{\text{up}}_t)$, by means of the dynamical phase transition. 
%%%%%%%%%%%%%%%%%%%%%%%%%
%%%%%%%%%%%%%%%%%%%%%%%%%%%%%%%%%%%%%%%%%%%%%%%%%%%%%%%%%%%%%%%%%%
\\\\ 
Sufficiently strengthening the third-order (over the second-order) interaction, we see that the one-qubit Purely-QMM-UE wavefunctions show the defining quality of the phase 3. For example, having $\lambda^I = 5$ fixed and raising $\eta$ above $\sim 10$ is sufficient to derive the system into the phase 3.\\\\ % $\eta \gtrsim 8$.\\   
%%%%%%%%%%%%%%%%%%%%%%%%%
\begin{figure}[t]
	\begin{subfigure}[b]{2.9in}
		\includegraphics[width=2.9in]{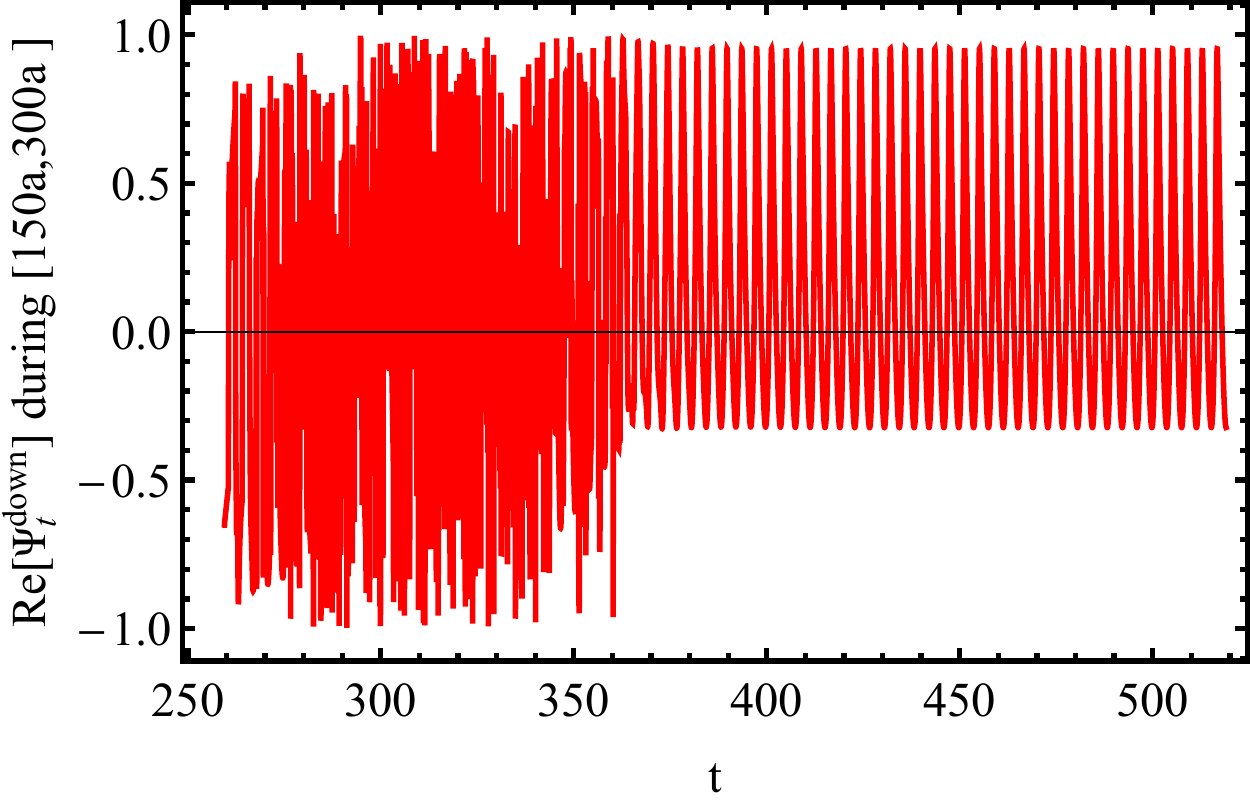}
		\caption{}
		\label{}
	\end{subfigure}
	\quad
	\begin{subfigure}[b]{2.9in}
		\includegraphics[width=2.9in]{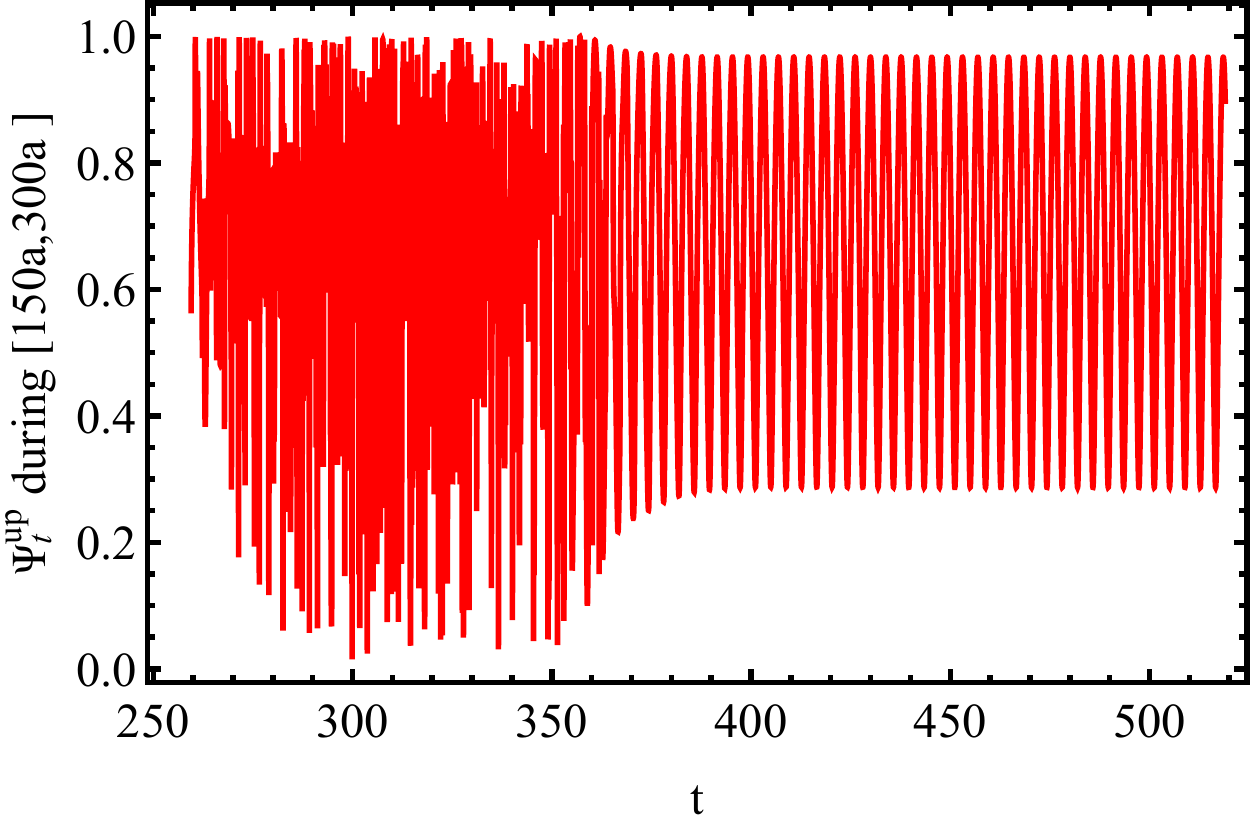}
		\caption{}
		\label{}
	\end{subfigure}
	\begin{center}
	\begin{subfigure}[b]{2.9in}
		\includegraphics[width=2.9in]{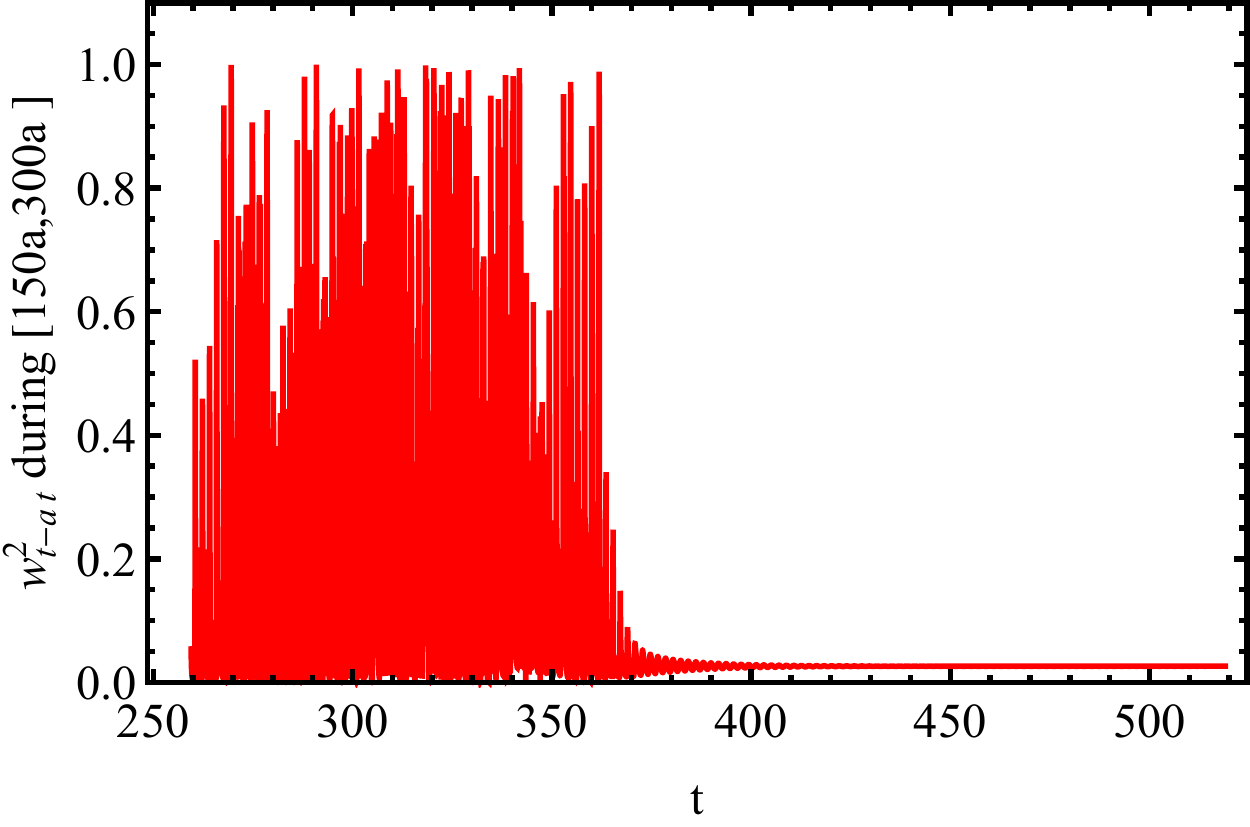}
		\caption{}
		\label{}
	\end{subfigure}
	\end{center}
	\caption{The Purely-QMM-UE described by (\ref{fgp},\ref{ate:ma},\ref{ate:mb},\ref{SE23POLAR}) with parameters:
		$(B_{Kicker}^{y} ; \theta_0; \phi_0 | \lambda^I, \eta | a; t_{completion}) = (7.5; 1.001, 0.089 | -5, 40.30 | 1.73; 3000 a)$.
		Dynamical Phase Transition of one-qubit (2,3)-Purely-QMM-UE: `Phase 3 to Phase 2'.}
	\label{fig-23-S2}
\end{figure}
\\\\ 
We now explore the distinct category of negative $\lambda^I$ in regard with the one-qubit phases of the (2,3)-Purely-QMM-UE.  Setting $\lambda^I = -5$, we increase $\eta$ from zero, with steps of $\sim 0.1$ or $0.01$, up to sufficiently large values, to identify and classify the corresponding phases. In the whole range of $\eta$ taking values from zero up to $\sim 50$, we observe the phase 2. At $\eta \sim 50$, where the third-order interaction has become dominant about one order of magnitude, the one-qubit purely-QMM-UE makes a crossover to irregularity: the phase 3 takes place beyond this soft border. Along the way to its crossover to irregularity, the one-qubit (2,3) system makes a unitary and purely-internal dynamical phase transition from a long-time-surviving phase 3 to an ultimate phase 2.  In fact, we observe this remarkable behavior in the whole range: $\lambda^I = -5, \; 40 \lesssim \eta \lesssim 50$, corresponding to the bound $\frac{\Delta \eta}{\eta} \sim 0.1 $. Fig. \ref{fig-23-S2} shows a clear representative numerical solution which features this dynamical phase transition. We highlight that, as it comes about in the next subsection, a phase-interpolating behavior of the same kind is seen in the one-qubit (3,3) system. This suggest that this dynamical phase transition is a signature of the third-order Purely-QMM interactions.
%%%%%%%%%%%%%%%%%%%%%%%%%
\begin{figure}[t]
	\begin{subfigure}[b]{2.9in}
		\includegraphics[width=2.9in]{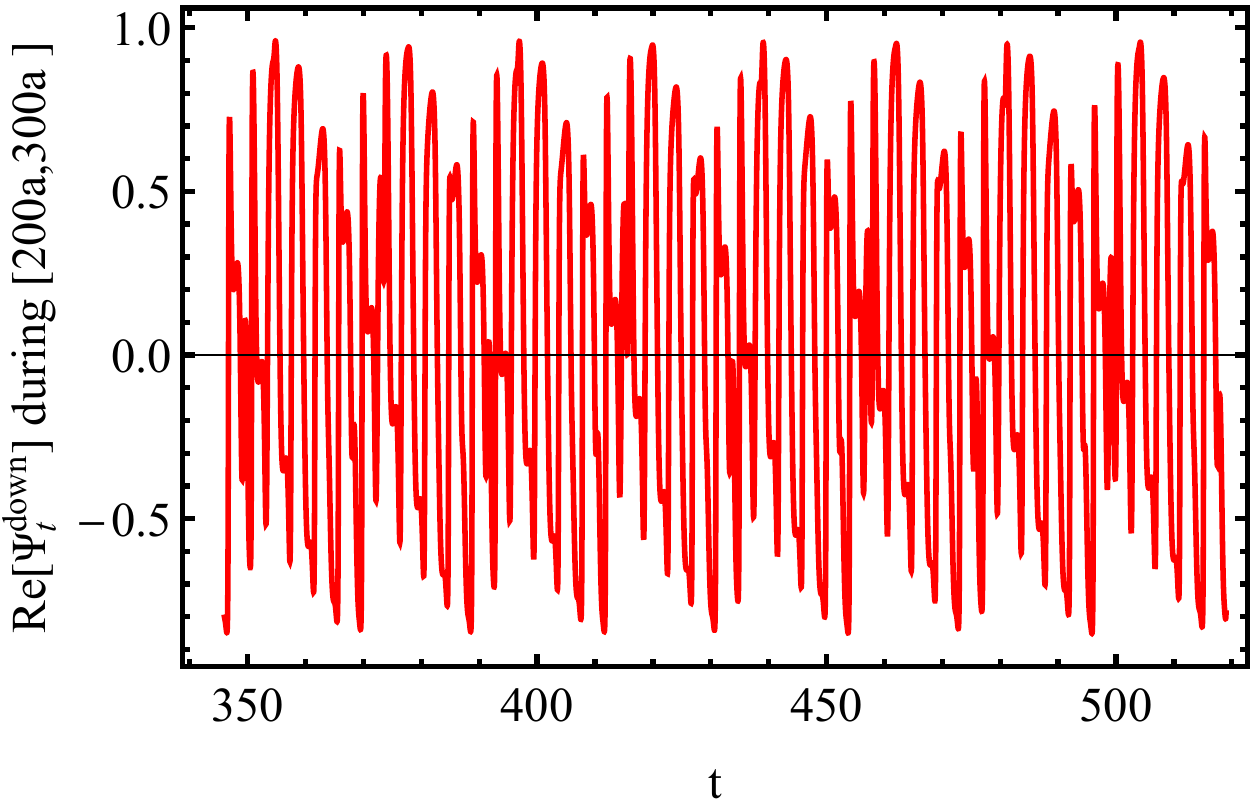}
		\caption{}
		\label{}
	\end{subfigure}
	\quad
	\begin{subfigure}[b]{2.9in}
		\includegraphics[width=2.9in]{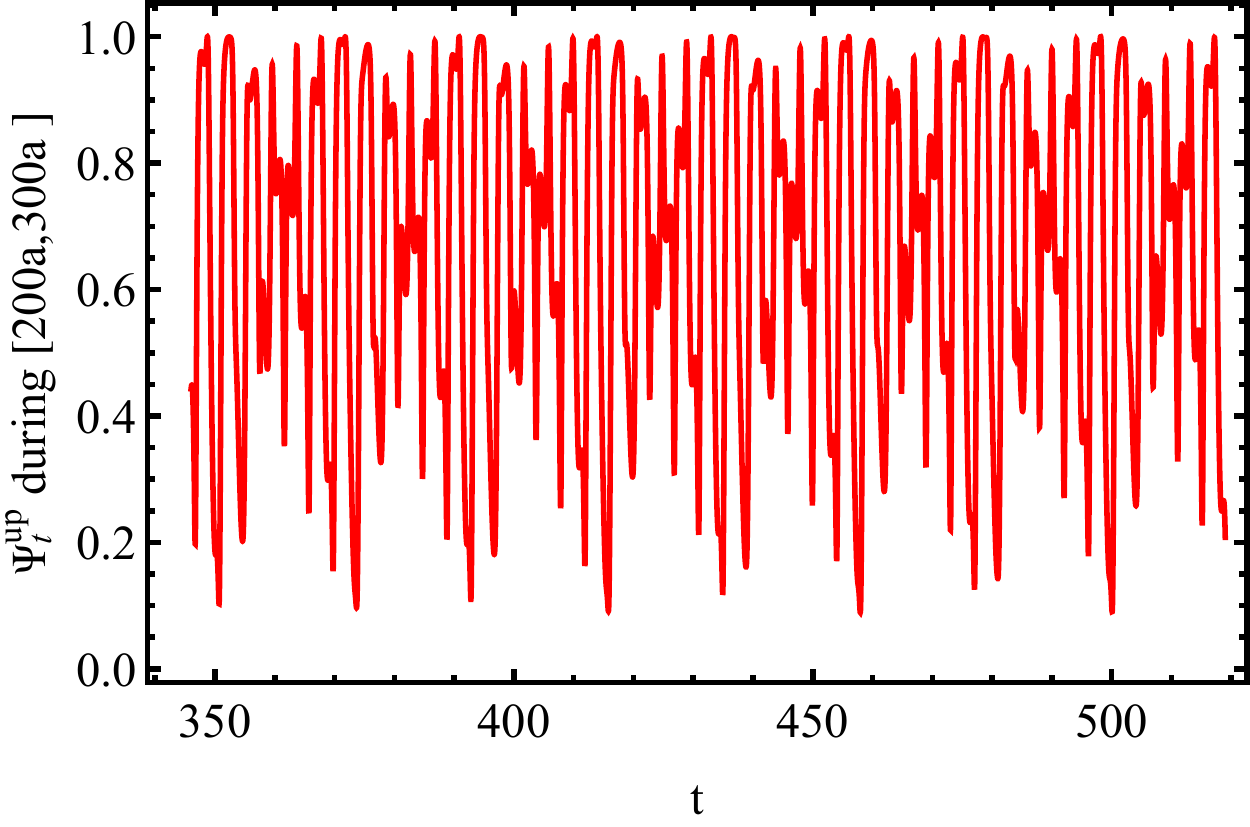}
		\caption{}
		\label{}
	\end{subfigure}
	\begin{center}
	\begin{subfigure}[b]{2.9in}
		\includegraphics[width=2.9in]{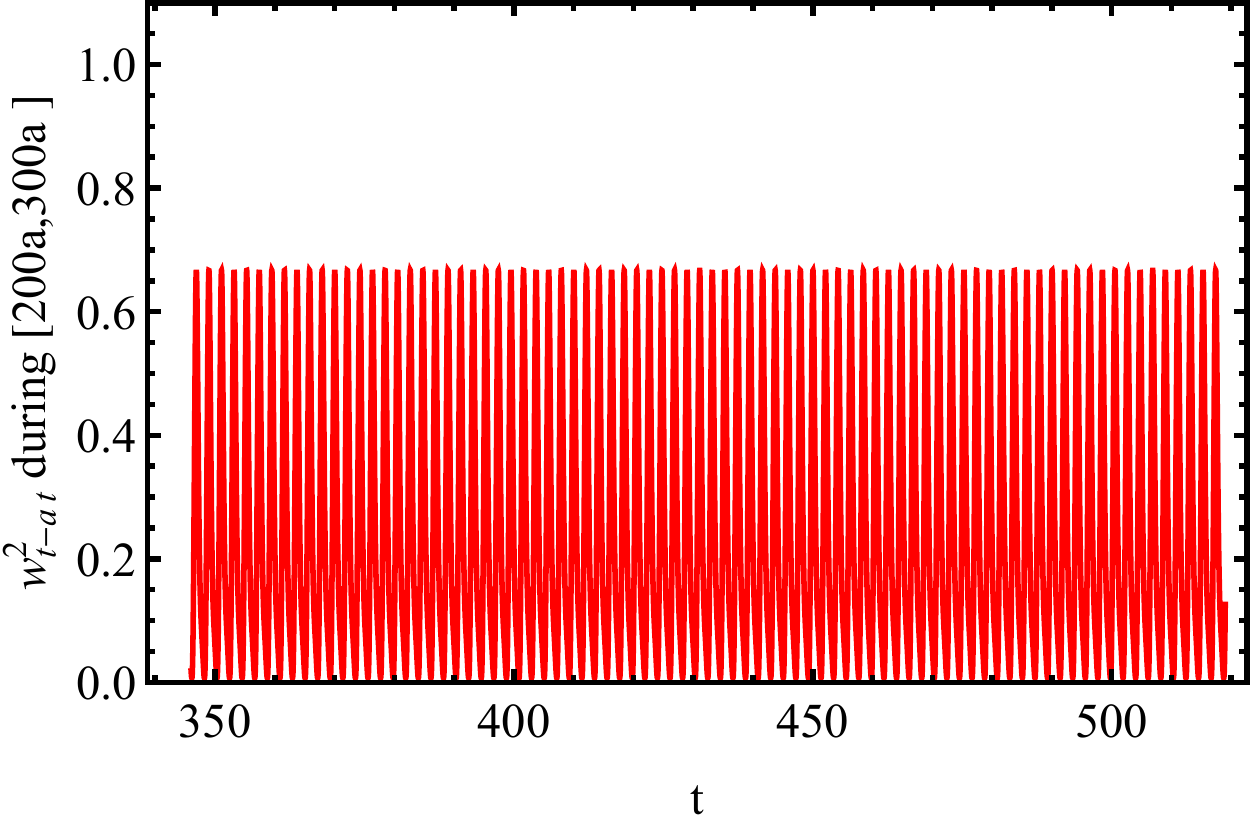}
		\caption{}
		\label{}
	\end{subfigure}
	\end{center}
	\caption{The Purely-QMM-UE described by (\ref{fgp},\ref{ate:ma},\ref{ate:mb},\ref{SE23POLAR}) with parameters:
		$(B_{Kicker}^{y} ; \theta_0; \phi_0 | \lambda^I, \eta | a; t_{completion}) = (7.5; 1.001, 0.089 | -1.30, 6.5 | 1.73; 3000 a)$.
		`Phase 4' of one-qubit (2,3)-Purely-QMM-UE.}
	\label{fig-23-4}
\end{figure}
%%%%%%%%%%%%%%%%%%%%%%%%%
\\\\ 
One-qubit (2,3) Purely-QMM-UE with negatively-signed second-order interactions can also form the phase 4. For example, realizations of phase 4 is obtained in the entire range: $\lambda^I = -1.3$ and $5.5 \lesssim \eta \lesssim 8$. Moreover, because now the third-order interactions are present in the QMM-H, quite non-generic versions of phase 4 are easily realizable. In these non-generic and refined versions, the structured modules of phase 4 feature significantly-enhanced orders in their patterns. A representative solution of the kind is given in Fig. \ref{fig-23-4} where, for visual clarity, only the window $[200\;a, 300\;a]$ is presented. Furthermore,
in narrow bounds of $(\lambda^I,\eta)$,
dynamical phase transitions from phase 3 to phase 4 is observed. Naturally, we expect that these alternative dynamical phase transitions can likewise occur without any fine-tuning of the couplings, in the presence of higher order interactions: $L > 3$.
%%%%%%%%%%%%%%%%%%%%%%%%%%%%%%%%%%%%%%%%%%
%%%%%%%%%%%%%%%%%%%%%%%%%%%%%%%%%%%%%%%%%%%

\subsection{The Phase Diagram of One-Qubit Wavefunctions Evolving Under A (3,3) 
Purely-QMM Hamiltonian}\label{SV-IV}
\begin{figure}
	%\centering
	\begin{subfigure}[b]{2.9in}
		\includegraphics[width=2.9in]{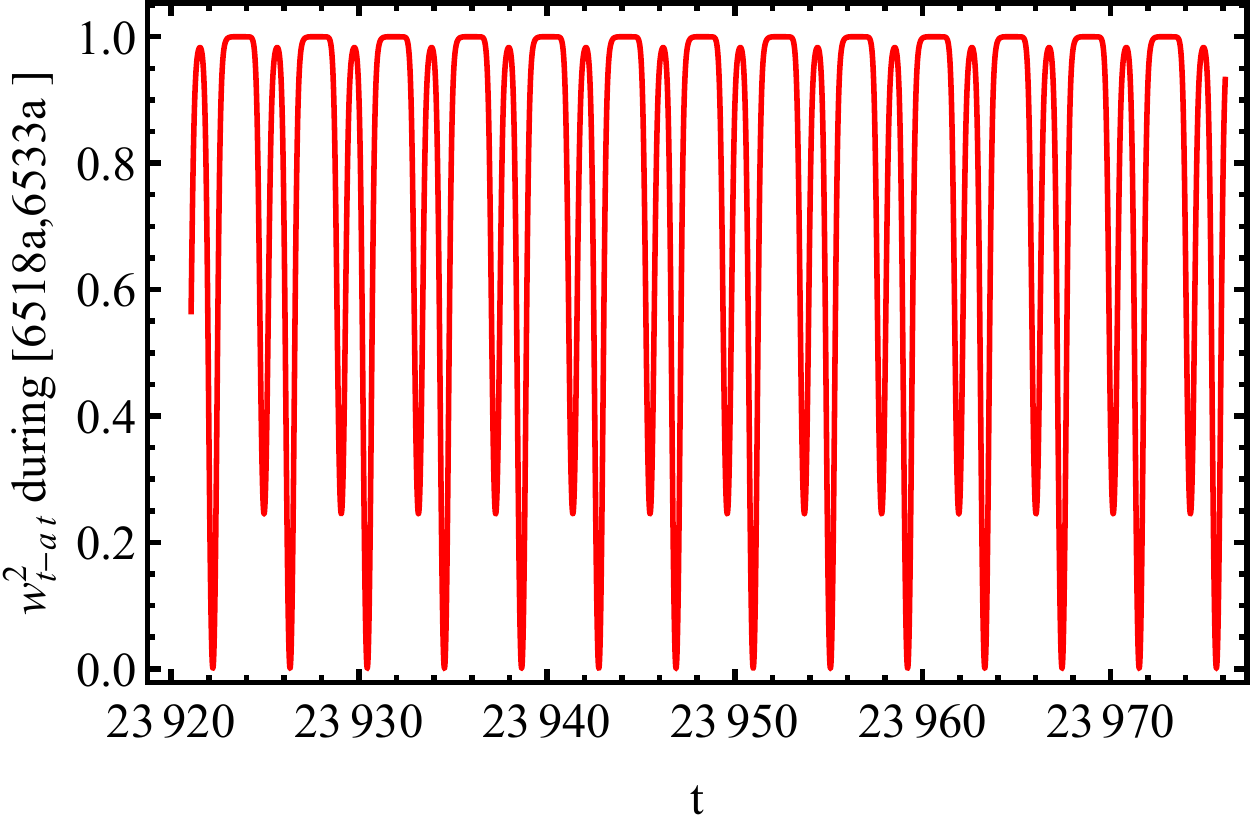}
		\caption{}
		\label{}
	\end{subfigure}
	\quad
	\begin{subfigure}[b]{2.9in}
		\includegraphics[width=2.9in]{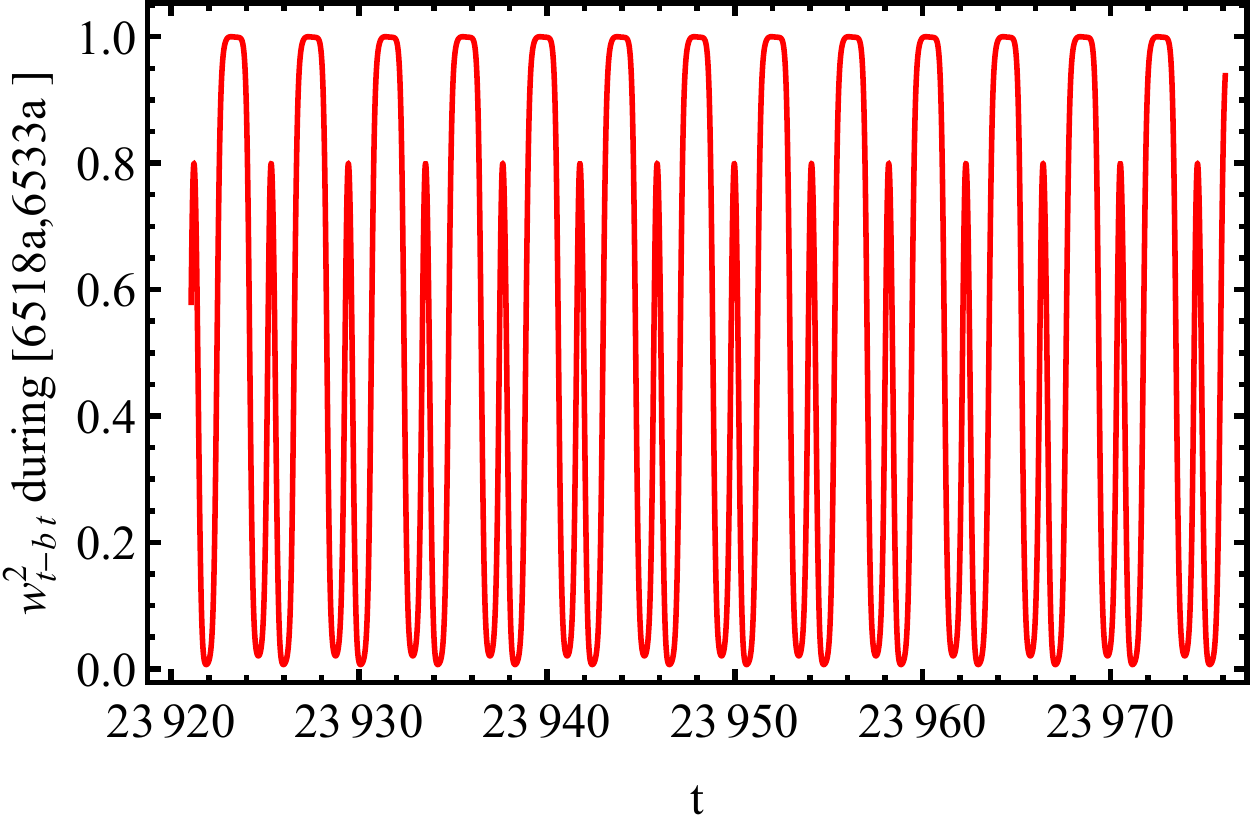}
		\caption{}
		\label{}
	\end{subfigure}
		\\
		\\
	\begin{subfigure}[b]{2.9in}
		\includegraphics[width=2.9in]{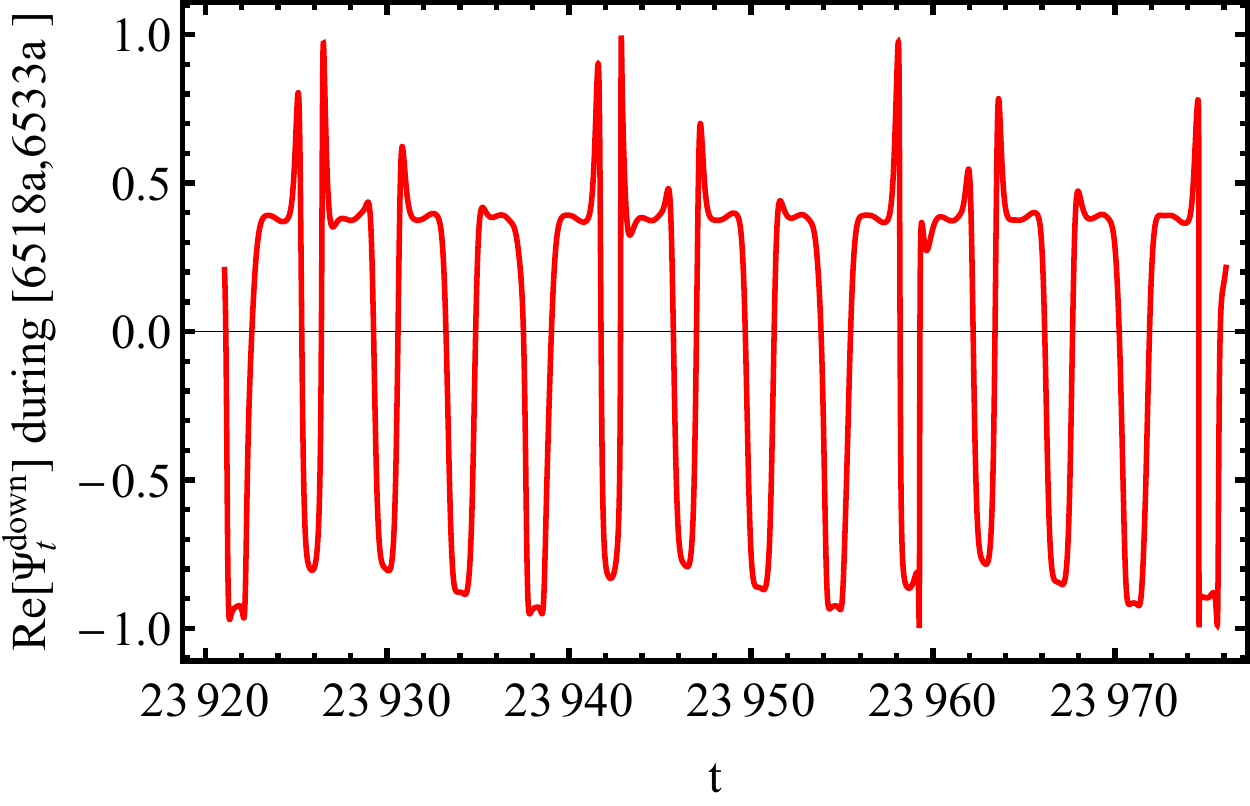}
		\caption{}
		\label{}
	\end{subfigure}
	\quad
	\begin{subfigure}[b]{3in}
		\includegraphics[width=2.9in]{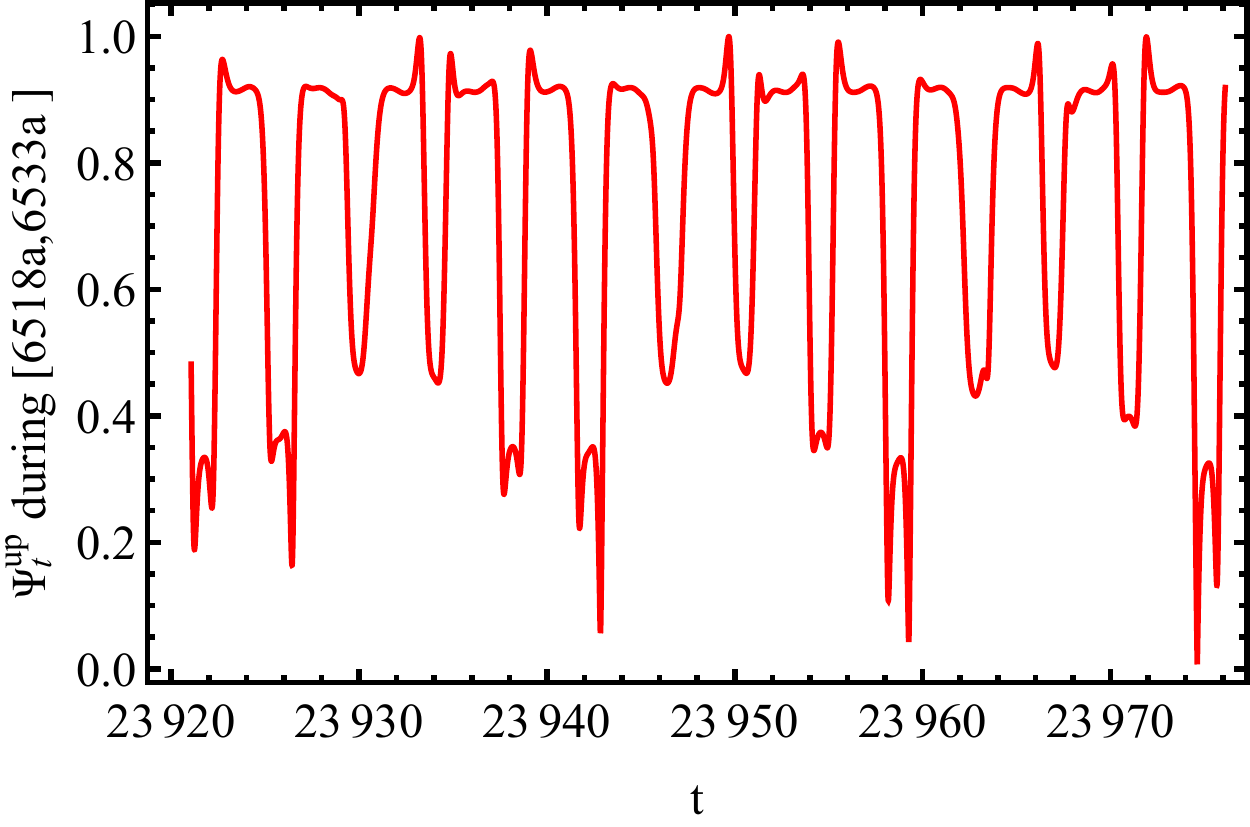}
		\caption{}
		\label{}
	\end{subfigure}
	\caption{Numerical Solution of The QMM-UE system (\ref{SE33POLAR},\ref{fgof33}) with parameters:\\ $r = 1.173$, ($B^y_{\text{Kicker}} ; \theta_0 , \phi_0 | \kappa^I , \kappa^R | a  ; t_{\text{completion}} $)=($35.5 ; 1.001 , 0.089 | 5.124 , 3.185 |  3.67 ; 6600\; a$).\\ A special phase of one-qubit (3,3)-Purely-QMM-UE: blending phases $5$ and $4$ of the one-qubit $N=2$ Theory.}
	\label{33_5R1_All}
\end{figure}

\begin{figure}
\label{33_5R2_All}
	\begin{subfigure}[b]{2.9in}
	\includegraphics[width=2.9in]{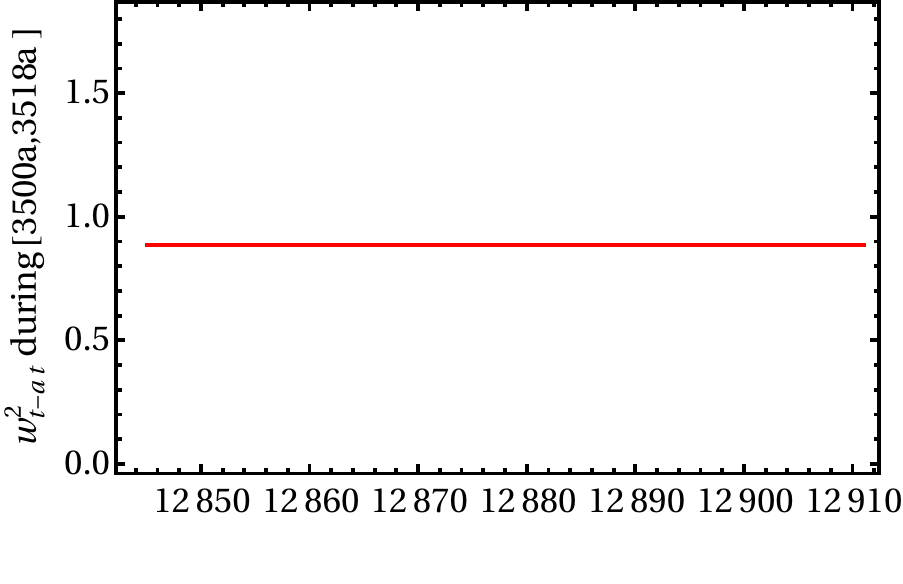}
	\end{subfigure}
	\quad
	\begin{subfigure}[b]{2.9in}
	\includegraphics[width=2.9in]{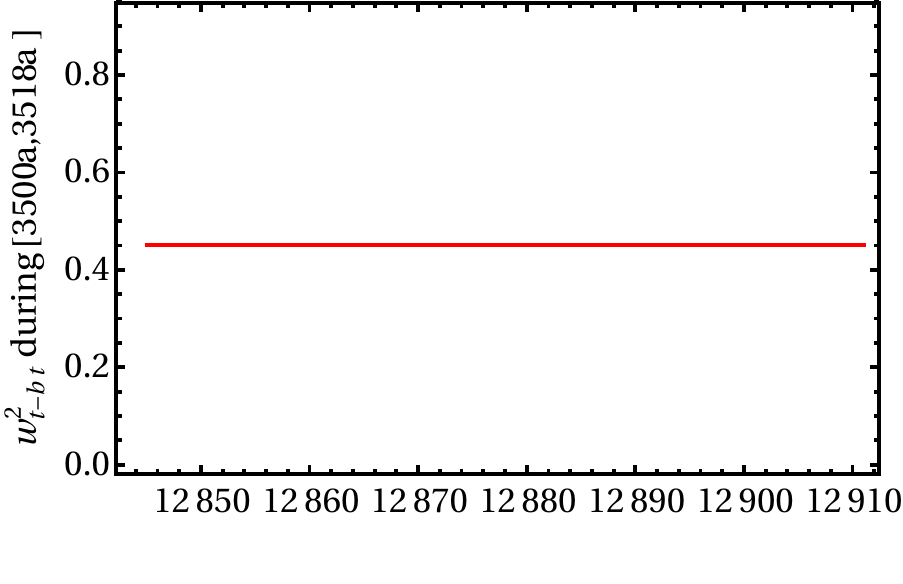}
	\end{subfigure}
		\\
		\\
	\begin{subfigure}[b]{2.9in}
	\includegraphics[width=2.9in]{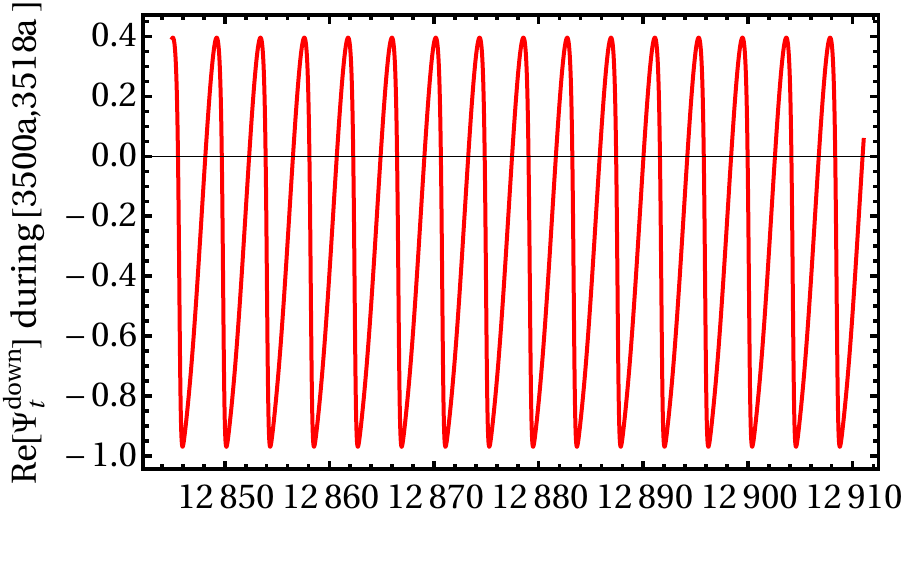}
	\end{subfigure}
	\quad
	\begin{subfigure}[b]{2.9in}
	\includegraphics[width=2.9in]{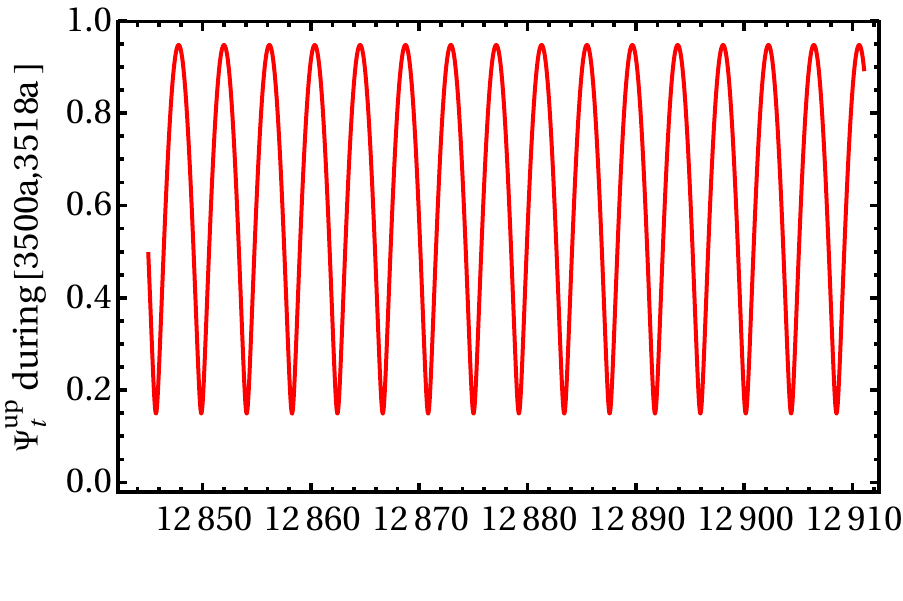}
	\end{subfigure}
	\caption{Numerical Solution of The QMM-UE system (\ref{SE33POLAR},\ref{fgof33}) with parameters: $r = 2.79$, ($B^y_{\text{Kicker}} ; \theta_0 , \phi_0 | \kappa^I , \kappa^R | a  ; t_{\text{completion}} $)=($35.5 ; 1.001 , 0.089 | 5.124 , 3.185 |  3.67 ; 3518\; a$). A special phase of one-qubit (3,3)-Purely-QMM-UE: fully-frozen version of phase $2$.}
\end{figure}

\begin{figure}
	%\centering
		\begin{subfigure}[b]{2.9in}
		\includegraphics[width=2.9in]{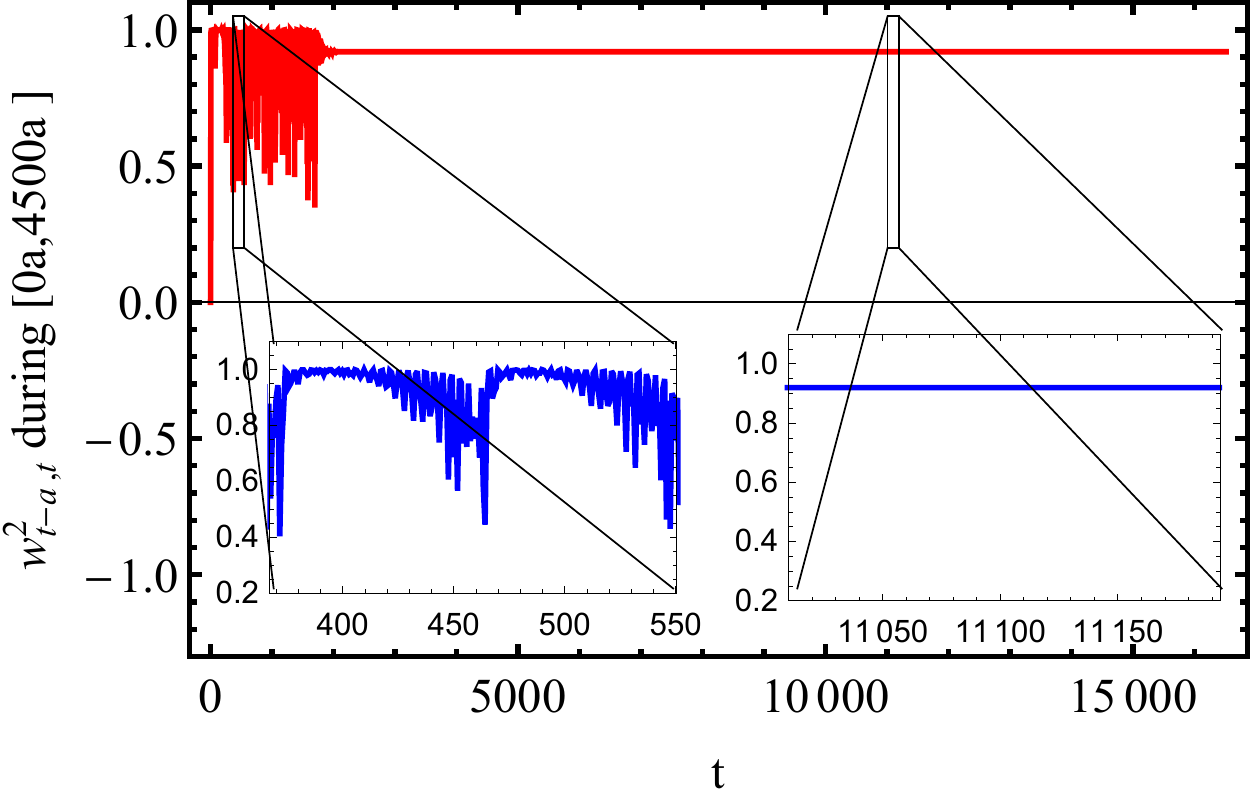}
		\caption{}
		\label{}
	\end{subfigure}
	\quad
	\begin{subfigure}[b]{2.9in}
		\includegraphics[width=2.9in]{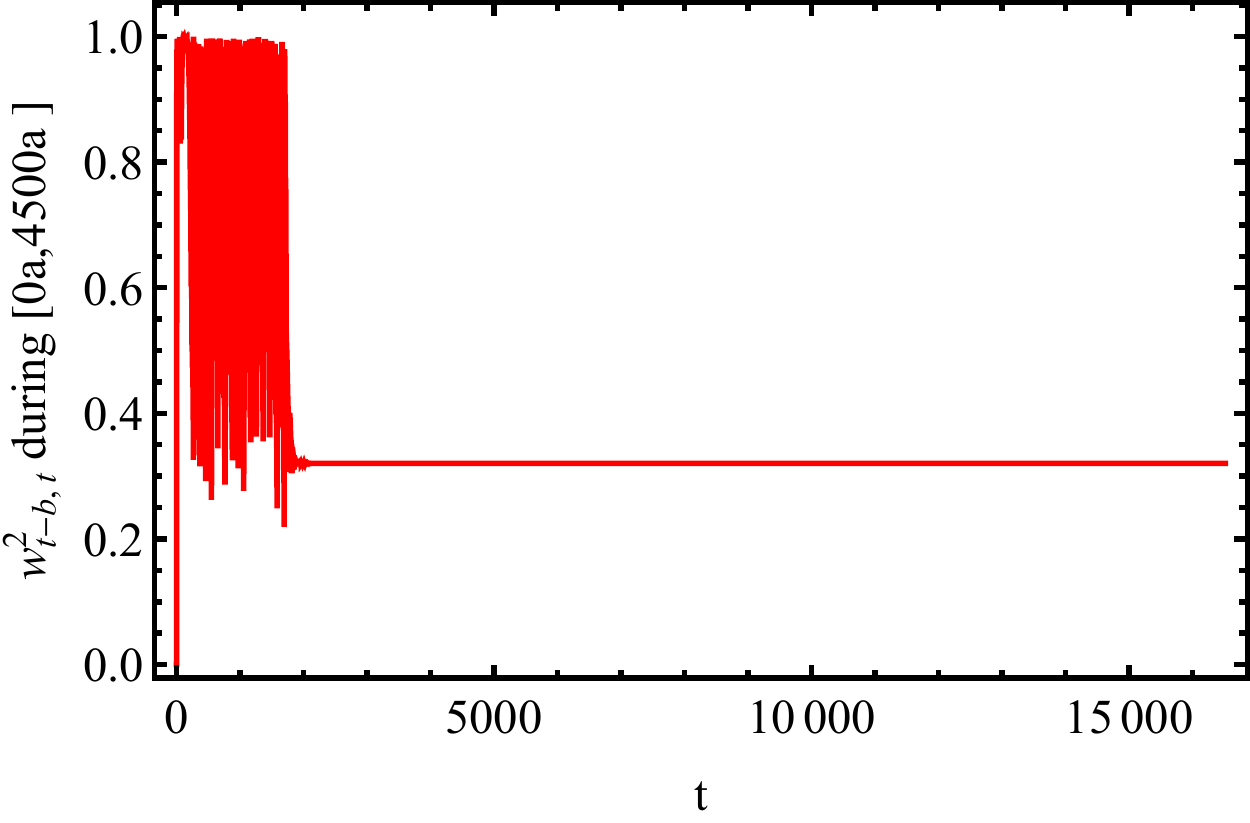}
		\caption{}
		\label{}
	\end{subfigure}
		\\
		\\
	\begin{subfigure}[b]{2.9in}
		\includegraphics[width=2.9in]{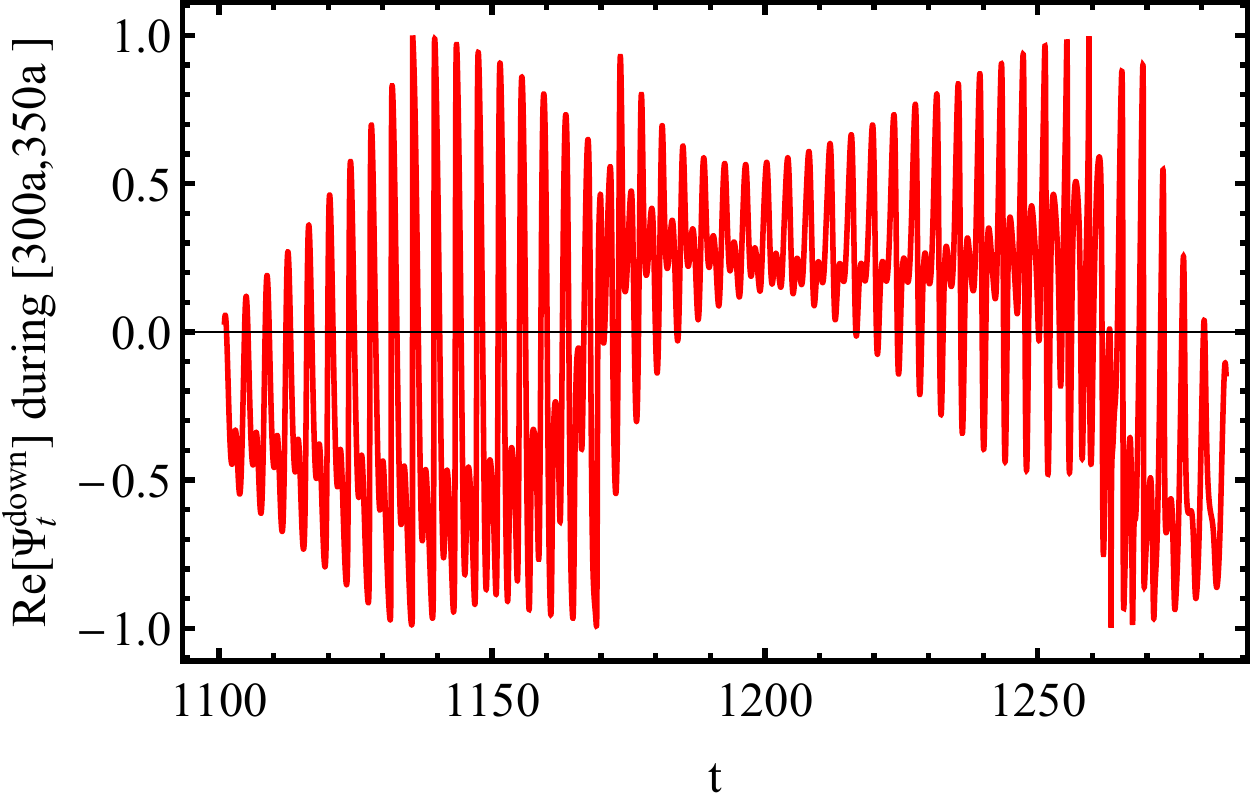}
		\caption{}
		\label{}
	\end{subfigure}
	\quad
	\begin{subfigure}[b]{2.9in}
		\includegraphics[width=2.9in]{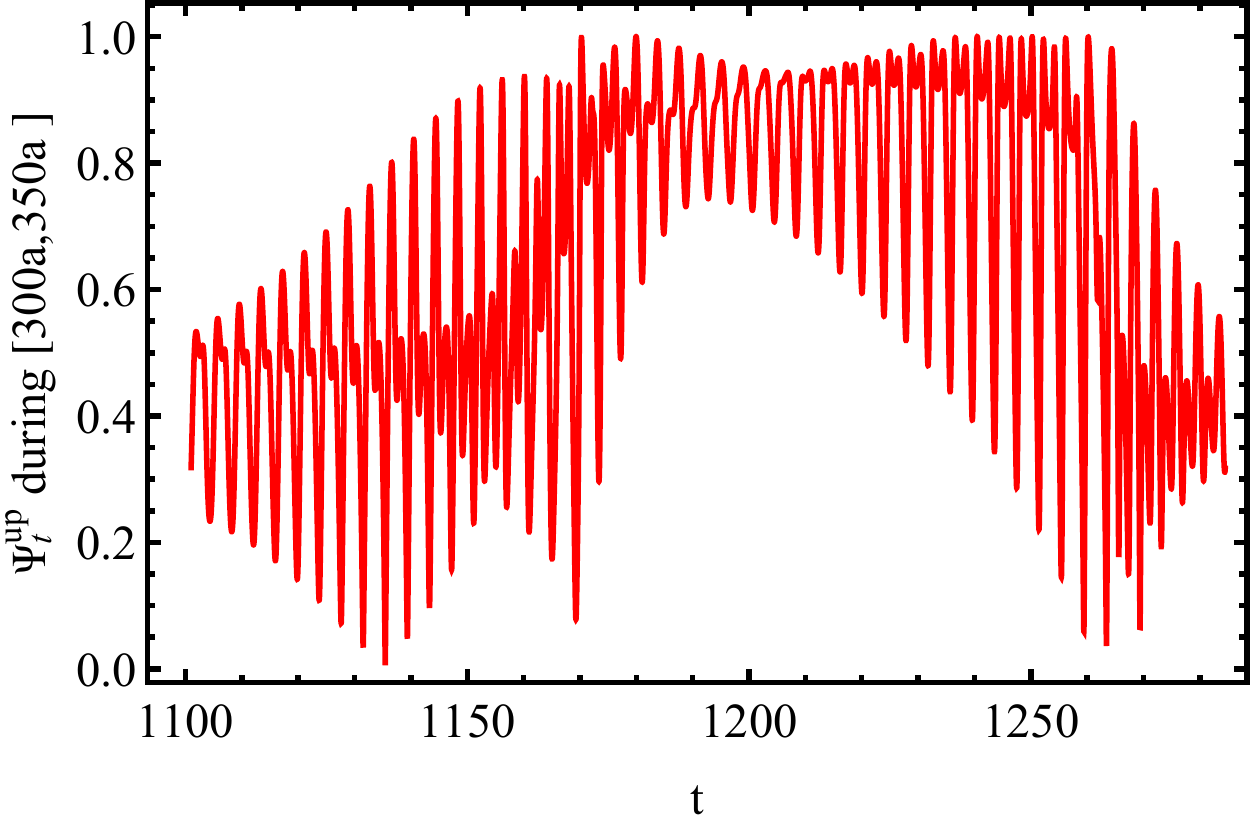}
		\caption{}
		\label{}
	\end{subfigure}
	\\
	\\
	\begin{subfigure}[b]{2.9in}
		\includegraphics[width=2.9in]{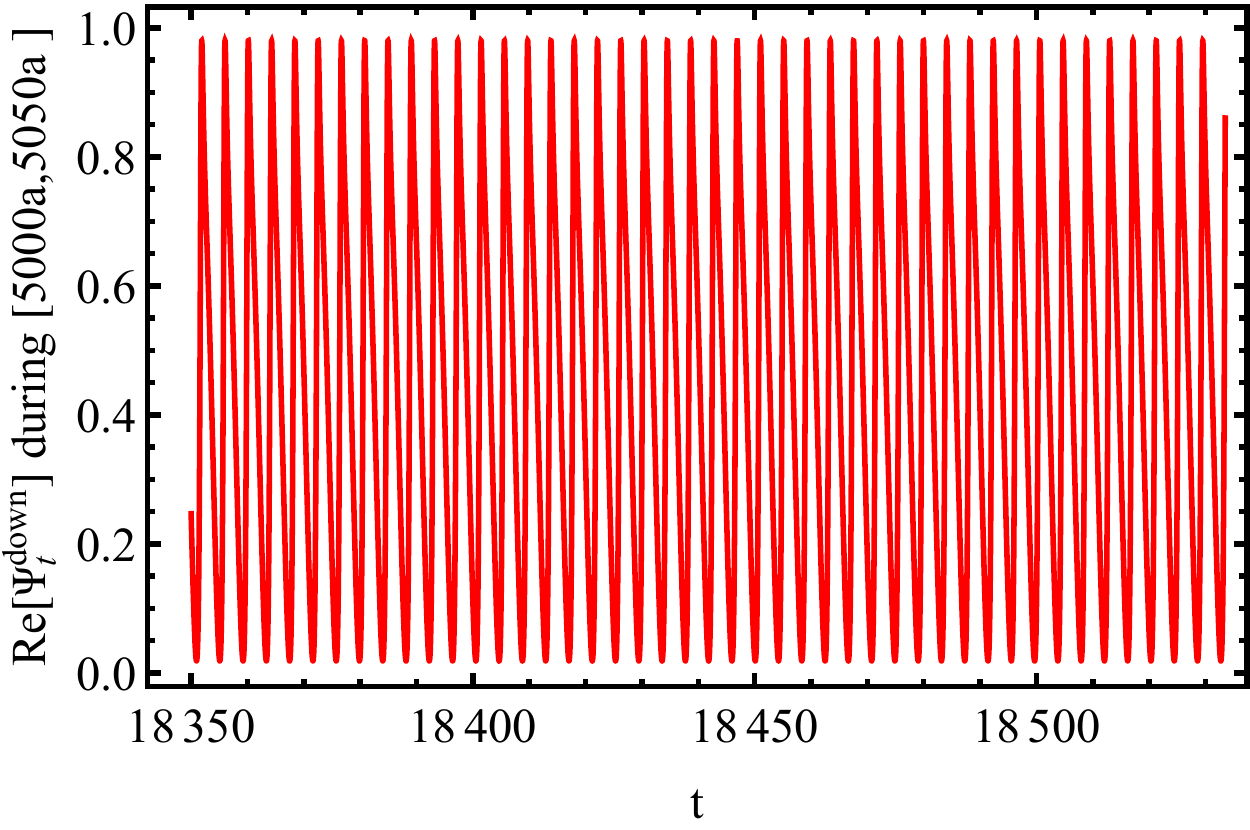}
		\caption{}
		\label{}
	\end{subfigure}
	\quad
	\begin{subfigure}[b]{2.9in}
		\includegraphics[width=2.9in]{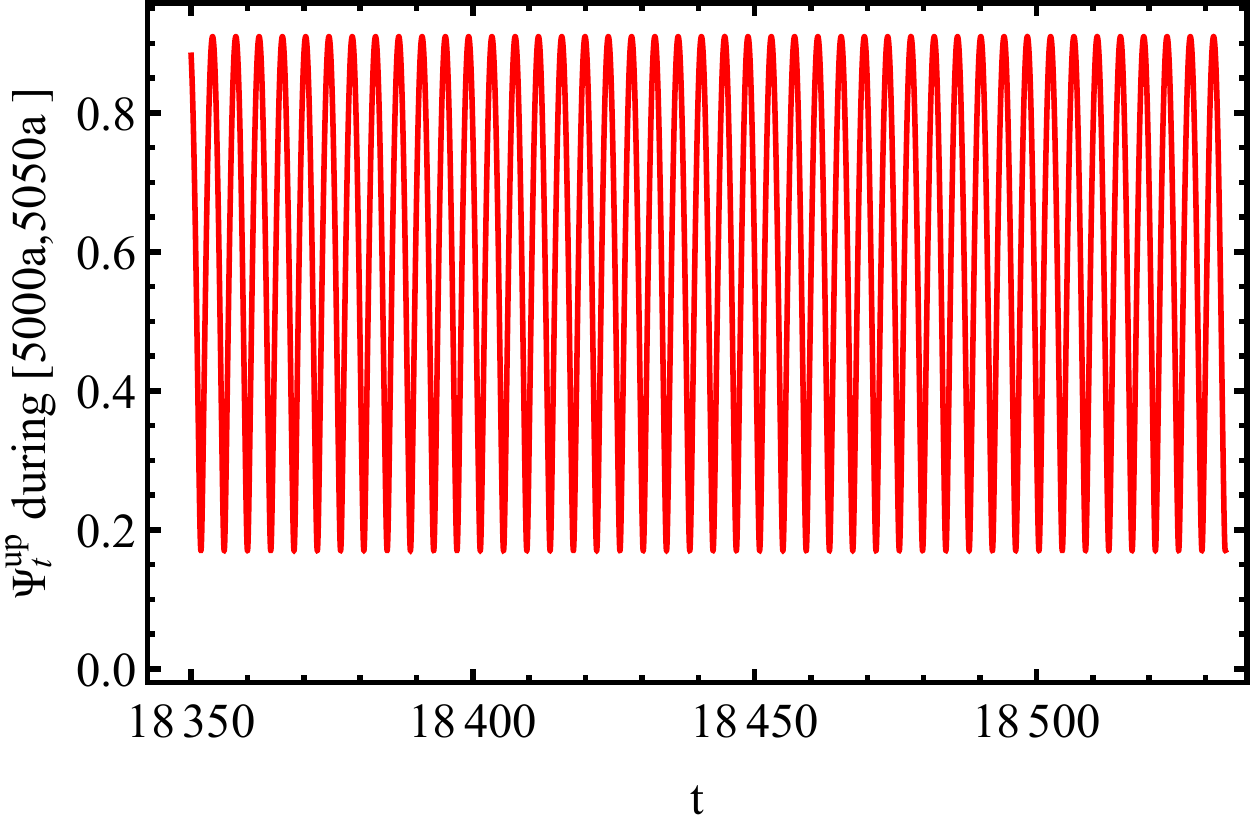}
		\caption{}
		\label{}
	\end{subfigure}
	\caption{Numerical Solution of The QMM-UE system (\ref{SE33POLAR},\ref{fgof33}) with parameters:\\ $r = 1.725$, ($B^y_{\text{Kicker}} ; \theta_0 , \phi_0 | \kappa^I , \kappa^R | a  ; t_{\text{completion}} $)=($35.5 ; 1.001 , 0.089 | 5.124 , 3.185 |  3.67 ; 6000\; a$).\\ A phase-interpolating state-history of one-qubit (3,3)-Purely-QMM-UE.}
	\label{33_5R3_All}
	\end{figure}

\begin{figure}
	%\centering
		\begin{subfigure}[b]{2.9in}
		\includegraphics[width=2.9in]{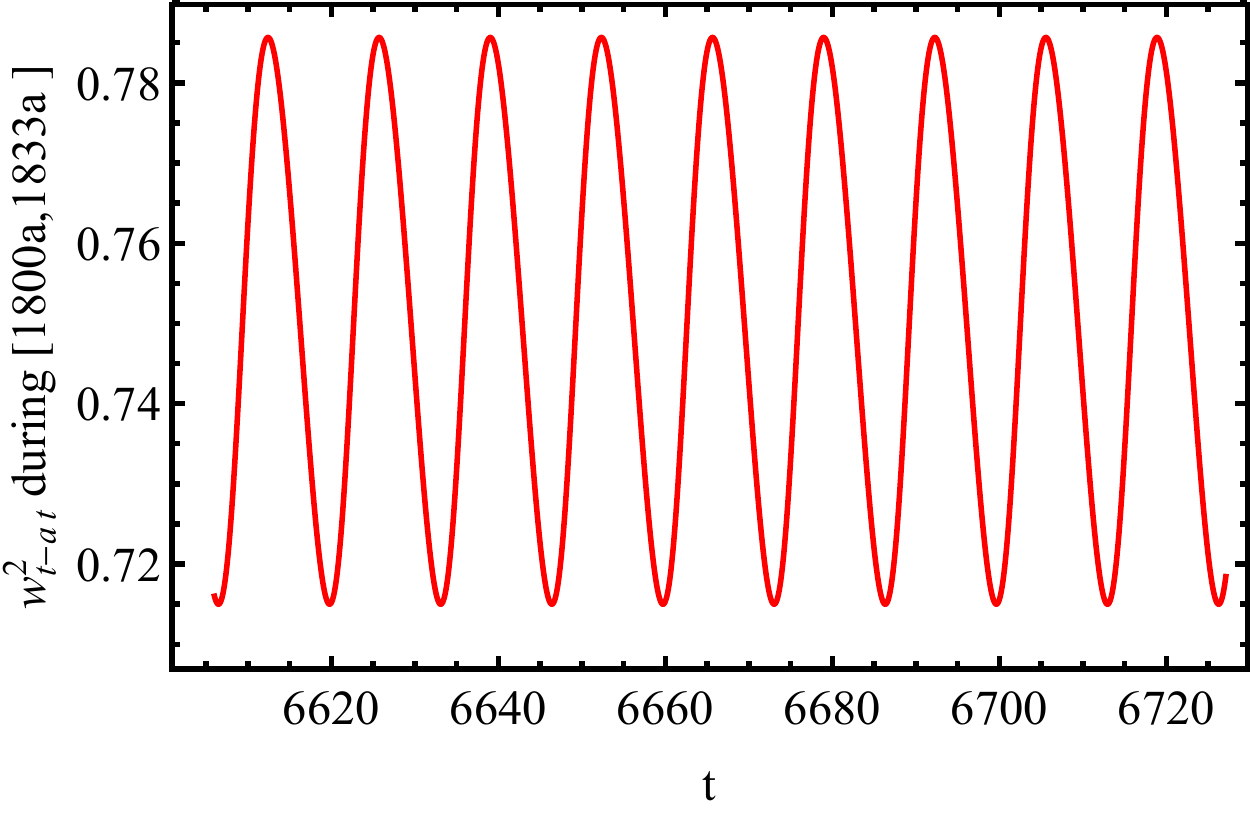}
		\caption{}
		\label{}
	\end{subfigure}
	\quad
	\begin{subfigure}[b]{2.9in}
		\includegraphics[width=2.9in]{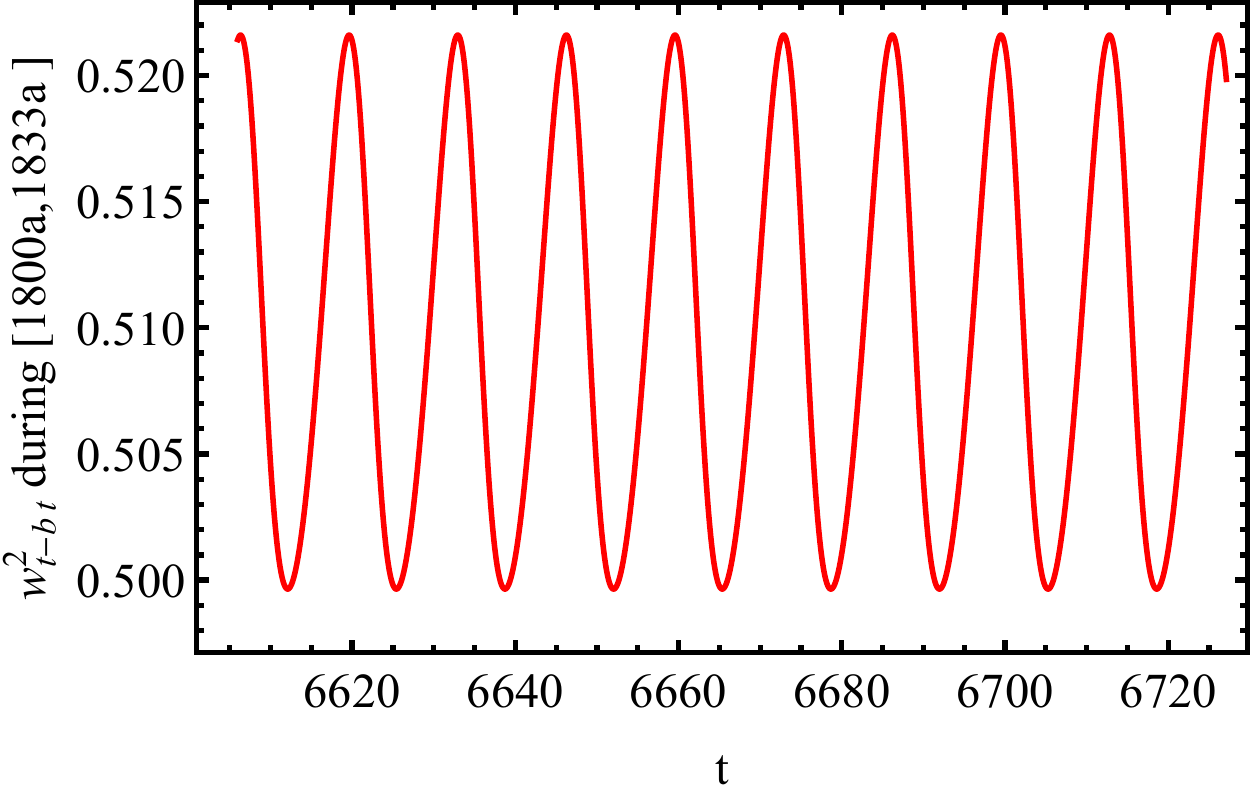}
		\caption{}
		\label{}
	\end{subfigure}
		\\
		\\
	\begin{subfigure}[b]{2.9in}
		\includegraphics[width=2.9in]{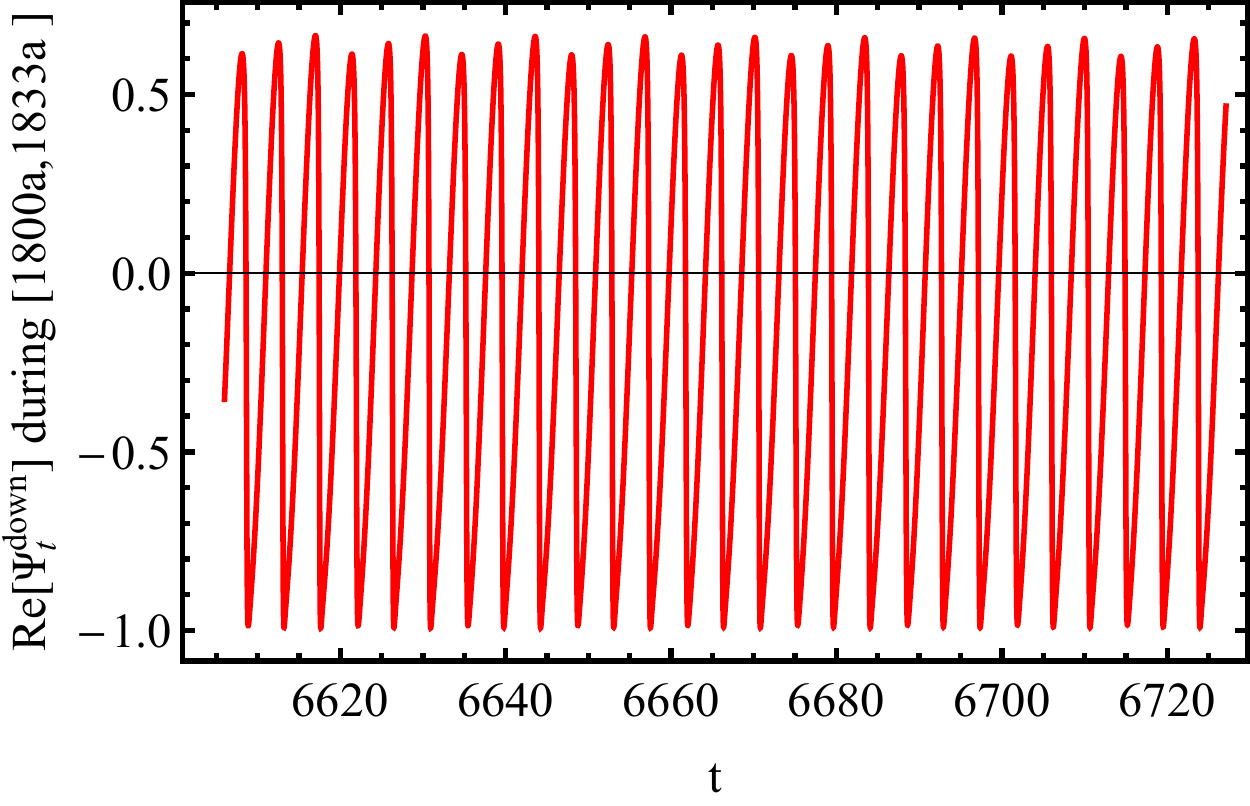}
		\caption{}
		\label{}
	\end{subfigure}
	\quad
	\begin{subfigure}[b]{2.9in}
		\includegraphics[width=2.9in]{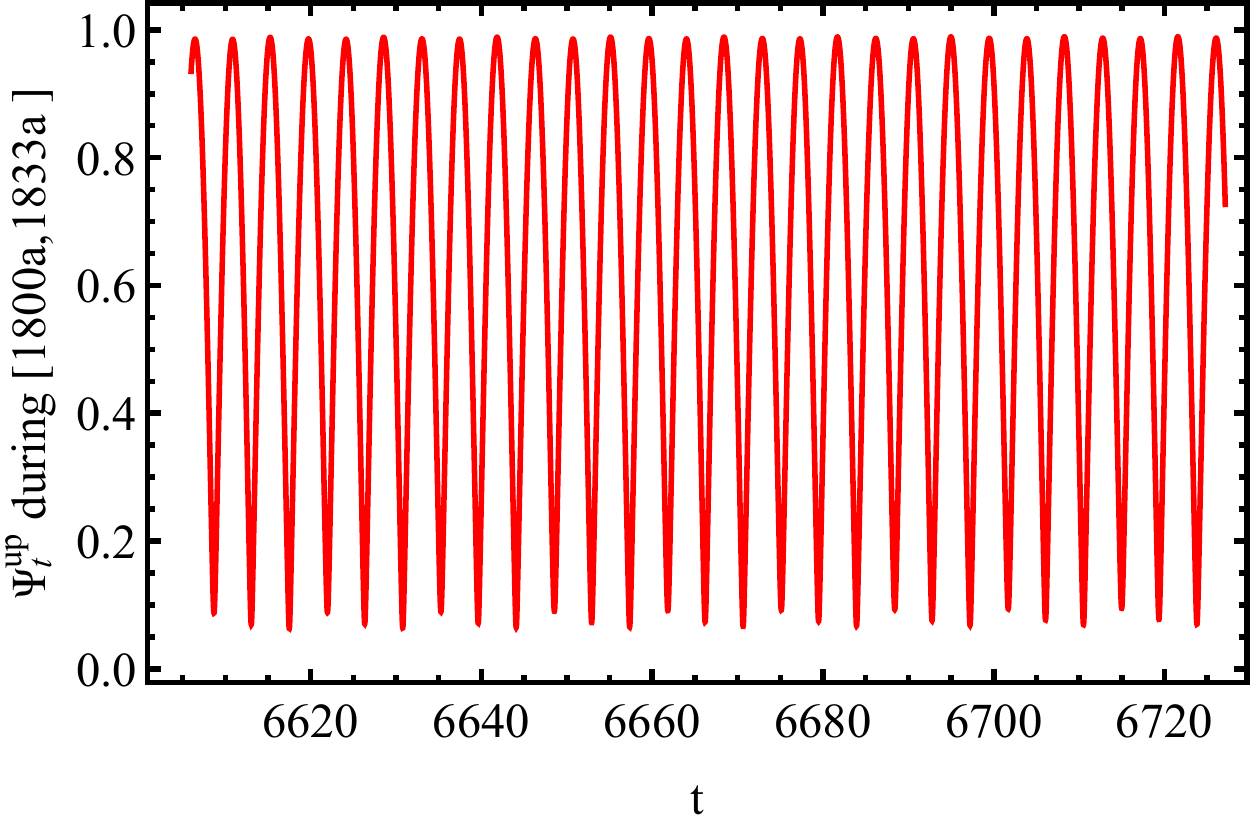}
		\caption{}
		\label{}
	\end{subfigure}
	\caption{Numerical Solution of The QMM-UE system (\ref{SE33POLAR},\ref{fgof33}) with parameters:\\ $r = 3.235$, ($B^y_{\text{Kicker}} ; \theta_0 , \phi_0 | \kappa^I , \kappa^R | a  ; t_{\text{completion}} $)=($35.5 ; 1.001 , 0.089 | 5.124 , 3.185 |  3.67 ; 6600\; a$).\\ Special  one-qubit (3,3)-Purely-QMM-UE, featuring Synchronized Simple Oscillations.}
	\label{33_5R4_All}
\end{figure}

\begin{figure}
	%\centering
		\begin{subfigure}[b]{2.9in}
		\includegraphics[width=2.9in]{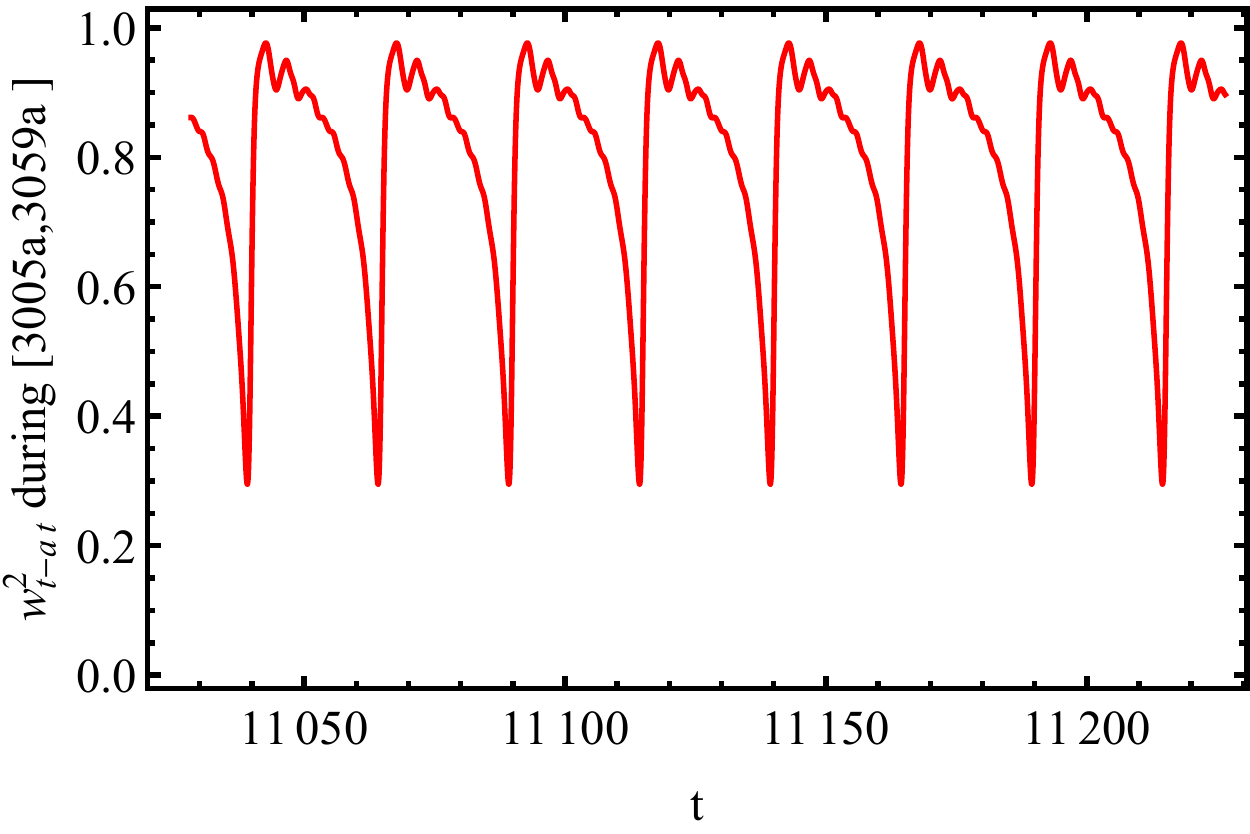}
		\caption{}
		\label{}
	\end{subfigure}
	\quad
	\begin{subfigure}[b]{2.9in}
		\includegraphics[width=2.9in]{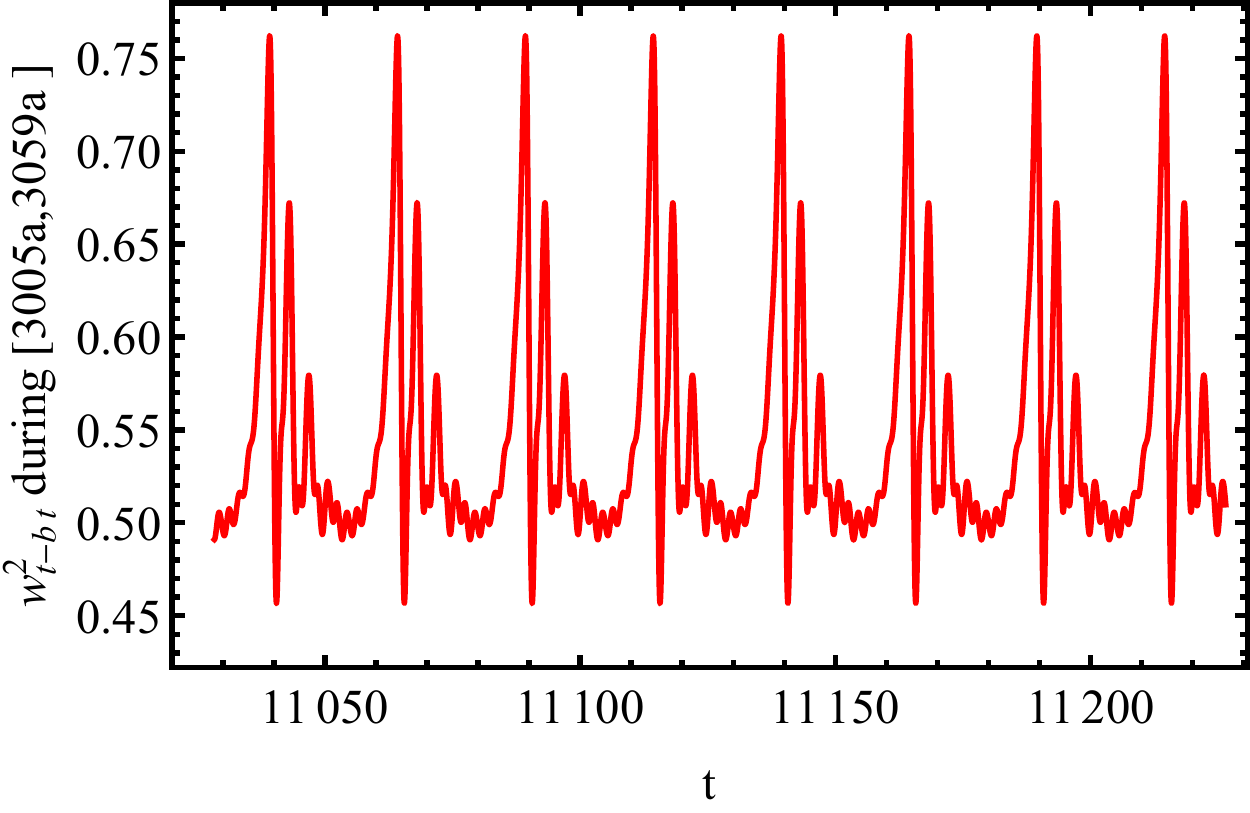}
		\caption{}
		\label{}
	\end{subfigure}
		\\
		\\
	\begin{subfigure}[b]{2.9in}
		\includegraphics[width=2.9in]{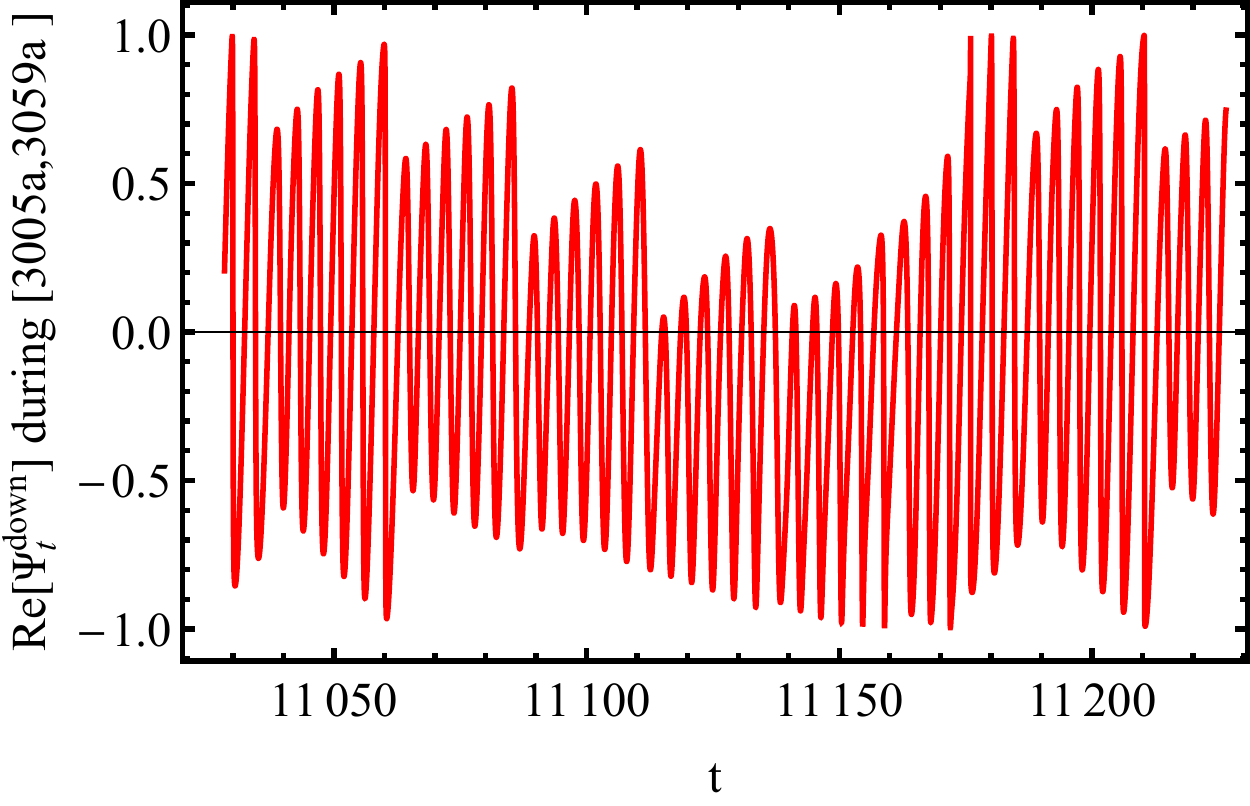}
		\caption{}
		\label{}
	\end{subfigure}
	\quad
	\begin{subfigure}[b]{2.9in}
		\includegraphics[width=2.9in]{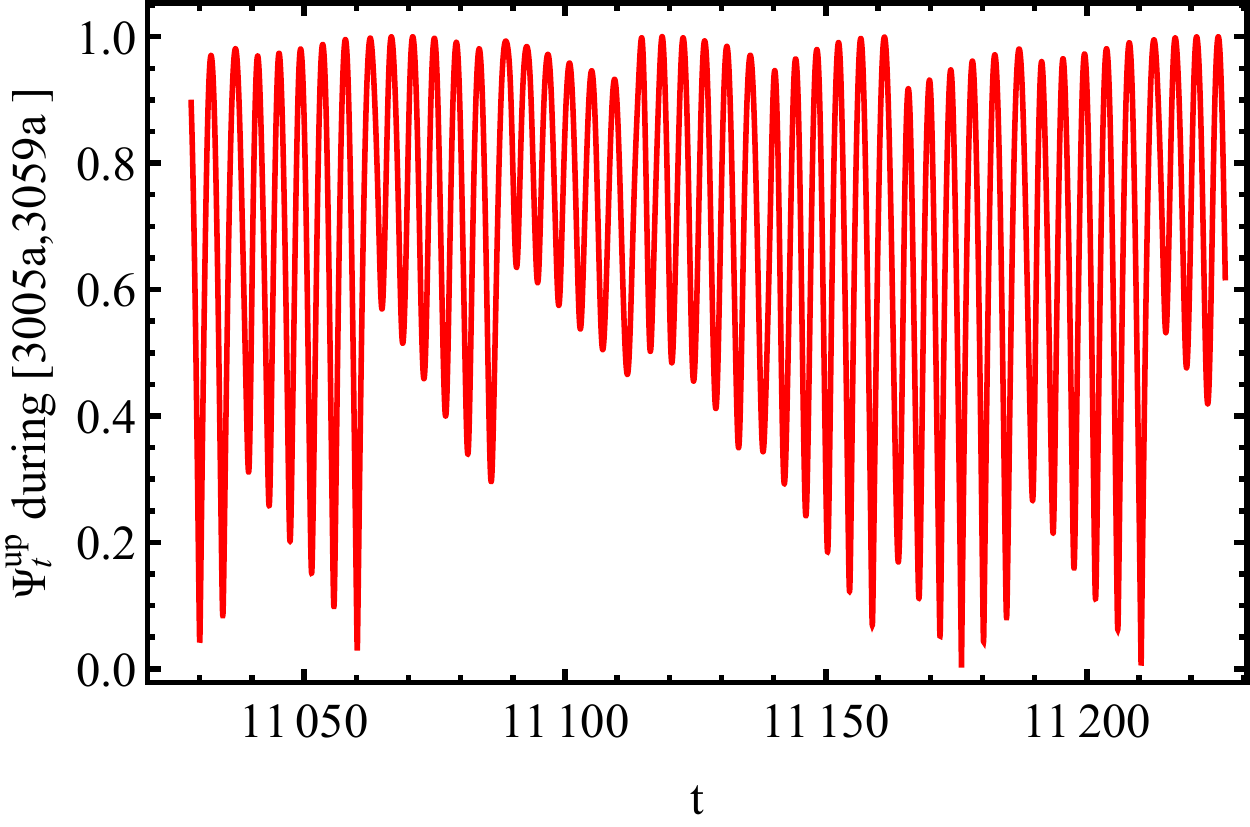}
		\caption{}
		\label{}
	\end{subfigure}
	\caption{Numerical Solution of The QMM-UE system (\ref{SE33POLAR},\ref{fgof33}) with parameters:\\ $r = 3.299$, ($B^y_{\text{Kicker}} ; \theta_0 , \phi_0 | \kappa^I , \kappa^R | a  ; t_{\text{completion}} $)=($35.5 ; 1.001 , 0.089 | 5.124 , 3.185 |  3.67 ; 3300\; a$).\\ It represents various intricately-patterned forms of one-qubit (3,3)-Purely-QMM-UEs, right before making the crossover from Phase $4$ to phase $5$.}
	\label{33_5R5_All}
\end{figure}

\begin{figure}
	%\centering
		\begin{subfigure}[b]{2.9in}
		\includegraphics[width=2.9in]{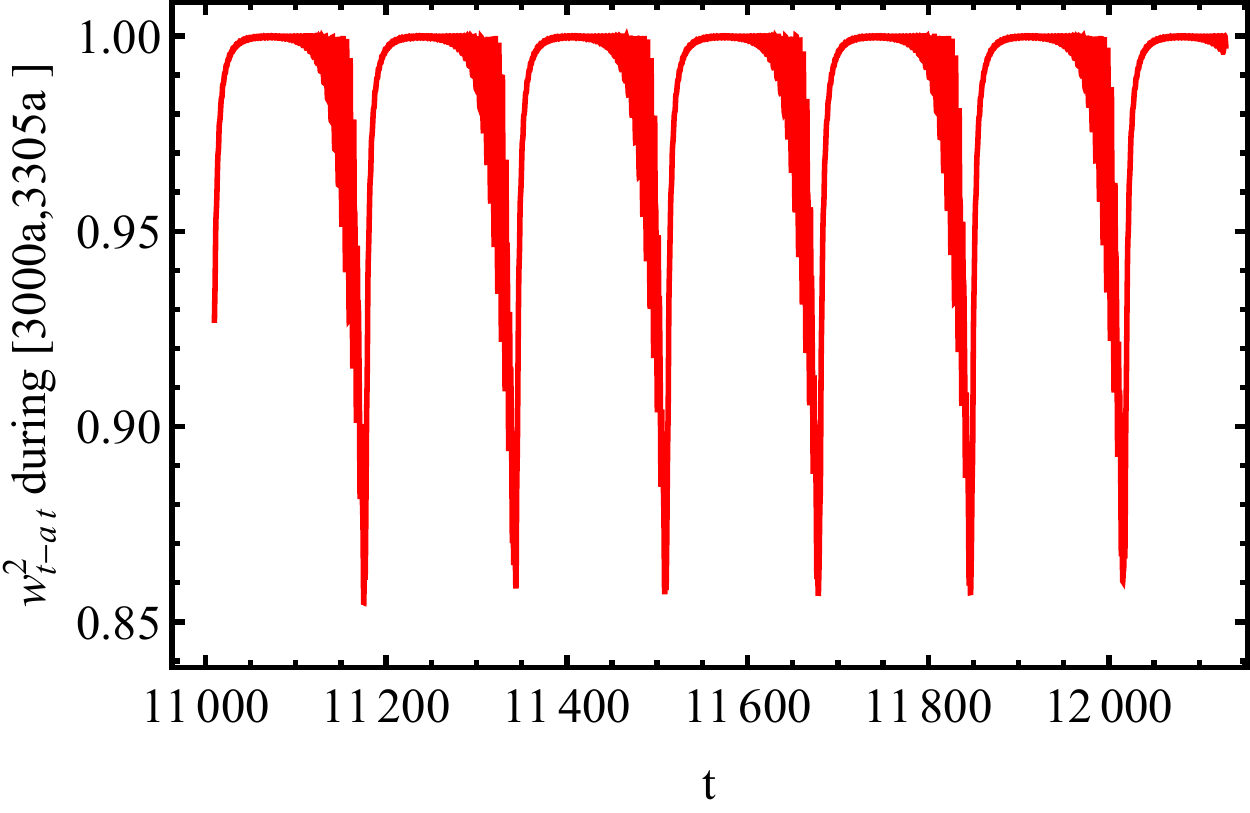}
		\caption{}
		\label{}
	\end{subfigure}
	\quad
	\begin{subfigure}[b]{2.9in}
		\includegraphics[width=2.9in]{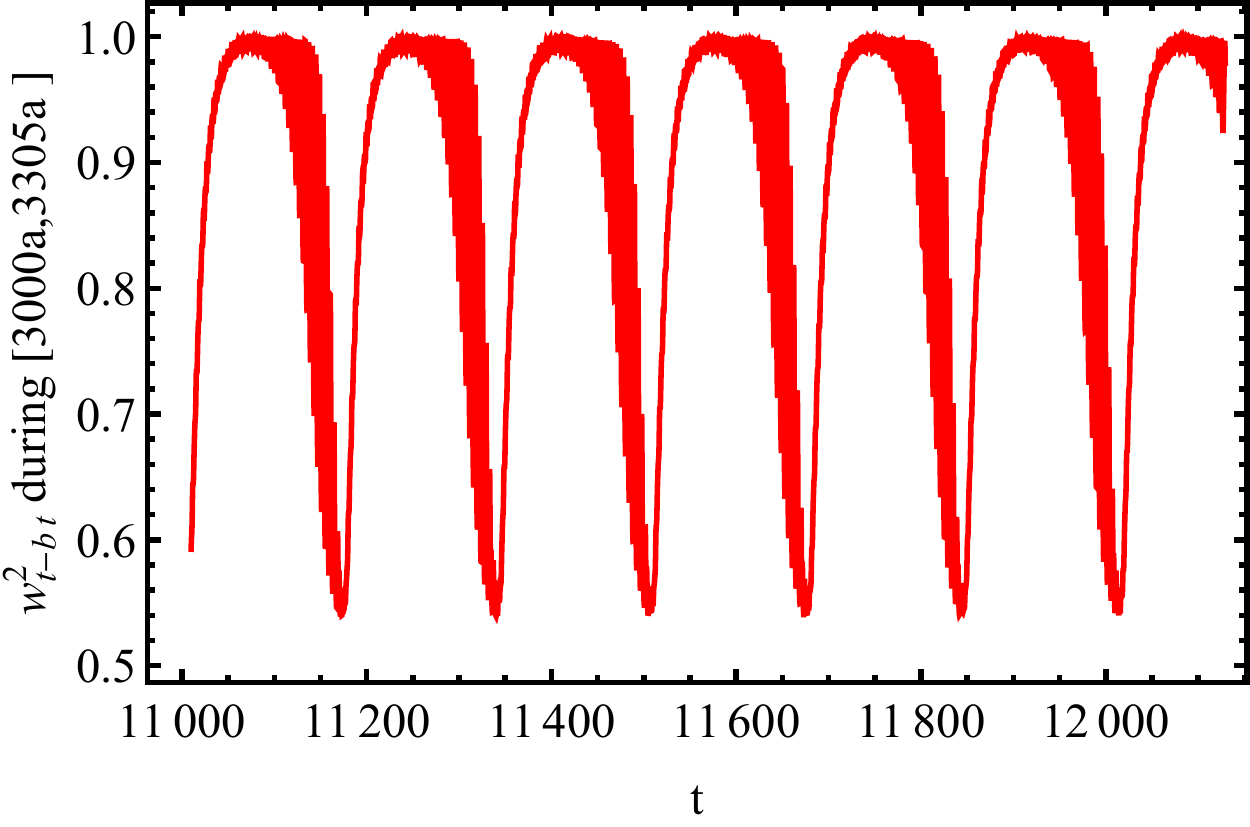}
		\caption{}
		\label{}
	\end{subfigure}
		\\
		\\
	\begin{subfigure}[b]{2.9in}
		\includegraphics[width=2.9in]{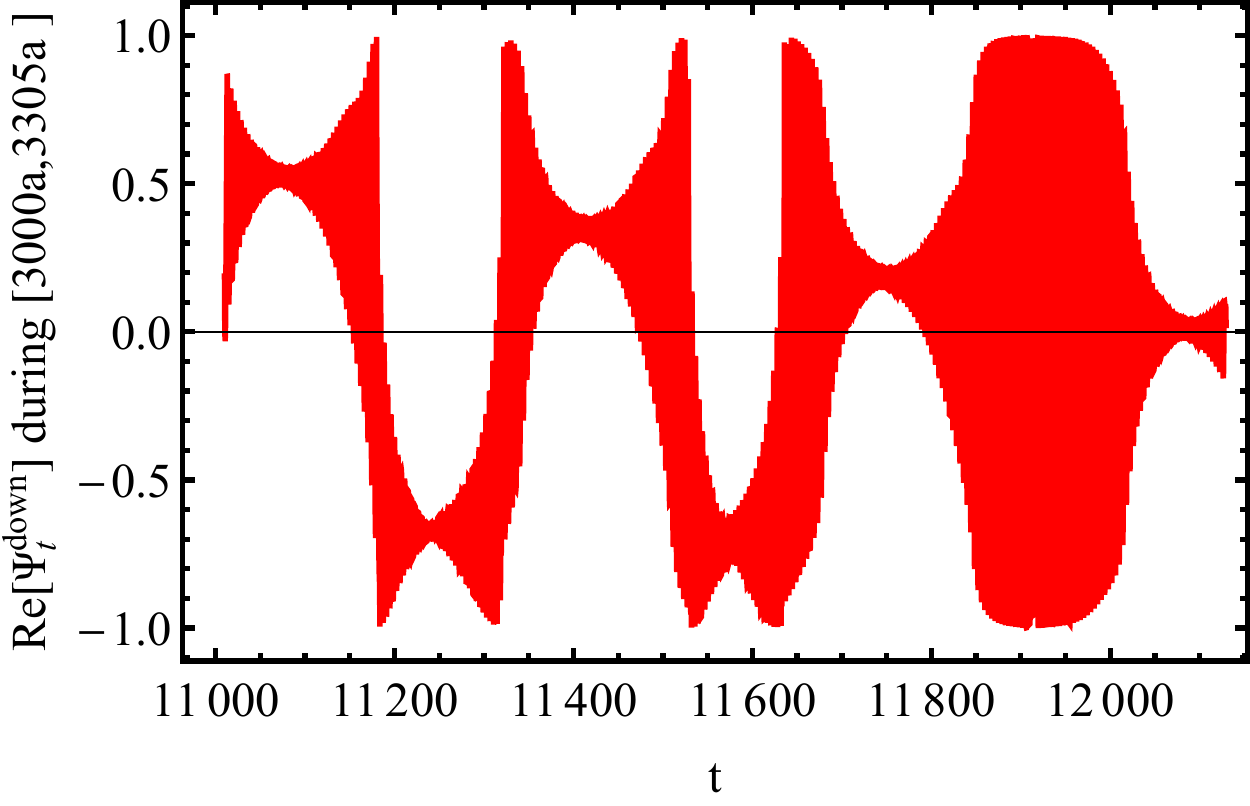}
		\caption{}
		\label{}
	\end{subfigure}
	\quad
	\begin{subfigure}[b]{2.9in}
		\includegraphics[width=2.9in]{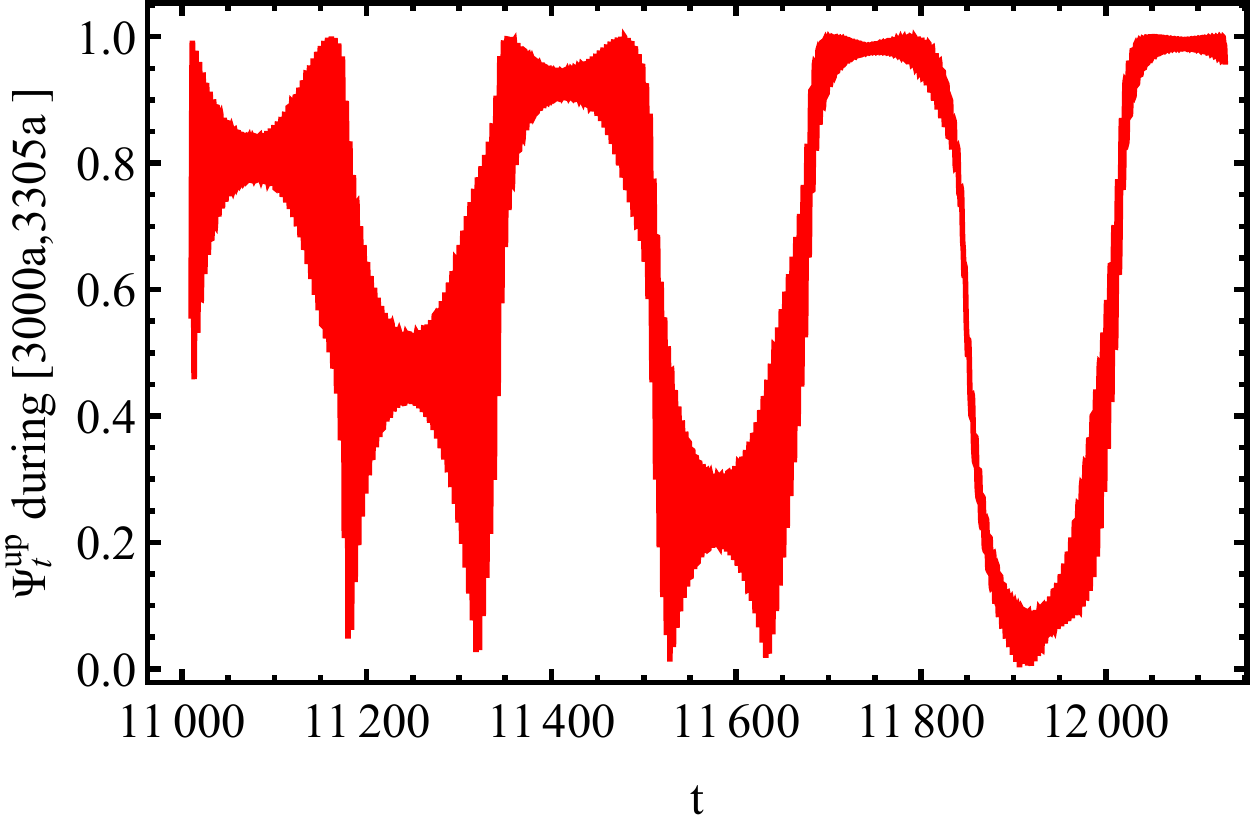}
		\caption{}
		\label{}
	\end{subfigure}
	\caption{Numerical Solution of The QMM-UE system (\ref{SE33POLAR},\ref{fgof33}) with parameters:\\ $r = 3.77$, ($B^y_{\text{Kicker}} ; \theta_0 , \phi_0 | \kappa^I , \kappa^R | a  ; t_{\text{completion}} $)=($35.5 ; 1.001 , 0.089 | 5.124 , 3.185 |  3.67 ; 3900\; a$).\\ One-qubit (3,3)-Purely-QMM-UE corresponding to morphologicaly-deformed phase $5$.}
	\label{33_5R6_All}
\end{figure}

Now we construct and investigate one-qubit state-histories for $(3,3)$-Purely-QMM-UE. We focus on the specific Hamiltonian \eqref{NL}, a simplified version of the most general case \eqref{MMHparts} one could take, as only one complex coupling is now turned on: $(\kappa^I,\kappa^R)$. However, the simplified (3,3) QMM-H \eqref{NL} is sufficiently complex, leading us to a whole variety of novel phases. The coupling $\kappa$ is the strength of \emph{the cubic hyper-link, $L = 3,$ between three specific quantum memories, $N=3$}. The quantum memories, reselected at every present moment $t$, correspond to these three wavefunctions of the one-qubit closed system: one is the qubit's pure state at the very now, $\rho_t$, one is qubit's pure state in its past moment with an unchanging temporal distance $b \in \mathbb{R}^+$ from the now, $\rho_{t - b}$, one is the third qubit's pure state, $\rho_{t-a}$, with $a>b$ being a free constant specifying a moment further in the past from the time now. Hence, besides the $(B^y_{\text{Kicker}}; \theta_0 , \phi_0)$ triplet of initial data, the one-qubit $(3,3)$-QMM-UE generated by the QMM-H \eqref{NL} has four more real valued control parameters which we identify as follows: the couplings $(\kappa^I,\kappa^R)$, the largest QMD $a$, and the ratio of the two past-time QMDs: $\frac{a}{b} \equiv r \in [1,\infty]$. In fact, we numerically investigate and establish that this \emph{seventh control parameter} $r$, despite being totally independent of the couplings of third order interactions $(\kappa^I,\kappa^R)$, plays significant role in deciding what particular phase of the phase diagram is formed by a given one-qubit $(3,3)$-Purely-QMM-UE. Moreover, as there are three quantum memories participating in the generation of unitary time evolution, each one state-history is marked with three QM-TPFs \eqref{ctmcs}: $w_{t-a,t-b}$, $w_{t-b,t}$ and $w_{t-a,t}$.\\\\ 
The QMM Schr\"odinger equation \eqref{pmmscheqwmctc} encoding the above unitary evolution is \eqref{SE33VECTOR} for the dynamics of the one-qubit Bloch vector, or equivalently is the coupled system of delay-differential equations (\ref{SE33POLAR},\ref{fgof33}) for the dynamics of the one-qubit Bloch angles. As before, we construct the unitary state-histories under consideration by solving the polar version of the QMM Schr\"odinger equation, namely (\ref{SE33POLAR},\ref{fgof33}) by means of the same numerical discipline explained and used in previous subsections. However, given that the matching of the two methods have been established in the previous examples, it suffices here to implementing the first method. We have seen by extensive numerical investigations that the complete phase diagram of $(3,3)$-Purely-QMM-UE happens to be \emph{much more complex} than that of its counterpart with only two quantum memories, even in as elementary quantum setting as the one-qubit closed system. %(\ref{SE33POLAR},\ref{fgof33}). 
Because of this, the forthcoming presentation of one-qbit unitary state-histories with $N=3$ and $L=3$ is cut down to a sufficiently-rich variety of numerical solutions to (\ref{SE33POLAR},\ref{fgof33}) which we have found to be most remarkable qualitatively.\\\\ 
Besides few exceptions, for every one-qubit QMM unitary state-history of the $(3,3)$ theory which is shown in this subsection, these \emph{four plots} are presented: $\text{Re}[\Psi_t^{\text{down}}], \Psi_t^{\text{up}}$ and also $w_{ t-a\;t}^2$, as were presented for all the QMM unitary state-histories with $N=1$ and $N=2$, together with only one (among the two) of the other QM-TPFs: $w_{ t-b\;t}^2$. Note that there still remains one more QM-TPF, $w_{ t-a\;t-b}^2$, whose plots we ignore to show. This is mainly because it develops the same qualitative profile as the two other QM-TPFs, in all the many numerical solutions we have constructed. This also prevents further enlargement of the volume of the paper, and the population of the plots, which are both already large enough. In addition to several numerical solutions singled out for demonstration in figures, we present brief explanations about all the phases of the one-qubit (3,3)-Purely-OMM-UE which we have seen in the course of our explorations of its (large and complex) phase diagram. We first explain the method with which we have organized our extensive (but not exhaustive) exploration of the phase diagram.\\\\ 
Geometrically, the phase diagram of the unitary state-histories whose generator is the $(3,3)$-QMM-H \eqref{NL} lives in one dimension higher than that corresponding to the $(2,2)$-QMM-H \eqref{QMMH22}. One can identify this additional dimension with geometrization of the ratio of the two larger QMDs, $b$ and $a$, that is (as introduced before) with the extra control parameter $r$. All the rest of their control parameters are in a one-to-one correspondence. In particular, the defining coupling constants of both QMM-Hs form $\mathbb{C} = \mathbb{R}^2$. Moreover, by purity of the unitary state-histories \eqref{hps}, and via a mapping $(\hat{\mu}_{t-a}, \lambda^I) \leftrightarrow (\kappa^R, \kappa^I)$, the total phase diagram of the $(2,2)$ theory corresponds to the following codimension-one hypersurfaces of the phase diagram of the $(3,3)$-theory: one is the boundary hypersurface  $r=1$, and the other is the asymptotic boundary $r = \infty$. In conclusion, $r \in \mathbb{S}^1$. Given geometric reconstruction as such, and knowing the five $N=2$ phases, we have extensively explored the phases of the $(3,3)$-Purely-QMM-UE.  Focusing in turn on every one of the five phases identified in Subsection \ref{SV-II}, we first obtain a handful of solutions to the QMM Schr\"odinger equations (\ref{SE33POLAR},\ref{fgof33}) at $r=1$, belonging to that phase. Next, we gradually raise the (3,3)-parameter $r$ above $r = 1$, having kept $(B^y_{\text{Kicker}} | \theta_0 ; \phi_0 | \kappa^R ; \kappa^I)$ and the largest QMD $a$ fixed, re-solve the system (\ref{SE33POLAR},\ref{fgof33}), and identify what qualitatively-distinct phases are formed by the resulted $r$-deformed solutions. We go on with this procedure up until hitting a sufficiently-large $r$ which (with acceptable approximation) reproduces the initial solution with $r = 1$.  Put differently, we systematically generate and single out the distinct phases of the $N = 3$ Purely-QMM theory as the higher-dimensional $r$-deformations of the five phases of the counterpart $N = 2$ theory.\\\\ 
\underline{\emph{Mapping the $r$-deformations of the Phase 1}}\\
Numerical investigations establish that the phase $1$ of $N = 2$ is invariant under the (3,3) $r$-deformations. All Purely-QMM unitarily-evolving one-qubit wavefunctions which solve the $(3,3)$ dynamical system (\ref{SE33POLAR},\ref{fgof33}) within the coupling range $0 \leq |\frac{\kappa^R}{\kappa^I}| \ll 1$, develop at sufficiently-late times \emph{fixed-point attractors}.\\\\
\underline{\emph{Mapping the $r$-deformations of the Phases 2,3,4}}\\
The (3,3) phases found by applying the $r$-deformation on the phases 2,3 and 4 of the one-qubit $N=2$ theory, are essentially accommodated in the total landscape of the $r$-deformations of the phase 5. A rich landscape to which we now turn. \\\\
\underline{\emph{Mapping the $r$-deformations of The Phase 5}}\\
Now, we simulate and classify the (3,3) solutions, by moving on the $S^1$ deformation, from within the original phase $5$ of the one-qubit $N=2$ theory. We know that at $r = 1$, the initial point of the deformations, the behavior of the one-qubit Purely-QMM-UE is the everlasting switching, under sharp transitions, between two meta-stable states. Concretely, we classify the numerical solutions obtained by varying the $r$ parameter along its $S^1$, with the other control parameters being chosen and kept fixed as follows: $(B^y_{\text{Kicker}} ; \theta_0 , \phi_0 | \lambda^I , \hat{\mu}_{t-a} | a ) =(35.5 ; 1.001 , 0.089 | 5.124 , 3.185 | 3.67)$. Beginning from $r=1$, increasing $r$-deformations are implemented in sufficiently-small steps, and the system (\ref{SE33POLAR},\ref{fgof33}) is solved numerically. With the choice of $r=1$: $\Delta^{(t)}_1 = \Delta^{(t)}_2 = a$, and we do obtain a numerical solution whose profile is of the form presented in Fig. \ref{fig-phase5}. We expect that within a not-too-wide band above $r =1$, the system must stay in the phase $5$. Indeed, numerical solutions confirm the stability of the phase $5$ under $r$-deformation in the range: $1 \leq r \lesssim 1.15$. Within this range, the only effect of increasing $r$-deformations is that all the three simultaneous stages of $w_{t_i t_j} = 1$ shorten monotonically. Moreover, one sees that the defining feature of bistate metastability is taken away gradually.\\\\ 
Making a \emph{crossover} at $r \sim 1.15$, one-qubit Purely-QMM-UEs enter a \emph{special phase}. This special phase, being formed everywhere in the following $r$-deformation range: $1.15 \lesssim r \lesssim 1.632$, is a \emph{finely-balanced blend of the phases $5$ and $4$}. A representative unitary state-history in this special phase is presented in Fig. \ref{33_5R1_All}. See the distinctive orders, both in the oscillations of the one-qubit wavefunctions, and in the clear patterns of the QM-TPFs. It marks a very non-generic sub-class of the phase $4$, in the sense of its built-in structures. Being formed as a phase-$5$ deformation, however, the norms of the three QM-TPs $(w_{t-a \; t} , w_{t-b \; t} , w_{t-a \; t-b})$ feature the following pattern. Generically, between the neighbouring pairs of the $w_{t-a \; t} =  w_{t-b \; t} = w_{t-a \; t-b} = 1$ stages, which are shortened by now about two orders of magnitude compared to the corresponding stage in the $r = 1$ phase $5$, one narrow bump is formed which contains three sharp extrema: two unequal minima and one maximum. Interestingly enough, the dynamics of the three QM-TPFs are finely synchronized, such that the $w_{t_i t_j} = 1$ stages and the aforementioned three extrema all happen simultaneously. %Fig. \ref{fig33-phase5}.
\\\\ 
Surpassing $r \gtrsim 1.632$, the $(3,3)$ Purely-OMM-UE crossovers to the general realm of the phase 2, and stays there up until $r \lesssim 3.23$. One sees, in a considerable sub-interval of this range, two deformed variants of the phase 2, in particular. Qualitatively, they are two possible `amplifications' of the phase 2: \emph{the `semi-frozen' and `fully-frozen' versions of phase 2}. In these remarkable variants, one, two or all of three $w_{t_i t_j}$s become totally frozen, with the wavefunction oscillations becoming perfectly regular. See Fig. \ref{33_5R2_All} for a representative solution. Moreover, depending on the value of $r$, other non-generic versions of the phase $2$ are formed.
\\\\ 
More remarkably, one observes that at $r \sim 1.725$, within the above range, the $(3,3)$ Purely-QMM-UE experiences a dynamical phase transition under which the one-qubit wavefunction changes a long-lasting phase 3 with the stable fully-frozen phase 2. The phenomenon is of the same nature observed and described in Subsection \ref{SV-III}: a unitary purely-internal dynamical transition, resembling a phase-interpolating time-like kink, caused by the third-order interactions between the chosen QMs of the closed system. The six plots of Fig. \ref{33_5R3_All} present a representative solution of the corresponding QMM Schr\"odinger equations, given by (\ref{SE33POLAR},\ref{fgof33}). In particular, to further enhance clarity, one of the QM-TPs has been zoomed in within two largely-separated time windows.  As we see, the dynamical phase transition has occurred in $\sim 920\;a$ after the outset of the (3,3) Purely-QMM-UE. 
\\\\ 
Walking the direction of $r$-deformations ahead, the one-qubit closed system makes crossover to `ordered behavioral patterns': the general realm of the phase  $4$ of the N=2 theory. Right before it, we see a special profile in a narrow band about $r = 3.235$: Purely-QMM-UEs are simple synchronized oscillations. Fig. \ref{33_5R4_All} shows clearly the highly nongeneric, regular, simple configurations which are formed in the aforementioned range. As we see in the representative solution presented in Fig. \ref{33_5R4_All}, all the three system's QM-TPFs, and its wavefunction experience stable, almost-perfectly simple oscillations. Moreover, the oscillations of $(w_{t-a\;t}, w_{t-b\;t})$ are oppositely synchronized, as precise examination of their maxima/minima shows. After this, the system moves into the ordered-but-complex oscillations of the phase 4, up until $r \lesssim 3.3$. 
\\\\ 
A remarkable family of one-qubit (3,3) Purely-QMM-UEs which structurally are highly `intricate and exotic' are formed, within a narrow band of $r$-deformations at the verge of system's crossovering from the phase $4$ to the phase $5$. Fig. \ref{33_5R5_All} shows a numerical solution which represents this intricately-patterned family of unitary state-histories, corresponding to $r = 3.299$. One can examine and confirm that the QM-TPFs $w_{t-a\;t}$ and $w_{t-b\;t}$ are, once more, oppositely synchronized: their local extrema inside every structural module, being a handful in population, happen in reverse orders at almost simultaneous moments.
\\\\ 
Finally, for $r \gtrsim 3.5$, the one-qubit closed system develops $(3,3)$ Purely-QMM-UEs which feature ongoing two-state-switching metastability, corresponding to the phase $5$, including \emph{morphological deformations}. See the representative of Fig. \ref{33_5R6_All} at $r = 3.77$.
\\\\ 
Likewise, we have analyzed the resulted $r$-deformations of the $r=1$ phase $5$ for a handful of independent examples, classified their phases, and found them qualitatively equivalent with those obtained based on the solutions above. But, we highlight that our extensive explorations do not necessarily exhaust the entirety of the large-and-complex spectrum of all possible $r$-deformations of the phase $5$. Nevertheless, we regard that the presented spectroscopy of one-qubit (3,3) Purely-QMM-UEs is inclusive enough.
%%%%%%%%%%%%%%%%%%%%%%%%%%%%%%%%%%%%%%
%%%%%%%%%%%%%%%%%%%%%%%%%%%%%%%%%%%%%%%
\subsection{Selected General Highlights On One-Qubit Purely-QMM-UEs}
\begin{figure}
	%\centering
	\begin{subfigure}[b]{2.9in}
		\includegraphics[width=2.9in]{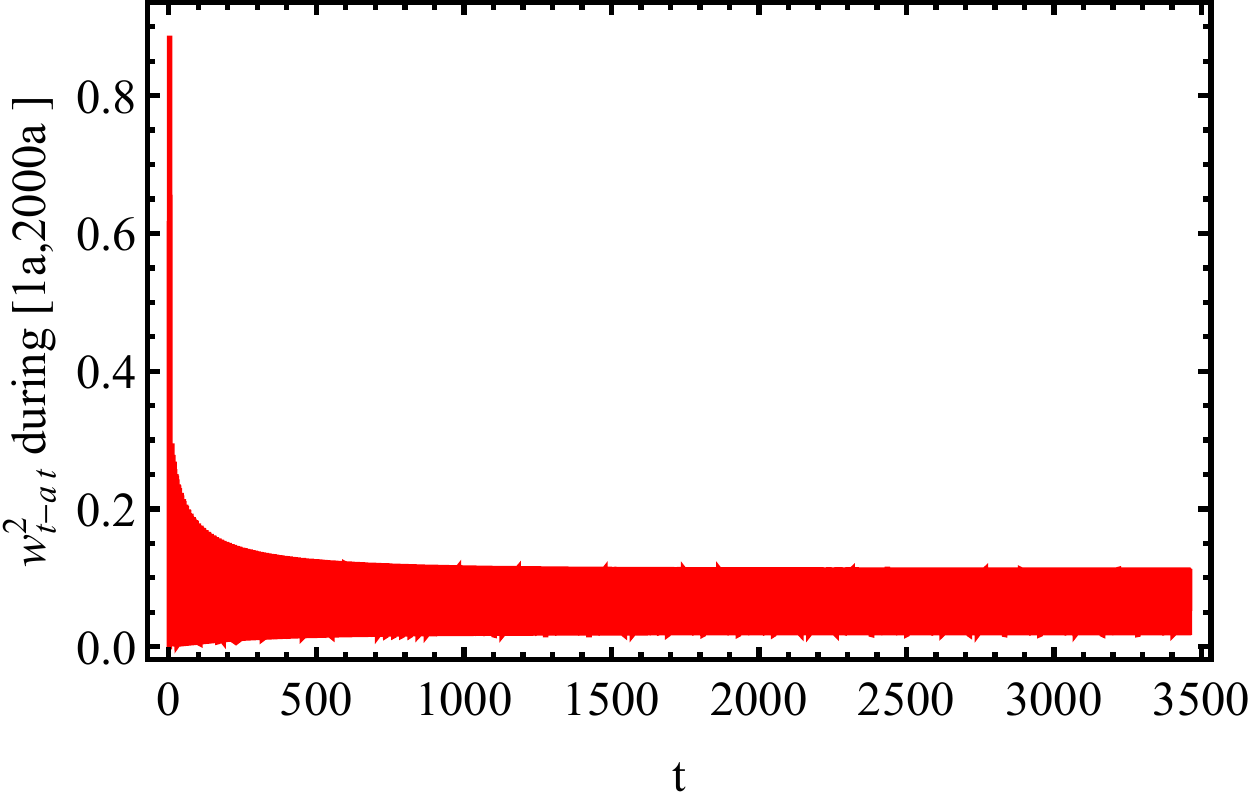}
		\caption{}
		\label{}
	\end{subfigure}
	\quad
	\begin{subfigure}[b]{2.9in}
		\includegraphics[width=2.9in]{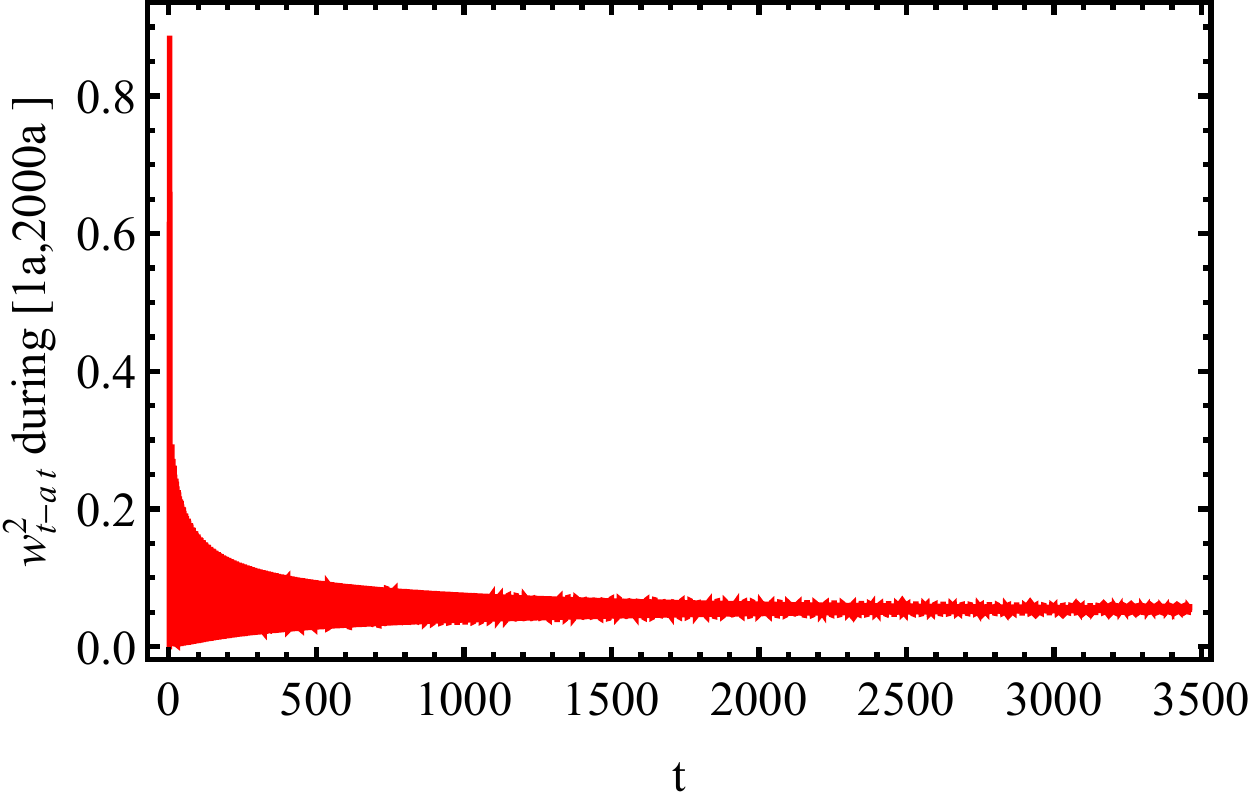}
		\caption{}
		\label{}
	\end{subfigure}
	\caption{Crossover by the numerical solution from Phase 4 (a) to Phase 2 (b) of the one-qubit (2,2)-Purely-QMM-UE, demonstrated by $w^2_{t-a\; t}$, 
	upon a shift: $\delta \hat{\mu}_{t-a} = 0.018$.\\ (a): ($B^y_{\text{Kicker}} ; \theta_0 , \phi_0 | \lambda^I , \hat{\mu}_{t-a} | a  ; t_{\text{completion}} $)=($7.5 ; 1.001 , 0.089 | -2 , 3.01 |  1.73 ; 3600\; a$).\\ (b): ($B^y_{\text{Kicker}} ; \theta_0 , \phi_0 | \lambda^I , \hat{\mu}_{t-a} | a  ; t_{\text{completion}} $)=($7.5 ; 1.001 , 0.089 | -2 , 2.992 |  1.73 ; 3600\; a$).}
	\label{CrossoverExampleAmongThe5Phases}
\end{figure}

Having known the distinct phases of the $(N \leq 3,L \leq3 )$ Purely-QMM-UE of the one-qubit closed system, 
we now want to highlight few general remarks about the phase diagram. Although these points were essentially brought about in the previous subsections given the context, these five remarks can serve the presented results as a supplement.\\\\
One-qubit Purely-QMM-UEs in which the wavefunction experiences ongoing disordered or structured oscillations are realizable \emph{if} (at least) one \emph{Hermitian} QMM monomial is turned on in the Hamiltonian. Hence, $(1,1)$-or-$(N=2)_{|_{\lambda^I=0}}$ QMM-Hs are sufficient.\\\\ 
One-qubit Purely-QMM-UE attains fixed-point attractor, along with the everlasting behavior of switching between two metastable attractors, the phases `one and five', \emph{on the condition} that one turns on interaction operators given by \emph{anti-Hermitian} QMM monomials. The simplest Hamiltonian to realize those two phases is (\ref{QMMH22}) in which the   \emph{commutators} of two quantum memories, $\mathcal{O}^{(\dagger)}_{a} = [\rho_{t-a}, \rho_{t-a}]^{(\dagger)}$, run the above-mentioned interactions. Moreover, being intuitive, the formation of the two-attractor meta-stable dynamics of the phase five demands that the strengths of the couplings $\lambda^I$ and $\hat{\mu}_{t-a}$, which compete to drive the system to oscillatory versus fixed-point-attractor behavior, are made comparable to one another.\\\\
The only kind of transitions that one observes between the distinct phases of one-qubit Purely-QMM-UEs, as a consequence of varying the QM-couplings $(\mu_{t-a\;t}; \lambda^I, \lambda^R; \eta; \kappa)$, are \emph{crossovers}. Turning on flows of QM-couplings causes continuous changes. Fig. \ref{CrossoverExampleAmongThe5Phases} shows a clear example of the inter-phase crossovers. Plot (a) shows a solution in phase four. With $\lambda^I = -2$ and all other parameters fixed, $\hat{\mu}_{t-a\;t}$ is lowered from $3.01$ to $2.992$. The new solution is in phase two. Gradually making the above shift, $\delta \hat{\mu}_{t-a\;t}  = 0.18$, and classifying the solutions, we find that the crossover has occurred at $\hat{\mu}_{t-a\;t} \cong 3$.
%MATCHING WITH SPECTROSCOPY
\\\\ 
Purely-QMM-UEs realize dynamical phase transitions which, resembling time-like kinks, switch between a pair of the one-qubit phases. A generic solution of the kind has a long lasting stay in one phase, but makes a sharp transition which stabilizes it in one other phase. Interestingly, they can switch between various (unstructured or structured) oscillatory phases and the (fixed-point or multi) attractor phases. They take place without external influences, coupling variations, or quenched disorders. These unitary transitions are caused internally, by the higher order interactions, $L \geq 3$, between the $N \geq 2$ QMs which compose the Hamiltonian. Finally, these unitary dynamical phase transitions are realized, in diverse forms, in various regions of the interaction-coupling space whose volumes increase by raising the degree of QM monomial interactions, but are considerable even at $L=3$.\\\\
Overall, the one-qubit closed system endowed with first, second or third order monomial interactions between its past-or-present quantum states can develop five principle kinds of unitary time evolutions, along with time-like kink behavior interpolating between them: fixed-point attractor behavior (phase one), three distinct kinds of everlasting oscillatory behavior (phases two, three and four), and everlasting swapping behavior between two meta-stable attractors (phase five). That, the Purely-QM-UEs of the most elementary quantum setup is capable of such surprising behavior is remarkable to us. These findings seem to be truly promising for the visions we bring about in Section \ref{visions}.
%%%%%%%%%%%%%%%%%%%%%%%%%%%%%%%%%%%%%%
%%%%%   6 .  Visions
%%%%%%%%%%%%%%%%%%%%%%%%%%%%%%%%%%%%%%%

\section{QMM-UEs: Outlook, Open Questions And Visions On \\ `Experience Centric Quantum Theory'}\label{visions}
\subsection{A Conceptual Overview And  Selected Elucidations}\label{elucidations}
% One 
All fundamental laws of nature as we have discovered are successfully formulated by means of microscopic interaction operators, Lagrangians and Hamiltonians which are unchangeably `enforced' on closed systems, in the sense that they 
are, and so remain, intrinsically indifferent to the accumulating-and-evolving `experiences' that the total closed system and its constituent parts develop in the course of their time evolutions. Mathematically, they are given by
local-in-time operators on the Hilbert space of a closed system that are defined with no dependencies on the system's past-or-present quantum states. Indeed, a discipline as such establishes a strictly-one-way relationship between states and their evolutions. Hamiltonians and their constituting interactions make the states moment-by-moment evolve unitarily, however, do not `tolerate' (any structural kind of) back reaction from them: they receive no impact, hence no update, from the past-or-present states of the closed system. The standard model (= theory) of particle physics is our best theoretical and experimented validation of how successful the assumption of the one-way relationship between the closed-system states and their unitary evolution has (perhaps astonishingly) turned out to be.\\\\ 
Nevertheless, a paradigm shift is necessary if we aim for a general sensible theory of \emph{quantum intelligence}. What we mean by quantum intelligence is the ability of a system to generate, organize and output a sufficiently-rich spectrum of behaviors which are general quantum behavior on the one hand, and general intelligent behavior on the other hand. In other words, if one aims to bring about a general synthesis and merge of quantum behavior and intelligent behavior, within a sound and successful theoretical framework, a novel approach, featuring the alteration of some conventional (and `less' fundamental) assumptions, must be adapted. Specially, it is in the outset crystal clear that the aforementioned one-way relationship between unitary evolutions and the system's quantum states is at complete odds with \emph{the most (or one of the most) fundamental characteristic(s) of intelligent behavior}. Any possible form of intelligent system must output manifold behaviors whose dynamical profiles undergo recurrent remarkable updates which are directly originated from its state-history experiences. In particular, the very time evolution of the states must receive moment-by-moment revisitations, and recurrent updates, based on the relevant sub-histories of the system. For closed systems which are the focus of the present work (as we shall argue soon), the mere possibility of intelligent behavior requires that the Hamiltonian generating the unitary evolution operator, and hence the system's internal interactions which source it, can-and-must-typically receive \emph{adequate `structural' updates} which are driven by the system's quantum-state-history. In other words, \emph{already in the primary context of closed evolving systems, we need quantum memory effects which manifest themselves as compositional nonlocal-in-time dependencies of the unitary time-evolution operator and its sourcing interactions on the (relevant information content of the) state-history}, as one necessary condition for the system's capability to develop behavior which are, at the same time, quantum and intelligent.\\\\ %%%%%%%%%%%%%%%%%%%
In this regards, it is highlighted that the quantum memory effects induced or mediated by the system-environment coupling are ubiquitous and cause a landscape of Non-Markovian time evolutions in open quantum systems. These effects are quantified, and also measured indirectly, using independent characterizations such as information backflow or lack of CP-divisibility. \emph{But}, environmental Non-Markovianities come generically with dissipation, decoherence, and time evolution irreversibility: they source evolutions which are \emph{Non-Unitary}. Moreover, the strength-or-largeness of such memory effects \emph{follows dynamically} the strength-or-largeness of the overall system-environment coupling, and hence generically the correlated strength-or-largeness of the induced Non-Unitarity of the system's evolution.\\\\ 
\emph{On the contrary}, as various thought experiments can easily show, \emph{no such monotonic or dynamical binding between the Non-Markovianity and the Non-Unitarity of the time evolution should-or-is required by quantum intelligent systems in the general context}. Clearly enough, general intelligent behaviors neither should, nor actually do, bound themselves to any inevitable-or-monotonic dynamical non-unitarity cost \emph{in exchange for} required memory dependencies as their experience resource. We must highlight that the above statement stays valid even when the initial formation of learning and intelligence would require coupling the system to an interacting environment for sufficiently-long times. Indeed, an intelligent system which `lives' in an interactive environment and respond to it, undergos generically some form of Non-Markovian evolution of the non-unitary kind. Nevertheless, if the system with (a degree, kind or level of) \emph{emerged} intelligent behavior becomes (even if nearly) isolated, its time evolution should become (nearly) unitary, but now in some non-conventional form which has sufficiently-strong dependencies on its in-the-making state-history. The three points in (conceptual and technical) relation with the above highlight are now summarized:\\\\ 
\emph{(1)} Firstly, there is no fundamental argument against the remarkable possibility that, even in the total absence of initial environmental interactions, a sufficiently-complex sufficiently-aged system can \emph{self-organize in isolation} (some degrees, kinds, or levels of) \emph{emergent intelligent goal-oriented behavior}. \emph{(2)} Even if we assume that the collective development of intelligence, as a sustainable behavioral \emph{emergent} quality of a complex (closed or open) system across a proper subset of all its phases, needs an initial history of sufficient environmental interactions, once the environmental couplings are made negligible, the corresponding (dominantly-) unitary evolution must be generated by a nonlocal-in-time Hamiltonian operator which is state-history-based `constructionally'. \emph{(3)} Now, if we consider \emph{open} intelligent systems coupled to an interactive environment, the intelligence-required Non-Markovianity of the system's evolution does not need to follow a corresponding scale of evolution Non-Unitarity. That is, the general dynamics of open intelligent systems should not necessarily be of the conventional Non-Markovian kind, namely, of the kind deducible from the conventional Hamiltonian evolution of an encompassing closed system. Indeed, as in the case of conventional quantum systems, a general deduction should instead follow a `closed $\to$ open' direction as follows. \emph{General Non-Unitary Non-Markovian evolutions of intelligent open systems} %ir
are deducible from Unitary Non-Markovian evolutions, such as those in the present work, of encompassing intelligent closed systems.\\\\
\emph{Above all, it is within the broader perspective of intelligent quantum behavior} that we have proposed, formulated and primarily investigated a complementary distinctive way of memory participations in the system's internal interactions and  unitary evolution. Quantum memory roles and effects come with three distinctive qualities in this novel way: \emph{(a)} they can be (completely or partially) \emph{Non-Environmental}, namely, \emph{Internal}; \emph{(b)} they are \emph{`constructive'}, serving the (re-)organization of the system's behavior, by being (some of the) direct building-block operators of the updating Hamiltonian; \emph{(c)} In principle, their compositional role in the system's evolution are pushable all the way to the \emph{extreme}, that is, the Hamiltonian operator can be entirely made from the system's density operators as developed in its state-history. Specifically, the focus of our work has been the central role that the quantum-state-history of a closed system can play in sourcing, updating and directing its internal interactions and unitary evolution.\\\\ 
The novel family of Non-Markovian quantum processes proposed by the present work are completely coherent: their quantum-state-history dependencies are developed free from any dissipative effects (at least ideally) and they remain so, therefore, preserving the quantum information in the entire state-history-based evolution. Indeed, despite being Non-Markovian, the proposed evolutions are (by construction) \emph{perfectly unitary}. The perfect unitarity of these evolutions is directly implied by the built-in \emph{Hermicity} of a moment-by-moment Quantum-Memory-Made operator which generates it, namely $H_t^{(\text{QMM})}$ which serves the closed system as its nonlocal-in-time Hamiltonian. Moreover, the unitarity of these QMM processes has been shown crystal clearly in \emph{the explicit construction} of their time-evolution operator presented in \eqref{mue}, using (the continual sequence of) momentary QMM-Hs generally defined in \eqref{VMV} and formulated in (\ref{QMMmonml},\ref{MMH}) specifically. Furthermore, we remark that in all one-qubit realizations presented in the Sections \ref{888}, \ref{iml} and \ref{SV}, the time-evolution unitarity has been tested and re-confirmed in several independent ways.\\\\
About the relation between standard quantum mechanics and the present work, namely our generalizing framework and the novel systems formulated within it, there are some clarifying points  which we now come to highlight. First and foremost, the QMM-UEs defined in (\ref{VMV}) and specified in (\ref{QMMmonml},\ref{MMH}) and in their next-subsection generalizations, \emph{do not necessitate any underlying form of the so-called `nonlinear quantum mechanics'}. The statement holds despite the fact that all QMM-UE must be unavoidably nonlinear and nonlocal-in-time, due to their compositional dependencies on the past-to-present state-history of the closed system, as QMM Schr\"odinger equations manifest. The point is, as explained above, the QMM-Hs ((\ref{VMV});(\ref{QMMmonml},\ref{MMH});(\ref{bcnqmbs:mbo.118})) are proposed \emph{foremostly} as \emph{emergent Hamiltonians} in any complex system that describe the (sufficiently-developed `higher-level') \emph{phases which possess (the primary characteristics of) quantum intelligent behavior}. The system itself, however, can be a completely-standard quantum system at its microscopic level of formulation, and likewise at all its `lower' levels. In fact, the system does not even need to be necessarily quantum-mechanical at all: it can be a classical complex many-body-interacting system which develops at functional collective levels a form of emergent quantum behavior. Moreover, \emph{intelligence-aside}, QMM-UEs and their non-unitary open-system descendents provide effective descriptions of new emergent phases of microscopically-standard quantum systems whose Non-Markovian behaviors feature significantly more coherent quantum memory effects. Nevertheless, one does not exclude, either, the intriguing possibility of \emph{the `orthogonal' branch} of the physical relevance of QMM-UEs, which deduction-wise goes the other way around: one can propose these generalized quantum evolutions to define more fundamental quantum theories, for example in the context of quantum gravity, out of which both standard quantum mechanics and geometric spacetime emerge. We shall present more detailed explanations on \emph{the above-mentioned points} in Subsection \ref{pip}.\\\\
The proposed unification of time evolution Unitarity and Non-Markovianity for closed quantum systems led us to observe, already in the most minimal context of a one-qubit closed system evolving by means of the lowest-order interactions between its quantum memories, a rich landscape of unprecedented behavioral phases and dynamical phase transitions. Mathematically, they arise from the inseparably-entangled nonlinearity and nonlocality-in-time of the corresponding QMM Schr\"odinger and Von Neumann equations presented in Sections \ref{gtp} and \ref{888}. The physical origin of these novel kinds of quantum behavior and purely-internal behavioral transitions, however, is \emph{the merge between state-history participations and a `renewable'  operator of evolution Unitarity.} The surprising characteristics of the behavioral landscape of the one-qubit QMM-UEs which were detailed and explained in Section \ref{SV}, do clearly indicate to us that these novel systems can naturally find promising applications in several envisioned domains and disciplines which shall be mentioned in the next subsection.\\\\
%%%%%%%%%\ref{MM-evo-eqn3}. %%%%%%%%%%%%%
\emph{Independently}, we have advanced towards a speculative idealized quantum scheme in which \emph{states, interactions and evolutions are all unified}. The extreme message of such a unified scheme is most clearly conveyed by the scenario \eqref{bcnqmbs:mbo.118}. Taking the subsystem decomposition in the recipe \eqref{bcnqmbs:mbo.118} all the way down to its ultimate level of resolution, a radical perspective of a consistently-circular interplay manifests. It speculates a closed universe in which all many-body interactions and all evolutionary processes are remade continually from within its internal experience(s), \emph{and vice versa}.
%%%%%%%%%%%%%%%%%%%%%%%%%%.%%%%%%%%%%%%%%%%%%%%%%%%%%%%%
\subsection{Forward Views: A Generalization, Selected Open Questions And The Envisioned Applications In Fundamental And Applied Domains}\label{pip}
QMM-UEs can naturally find promising applications in several distinct domains, as we envision and conjecture. Clearly, there are also many directions to advance the work of the present paper, and there are many interesting open questions to be addressed.  We now elaborate, on a selective basis, on some open questions, future advancements, and envisioned applications of QMM-UEs.\\\\
Firstly, we present an alternative more general recipe to formulate the QMM-UEs for a general closed system, denoted by $\mathcal{S}$. Indeed, it is within the \emph{very same} paradigm we conceptualized in Section \ref{gtp} that this second recipe is now proposed. Nevertheless, it generalizes the formulation in (\ref{QMMmonml},\ref{MMH}). Consider a partitioning of $\mathcal{S}$ into $M$ interacting subsystems $\{\mathcal{S}_i\}_{i=1}^M$, mirroring an arbitrary Hilbert space factorization, $\mathcal{H}^{(\mathcal{S})} = \otimes_{i = 1}^{M} \mathcal{H}_i$. \emph{The second recipe} is to define the system's total QMM-H, $H_{(\mathcal{S})}^{\text{QMM}}$, directly based on the chosen subsystem decomposition, using the past-or-present states of the subsystems $\rho_i(t_i)$, defined by the complementary-space traces $\rho_i(t_i) \equiv \text{Tr}_{ \mathcal{H}_{\bar{i}}} [\rho^{(\mathcal{S})}(t_i)]$. One chooses a set of QMM $r$-body operators interlinking the $M$ $\mathcal{S}_i$'s, $\mathcal{O}^{\text{QMM}}_{i_ 1 \cdots i_r | t_{i_1} \cdots t_{i_r} } [\;\rho_{i_1}(t_{i_1}),\cdots,\rho_{i_r}(t_{i_r})\;]$, and packs them into a Hermitian $H_{(\mathcal{S})}^{\text{QMM}}$. For example, one can form such a $H_{(\mathcal{S})}^{\text{QMM}}$, by introducing $\mathcal{O}^{\text{QMM}}_{i_ 1 \cdots i_r | t_{i_1} \cdots t_{i_r} }$ with $r \leq 3$, together with the corresponding collections of state-history-independent operators and the $\lambda$-couplings $(\mathcal{A} , \Lambda)$, likewise (\ref{QMMmonml},\ref{MMH}), as,
\begin{equation} \label{bcnqmbs:mbo.118} \begin{split} & H_{(\mathcal{S})}^{\text{QMM}}(t) \;\equiv \; H_{(\text{Polynomial},\;\text{Exponential},\;\cdots )}^{(\text{Hermitian})}[\;  \{\; \mathcal{O}_{i | t_i}^{\text{QMM}}\; \}\; ,\; \{\; \mathcal{O}^{\text{QMM}}_{i j | t_i t_j} \; \}  \; , \; \{\;  \mathcal{O}^{\text{QMM}}_{i j k | t_i t_j t_k} \; \}\;|\; ( \mathcal{A} \;,\; \Lambda ) \; ] ,\\ & \mathcal{O}_{i | t_i}^{\text{QMM}} \equiv \rho_i(t_i) \otimes \mathbb{1}_{\bar{i}} \;\;;\;\; \mathcal{O}^{\text{QMM}}_{i j | t_i t_j}  \equiv \rho_i(t_i) \rho_j(t_j)  \otimes \mathbb{1}_{\overline{ij}}\;\; ; \;\; \mathcal{O}^{\text{QMM}}_{i j k | t_i t_j t_k} \equiv \rho_i(t_i) \rho_j(t_j) \rho_k(t_k) \otimes \mathbb{1}_{\overline{ijk}} ,\\ & \rho_i(t_i) = \text{Tr}_{\mathcal{H}_{\bar{i}}} [\rho^{(\mathcal{S})}(t_i)] \;;\; \mathcal{H}^{(\mathcal{S})} = \mathcal{H}_i \otimes \mathcal{H}_{\bar{i}}\;,\; \hspace{.65cm} \forall \; t, t_i, t_j, t_k \in \; [t_*, t]\;,\; \forall \;i,j,k \in \; 1, \cdots, M  .
\end{split}\end{equation} For $M=1$ and Polynomial QMM-Hs, the above recipe reduces to the recipe presented in (\ref{QMMmonml},\ref{MMH}). Nevertheless, they represent two distinctive scenarios to form QMM-UEs. In the first scenario the QMM-UE of a closed system is directly driven from the history of its total-or-`global' quantum states. In the second scenario, however, a closed system interconnects the state-histories of its chosen subsystems, to structure its QMM-UE.\\\\  
Regarding analysis extensions, an obvious direction is studying the closed system of $M$ interacting qubits, for small $M$ and $M \to \infty$, and $M$ quantum fields (of various kinds and symmetries) endowed with QMM many-body interactions. We have focused on a single qubit, but the Hubbard-model-like reformulation of the $N=2$ one-qubit system in (\ref{hubbard33},\ref{hubbardcouplingsare}), manifests possible forms of two-body and higher $r$-body QMM interactions. It is straightforward to extract these interactions out of the two recipes, likewise one-qubit cases in Section \ref{888}. Investigating the behaviors of the QMM variants of quantum spin systems, Hubbard models, and various QFTs can be insightful.\\\\
Another interesting direction of furthering our analyses corresponds to QMM-UEs sourced by interacting quantum memories which fill up a continuum state-(sub)history. One re-introduces the $H_{\text{(QMM)}}^{(N;L)}$'s defined in (\ref{QMMmonml},\ref{MMH},\ref{MMHparts}) in their $N \to \infty$ QM continua, integrating degree-$l^{(\leq L)}$ monomials of the past-or-present density operators over entire participating subsets of $[t_*,t]$.
\emph{For example}, choosing a `QM-width' parameter $a \in [t_*,t]$ and taking $L=2$, one intuitive QMM-H made by all quantum states in the history window $[t-a,t]$, $H_{\text{QMM}}^{(\infty|2|a)}(t)$, can be formulated as the following reformation of \eqref{NL},  
\begin{equation}\label{218}\begin{split} & H_{\text{QMM}}^{(\infty|2|a)}(t) \;\equiv \; \\ & \equiv \; \int_{t - a}^t d t'\; f_{t' t}\;\big[\; \mu_{t'} (\mathcal{P}^{{\rm chosen}}_{[t-a , t]}) \; \rho_{t'} \;+\; i\; \lambda^I_{t' t}(\mathcal{P}^{{\rm chosen}}_{[t-a , t]} )\; [\; \rho_{ t'} \;,\; \rho_{t} \;] \;+\; \lambda_{t' t}^R (\mathcal{P}^{{\rm chosen}}_{[t-a , t]}) \; 
\{\; \rho_{t'} \;,\; \rho_{t}\; \}\;\big] . \;\end{split} \end{equation}
The temporal functions $f_{t' t}$ multiply the QM-couplings $\big(\mu_{t'} , \lambda_{t't}^I , \lambda_{t' t}^R \big)$ in order to weight the participation of those quantum memories. One can choose a monotonically-decreasing tunable function $f_{t' t} = f(t-t')$. Moreover, one can choose the QM-couplings $(\lambda_{t' t}^{I,R} , \mu_{t'})$ to be themselves functions of the quantum-memory fidelities or various other kinds of state-history-dependent scaler observables, incorporating higher-oder interactions and quantum memory effects already at the $L=2$ level.\\\\
One can investigate statistical behavioral effects of `memory-participation randomness' in QMM-UEs, using the same general recipe of QMM-Hs formulated in (\ref{ds}, \ref{cps},\ref{QMMmonml},\ref{MMH}).\\ For example, considering a probability distribution $\mathcal{A}$ with mean value $a$, variance $\alpha^2$, and support $D_{{}_\mathcal{A}} \subset \mathbb{R}^+$, a QM-random $H^{(2|2)}_{{\rm QMM}}(t)$ is formulated by imposing that the participating past-states are chosen moment by moment, based on a random-variable QM distance drawn from $\mathcal{A}$. The QMM-H reads as,
\begin{equation}\label{318} H^{(2|2)}_{{\rm QMM}}(t|\mathcal{A}) \; = \; \mu \; \rho_{t -  a_t} + \;i \lambda^i \;[\; \rho_{t -  a_t} \;,\; \rho_{t} \;] +  \lambda^R\;\{\;\rho_{t-a_t} \;,\; \rho_{t} \;\}\;\;\;;\;\;\; a_t^{(\in\; \mathbb{R}^+)}\;\sim\;\mathcal{A} (\;a , \alpha^2 \;) . \end{equation}
Moreover, one can take some or all independent dimensionless couplings to be drawn from probability distributions. Mapping disorder-averaged phase diagrams of random QMM-UEs \eqref{318} and their generalizations is interesting. In the context of their spin models, it leads us to various novel families of quantum spin glasses endowed with QMM random interactions.\\\\
We find it natural, moreover theoretically elegant, to reformulate general QMM-UEs in the `History Hilbert space' frameworks such as \cite{CW2016,NCH2018,C2021}, using higher quantizations and the postulated algebras of quantum-memory fundamental operators. It can serve as a useful abstract language to investigate and establish general characterizations of QM many body interactions. Independently, another advancement of our analyses is to explore QMM-UEs made of general mixed quantum states, mapping their richer behavioral phases and dynamical phase transitions.\\\\  
The hybrid family of QMM-UEs, generated by combining conventional (time-independent or local-in-time) Hamilonians and Purely-QMM Hamiltonians, constitute novel closed quantum systems of special importance to be investigated extensively. The importance is both theoretical and observational. Because the limit of QMM-UEs with extreme memory dependency, namely the case of Purely-QMM-Hs, was totally unexplored while quite interesting and sufficiently complex, it constituted the bulk of our analyses in the present work. However, the analysis of Subsection \ref{the hybrid versions} shows that the hybridization is qualitatively nontrivial: some principal features of the pure limit do remain stable, other behavioral aspects undergo interesting deformations. Hence, the phase diagram of hybrid QMM-UEs is a matter for careful explorations. Phenomenologically, hybrid Hamiltonians which are dominantly conventional, that is coming from perturbing the conventional Hamiltonians by means of QMM operators, are specially interesting. This is because, in the context of closed quantum systems in nature, even if one finds some structural Non-Markovianity in their unitary evolutions, the observational effects likely come from some very-weakly-coupled perturbative QMM operators \cite{188}. Nevertheless, because of aspects such as `mobility phenomenon' \cite{M1980,M1985,C1993}, perturbative couplings can be impactful. Hence, one specially interesting question for future works is to pinpoint robust fingerprints of perturbatively-coupled QMM operators on two-or-higher point functions of observables in cosmological settings and laboratory-setup closed quantum matter. Indeed, \emph{investigating and characterizing such robust traces should be a crucial advancement toward discovering phenomenological landmarks of QMM-UEs }\cite{188}.\\\\ 
The capacity of our model to explore a wide range of regimes, from a completely memory-less all the way to a heavily non-Markovian regime, places our work naturally in the stream of research studying the crossover from a Markovian to a non-Markovian dynamics, explored previously on the theory side in works such as~\cite{Wang17}, \cite{Zhang} and \cite{Ma14}, and on the experimental side in~\cite{LiuBreuer11} in which full control over the (non-)Markovian character of the dynamics could be achieved in an all-optical setup.
Moreover, the ability to access explicitly the past states of the system can potentially shed more light on the connection between memory in quantum dynamics and quantum correlations, a topic already explored for example in~\cite{Mazzola12,Smirne13,fanchini14,DArrigo14,Campbell18}. The main focus of these works is on the momentary system-environment correlations, while we can look instead at time correlations of the system with itself. In the common scenario where memory effects are induced by the system-environment correlations, the environment plays the role of an intermediary, storing and carrying over traces of the system's history. In a way, our approach is more fundamental, since we wish to look instead at a purely-internal source of participating quantum memories: the closed quantum system itself.\\\\
\emph{The experimental realization of the QMM-UEs} defined in \eqref{VMV} is a subtle exciting open question. One can first consider the scenarios (\ref{QMMmonml},\ref{MMH},\ref{MMHparts}) and \eqref{bcnqmbs:mbo.118} which are two specific ways of structuring \eqref{VMV} as \emph{the} general definition of QMM-UEs. In either scenario, a closed system must select and connect its available `internal experiences', as contained in its quantum-state-history, to update its momentary unitary evolution. They however differ in the way that this central point is structured. In (\ref{QMMmonml},\ref{MMH},\ref{MMHparts}), the available memories of the total system link up to update the system's Hamiltonian. In scenario \eqref{bcnqmbs:mbo.118}, a system updates its Hamiltonian by joining accessible memories of its subsystems. \emph{Primarily, we suggest} that the second scenario offers a flexible framework for the experimental realization of 
(at least well-approximated) QMM-UEs, via befitting alterations of \emph{platforms such as} \cite{Mataloni19,Bernardes15,Schmidt14,PekolaMani18}, of course combined with necessary novel ideas. One can also devise QMM-UEs \eqref{VMV} using partial content of the (sub)system states. Finally, we highlight that, as the intriguing generalizing perspectives and formalisms such as the `Quantum Cognition' \cite{AA20515,QCB,FWHSY2016,MSB2016,K2016,BZBW2020,LDW2020,OB2021,Love2022} or \cite{V2021A} suggest us, a (rather surprising but) natural way to experimentally realize the proposed QMM-UEs may be paved by \emph{the emergent functionally-quantum phases} of classical complex neural networks, even \emph{the brain itself}.\\\\ 
%may qubits~\cite{PekolaMani18}.\\
As explicitly derived in Section \ref{gtp}, sate-history-driven unitary evolutions are described by QMM von Neumann and Schr\"odinger equations whose nonlinearities and temporal nonlocalities are coupled strongly. \emph{Nonlinear quantum dynamics} has a long history of unsettled conceptual, theoretical and phenomenological investigations. In particular, unitary nonlinear evolutions, if being fully deterministic, can result in superluminal signalling, within the standard settings \cite{G84,G89,W89}. One can formulate, however, diverse distinct forms of nonlinear deformations which \emph{safeguard causality} either by `construct' or by the `context'. The former case includes, alongside stochastic nonlinearities and the (physical or mathematical) generalized frameworks \cite{P91,J1993,CD2002,K2005,ACD2002,BMHin2019}, various causal families of \emph{constrained or finely-structured} nonlinear quantum evolutions \cite{K1978,FSS2004,RC2020,KR2021}. Moreover, broad classes of nonlinear quantum evolutions arise in the \emph{effective} contexts where the indicated concerns, such as the superluminal signalling, become irrelevant naturally \cite{acronlse:a}. Finally and importantly, we must highlight that state-dependent Hamiltonians and nonlinear Schrodinger equations are \emph{ubiquitous} in the whole domain of `emergent quantum theories', specially the diverse distinct scenarios where quantum dynamics emerges either as an effective description, or an alternative equivalent formulation, in an \emph{underlying `non-quantum' system} which can even be classical \cite{R2005,tH2007,CBR2015,MP2016,MAN81820199,AdlerEQat2012,V2020,V2021A}. \\\\ 
Another principal category of the above-mentioned theories are `topological quantum theories', and beyond them, the fundamental theories formulating quantum systems underlying the emergent spacetime, such as the $(0 + 1)$-dimensional matrix-or-network formulations of pregeometric quantum gravity, in the spirit of the proposals \cite{BFSS97,KMS2008}. Furthermore, nonlinear unitary evolutions can find applications in modeling black holes \cite{HM2004, HY2005, LP2014}.  Characterizing causality constraints such as those in \cite{FSS2004,RC2020} for QMM-Hs, where the temporal nonlocalities and nonlinearities are entangled, is \emph{an open question}. Independently, we like to state the envisioned possible applications of QMM-UEs in the fundamental domain of pregeometric \emph{quantum gravity}. Revisiting proposals-and-models \cite{WheelerIQ, S2005,BFSS97,KMS2008,L2018,L2020,V2020,V2021B,ACLSSTW2021}, with similar objectives and within similar perspectives, \emph{we conjecture one can formulate novel $(0 + 1)$-dimensional unitarily-evolving information-theoretic quantum many body systems from which phenomenologically-plausible dynamical local geometric spacetime and interacting quantum matter can emerge via the `state-history driven' mechanisms of self-organization which are processed by means of the QMM-UEs of the kind (\ref{QMMmonml},\ref{MMH}),(\ref{bcnqmbs:mbo.118},\ref{H-CQS})}.\\\\
%%%%%%%%%%%%%%%%%%%%%%%%%%%%%%%%%%%%%%%%%%%%%%%%
One principal realm where the proposed \emph{`Experience Centric Unitary Evolutions'} (\ref{H-CQS}), structured as (\ref{QMMmonml},\ref{MMH}) or (\ref{bcnqmbs:mbo.118}), can have natural, diverse and promising applications is \emph{Quantum Intelligent Systems}, as generally highlighted in (the previous subsection) \ref{elucidations}. We conjecture, and aim to show in the future works, that one can come up with novel advantageous systems and mechanisms of \emph{Quantum Neural Computation}, and broadly with \emph{Interacting Systems Developing Quantum Intelligent Behaviour}, using QMM-UEs. In the former case, we envision developing advantageous qualitatively-novel kinds of Quantum Artificial Neurons and Quantum Neural Networks, including the data-driven context of Quantum Machine Learning \cite{K95, SSP2014,BWPRWL2017,DB2018,W2014,KDWAGZ2021,CHAPRSW2018,AW2017,CCCDSTVMZ2019,GSS2020,SR2020,ASZLFW2021}. In the much-broader latter category, we envision their natural applications in developing or modeling diverse (quantum or classical) interacting many-body systems which, without (any underlying or explicit kind) of neural network structure, develop and feature various forms and hierarchical levels of intelligent quantum behavior.\\\\ 
As such, we envision and highlight the following application categories for QMM-UEs: \emph{(a)} The Agent-Environment-Interacting Systems of Quantum Artificial Intelligence, including specially the broadly-defined context of Quantum Reinforcement Learning \cite{C2019, SAH2021, DB2018,DCLT2008}, \emph{(b)} The alternative frameworks of Quantum Decision-Making Models \cite{BBA2014,BHK2017,Y2020,OB2021}, and Computational Quantum Modeling of Noisy Decision Making of classical subjects \cite{AA20515,FWHSY2016,MSB2016, BZBW2020, LDW2020,OB2021}, specially engaged with Complex Bayesian Inference. \emph{(c)} Closed quantum systems performing intelligent tasks by means of their Many Body Interactions which can be task-engineered \cite{AA2003,T2001,SSB2017}, or are emergent via self-organization. \emph{(d)} 
On partially-common technical grounds, we suggest QMM-UEs can be utilized in the Quantum Modeling of General Stochastic Processes, specially as advantageous variants of 
Quantum Hidden Markov Models and their close associates \cite{MBW2011,CHBB2015,HGE2020}. \emph{(e)} Moreover, we must emphasize `Cognitive Complex Systems' formulated, modeled and investigated in their broadest perspective, specially in the context of `Quantum Cognition' and its associates  \cite{AA20515,QCB,FWHSY2016,MSB2016,K2016,BZBW2020,LDW2020,OB2021,Love2022}.\\\\ 
General reasons motivating advantageous applications of QMM-UEs in formulating Quantum Intelligent Systems, as suggested above, are manifold. We outline them.
The state evolution of every sensible neural circuit is intrinsically nonlinear by virtue of the activation functions of its computing units. Functionally, evolution nonlinearities are essential for neural computations, for example central to their attractor behaviors. Quantum computing circuits, however, are structured based on linear unitary gates, albeit up to quantum measurement effects or engineered environment-coupling effects. Therefore, quantum implementations of computational neural nonlinearities have been limited to designing sequential-or-recurrent quantum measurement actions or the open system environmental interactions or correlations. The resulted non-unitary quantum neural processes, however, experience general drawbacks such as unwanted dissipations, reduced or inflexible Controllability, and limitations on the learning-task generalization. Reliable quantum neural nonlinearity has been elusive. QMM-UEs can yield unitary nonlinear quantum gates which are free from dissipation and can be robustly flexible. Hence, advantageous Quantum Artificial Neurons \cite{KDWAGZ2021,GSS2020,TMGB2019,PLOS2019,YQC2020} can be naturally realized by means of QMM-UEs.\\\\
Compared to the standard linear quantum algorithms, significant manifold benefits of nonlinear quantum computations has been established in various old and recent works. Nonlinear quantum algorithms, based on (even using few) nonlinear gates, can provide exponential superiorities across a wide range of realistic computational problems, such as NP-complete and $\#P$ problems, state discrimination and unstructured search, and the Neural-Network-Architecture learning \cite{AL1998,C1998lm,A2005,PM2009,PM2011,MW2013,MW2014,SLO2016,CY2016,G2021}. Indeed, the family of QMM-UEs can serve as a large flexible resource for implementing these superiorities unitarily.\\\\
QMM-UEs, by the virtue of their state-history compositeness, offer other significant advantages, alongside flexible unitary nonlinearity, for quantum neural computation. Using quantum memories in quantum learning algorithms is exponentially gainful \cite{CCHL2021}. The general formulation of QMM-UEs, specially its subsystem-based prescription \eqref{bcnqmbs:mbo.118} host a wealth of possibilities to design closed systems of quantum neural networks in which computations can be processed on-and-utilizing unitarily-stored quantum data. Independently, classical networks whose synaptic connections experience (time-)delay interactions have been investigated, already from the outset of neural computations. Indeed, the effects of nonlocal-in-time synaptic interactions in classical neural networks, even in the simplest models of associative memory, are significant and diverse, sourcing behavioral richness, functional advantages and broader range of applications \cite{SK1986,HKP1991,BA1994}. The intrinsic time-nonlocality of QMM-UEs and their generating interactions provide the natural setting to bring these functional features to the quantum realm, as unitary time-delay quantum neural networks.\\\\ 
Moreover, as our extensive analyses showed in great details, the intrinsic interplays of delayed response effects and nonlinearity effects in QMM-UEs naturally enable a closed system to develop behavioral phases whose usefulness is known for the purpose of neural computations. For instance, the classical analogues of dynamical phases such as the phase `five' of one-qubit QMM-UEs, featuring robust patterns of multistable attractor behavior, play important roles in brain neural computations, see \cite{LZ2021} as an example. It is worthwhile to highlight that stable and multistable attractors of QMM-UEs not only occur in the smallest quantum system with simplest interactions between only two arbitrary quantum memories, but also take place in the complete absence of any kind of stochasticity. Furthermore, straightforward generalization of our behavioral analyses in Sections \ref{iml} and \ref{SV} to larger systems should reveal much richer kinds of
multi-attractor and dynamic-attractor phases, such as sequential limit cycles and higher dimensional attractors in biologically-inspired neural networks \cite{SCT2021,CC2021}. QMM-UEs are natural to realize these behaviors in quantum neural computations.\\\\  
The first signatures of the QMM-UE advantages for quantum neural computations, specially in supervised and unsupervised learning, should likely appear in the quantum spin glass category of the learning machines. The first steps forward could naturally be the QMM-UE reformulations of Quantum Hopfield Neural Networks \cite{RBWL2018,STT2020,MS2017,PMGLM2018}, and Quantum Restricted Boltzmann Machines \cite{AARKM2018,ZLW2021}. The Hopfield class includes the 
QMM-UE reformulation of their alternative variants, for example \cite{AA2003,T2001}, likewise. Specially, the free-energy-minima landscape, quantum-enhanced memory capacity and memory retrieval processing of renovated Quantum Hopfield Neural Networks must be compared with those of all the earlier quantum models, see for example \cite{SSB2017}. Moreover, quantum versions of $p$-spin-glass and `Modern' Hopfield networks \cite{FAA2018,ayn2020,KH2020} should be formulated using QMM-UEs. Obvious further advancements include the reformulation of Quantum Convolutional Networks \cite{CCL2019}. In brief, we suggest that QMM-UEs (\ref{H-CQS}) can offer a novel discipline of \emph{state-history-driven quantum neural computation}.\\\\ 
In the realm of Quantum Intelligent Systems,  the envisioned utilities of QMM-UEs should naturally go beyond quantum neural computation and data centric learning. The expectation is, in fact, natural. Alongside those reasons in common with the aforementioned applications, it has to do, first and foremost, with the intrinsic aspects\\ of a unitary quantum system which realizes (\ref{H-CQS}), specially in the general form of  (\ref{bcnqmbs:mbo.118}). \emph{These characteristics naturally enable a network of quantum agents living inside an encompassing quantum system to pick up, interjoin and use their experiences, their accessible histories of states-actions-outcomes, in order to (continually or recurrently) redirect their momentary many-body interactions and interdependent time evolutions, including policy updatations, on a goal-driven basis}. Clearly, generic problems of (single and multi-agent) Quantum Reinforcement Learning \cite{DB2018} are flexibly programmable in the contextual formulations of the QMM-UEs (\ref{bcnqmbs:mbo.118},\ref{H-CQS}). Moreover, the formalism can lead to novel-and-flexible quantum models and quantum algorithms in the context of the various conceptual and mathematical generalizations of Reinforcement Learning \cite{DB2018}. One can begin with a comparative reformulation of the Projective-Simulation framework of \cite{BC2012} using QMM-UEs, and further advance to program evolving networks of quantum agents being endowed with the vital faculties of Generalization and Meta Learning \cite{AMH2017}.\\\\
%. hierarchical vs anarchical\\ %INTUITIVE 
Beyond the framework of Reinforcement Learning and its straightforward extensions, we also envision that QMM many body interactions and unitary evolutions (\ref{QMMmonml},\ref{MMH};\ref{bcnqmbs:mbo.118}) and (\ref{H-CQS}) can find applications in the behavioral modeling and physical formulation of Complex Quantum Systems featuring various kinds and levels of Emergent Intelligence. 
Indeed, the generality of our proposed framework, and its mathematical-and-physical formulation, should enable us to systematically come up with unprecedented diverse kinds of many body systems which feature behaviorally-superior quantum intelligence. Complex systems which ongoingly exploit the quantum-behavior experience, as the compositional resource of a \emph{a re-emergence flow}, to reinvent and redirect themselves. Their many-body interactions and time evolution operators are \emph{internally} re-designed and re-constructed on a moment-by-moment basis or recurrently, utilizing the chosen `modules' of their quantum-state history. Based on our extensive investigations of the one-qubit closed systems driven by first, second and third order QMM interactions, it is crystal clear that the many-body behavioral phases of the resulted quantum-intelligent systems should be outstandingly rich and largely promising.
\subsection{Conclusion}
We have taken only the first step in formulating, investigating and understanding the innermost interactions and the unification of time evolution unitarity in closed quantum systems and one fundamental characteristic of intelligent behavior, being the impactful exploitations of the `experience' of the system, namely its state history, to reorganize and update its quantum behavior. 
The further details, the broader perspectives, and manifold outcomes of the resulted theory, \emph{`Experience Centric Quantum Theory'}, is to be investigated and disclosed in future works.
%%%%% Qpen question, how of the emergence!

\section{Dedication}
\emph{Alireza Tavanfar dedicates the present paper to his father, Manouchehr Tavanfar, who always inspired and encouraged him on the marvelous path of research and discovery in natural sciences, specially in the deeper contexts of fundamental questions.} 

%%%%%%%%%%%%%%%%%%%%%%%%%%%%% 
%present work represents a first step\\ %. brought indiscriminate mentioning because gneral reasons applying to them collectively % classical-limit-driven novel classical systems, the baez work for the classical via QFT \\\\
%%%%%%%%%%%%%%%%%%%%%%%%%%%%%%%
%%%%%%%%%%%%%%%%%%%%%%%%%%%%%%%%%%%%%% %%%%%   7 .  Acknowledgement %%%%%%%%%%%%%%%%%%%%%%%%%%%%%%%%%%%%%%%
\section{Acknowledgments}
The authors thank, in alphabetic order, Amirfarshad Bahrehbakhsh, Miguel I. Berganza, Daniel Burgarth, Mateo Casariego, Tiago Costa, Roohollah Ghobadi, Mohammad Hafezi, Seyyed Ali Hosseini Mansoori, Javier R. Laguna, Javier M. Magan, Duarte  Magano, Luca Mazzucato, Thomas S. Mullen, James Murray, Mohammad Nouri Zonoz, Yasser Omar, Simone Paganelli, Kyriakos Papadopoulos, Rui A. P. Perdigão, Gonzalo Polavieja, Morteza Rafiee,  Seyyed Nariman Saadatmand, Hamid Reza Sepangi, Seyyed Hamed Shaker, Hesam Soltanpanahi and Behrad Taghavi for discussions, comments and suggestions.\\
Ali Parvizi thanks the Institute of Theoretical Physics, University of Wroclaw for their kind hospitality during his visit when part of this project was carried out.\\ Alireza Tavanfar thanks the kind hospitality of Physics of Information and Quantum Technologies Group of Instituto de Telecomunicações at Lisbon in the context of his visit during part of the project. Alireza Tavanfar wishes to acknowledge his research funding through grants to Zachary F. Mainen from the Fundação para a Ciência e a Tecnologia (FCT-PTDC/MED-NEU/32068/2017) and the Champalimaud Foundation thankfully.

% one

%%%%%%%%%%%%%%%%%%%%%%%%%%%%%
%%%%% 8 . bibliography
%%%%%%%%%%%%%%%%%%%%%%%%%%%%%%
%\bibliographystyle{utcaps}
%\bibliography{all}

\newpage

\appendix

\section{Purely-QMM Kicker Hamiltonians}
\label{appendix1}
Here we ask if one can construct diverse and natural QMM realizations of (\ref{VMV}) to serve us as the Kicker part of total Hamiltonian (\ref{MMHpure}). In other words, the question we address is the following. \emph{Can $H^{({\rm Kicker})}(t)$ itself be a construct of the state history of the closed system, and (in the most extreme case) nothing else?}  The question is  interesting conceptually, because such Purely-QMM realizations of $H^{({\rm Kicker})}(t)$ secure the `self-sufficiency' of QMM-UEs as a \emph{dynamically complete theory}, that is, a theory free from any conventional Hamiltonian. \\\\
The answer to the above question is \emph{affirmative}, with various ways of realizing it. The point is, for a Hamiltonian to be QMM, it does not need to be made from the system's density operators \emph{themselves}, as for example implemented in the recipe (\ref{MMH}). Rather, an operator is QMM if, by any means, it is made (partially or entirely) from the quantum information which is encoded in the state history of the closed system. For example, as building blocks of QMM operators, one can break apart momentary density operators of the system, and re-assemble them as various deformed operators which nevertheless contain (at least partially) the qubit memories of the closed system. Employing \emph{state alterations} as such, we can come up with infinitely many versions of Purely-QMM Kicker Hamiltonians, $H_{{\rm Kicker}}(t)$, without the use of any conventional Hamiltonian. For example, it does suffice to introduce $H^{({\rm Kicker})}_{{\rm QMM}}(t)$ at any moment $t \in [t_0,t_*]$ which is a Hermitian `alteration' of $\rho_{t}$, call it $\tilde{\rho}_{t}$, satisfying $[\tilde{\rho}_{t}, \rho_{t}] \neq 0$, hence triggering the momentary evolution of the system.\\\\ 
One good recipe, among many others, goes as follows. Bipartitioning the total Hilbert space of the total system into arbitrary sub-Hilbert spaces, call them Left and Right, $\mathcal{H}^{(\text{total})}\;=\;\mathcal{H}_L \; \otimes \mathcal{H}_R$, and using any doublet $(A^{(\text{fixed})}_L, A^{(\text{fixed})}_R)$  of QM-independent Hermitian operators, an Purely-QMM Kicker Hamiltonian can be introduced as such,
\begin{equation}
\label{adi_1}
\begin{split} 
& \hspace{.0000199 cm}
H^{({\rm Kicker})}_{{\rm QMM}}(t_0)\;\equiv\;\lambda_1\;\rho^{(\text{altered})}_{t_0}\;\;\;;\;\;\; \rho^{(\text{altered})}_{t_0}\;\equiv\; a_L \; \rho^{L}_{t_0} \; \otimes \; A^{(\text{fixed})}_R \;+\;a_R \; A^{(\text{fixed})}_L \; \otimes \; \rho^R_{t_0} \\
&  \rho^L_{t_0} \;\equiv\; \text{Tr}_{\mathcal{H}_R}\; \rho_{t_0} \;\;\;;\;\;\;\; \rho^R_{t_1} \;\equiv\; \text{Tr}_{\mathcal{H}_L}\; \rho_{t_1} \;\;\;;\;\;\; \;\;\; (a_L,a_R) \in \mathbb{R}^2.  
\end{split} 
\end{equation}
Indeed, the idea of employing the states of the chosen subsystems of the closed system to construct its Purely-QMM Kicker Hamiltonian  conceptually connects with generalized QMM-Hs presented in the recipe  \eqref{bcnqmbs:mbo.118}. Moreover, alternative forms of Purely-QMM Kicker Hamiltonians can be introduced which do not require a tensorial decomposition of the Hilbert space. For example, employing any state-history-independent Hermitian operator $A$ acting on the Hilbert space of the total system, one can introduce,
\begin{equation}
\label{adi_2}
H^{({\rm Total})}_{({\rm QMM}|\mathcal{A})}(t) \in \Big\{\; \{ \rho_t, A \} \;,\; i [ \rho_t, A ] \;,\; A \rho_t A \;  \Big\}.
\end{equation} 
The constructions \eqref{adi_1} or \eqref{adi_2} still require state-history-independent operators to be used. In the terminology of the paper, those Kicker Hamiltonians are Purely-QMM, however they are not Entirely-QMM. Now we highlight that one can introduce Kicker Hamiltonians which also satisfy the stronger condition, that is being Entirely-QMM. 
For example, we can use operations such as the transposition or time derivatives on the momentary density operator of the closed system, to construct Kicker Hamiltonians,  
\begin{equation}
\label{adi_3}
H^{({\rm Total})}_{({\rm QMM}|\mathcal{A})}(t) \propto \rho_t^T \;\;\; \text{or} \;\;\; H^{({\rm Total})}_{({\rm QMM}|\mathcal{A})}(t) \propto \dot{\rho}_t \;.
\end{equation} 

\end{document}